%% file: thesis.tex
\documentclass
[
    twoside,                 
    openright,               
    cleardoublepage = empty, 
    fontsize = 11 pt,        
    american,                
    captions = tableheading, 
    numbers = noenddot,      
    footheight = 35 pt,      
]
{scrbook}

\input{core/preamble/document_settings} 

    \extraBorder{3 mm}

    \bindingCorrection{10 mm}

    \myTitle{A Fully Anisotropic Formulation of Stochastic Cell Rescaling}{\empty}

    \myName{Vittorio Del Tatto}

    \myProgram{IT Systems Engineering}

    \dateOfHandingIn{30. Februar 20XX}

    \nameOfMySupervisor{Prof.\,Dr. Giovanni Bussi\\Prof.\,Dr. Raffaello Potestio}

    \additionalExaminers{A Postdoc\newline A PhD student}

    \mySubject{A cool bachelor/master thesis.}

    \myKeywords{bachelor--master thesis | world-changing | very important | please like and share and subscribe}

\input{core/preamble/format}                        
\input{core/bibliography/bibliography_format}       
\input{packages_and_commands/standard_packages}     
\input{packages_and_commands/general_commands}      

\usepackage{amsmath,amsfonts,amsthm,bm} 
\usepackage[ruled,vlined]{algorithm2e}
\usepackage{thmtools}
\usepackage[toc,page]{appendix}
\usepackage{graphicx}
\usepackage{xpatch}
\usepackage{float}
\usepackage{ragged2e}
\xpretocmd{\appendixpagename}{\sffamily}{}{}
\newcommand{\overbar}[1]{\mkern 1.5mu\overline{\mkern-1.5mu#1\mkern-1.5mu}\mkern 1.5mu}
\DeclareMathAlphabet\mathbfcal{OMS}{cmsy}{b}{n}

\usepackage{booktabs}
\usepackage[referable]{threeparttablex}
\renewlist{tablenotes}{enumerate}{1}
\makeatletter
\setlist[tablenotes]{label=\tnote{\alph*},ref=\alph*,itemsep=\z@,topsep=\z@skip,partopsep=\z@skip,parsep=\z@,itemindent=\z@,labelindent=\tabcolsep,labelsep=.2em,leftmargin=*,align=left,before={\footnotesize}}
\makeatother

\makeatletter
\let\blx@rerun@biber\relax
\makeatother

\usepackage{scalerel}

\def\vecsign#1{\rule[1.388\LMex]{\dimexpr#1-2.5pt}{.26\LMpt}%
  \kern-4.0\LMpt\mathchar"017E}

\usepackage{stackengine}
\stackMath

\declaretheorem[numbered=no,name=Ergodic hypothesis]{ergodic_hyp}
\declaretheorem[numbered=no,name=Liouville's theorem]{liouville_th}
\declaretheorem[numbered=no,name=Principle of equal a propri probabilities]{equiprobab_pr}

\declaretheorem[numbered=no,name=Ito's lemma]{ito_lemma}
\declaretheorem[numbered=no,name=Ito's lemma (multidimensional case)]{ito_lemma_multidim}

\crefformat{pluralequation}{#2eqs.~(#1)#3}
\Crefformat{pluralequation}{#2Eqs.~(#1)#3}

\begin{document}

    \frontmatter
    \include{core/title_page/title_page}

    \pagestyle{plain}
    
    \addchap*{Note on the content}
    This work is also deposited in the thesis catalogue of the University of Trento and it is available for consultation according to the procedures established by the University regulations\footnote{\href{https://www.biblioteca.unitn.it/446/regolamento-per-la-consultazione-delle-tesi-di-laurea}{https://www.biblioteca.unitn.it/446/regolamento-per-la-consultazione-delle-tesi-di-laurea}}. The present version includes corrections and additions compared to the deposited one.

    \addchap{Introduction}
    Molecular Dynamics (MD) simulations are a set of computational techniques that allow to simulate the physical motion of atoms and molecules, with the purpose of inspecting dynamical properties of microscopic systems. MD simulations are widely employed in biophysics, chemical physics and material science, both to validate theoretical models and to guide future experiments. Furthermore, MD simulations allow to probe atomistic details that would be inaccessible in real experiments, and for this reason they have been metaphorically described as a "computational microscope" \cite{microscope}.

    When a classical system is isolated, its dynamical evolution is completely determined by Hamilton's equations of motion and its total energy is conserved. However, in several situations of practical interest the system is at equilibrium with an external bath and can exchange energy with it in form of heat (isothermal equilibrium) or as mechanical work (isobaric equilibrium), also allowing for volume fluctuations. Temperature and pressure are the thermodynamic variables that control respectively these energy exchanges, and the MD algorithms employed to simulate systems in these equilibrium conditions are called thermostats and barostats. In particular, the proper coupling of a thermostat with a barostat allows sampling the isothermal-isobaric distribution, which describes the statistical behaviour of a system in the typical conditions of real experiments.
    
    One of the most popular pressure coupling methods, based on a deterministic first-order equation, is the Berendsen barostat \cite{berendsen}. This algorithm can correctly reproduce the average volume of the system but not its statistical fluctuations; it is however very efficient in the equilibration phase and contains a single parameter, i.e. the volume relaxation time, which is easy to tune and to interpret. The traditional MD algorithms that produce the correct volume fluctuations are methods based on second-order equations \cite{andersen,langevin_piston,mtk,parr-rahman,shinoda}; an isolated exception is the Monte Carlo barostat \cite{MC_barostat_limitations}, which is simpler to implement but sometimes less efficient than second-order methods. Also second-order barostats show some relevant drawbacks. First, they are known to be less efficient than the Berendsen barostat in the equilibration phase; in particular, they might show instabilities and slowly damped oscillations of the volume when the system is initialized far from equilibrium. Moreover, these algorithms are typically formulated in terms of an input parameter called barostat mass, which is related in a nonlinear manner to the actual volume relaxation time and whose interpretation is not straightforward. In principle, a decrease in the numerical value of this parameter results in a decreased period of the oscillations around the equilibrium value of the volume. However, the relation between the damping period of these oscillations and the barostat mass is not trivial, so that a too small value of the latter can actually result in an increase of the former. As a consequence, an optimal and system-dependent value of the barostat mass always exists, but no general receipe allows to find it \emph{a priori}. Stochastic Cell Rescaling (SCR) \cite{crescale_iso} is a recently proposed first-order barostat that samples the correct volume fluctuations, employing a Berendsen-like deterministic part and a suitable noise term. Due to his formulation, SCR employs the same input parameter of the Berendsen algorithm, avoiding the mentioned drawbacks of second-order methods. This algorithm, which can also be interpreted as a high-friction variant of the Langevin piston barostat \cite{langevin_piston}, has been designed so far in its isotropic and semi-isotropic versions, which employ respectively one and two stochastic equations. Tests performed on Lennard-Jones fluid, water and membrane simulation have shown that SCR can be effectively used in both the equilibration and the production phases.
    \begin{figure}[h!]
        \centering
        \includegraphics[width = 0.55\textwidth]{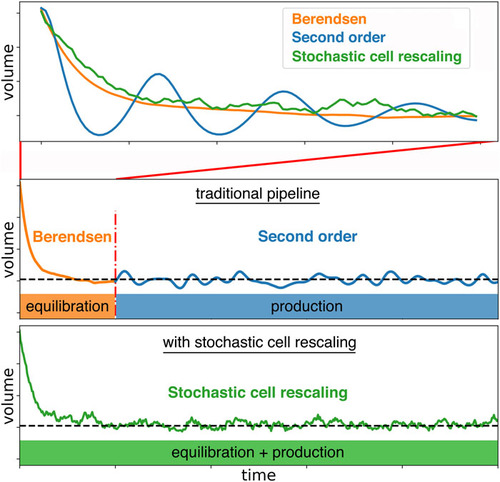}
        \caption{The use of second-order barostats on non-equilibrated systems can lead to oscillations and instabilities (upper panel). Therefore, in the traditional pipeline for constant pressure simulations the first-order Berendsen barostat is employed in the equilibration run and a second-order barostat is used in the production run (middle panel). SCR can be efficiently applied in both phases (lower panel). Source: \cite{crescale_iso}}
        \label{fig:pipeline}
    \end{figure}

    There are several scenarios where the assumption of isotropic volume fluctuations is no more a good approximation to describe the dynamics of a system. These scenarios include crystal systems, whose ordered structure intrinsically breaks the concept of "average isotropic behaviour" holding for liquids, and systems with an anisotropic external pressure, e.g. when a shear stress is applied. Moreover, it has been shown that systems of biological interest, for instance membrane environments, can also benefit from the use of anisotropic pressure coupling methods \cite{aniso_bio}. 

    This work includes the derivation, implementation and test of the anisotropic version of SCR, where volume is supposed to fluctuate with no constraints on the shape of the simulation box. In that regard, the isotropic SCR equation is generalized into nine first-order and coupled stochastic equations for the components of the so-called cell matrix, representing the box in which the system is confined. As in the isotropic case, a first motivation of the work is the creation of an algorithm that generates the correct (anisotropic) isothermal-isobaric ensemble and which is also efficient in the equilibration phase. Moreover, the algorithm preserves the same parameter employed in the Berendsen barostat, which is not only easier to set than the input parameters of second-order methods, but also carries relevant \emph{a priori} information on the dynamics of the volume. 
    
    The elaborate is structured as it follows:
    \begin{itemize}
        \item the first chapter aims to provide an introduction to the field of MD simulations, with a review of fundamental concepts of Statistical Mechanics and a parallel discussion of the main algorithms to simulate the dynamics of classical physical systems;
        \item the second chapter reports the theoretical formulation of the anisotropic SCR equations and the details on their numerical integration;
        \item the third chapter contains further details on the practical implementations of the algorithm and illustrates the tests carried out for its validation;
        \item the appendices contain insights concerning all three chapters, including the complete derivations and the mathematical details not reported in the main text.
    \end{itemize}

    \selectlanguage{american}

    \addchap{Acknowledgments}
    \input{preface/acknowledgments}

    \setuptoc{toc}{totoc}
    \renewcommand{\contentsname}{Contents}
    \tableofcontents

    \pagestyle{headings}
    \mainmatter




    \chapter{Introduction to Molecular Dynamics}
    \renewcommand{\chaptername}{Chapter}
    The computational methodology carrying the name of Molecular Dynamics (MD) is entirely built on the laws of classical Statistical Mechanics, that allow to describe with a surprisingly good approximation a large class of microscopic system, without resorting to the tools of Quantum Mechanics. One of the founding concepts of Statistical Mechanics is the notion of \emph{statistical ensemble}, which can be defined as a set of infinite copies of a system with different \emph{microscopic} states but sharing the same \emph{macroscopic} state. The macroscopic state is defined by fixing a few thermodynamic quantities, which univocally identify the corresponding statistical ensemble (e.g. number of particles $N$, volume $V$ and internal energy $E$ in the microcanonical ensemble), while the microscopic state includes the particles' positions $\{\mathbf{q}_i\}$ and momenta $\{\mathbf{p}_i\}$ ($i=1,...,N$), which compose the $6N$-dimensional phase space $\Omega$. Importantly, the algorithms employed in a MD simulation depend on the statistical ensemble the system belongs to. Among all the possible statistical ensembles, the most common ones in MD simulations are the microcanonical, the canonical and the isothermal-isobaric ensembles. In the following, these three ensembles are reviewed in their fundamental concepts, together with the most popular MD algorithms that have been developed over the years to reproduce them in computer simulations.

    \section{Microcanonical ensemble (NVE)}\label{hamilton}
    When a system is isolated, its dynamical evolution is completely determined by the Hamilton's equations of motion,
    \begin{subequations}\label[pluralequation]{hamiltons_eqs}
    \begin{align}
        \label{hamilton_eqs1}
        \text{d}\mathbf{q}_i &= \frac{\mathbf{p}_i}{m_i}\text{d}t\,, \\ 
        \label{hamilton_eqs2}
        \text{d}\mathbf{p}_i &= -\frac{\partial U}{\partial\mathbf{q}_i}\text{d}t\,,
    \end{align}
    \end{subequations}
    where $U=U\big(\{\mathbf{q}_i\}\big)$ is the potential energy of the system, $m_i$ is the mass of the particle $i$ and $t$ is the time. It is possible to prove that \cref{hamiltons_eqs} have a conserved quantity $\mathcal{H}$, which is called Hamiltonian or energy and whose value only depends on the initial conditions $\{\mathbf{q}_i(t_0),\mathbf{p}_i(t_0)\}$:
    \begin{equation}
        \mathcal{H}\big(\{\mathbf{q}_i,\mathbf{p}_i\}\big) \equiv \sum_i \frac{|\mathbf{p}_i|^2}{2 m_i} + U\big(\{\mathbf{q}_i\}\big) 
        = \mathcal{H}\big(\{\mathbf{q}_i(t_0),\mathbf{p}_i(t_0)\}\big) \equiv E\,.
    \end{equation}
    The term $K\big(\{\mathbf{p}_i\}\big) = \sum_i |\mathbf{p}_i|^2/(2 m_i)$ is called kinetic energy of the system.
    
    As an additional property, \cref{hamiltons_eqs} are time-reversible, meaning that if the trajectory $\{\mathbf{q}_i(t),\mathbf{p}_i(t)\}$ is a solution of Hamilton's equations, then also the \emph{time-reversed trajectory} 
    \begin{equation}
    \{\mathbf{q^*}_i(t),\mathbf{p^*}_i(t)\} \equiv \{\mathbf{q}_i(-t),\mathbf{p}_i(-t)\}
    \end{equation}
    is a solution.
    
    The bridge between the deterministic Hamiltonian mechanics and the statistical approach embedded in the microcanonical ensemble can be constructed with three fundamental steps, i.e. the ergodic hypothesis, the Liouville's theorem and the principle of equal a priori probabilities.
    
    \begin{ergodic_hyp}
    Given a region d$\Omega$ of the phase space, over large times the amount of time d$t$ that the system spends in d$\Omega$ is proportional to the volume of this region:
    \begin{equation}
        \label{erg_hyp}
        \lim_{\tau\rightarrow\infty}\frac{\text{d}t}{\tau} = \rho\big(\{\mathbf{q}_i,\mathbf{p}_i\}\big)\, \text{d}\Omega\,.
    \end{equation}
    \end{ergodic_hyp}
    \noindent The quantity $\rho$ that appears in \cref{erg_hyp} is called \emph{phase space density}, and if properly normalized it can be interpreted as the probability (density) of finding the system in a given point of phase space. This allows to replace time averages of physical observables with ensemble averages over the phase space: if $a\big(\{\mathbf{q}_i,\mathbf{p}_i\}\big)$ is the instantaneous microscopical estimator of a macroscopic quantity $A$, i.e.
    \begin{equation}
        A = \lim_{\tau\rightarrow\infty}\frac{1}{\tau}\int_t^{t+\tau}\text{d}t'\, a\big(\{\mathbf{q}_i(t'),\mathbf{p}_i(t')\}\big)\, ,
    \end{equation}
    by applying the ergodic hypothesis we can replace the integral over time with an integral over the phase space, weighted by the density $\rho$:
    \begin{align}
        A &= \lim_{\tau\rightarrow\infty}\int_t^{t+\tau}\frac{\text{d}t'}{\tau}\, a\big(\{\mathbf{q}_i(t'),\mathbf{p}_i(t')\}\big) \\
        &= \int \text{d}\Omega\, \rho\big(\{\mathbf{q}_i,\mathbf{p}_i\}\big) \, a\big(\{\mathbf{q}_i,\mathbf{p}_i\}\big) \,.
        \label{ens_average}
    \end{align}

\begin{figure}[h!]
        \centering
        \includegraphics[width = 0.75\textwidth]{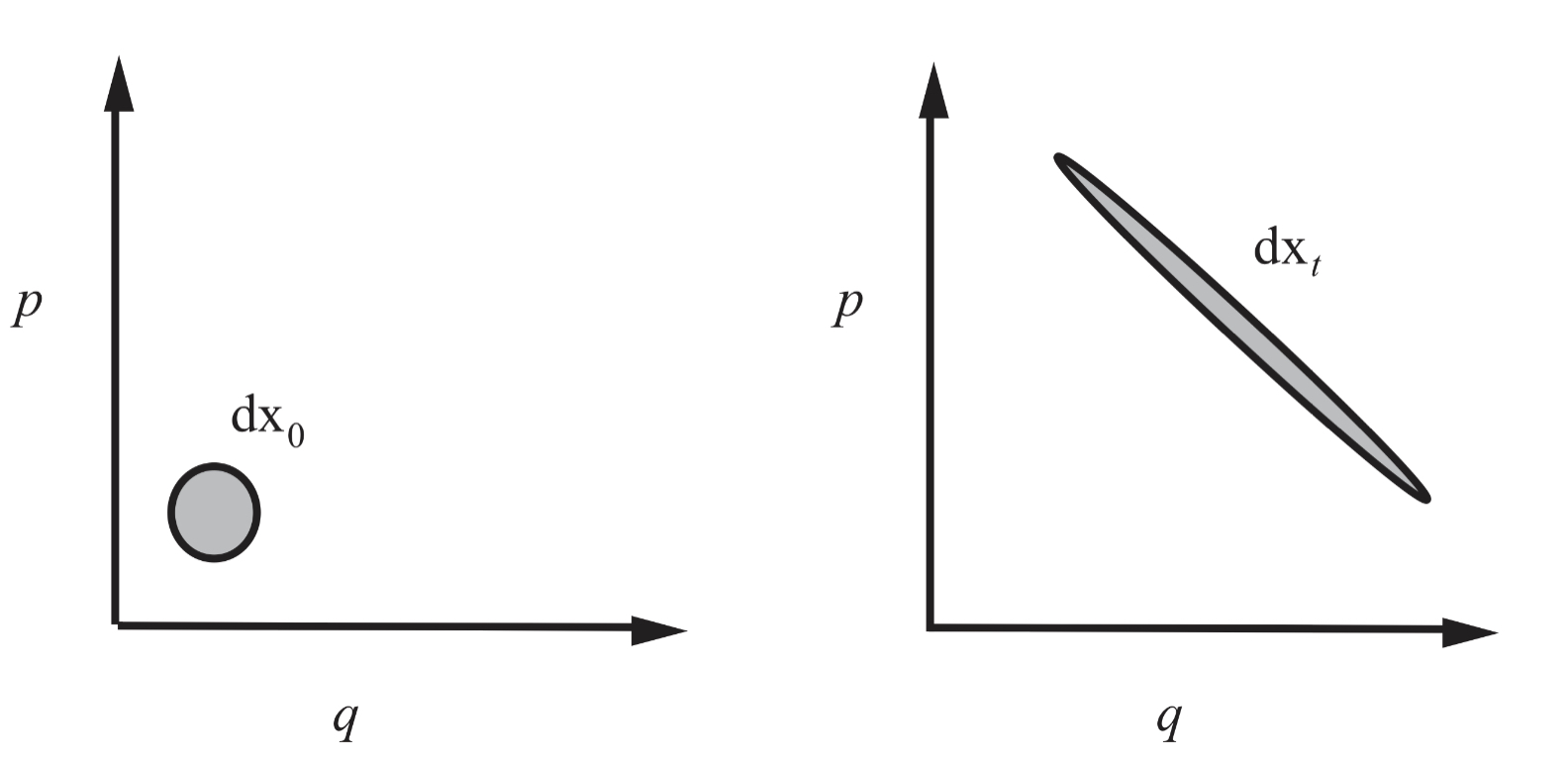}
        \caption{Graphical representation of the phase space \emph{volume conservation} prescribed by Liouville's theorem for a one-dimensional system. Source: \cite{tuckerman}.}
        \label{fig:liouville}
    \end{figure}

    \begin{liouville_th}
    The phase space density $\rho$ is conserved: 
    \begin{equation}\label{liouv_th}
        \frac{\text{d}\rho}{\text{d}t}=0\,.
    \end{equation}
    \end{liouville_th}
    \noindent The so-called Liouville's equation includes a total time derivative, which describes how $\rho$ changes in time as a function of how positions and momenta change in time. Considering also an explicit time dependence, i.e. $\rho = \rho\big(\{\mathbf{q}_i,\mathbf{p}_i\},t\big)$, \cref{liouv_th} can be rewritten by applying the chain rule for derivatives, namely
    \begin{equation}\label{chain_rule}
        \frac{\text{d}\rho}{\text{d}t} = \frac{\partial\rho}{\partial t} + \sum_{i,\alpha} \Big(\frac{\partial\rho}{\partial q_{i}^{\alpha}}\,\dot{q}_{i}^{\alpha} + \frac{\partial\rho}{\partial p_{i}^{\alpha}}\,\dot{p}_{i}^{\alpha} \Big)\,.
    \end{equation}
    Defining the vector of all phase space coordinates as $\mathbf{x} = \left(q_1^x,q_1^y,q_1^z,...\,,p_N^x,p_N^y,p_N^z\right)$, this condition turns out to be equivalent to 
    \begin{equation}\label{incompressibility}
        \nabla_{\mathbf{x}}\cdot\dot{\mathbf{x}} = 0\,,
    \end{equation}
    where $\nabla_{\mathbf{x}} = \left(\partial/\partial q_1^x,...\,,\partial/\partial p_N^z\right)$. For this reason Liouville's theorem is also said to describe the property of \emph{phase space incompressibility}, since \cref{incompressibility} implies the absence of sources or sinks for a fluid with velocity flow field $\dot{\mathbf{x}}$. Providing a geometrical interpretation, this condition implies that the copies of the system initialized within the phase space element $\text{d}\mathbf{x}_0$ will evolve in a phase space element $\text{d}\mathbf{x}_t$ with the same volume, at any future time $t$ (see \cref{fig:liouville}). 
    
    \noindent Using \cref{chain_rule} together with Hamilton's equations, \cref{liouv_th} can be rewritten as:
    \begin{equation}
        0 = \frac{\partial\rho}{\partial t} + \sum_{i,\alpha} \Big(\frac{\partial\rho}{\partial q_{i}^{\alpha}}\,\frac{\partial\mathcal{H}}{\partial p_{i}^{\alpha}} - \frac{\partial\rho}{\partial p_{i}^{\alpha}}\,\frac{\partial\mathcal{H}}{\partial q_{i}^{\alpha}} \Big)
        = \frac{\partial\rho}{\partial t} + \Big\{\rho,\,\mathcal{H}\Big\}\,.
        \label{liouv}
    \end{equation}
    Here $\alpha$ identifies the cartesian component ($\alpha=x,y,z$), the dot notation is used for the total time derivative and $\{\cdot\,,\cdot\}$ define the Poisson brackets. Therefore in a system at equilibrium, for which 
    \begin{equation}
        \frac{\partial\rho}{\partial t} = 0\,,
    \end{equation}
    \cref{liouv} brings to $\{\rho,\,\mathcal{H}\} = 0$, which implies that the density $\rho$ can only depend on the phase space coordinates $\{\mathbf{q}_i,\mathbf{p}_i\}$ through the Hamiltonian $\mathcal{H}$:
    \begin{equation}
        \rho = \rho \Big(\mathcal{H}\big(\{\mathbf{q}_i,\mathbf{p}_i\}\big)\Big)\,.
    \end{equation}
    \noindent Among all the possible functional forms satisfying \cref{rho_H}, the one that defines the density $\rho$ in the microcanonical ensemble is fixed by the following principle.
    \begin{equiprobab_pr}
    The probability associated to each microstate, i.e. to each point of phase space, is the same for all the microstates compatible with the energy $E$ of the system, fixed by the initial conditions.
    \end{equiprobab_pr}
    \noindent In other words, the probability distribution of the microcanonical ensemble can be written as a $\delta$-function:
    \begin{equation}
        \mathcal{P}_{NVE}\big(\{\mathbf{q}_i,\mathbf{p}_i\}\big) = \frac{1}{\Gamma}\, \delta\Big(\mathcal{H}\big(\{\mathbf{q}_i,\mathbf{p}_i\}\big) - E\Big)\,,
        \label{rho_H}
    \end{equation}
    where $\Gamma = \Gamma(N,V,E)$ is the normalization factor of the distribution, fixed by the constraint
    \begin{equation}
        \int_{\Omega} \Big(\prod_{i=1}^N d\mathbf{q}_id\mathbf{p}_i\Big)\, \mathcal{P}\big(\{\mathbf{q}_i,\mathbf{p}_i\}\big) = 1\,.
    \end{equation}

    \subsection{Velocity Verlet and leapfrog algorithms}\label{velocity_verlet}
    Since Liouville's theorem is derived by means of Hamilton's equations, the dynamical evolution of the system can be obtained by solving \cref{liouv_th}, that can be alternatively written as
    \begin{equation}\label{liouvilleq}
        \frac{\partial\rho}{\partial t} = -i\mathcal{\hat{L}}\,\rho\,,
    \end{equation}
    where $\mathcal{\hat{L}}$, called \emph{Liouville operator}, is typically split in the \emph{drift operator} $\mathcal{\hat{L}}_q$ and the \emph{kick operator} $\mathcal{\hat{L}}_p$:
    \begin{equation}
    i\mathcal{\hat{L}}\equiv i\mathcal{\hat{L}}_q + i\mathcal{\hat{L}}_p  \equiv\sum_{i,\alpha} \left(\frac{\partial \mathcal{H}}{\partial p_{i}^{\alpha}}\,\frac{\partial}{\partial q_{i}^{\alpha}}-\frac{\partial \mathcal{H}}{\partial q_{i}^{\alpha}}\,\frac{\partial }{\partial p_{i}^{\alpha}}\right)\,.
    \end{equation}
    Isolating the action of $\mathcal{\hat{L}}_q$ and $\mathcal{\hat{L}}_p$ separately, it is easy to show that the two following relations hold exactly:
    \begin{subequations}
    \begin{align}
        \label{Lq}
        e^{-i \Delta t\mathcal{\hat{L}}_q}\rho(\mathbf{q}_i,\mathbf{p_i},t) &= \rho\left(\mathbf{q}_i -\frac{\mathbf{p}_i}{m_i}\Delta t,\mathbf{p_i},t\right)\,, \\
        \label{Lp}
        e^{-i \Delta t\mathcal{\hat{L}}_p}\rho(\mathbf{q}_i,\mathbf{p_i},t) &= \rho\left(\mathbf{q}_i,\mathbf{p_i}-\mathbf{F}_i\Delta t,t\right)\,,
    \end{align}
    \end{subequations}
    where $\mathbf{F}_i=-\frac{\partial U}{\partial \mathbf{q}_i}$ is the force acting on atom $i$ and $\Delta t$ is the integration time step.
    However, the formal solution of \cref{liouvilleq},
    \begin{equation}
    \label{exact_sol}
    \rho(\mathbf{q}_i,\mathbf{p_i},t+\Delta t) = e^{-i \Delta t\mathcal{\hat{L}}}\rho(\mathbf{q}_i,\mathbf{p_i},t)\,,
    \end{equation}
    includes the exponential of a sum of two non-commuting operators and cannot be rewritten separating the action of the two operators. A popular approximation of \cref{exact_sol}, which is correct up to second-order in $\Delta t$, employs the so-called \emph{Trotter splitting}:
    \begin{equation}\label{vv_trotter}
        \exp\left(-i \Delta t\mathcal{\hat{L}}\right) = \exp\left(-i \frac{\Delta t}{2}\mathcal{\hat{L}}_p\right) \exp\left(-i \Delta t\mathcal{\hat{L}}_q\right) \exp\left(-i \frac{\Delta t}{2}\mathcal{\hat{L}}_p\right) + \mathcal{O}(\Delta t^3)
    \end{equation}
    In words, \cref{vv_trotter} gives a simple recipe to evolve the system according to Hamilton's equations, with an accuracy depending on the timestep $\Delta t$. A single iteration of the algorithm for atom $i$ is reported below.\\
    
    \begin{algorithm}[H]\label{velocity_verlet_alg}
    \SetAlgoLined
     $\mathbf{p}_i(t+\Delta t/2) = \mathbf{p}_i(t) + \mathbf{F}_i(t) \Delta t/2$\;
     $\mathbf{q}_i(t+\Delta t) = \mathbf{q}_i + \frac{\mathbf{p}_i}{m_i}(t+\Delta t/2) \Delta t$\;
     recompute forces: $\mathbf{F}_i(t+\Delta t) \longleftarrow\mathbf{F}_i(t)$\;
     $\mathbf{p}_i(t+\Delta t) = \mathbf{p}_i(t+\Delta t/2) + \mathbf{F}_i(t+\Delta t) \Delta t/2$\;
     \caption{Velocity Verlet}
    \end{algorithm}
    \vspace{0.4cm}
    \noindent This integration scheme is named \emph{velocity Verlet}, as opposed to the algorithm where drift and kick operators are exchanged in the Trotter splitting (called \emph{position Verlet}), and it's easy to show that it satisfies both the time-reversibility of Hamilton's equations and the property of density conservation in phase space, but clearly violates the conservation of energy.
    
    Since the last momentum update and the first one in the next iteration employ the same forces, these two steps can be combined in order to improve the efficiency of the algorithm. This approach defines the \emph{leapfrog integrator}, where positions and momenta are propagated with a time lag of half timestep:\\
    
    \begin{algorithm}[H]
    \SetAlgoLined
     $\mathbf{p}_i(t+\Delta t/2) = \mathbf{p}_i(t-\Delta t/2) + \mathbf{F}_i(t) \Delta t$\;
     $\mathbf{q}_i(t+\Delta t) = \mathbf{q}_i + \frac{\mathbf{p}_i}{m_i}(t+\Delta t/2) \Delta t$\;
     recompute forces: $\mathbf{F}_i(t+\Delta t) \longleftarrow\mathbf{F}_i(t)$\;
     \caption{Leapfrog integrator}
    \end{algorithm}

    \section{Canonical ensemble (NVT)}
    A physical system belongs to the canonical ensemble if it is closed but can exchange energy with the external environment in the form of heat, i.e. if its number of particles $N$, its volume $V$ and its temperature $T$ are fixed. In this context, the system is typically said to be coupled with an external bath at temperature $T$, which is the intensive thermodynamic quantity controlling the energy fluctuations. Indeed, by considering both the system and the external bath as a single system belonging to the microcanonical ensemble, i.e. with fixed total energy, it is possible to prove that the probability distribution as a function of the phase space coordinates takes the form
    \begin{equation}
        \label{boltzmann}
        \mathcal{P}_{NVT}\big(\{\mathbf{q}_i,\mathbf{p}_i\}\big) = \frac{1}{\mathcal{Z}} \exp\left[ -\frac{1}{k_B T}\mathcal{H}\big(\{\mathbf{q}_i,\mathbf{p}_i\}\big)\right]\,,
    \end{equation}
    where $k_B$ is the Boltzmann constant and $\mathcal{Z} = \mathcal{Z}(N,V,T)$ is the normalization factor, called \emph{canonical partition function}. This functional form defines what is called the \emph{Boltzmann distribution} (or canonical distribution), in term of which the generic ensemble average of the physical observable $a$ can be written as: 
    \begin{equation}
        \label{ens_average_canonical}
        \langle a \rangle = \frac{1}{\mathcal{Z}}\int_{\Omega} \Big(\prod_{i=1}^N d\mathbf{q}_id\mathbf{p}_i\Big)\, \exp\left[ -\frac{1}{k_B T}\mathcal{H}\big(\{\mathbf{q}_i,\mathbf{p}_i\}\big)\right]\,a\big(\{\mathbf{q}_i,\mathbf{p}_i\}\big)\,,
    \end{equation}
    Some relevant relations following from \cref{boltzmann} and \cref{ens_average_canonical} are the expression for the average energy of the system, 
    \begin{equation}
        \langle \mathcal{H} \rangle = -\frac{\partial \log \mathcal{Z}}{\partial\beta}\,,
    \end{equation}
    where $\beta=1/(k_B T)$, and its relation with the temperature and the standard deviation $\sigma_{\mathcal{H}}$ of the energy distribution:
    \begin{equation}
        \frac{\partial \langle \mathcal{H} \rangle}{\partial T} = \frac{1}{k_B T^2}\sigma_{\mathcal{H}}^2\,.
    \end{equation}
    As a function of the kinetic energy $K$ only, the canonical distribution becomes a gamma distribution:
    \begin{equation}\label{kineng_distr}
        \mathcal{P}_{NVT}(K) \propto K^{\frac{N_f}{2}-1}\,e^{-\frac{K}{k_B T}}\,.
    \end{equation}
    Here $N_f$ is the number of degrees of freedom in the system. As a consequence, the average kinetic energy $\langle K\rangle$ and its standard deviation $\sigma_K$ are:
    \begin{subequations}
    \begin{align}
        \langle K \rangle &= \frac{1}{2}N_f k_B T \,,\label{avg_kin}\\
        \sigma_K &= \langle K \rangle \sqrt{\frac{2}{N_f}}\,.
    \end{align}
    \end{subequations}
    It is worth observing that \cref{avg_kin} is in perfect agreement with the classical equipartition theorem. Instead of studying the behaviour of the instantaneous kinetic energy $K$, an equivalent description can be given in terms of the instantaneous temperature $T_s$, defined as:
    \begin{equation}\label{instantaneous_temp}
        \frac{1}{2}N_d k_B T_s = K\,.
    \end{equation}

    \section{Thermostats}
    Simulating a system in the canonical ensemble means generating a sequence of samplings from the kinetic energy distribution defined in \cref{kineng_distr}, and the algorithms to accomplish this task are called \emph{thermostats} in the MD language. Thermostats can be either \emph{global}, when all the single atom kinetic energies are rescaled with the same factor - the same used to rescale the total kinetic energy - or \emph{local}, when the change in the total kinetic energy is a result of a different rescaling for each atom \cite{global_vs_local}. A second distinction is between \emph{deterministic} and \emph{stochastic} thermostats, where the second category employs random number generators. 
    
    \subsection{Properties of sampling algorithms}\label{properties_sampling}
    A good thermostat should have at least three properties, which are stated below for a generic sampling algorithm and a generic probability distribution.
    
    \begin{enumerate}[label=\textbf{\arabic*}, align = left, labelwidth = 2 em, labelsep = 0 em]
        \item \textbf{Stationarity of the target distribution (or balance).} A sampling algorithm S is said to satisfy the balance condition with respect to the probability distribution $\mathcal{P}(x)$ if the application of S on a set of $N$ samplings $\{x_i\}_{i=1}^N$ independently drawn from $\mathcal{P}(x)$ produces a new set of samplings $\{x'_i\}_{i=1}^N$ that are still independently drawn from $\mathcal{P}(x)$, in the limit of large $N$. Equivalently, $\mathcal{P}(x)$ is said to be stationary with respect to S.
        
        \item \textbf{Ergodicity.} A sampling algorithm S is said to be ergodic with respect to its stationary distribution $\mathcal{P}(x)$ if, starting from any point $x_0$ in the domain of $\mathcal{P}$, it allows to reach in a finite number of steps any point $\overbar{x}$ such that $\mathcal{P}(\overbar{x}) \neq 0$.
        
        It is possible to prove that a necessay condition for the ergodicity of a sampling algorithm is the existence of a single stationary distribution.
        
        \item \textbf{Fast decorrelation of the samplings.} An ideal sampling algorithm should generate independent samplings of the target distribution. The degree of correlation after a time lag $\tau$ can be measured via the autocorrelation function
        \begin{equation}
            C(\tau)=\frac{\langle x_t x_{t+\tau}\rangle - \langle x\rangle^2}{\sigma_x^2}\,,
        \end{equation}
        which has the properties $C(\tau) \in [-1,1]$, $C(0) = 1$  and for which $C(\tau) = 0$ denotes independent samplings after a time lag $\tau$. 
    \end{enumerate}

    \subsection{Berendsen thermostat}
    The Berendsen thermostat \cite{berendsen} is a global and first-order deterministic thermostat, i.e. it can be formulated in terms of a first-order differential equation that evolves the total kinetic energy $K$ towards the target value $\overbar{K} = \frac{1}{2}N_f k_B T$:
    \begin{equation}
        \label{berendsen_eq}
        \text{d}K = \frac{\overbar{K}-K}{\tau}\,\text{d}t
    \end{equation}
    The parameter $\tau$ plays the role of the relaxation time of the kinetic energy and it is also related to the autocorrelation time of the samplings; with the proper timestep it can be chosen arbitrarily small without any drawback. 
    
    The implementation of the Berendsen thermostat simply consists in the following momentum rescaling, applied to each atom $i$:
    \begin{equation}
        \label{berendsen_alg}
        \mathbf{p}_i(t+\Delta t) = \mathbf{p}_i(t)\,\sqrt{\frac{e^{-\Delta t/\tau}\,K + (1-e^{-\Delta t/\tau})\,\overbar{K}}{K}}\,.
    \end{equation}
    In fact, it is possible to show that \cref{berendsen_alg} brings to \cref{berendsen_eq} in the limit $\Delta t/\tau\rightarrow 0$.
    
    Although the Berendsen thermostat is efficient in the equilibration of the system it is typically not employed in the production phase, since it reproduces the correct average of the kinetic energy but not the higher order moments of its canonical distribution, i.e. it targets a probability distribution which is not the one defined in \cref{kineng_distr}. 
    
    \subsection{Andersen thermostat}
    The temperature coupling method developed by Andersen \emph{et al.} \cite{andersen} consists in a local and stochastic thermostat that does not admit a continuous formulation, i.e. it cannot be described in terms of a differential equation for the kinetic energy or the momenta. In its easiest formulation, the idea of the algorithm is to propagate the system at constant energy, i.e. in the microcanonical ensemble, and to redefine the momentum of each atom once every $n_s$ steps by extracting its components $p_i^{\alpha}$ from their reference distribution in the canonical ensemble,
    \begin{equation}\label{gaussian_momentum}
        \mathcal{P}_{NVT}(p_i^{\alpha}) = \frac{1}{\sqrt{2\pi m_i k_B T}}\,\exp\left(-\frac{(p_i^{\alpha})^2}{2 m_i k_B T}\right)\,,
    \end{equation}
    which can be obtained from \cref{boltzmann} by marginalizing the Boltzmann distribution over all the atom positions $\{\mathbf{q}_i\}$ and the remaining momenta $\{\mathbf{p}_j\}_{j\neq i}$. The parameter $n_s$ is called \emph{stride} of the thermostat, and the resampling can also be applied using a different stride for each atom or choosing randomly a different atom at each step. Thus, the algorithm can be easily implemented using a Gaussian random number generator. Using the formulation with a common stride for all the $N$ atoms, the scheme of the algorithm is the following: \\
    
    \begin{algorithm}[H]
    \SetAlgoLined
    velocity Verlet (or leapfrog integrator) for $n_s$ steps\;
    \For{$i=1$ \KwTo $N$}{
        \For{$\alpha = x,y,z$} {
            extract $\mathcal{R}_{i}^{\alpha}\sim\mathcal{N}(0,1)$\;
            update momentum component: $p_{i}^{\alpha} \gets \sqrt{m_i\, k_B T}\,\mathcal{R}_{i}^{\alpha}$\;
        }
    }
    \caption{Andersen thermostat}
    \label{andersen_alg}
    \end{algorithm}
    \vspace{0.4cm}
    It is worth underlining that in the original formulation of the Andersen algorithm \cite{andersen} particles' momenta are not randomized with a fixed stride; instead, for each particle the time intervals $\Delta t_j$ between successive collisions with the bath are extracted from the distribution
    \begin{equation}
        \mathcal{P}(\Delta t_j) = \nu e^{-\nu \Delta t_j}\,,
    \end{equation}
    where the input parameter $\nu$ is the the mean rate of the collisions.

    \subsection{Langevin thermostat}\label{langevin_thermo}
    The Langevin thermostat \cite{langevin_bussi} evolves each momentum $\mathbf{p_i}$ according to the stochastic equation (see Appendix~\ref{appendix_stocdiffeq})
    \begin{equation}
        \label{langevin_eq}
        \text{d}\mathbf{p}_i = -\gamma \mathbf{p}_i\text{d}t + \sqrt{2\gamma m_i\,k_B T}\text{d}\mathbf{W}_i\,,
    \end{equation}
    where the scalar parameter $\gamma$ is called \emph{friction}. Since \cref{langevin_eq} includes a noise term and is referred to the momentum of the single atom $i$, this thermostat is stochastic and local. It is possible to interpret the Langevin thermostat as a continuous version of the Andersen one shown in \cref{andersen_alg}, where the exact resampling from the target distribution is replaced by the "smoother" update 
    \begin{equation}
        \label{mom_update_andersen}
        \mathbf{p}_i (t+\Delta t) = e^{-\Delta t/\tau}\mathbf{p}_i(t) + \sqrt{\left(1-e^{-2\Delta t/\tau} \right)m_i\, k_B T}\,\mathbfcal{R}_i\,.
    \end{equation}
    In fact, by defining $\gamma\equiv 1/\tau$ it is possible to recover \cref{langevin_eq} from \cref{mom_update_andersen} in the limit $\Delta t/\tau\rightarrow 0$. By coupling the Langevin thermostat with Hamilton's equations one gets the \emph{underdamped Langevin equations}
    \begin{subequations}
    \begin{align}
        \text{d}\mathbf{q}_i &= \frac{\mathbf{p}_i}{m_i}\text{d}t\,, \\
        \text{d}\mathbf{p}_i &= \mathbf{F}_i\text{d}t - \gamma \mathbf{p}_i\text{d}t + \sqrt{2\gamma m_i\,k_B T}\text{d}\mathbf{W}_i\,,
    \end{align}
    \end{subequations}
    which can be integrated in a time-reversible way with the following scheme:\\
    
    \begin{algorithm}[H]
    \SetAlgoLined
    \For{$i=1$ \KwTo $N$}{
        extract $\mathbf{\mathbfcal{R}}_{i}\sim\mathcal{N}_3\left((0,0,0),(1,1,1)\right)$\;
        update $\mathbf{p}_i$ with time step $\frac{\Delta t}{2}$: \indent$\hspace{1cm}\mathbf{p}_i \gets e^{-\Delta t/(2\tau)}\mathbf{p}_i + \sqrt{\left(1-e^{-\Delta t/\tau} \right)m_i\, k_B T}\,\mathbf{\mathbfcal{R}}_i$\;
    }
    Velocity Verlet (or leapfrog integrator)\;
    \For{$i=1$ \KwTo $N$}{
        extract $\mathbf{\mathbfcal{R}}_{i}\sim\mathcal{N}_3\left((0,0,0),(1,1,1)\right)$\;
        update $\mathbf{p}_i$ with time step $\frac{\Delta t}{2}$: \indent$\hspace{1cm}\mathbf{p}_i \gets e^{-\Delta t/(2\tau)}\mathbf{p}_i + \sqrt{\left(1-e^{-\Delta t/\tau} \right)m_i\, k_B T}\,\mathbf{\mathbfcal{R}}_i$\;
    }
    \caption{Langevin thermostat}
    \label{langevin_alg}
    \end{algorithm}
    \vspace{0.4cm}
    \noindent  In the Langevin thermostat the value of the friction $\gamma$ affects in a non-trivial way the efficiency of the algorithm, especially when the system has to be equilibrated. Since a large friction accounts for a strong coupling with the external bath, one could think that increasing $\gamma$ always leads to accelerate the thermalization of the system. Conversely, when $\gamma$ is chosen too large the equilibration time actually increases \cite{global_vs_local}; as a consequence, for each system it exists an optimal value of $\gamma$ that minimizes the relaxation time, but there is no general recipe to set it a priori. The physical interpretation of this behaviour is that when the friction is too large the "collisions" of the particles with the bath become so frequent that they suppress the collisions of the particles among themselves, reducing the momentum exchanges and slowing down the dynamics of the system. Hence, since different choices of $\gamma$ can alter significantly the Hamiltonian dynamics, this algorithm cannot be used to compute dynamical properties, unless an extremely small friction is used \cite{global_vs_local}.
    
    The Langevin thermostat can be shown to satisfy the \emph{detailed balance} condition (see Appendix~\ref{appendix_fp}) with respect to the canonical distribution in \cref{boltzmann} and it allows to calculate the \emph{effective energy drift} (see Appendix~\ref{appendix_effenergy}), which quantifies the detailed balance violations \cite{langevin_bussi}.
    
    \subsection{Nosé-Hoover thermostat}
    The Nosé-Hoover method \cite{nose-original,hoover} is a global thermostat based on a second-order deterministic equation, which can be derived from an extended Hamiltonian that includes a new degree of freedom related to the bath coupling. Split into two first-order equations, the Nosé-Hoover dynamics reads
    \begin{subequations}
    \begin{align}
        \text{d}K &= -2 \gamma K \text{d}t\,, \\
        \text{d} \gamma &= \frac{2}{M}\left(K -\overbar{K} \right)\text{d}t\,,
    \end{align}
    \end{subequations}
    where $\gamma$ is a time-dependent friction and $M$ - usually called \emph{thermostat mass} - is a non-trivial parameter which affects the behaviour in the equilibration phase as well as the autocorrelation time of the samplings. In particular, a non-optimal choice of $M$ can bring to large and slow oscillations when the system is initialized far from equilibrium; for this reason, first-order thermostats are in general more efficient when a system needs to be thermalized.  
    
    Using the formalism of Fokker-Planck equations (see Appendix~\ref{appendix_fp}) it is possible to show that the canonical distribution defined in \cref{kineng_distr} satisfies the condition of stationarity described in \cref{properties_sampling}. However the Nosé-Hoover algorithm is not ergodic, especially for small and stiff systems and harmonic oscillators. Ergodicity can be achieved by coupling the friction variable to a second external bath, typically implemented as a new Nosé-Hoover thermostat with its own friction. This coupling procedure can be repeated several times, in a scheme named \emph{Nosé-Hoover chains} \cite{nosehoover_chains}. If the additional coupling is performed with of a Langevin thermostat, the temperature coupling method is called \emph{Nosé-Hoover-Langevin thermostat} \cite{nose-hoover-langevin}.

    \subsection{Stochastic velocity rescaling}\label{SVR}
    Stochastic velocity rescaling (SVR), also known as Bussi-Donadio-Parrinello thermostat \cite{SVR}, is a global and first-order algorithm based on the following stochastic equation:
    \begin{equation}
        \text{d}K=-(K-\overbar{K})\frac{\text{d}t}{\tau} + 2\sqrt{\frac{K\overbar{K}}{\tau N_f}}\text{d}W\,,   
    \end{equation}
    where $N_f$ is the number of degrees of freedom and $\overbar{K}$ is the target kinetic energy, related to the external temperature via the usual equipartition theorem $\overbar{K}=N_f k_B T/2$. It is worth noting that the deterministic part of the equation is exactly the Berendsen thermostat; hence the time constant $\tau$ can be interpreted also in this case as the relaxation time of the system and the autocorrelation time of the samplings. The additional noise term brings to a Fokker-Planck equation that can be shown to satisfy the \emph{detailed balance} condition (see Appendix \cref{appendix_fp}) with respect to the canonical distribution in \cref{kineng_distr}, meaning that SVR generates the correct canonical ensemble. The algorithm can be formulated in a time-reversible way that allows the calculation of the effective energy drift (see Appendix \ref{appendix_effenergy}).

    \subsection{Monte Carlo thermostat}\label{MC_thermostat}
    Canonical sampling can be also achieved through the popular technique of \emph{Markov chain Monte Carlo} (MCMC), that will be here discussed in a general framework and considering a discrete sample space. The aim of MCMC is to achieve the ergodic sampling of a probability distribution $\mathcal{P}(x)$ by repeatedly applying a stochastic rule embedded in the transition matrix $\Pi_{x,x'} \equiv \Pi(x\rightarrow x')$, which represents the conditional probability of sampling $x'$ starting from the previous sample $x$. The fact that this conditional probability only depends on the previous step is called \emph{Markov property}. $\Pi(x\rightarrow x')$ is a stochastic matrix, i.e. it satisfies the two properties
    \begin{subequations}
    \begin{align}
        \label{normaliz_trmatrix}
        \sum_{x'}\Pi(x\rightarrow x') &= 1\,,\\
        \Pi(x\rightarrow x') &\geq 0\,,
    \end{align}
    \end{subequations}
    and it can be used to relate the marginal probabilities of consecutive steps, via the so-called \emph{Master equation}:
    \begin{equation}\label{master_eq_mcmc}
        \mathcal{P}_{i+1}(x) = \mathcal{P}_{i}(x) + \sum_{x'\neq x} \mathcal{P}_{i}(x') \Pi(x'\rightarrow x) - \sum_{x'\neq x} \mathcal{P}_{i}(x) \Pi(x\rightarrow x')\,.
    \end{equation}
    In \cref{master_eq_mcmc} the index $i$ represents the discrete time step in the sampling chain, and the last two terms in the RHS of the equation are called \emph{gain} and \emph{loss} terms respectively. The stationarity or balance condition reads $\mathcal{P}_{i+1}(x) = \mathcal{P}_i(x)$, hence it brings to the equation
    \begin{equation}\label{balance_MC}
        \sum_{x'} \mathcal{P}(x') \Pi(x'\rightarrow x) = \sum_{x'} \mathcal{P}(x) \Pi(x\rightarrow x')\,,
    \end{equation}
    where the case $x'= x$ is now included in the sums since it gives the same contribution to the RHS and LHS of the equation. By applying the property in \cref{normaliz_trmatrix} this condition becomes
    \begin{equation}\label{lefteig}
        \sum_{x'} \mathcal{P}(x') \Pi(x'\rightarrow x) = \mathcal{P}(x) \,.
    \end{equation}
    In words, \cref{lefteig} tells that $\mathcal{P}$ is a stationary distribution with respect to the sampling algorithm embedded in $\Pi$ if $\mathcal{P}$ is a a left eigenvector of $\Pi$ with eigenvalue equal to 1. The existence of such a distribution is a consequence of the Perron-Frobenius theorem, and the ergodicity of the sampling algorithm can only be achieved if the left eigenvector of $\Pi$ is unique. A simpler condition that implies the stationarity of $\mathcal{P}$ in \cref{balance_MC} is the so-called \emph{detailed balance} or \emph{equilibrium} condition, that in the framework of MCMC reads:
    \begin{equation}\label{detailed_balance_MC}
        \mathcal{P}(x') \Pi(x'\rightarrow x) = \mathcal{P}(x) \Pi(x\rightarrow x')\,.
    \end{equation}
    This stricter condition is usually more employed than the balance one since it is generally easier to construct algorithms satisfying \cref{detailed_balance_MC} than \cref{balance_MC}; however, detailed balance is not a necessary condition for MCMC \cite{db_notnecessary}. The first step to create a rule satisfying the detailed balance condition is to split the transition matrix $\Pi$ into a \emph{proposal matrix} $M$ and an \emph{acceptance matrix} $\alpha$:
    \begin{equation}
        \Pi(x\rightarrow x') = M(x\rightarrow x') \,\alpha(x\rightarrow x')\,.
    \end{equation}
    By decomposing $\Pi$ in this way, \cref{detailed_balance_MC} can be rewritten as
    \begin{equation}\label{db_ratioacc}
        \frac{\alpha(x\rightarrow x')}{\alpha(x'\rightarrow x)} = \frac{\mathcal{P}(x')\,M(x'\rightarrow x)}{\mathcal{P}(x)\,M(x\rightarrow x')}\,.
    \end{equation}
    Among several possibilities to satify \cref{db_ratioacc}, the one giving the highest acceptance is the Metropolis-Hastings rule:
    \begin{equation}
        \alpha(x\rightarrow x') \equiv \min \left(1,\,\frac{\mathcal{P}(x')\,M(x'\rightarrow x)}{\mathcal{P}(x)\,M(x\rightarrow x')}\right)
    \end{equation}
    Although not strictly necessary, the proposal rule embedded in $M$ is typically constructed in order to make $M$ a symmetric matrix, so that the acceptance can be simply calculated as
    \begin{equation}
        \alpha(x\rightarrow x') \equiv \min \left(1,\,\frac{\mathcal{P}(x')}{\mathcal{P}(x)}\right)\,.
    \end{equation}
    Since the acceptance $\alpha$ is always calculated with ratios of probabilities, the Metropolis-Hastings rule allows to sample a generic distribution $\mathcal{P}$ without knowing its normalization, which can be hard to compute in a high-dimensional space. In general, the trial move embedded in the proposal matrix $M$ should be designed in order to obtain the highest possible value for the average value of $\alpha$ over the simulation. A scheme of the method is reported in \cref{MC_algorithm}.\\
    
    \begin{algorithm}[H]
    \SetAlgoLined
        Propose the move $x\rightarrow x'$\;
        Compute the acceptance: $\alpha=\min \left(1,\,\frac{\mathcal{P}(x')}{\mathcal{P}(x)}\right) $ \;
        Extract a uniform random number in $[0,1]$: $\mathcal{R}\sim U(0,1)$\;
        \eIf{$\alpha> \mathcal{R}$}{
            Accept the move: $x\gets x'$\;
        }{ 
            Refuse the move: $x\gets x$\;
        }
    \caption{Markov chain MC with Metropolis-Hastings}
    \label{MC_algorithm}
    \end{algorithm}
    \vspace{0.4cm}
    When a system is simulated in the canonical ensemble this scheme can be used to perform single-particle trial moves on their positions, where at each step the particle involved can be selected both with a deterministic \emph{sweep} strategy and with random selections \cite{sweep}. However, when constraints are present the construction of an efficient trial move becomes a complex task, and more efficient Monte Carlo strategies can be employed. As an additional limitation, standard MCMC applied to the canonical distribution carries no information about momentum variables, since the proposal move only involves atomic positions. \\

    \subsection{Hybrid Monte Carlo}\label{hybrid_MC}
    Hybrid (or smart) Monte Carlo \cite{hybrid_MC} is a technique to sample the canonical ensemble that employs a molecular dynamics algorithm in the microcanonical ensemble - such as velocity Verlet - to propose a new move, and the Metropolis-Hastings rule to accept it or refuse it.
    
    As a starting point to illustrate the method, it is possible to show that the canonical distribution, and more in general every distribution $\mathcal{P}$ depending on the phase space coordinates via the Hamiltonian $\mathcal{H}$, 
    \begin{equation}\label{p_of_H}
        \mathcal{P} = \mathcal{P}\Big(\mathcal{H}\big(\{\mathbf{q}_i,\mathbf{p}_i\}\big)\Big)\,,
    \end{equation}
    is stationary with respect to the Hamiltonian dynamics. Indeed the Fokker-Planck equation that one can write from \cref{hamilton_eqs1,hamilton_eqs2}, which does not include any diffusion term due to the deterministic nature of Hamilton's equations (see Appendix \ref{appendix_fp}), is
    \begin{equation}
        \frac{\partial \mathcal{P}}{\partial t} = - \sum_{i,\alpha} \frac{p_i^{\alpha}}{m_i} \frac{\partial\,\mathcal{P}}{\partial q_i^{\alpha}} - \sum_{i,\alpha}F_i^{\alpha}\frac{\partial\,\mathcal{P}}{\partial p_i^{\alpha}}\,
    \end{equation}
    and it is straightforward to show that $\frac{\partial \mathcal{P}}{\partial t} = 0$ if \cref{p_of_H} holds. Hence, the canonical distribution is stationary with respect to Hamilton's equations. Nevertheless this is not a sufficient condition to sample the canonical distribution, because the ergodicity condition is clearly not satisfied. The idea of Hybrid MC is to achieve ergodicity by employing a microcanonical integrator that exactly satisfies the time-reversibility property of Hamilton's equations (see \cref{hamilton}) but violates the energy conservation, so that energy variations can be accepted or refused according to the Metropolis-Hastings rule. In order to apply this rule, a detailed balance condition such as the one in \cref{detailed_balance_MC} should be satisfied. Calling a point in phase space $\mathbf{x} \equiv \big(\{\mathbf{q}_i,\mathbf{p}_i\}\big)$ and embedding the (deterministic) Hamiltonian dynamics in the matrix $\Pi$, it is clear that the standard detailed balance condition does not hold with respect to the canonical distribution, since in general $\Pi(\mathbf{x'}\rightarrow\mathbf{x}) = 0$ if $\Pi(\mathbf{x}\rightarrow\mathbf{x'}) \neq 0$.
    On the other hand, defining $\mathbf{x^*} \equiv \left(\{\mathbf{q}_i,-\mathbf{p}_i\}\right)$ and recalling the time-reversibility property of Hamilton's equations, the transition matrix $\Pi$ satisfies:
    \begin{equation}
        \Pi(\mathbf{x}\rightarrow\mathbf{x'}) = \Pi(\mathbf{x'^*}\rightarrow\mathbf{x^*})\,.
    \end{equation}
    Observing that the canonical distribution $\mathcal{P}_{NVT}$ in \cref{boltzmann} does not depend on the sign of momenta, it is possible to introduce a condition called \emph{generalized detailed balance}, 
    \begin{equation}\label{generalized_db}
        \mathcal{P}_{NVT}(\mathbf{x})\,\Pi(\mathbf{x}\rightarrow\mathbf{x'}) = \mathcal{P}_{NVT}(\mathbf{x'^*})\,\Pi(\mathbf{x'^*}\rightarrow\mathbf{x^*})\,,
    \end{equation}
    which justifies the use of the Metropolis-Hastings rule. In a schematic way, a single iteration of the hybrid MC method is reported in \cref{hybrid_MC_alg}.
    
    It is worth Note that the parameter $\Delta t$ employed in the proposal move acquires a whole new meaning in this context: in the propagation of a system in the microcanonical ensemble $\Delta t$ is the physical time step in the dynamics and it is related to the error made by the algorithm; in hybrid MC instead $\Delta t$ is a parameter which affects the average acceptance and the efficiency of the sampling, but not its correctness. In other words, hybrid MC (as well as standard MC) does not carry any information about the time dependence of the fluctuations, because the discrete "MC time" has no relation with the physical time in the dynamics of the system.
    
    \begin{algorithm}[h!]
    \SetAlgoLined
        Compute initial energy: $E=\mathcal{H}(\mathbf{x})$\;
        Propose new state with velocity Verlet (or leapfrog integrator): $\mathbf{x}\rightarrow\mathbf{x'}$\;
        Compute energy of proposed state: $E'=\mathcal{H}(\mathbf{x'})$\;
        Compute acceptance: $\alpha=\min \left[1,\,\exp{\left(-\frac{1}{k_B T}(E'-E)\right)}\right] $ \;
        Extract uniform random number: $\mathcal{R}\sim U(0,1)$\;
        \eIf{$\alpha> \mathcal{R}$}{
            Accept the move: $\mathbf{x}\gets\mathbf{x'}$ 
        }{ 
            Refuse the move: $\mathbf{x}\gets\mathbf{x}$ 
        }
    \caption{Hybrid Monte Carlo}
    \label{hybrid_MC_alg}
    \end{algorithm}

    \section{Isothermal-isobaric ensemble (NPT)}\label{npt_theory}
    A system in the isothermal-isobaric ensemble is defined by a fixing the number of particles $N$, the external pressure $P_0$ and the external temperature $T$. In the isotropic formulation of the ensemble the volume $V$ is allowed to fluctuate according to the probability distribution
    \begin{equation}\label{npt_isotropic}
        \mathcal{P}_{NP_0T}\big(\{\mathbf{q}_i,\mathbf{p}_i\},V\big) = \frac{1}{\Delta}\exp\left[-\frac{1}{k_B T}\Big(K\big(\{\mathbf{p}_i\}\big) + U\big(\{\mathbf{q}_i\}\big) + P_0 V\Big)\right]\,,
    \end{equation}
    where $\Delta=\Delta(N,P_0,T)$ is the isothermal-isobaric partition function. In the limit case of an ideal gas it is easy to show that the marginal distribution of the volume is reduced to a Gamma distribution, $\mathcal{P}_{NP_0T}^{\text{id}}(V) \propto V^N \exp \left(-\frac{1}{k_B T}P_0 V \right)$.
    
    As an instantaneous temperature was defined in the $NVT$ ensemble, it is also possible to define an instantaneous \emph{internal pressure} in the $NP_0 T$ ensemble, which is calculated via the Clausius virial theorem as
    \begin{equation}\label{internal_pressure}
        P_{\text{int}} = \frac{2 K}{3 V} - \frac{\partial U}{\partial V}\,.
    \end{equation}
    At a given temperature $T$ and number of particles $N$, the variation of the average volume with respect to the pressure, which is an intrinsic property of the system, is quantified by the \emph{isothermal compressibility}:
    \begin{equation}
        \beta_T = -\frac{1}{\langle V \rangle}\frac{\partial \langle V\rangle}{\partial P_0}\,,
    \end{equation}
    This thermodynamic quantity is tightly related to volume fluctuations, since it can be computed as
    \begin{equation}\label{beta_T_formula}
        \beta_T = \frac{k_B T}{\langle V\rangle}\sigma_V^2\,,
    \end{equation}
    where $\sigma_V^2 = \langle \left(V-\langle V\rangle  \right)^2\rangle$ is the variance of the volume distribution at pressure $P_0$. The same information carried by $\beta_T$ is sometimes expressed in terms its reciprocal, called bulk modulus:
    \begin{equation}
        k_T = \frac{1}{\beta_T}\,.
    \end{equation}
    
\begin{figure}[h!]
        \centering
        \includegraphics[width = 0.25\textwidth]{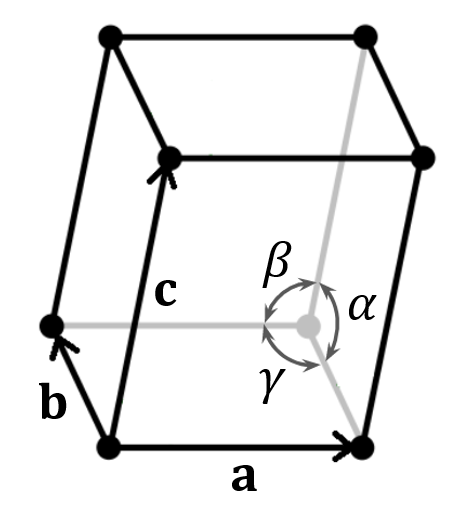}
        \caption{Construction of the box using the three cell vectors $\mathbf{a},\,\mathbf{b},\,\text{and}\,\mathbf{c}$.}
        \label{fig:box}
    \end{figure}

    While \cref{npt_isotropic} assumes that the volume fluctuates in a isotropic way, namely that the box where the system is confined changes its size but not its shape, there are several situations where anisotropic fluctuations can be relevant. For instance, in solid-state physics a fully flexible description of the system allows to predict crystal structures \cite{crystal_structure} and to study conformational transitions between them \cite{parr-rahman}. Moreover, semi-isotropic volume fluctuations can play a central role to study liquid-liquid interfaces \cite{constant_surface_tension} or to simulate membranes. In the fully flexible isothermal-isobaric ensemble the system is typically contained within a general parallelepiped, which represents the most general box shape and appears appropriate to describe, for example, solids whose unit cells are generally triclinic \cite{tuckerman}. Such a box can be described in terms of three vectors $\mathbf{a},\,\mathbf{b},\,\text{and}\,\mathbf{c}$ that lie along the three edges starting from a certain vertex (see \cref{fig:box}). Their nine components can be collected in the 3$\times$3 \emph{box matrix} (or \emph{cell matrix}) $\mathbf{h}$, which contains the three box vectors along its columns according to the convention adopted here:
    \begin{equation}
        \mathbf{h}=
        \begin{pmatrix}
        a_x & b_x & c_x\\
        a_y & b_y & c_y\\
        a_z & b_z & c_z
        \end{pmatrix}
    \end{equation}
    Using Greek and Latin letters to label the cartesian component and the number of the cell vector espectively, the elements of the box matrix will also be written as $h_{\alpha i}$ ($\alpha=x,y,z,\,i=1,2,3$). 
    
    \noindent The volume of the box is given by the triple product of the three cell vectors $\mathbf{a},\,\mathbf{b},\,\text{and}\,\mathbf{c}$, assuming that they form in that order a right-handed triad:
    \begin{equation}\label{volume_box}
        V = \mathbf{a}\cdot\mathbf{b}\times\mathbf{c} = \det \mathbf{h}\,.
    \end{equation}
    From \cref{fig:box} it is clear that the shape of the box can actually be described with six scalar numbers only, i.e. the moduli $|\mathbf{a}|,\,|\mathbf{b}|,\,|\mathbf{c}|$ of the three cell vectors and the angles $\alpha,\,\beta,\,\gamma$ between them. The three additional degrees of freedom in the box matrix $\mathbf{h}$ account for overall rotations of the cell, which leave both the moduli and the angles untouched. Since rotations of the entire system are not of interest in MD simulations, the three redundant degrees of freedom can be eliminated by employing different methods, which are discussed in \cref{eliminate_rotations}.
    
    In the anisotropic formulation of the isothermal-isobaric ensemble, the reference pressure $P_0$, also called external \emph{hydrostatic pressure}, is generalized to a symmetric 3$\times$3 tensor $\mathbf{S}$, where the off-diagonal components account for possible shear stresses applied to the system while the diagonal ones are related to the isotropic counterpart via the relation
    \begin{equation}\label{hydro_pressure}
        P_0 = \frac{\text{Tr}\left(\mathbf{S}\right)}{3}\,.
    \end{equation}
    $\mathbf{S}$ is commonly referred to as \emph{external pressure tensor} or \emph{stress tensor}, and the anisotropic iosthermal-isobaric ensemble is also denoted as $N\mathbf{S}T$ ensemble. Given \cref{hydro_pressure}, $\mathbf{S}$ can be split into the sum of a hydrostatic contribution and a trace-less \emph{deviatoric stress tensor} $\mathbf{S}_{\text{dev}}$:
    \begin{equation}
         \mathbf{S} = P_0\mathbf{I} + \mathbf{S}_{\text{dev}} \,,
    \end{equation}
    where $\mathbf{I}$ is the 3$\times$3 identity matrix.
    Similarly, the scalar internal pressure $P_{\text{int}}$ is generalized to a pressure tensor $\mathbf{P}_{\text{int}}$ that is calculated with a tensorial formulation of the virial theorem,
    \begin{equation}\label{internal_pressure_tensor}
        \mathbf{P}_{\text{int}} = \frac{2}{V}\left(\mathbf{K}-\Xi\right)\,,
    \end{equation}
    where $\mathbf{K}$ and $\Xi$ are called \emph{kinetic energy tensor} and \emph{virial tensor} respectively:
    \begin{subequations}
    \begin{align}
        K_{\alpha\beta} &= \sum_{i=1}^N \frac{p_i^{\alpha}p_i^{\beta}}{2m_i}\,, \\
        \Xi_{\alpha\beta} &= -\frac{1}{2}\sum_{i=1}^N F_i^{\alpha}\,q_i^{\beta}\,.
    \end{align}
    \end{subequations}
    \noindent The relations between these tensors and the corresponding scalar quantities are straightforward:
    \begin{subequations}
    \begin{align}
        K &= \frac{\text{Tr}\left(\mathbf{K}\right)}{3}\,, \\
        P_{\text{int}} &= \frac{\text{Tr}\left(\mathbf{P}_{\text{int}}\right)}{3}\,.
    \end{align}
    \end{subequations}
    Given these definitions, in presence of an external hydrostatic pressure - i.e. $\mathbf{S} = P_0 \mathbf{I}$ - the $N\mathbf{S}T$ ensemble is described by the following distribution:
    \begin{equation}\label{NPT_anisotropic}
        \mathcal{P}_{N\mathbf{S}T}\big(\{\mathbf{q}_i,\mathbf{p}_i\},\mathbf{h}\big) = \frac{\left(\det \mathbf{h}\right)^{-2}}{\Delta} \exp\left[-\frac{1}{k_B T}\Big(K + U + P_0 \det \mathbf{h}\Big)\right]\,.
    \end{equation}
    \noindent To derive this expression and explain the origin of the factor $\left(\det\mathbf{h}\right)^{-2}$ it is possible to start from the expression of the isothermal-isobaric partition function $\Delta(N,P_0,T)$ in terms of the isotropic $NP_0 T$ distribution:
    \begin{equation}
        \Delta(N,P_0,T) = \frac{1}{V_0}\int \text{d}V \exp \left(-\frac{P_0 V}{k_B T}\right)\,\mathcal{Z}(N,V,T)\,,
    \end{equation}
    where $V_0$ is a reference volume that is necessary to make the partition function dimensionless.
    Introducing the matrix $\mathbf{h}_u$ such that $\mathbf{h}=V^{1/3}\mathbf{h}_u$, $\mathcal{Z}(N,V,T)$ can be rewritten in terms of a constrained canonical ensemble - with partition function $\mathcal{Z}(N,V,T,\mathbf{h}_u)$ - where not only the volume of the system is fixed but also its shape:
    \begin{align}
        \mathcal{Z}(N,V,T) &= \frac{1}{V_0}\int \text{d}\mathbf{h}_u\, {Z}(N,V,T,\mathbf{h}_u)\,\delta(\det\mathbf{h}_u -1) \\
        &= \frac{1}{V_0}\int \text{d}\mathbf{h}_u \int\left(\prod_i\text{d}\mathbf{q}_i\text{d}\mathbf{p}_i\right)\,\exp\left(-\frac{1}{k_B T}\mathcal{H}\big(\{\mathbf{q}_i,\mathbf{p}_i\}\big) \right) \delta(\det\mathbf{h}_u -1) \nonumber
        \label{from_iso_to_flex}
    \end{align}
    Since $\text{d}\mathbf{h}_u = \text{d}\mathbf{h}/V^3$ and $\delta(\det\mathbf{h}_u -1) = V\delta(V-\det\mathbf{h})$, substituting in \cref{from_iso_to_flex} and integrating the $\delta$-function in $\text{d}V$ it is possible to express the isothermal-isobaric partition function as
    \begin{align}
        \Delta(N,P_0,T) = \frac{1}{V_0}\int& \text{d}\mathbf{h}\int\left(\prod_i\text{d}\mathbf{q}_i\text{d}\mathbf{p}_i\right)\, \left(\det\mathbf{h}\right)^{-2} \nonumber \\ 
        &\times\exp\left[-\frac{1}{k_B T}\Big(K\big(\{\mathbf{p}_i\}\big) + U\big(\{\mathbf{q}_i\}\big) + P_0 \det{\mathbf{h}}\Big)\right]\,,
    \end{align}
    which is consistent with the expression of the $N\mathbf{S}T$ distribution given in \cref{NPT_anisotropic}. In an arbitrary number $d$ of spatial dimensions, the relation between the original box and the unitary one is $\mathbf{h}=V^{1/d}\mathbf{h}_u$ and the factor $\left(\det\mathbf{h}\right)^{-2}$ is generalized to $\left(\det\mathbf{h}\right)^{1-d}$ \cite{tuckerman}.
    
    In presence of a generic external stress $\mathbf{S}$ - namely if $\mathbf{S}_{\text{dev}}\neq \mathbf{0}$ - it is necessary to introduce an additional \emph{strain energy}
    \begin{equation}\label{strain_energy}
        E_{\text{S}} = \frac{1}{2}\text{Tr}\left(\bm{\Sigma}\mathbf{G}\right)\,,
    \end{equation}
    where $\mathbf{G} = \mathbf{h}^T\mathbf{h}$ is called \emph{metric tensor} and the matrix $\bm{\Sigma}$ is defined with respect to a reference system with cell $\mathbf{h}_0$ and volume $V_0=\det\mathbf{h}_0$, which is typically identified with the initial system in MD simulations:
    \begin{equation}\label{sigma_matrix}
        \bm{\Sigma} = V_0\, \mathbf{h}_0^{-1}\,\mathbf{S}_{\text{dev}}\,\big(\mathbf{h}_0^T\big)^{-1} = V_0\, \mathbf{h}_0^{-1}\left(\mathbf{S}-P_0\mathbf{I} \right)\big(\mathbf{h}_0^T\big)^{-1}\,.
    \end{equation}
    This energy contribution can be derived in the framework of elasticity theory and was fist proposed by Parrinello and Rahman \cite{parr-rahman}. Hence the $N\mathbf{S}T$ distribution in presence of a generic external stress becomes:
    \begin{equation}\label{NPT_anisotropic_shear}
        \mathcal{P}_{N\mathbf{S}T} \propto V^{-2}\exp\left[-\frac{1}{k_B T}\Big(K + U + P_0 V + \frac{1}{2}\text{Tr}\left(\bm{\Sigma}\mathbf{G}\right)\Big)\right]\,.
    \end{equation}
    An intermediate case between the fully flexible and isotropic ensembles is given by the \emph{constant normal pressure and surface-tension} ensemble $NP_0^\perp\gamma_0 T$, where volume fluctuations result from only two decoupled degrees of freedom, i.e. the area $A$ of the simulation box in the $xy$-plane and its height $L$. Supposing that the system is contained in an orthorhombic box, i.e. that the box matrix $\mathbf{h}$ is diagonal, $A = h_{x1} h_{y2}$ and $L = h_{z3}$. In this case the external stress can be written as the diagonal tensor
    \begin{equation}
        \mathbf{S}=
        \begin{pmatrix}
        P_{0,xx} & 0 & 0\\
        0 & P_{0,yy} & 0\\
        0 & 0 & P_{0,zz}
        \end{pmatrix}\,,
    \end{equation}
    and the fixed control parameters of the ensemble, i.e. the normal pressure $P_0^\perp$ to the surface $A$ and the surface tension $\gamma$ (multiplied by the number of surfaces), are related to $\mathbf{S}$ by the following relations \cite{constant_surface_tension}:
    \begin{subequations}
    \begin{align}
        P_0^\perp &= P_{0,zz}\,, \\
        P_0^{\parallel} &= \frac{P_{0,xx} + P_{0,yy}}{2}\,,\\
        \gamma_0 &= L\left(P_0^\perp - P_0^{\parallel}\right)\,. \label{gamma_def}
    \end{align}
    \end{subequations}
    The additional degree of freedom brings a new energy contribution with respect to the isotropic case, resulting in the $NP_0^\perp\gamma_0 T$ probability distribution
    \begin{equation}\label{npt_constant_surface_tension}
        \mathcal{P}_{NP_0^\perp\gamma_0 T}\big(\{\mathbf{q}_i,\mathbf{p}_i\},A,L\big) \propto\exp\left[-\frac{1}{k_B T}\Big(K + U + P_0^\perp A L -\gamma_0 A\Big)\right]\,.
    \end{equation}

    \section{Barostats}
    In the MD language, the algorithms which allow to simulate volume fluctuations at constant pressure are called barostats. Common features of all these algorithms are that they must be coupled with a thermostat in order to reproduce the correct isothermal-isobaric ensemble, and they all rescale the particles' positions together with the volume; instead, rescaling of momenta is not in general a necessary feature. As in the case of thermostats, also barostats can be distinguished in deterministic and stochastic ones.
    
    \subsection{Monte Carlo barostat}\label{MC_barostat}
    The first strategy to simulate a system at constant pressure and with isotropic volume fluctuations is to employ a MCMC approach (see \cref{MC_thermostat}), namely to propose a move on the volume that is accepted or refused according to the Metropolis-Hastings rule. This kind of algorithm is called MC barostat \cite{MC_barostat1,MC_barostat2}. The proposal move is typically implemented as
    \begin{equation}
        V' = V + (2\,\mathcal{R}-1)\delta V\,,
    \end{equation}
    where $\mathcal{R}\sim U(0,1)$ and $\delta V$ is a parameter that quantifies the maximum volume variation achievable in a single MC step. Before computing the acceptance, all the particles' positions are rescaled as
    \begin{equation}\label{pos_rescaling_MC}
        \mathbf{q}_i' = \left(\frac{V'}{V}\right)^{1/3}\mathbf{q}_i\,,
    \end{equation}
    with the assumption that the system is confined in a cubic box. This step is necessary to avoid large modifications of the distance - and thus of the interaction - between each atom and the copies of the others (see Appendix \ref{PBCs}), which would drastically reduce the acceptance. Rescaling the physical positions as in \cref{pos_rescaling_MC} is equivalent to say that the volume is propagated at constant \emph{rescaled positions} 
    \begin{equation}\label{rescaled_pos_iso}
        \mathbf{s}_i \equiv \frac{\mathbf{q}_i}{V^{1/3}}\,.
    \end{equation}
    In terms of these variables, the $NP_0T$ distribution in \cref{npt_isotropic} becomes
    \begin{equation}
        \mathcal{P}_{NP_0T}\big(\{\mathbf{s}_i,\mathbf{p}_i\},V\big) = \frac{V^N}{\Delta}\exp\left[-\frac{1}{k_B T}\Big(K\big(\{\mathbf{p}_i\}\big) + U\big(\{\mathbf{s}_i\},V\big) + P_0 V\Big)\right]\,,
    \end{equation}
    therefore - when smaller than 1 - the acceptance $\alpha = \mathcal{P'}_{NP_0T}/ \mathcal{P_{NP_0T}}$ satisfies:
    \begin{equation}
        -k_B T \log\alpha = \Delta U + P_0\Delta V -N k_B T \Delta(\log V)\,, 
    \end{equation}
    where $\Delta U = U\big(\{\mathbf{s}_i\},V'\big)-U\big(\{\mathbf{s}_i\},V\big)$. Taking the limit $\Delta V = V'-V\rightarrow0$ one finds that $\alpha$ can be written as
    \begin{equation}\label{acceptance_MC_barostat}
        -k_B T \log\alpha = \Delta V \left(P_0-P_{\text{int}}^{\langle K \rangle}\right)\,,
    \end{equation}
    where $P_{\text{int}}^{\langle K \rangle}$ is the internal pressure defined in \cref{internal_pressure} but computed with the average kinetic energy instead of the instantaneous one: 
    \begin{equation}\label{internal_pressure_avgkin}
        P_{\text{int}}^{\langle K \rangle} = \frac{N k_B T}{V}-\frac{\partial U}{\partial V}\,.
    \end{equation}
    \noindent An alternative formulation of the MC barostat employs the additional rescaling
    \begin{equation}
        \mathbf{p}_i' = \left(\frac{V}{V'}\right)^{1/3}\mathbf{p}_i\,,
    \end{equation}
    namely the volume is propagated at constant \emph{rescaled momenta} 
    \begin{equation}\label{rescaled_mom_iso}
        \bm{\pi}_i = V^{1/3} \mathbf{p}_i\,.
    \end{equation}
    Since the rescaling factor in \cref{rescaled_mom_iso} is the inverse of the one in \cref{rescaled_pos_iso}, the $NP_0 T$ distribution as a function of $\{\mathbf{s}_i,\bm{\pi}_i\}$ gains no additional prefactors, i.e. the Jacobians of the two tranformations cancel each other. It is possible to show that in this case the acceptance can be computed as in \cref{acceptance_MC_barostat} but with the actual expression of the internal pressure, i.e. substituting the average kinetic energy with the instantaneous one.
    
    Since it does not require the calculation of the virial, the MC barostat is one of the easiest algorithms to control volume fluctuations at constant pressure; however there are situations where it appears less efficient than virial-based barostats \cite{MC_barostat_limitations}. Moreover, it does not allow to interpret the volume dynamics in a physical way, similarly to what was discussed in \cref{MC_thermostat}.
    
    \nopagebreak
    \subsection{Berendsen barostat}\label{berendsen_barostat}
    The Berendsen barostat \cite{berendsen} is a deterministic pressure coupling algorithm based on the following first-order differential equation:
    \begin{equation}\label{berendsen_barostat_iso}
        \dot{V} = -\frac{\beta_T V}{\tau_p} \left(P_0 - P_{\text{int}}\right)\,, 
    \end{equation}
    where $\tau_p$ and $\beta_T$, namely the relaxation time of the volume and the isothermal compressibility, are the input parameters of the barostat.
    Therefore, in a single step of the algorithm (with time step $\Delta t$) the box edges and the positions' components are rescaled by a factor
    \begin{equation}
        \mu = \left[1 - \frac{\beta_T \Delta t}{\tau_p}\left(P_0 -P_{\text{int}} \right)\right]^{1/3} \simeq 1 - \frac{\beta_T \Delta t}{3\tau_p}\left(P_0 -P_{\text{int}} \right)\,.
    \end{equation}
    
    Although the Berendsen barostat allows to sample the correct average volume, it does not reproduce the correct volume fluctuations; in other words, the stationary distribution that it samples is not the isothermal-isobaric one. Anyway, since it does not produce instabilities when the system is initialized far from equilibrium, the Berendsen algorithm is largely employed in the first part of constant pressure simulations, i.e. in the equilibration phase.
    
    Since $\tau_p$ only appears in $\beta_T/\tau_p$, this ratio is the only effective input parameter of the barostat, and an error in the isothermal compressibility - which may not be accurately known - only influences the accuracy of the time constant $\tau_p$ but has no consequence for the dynamics.
    As a consequence, it is sufficient to use as input a rough estimate of $\beta_T$; then, after computing the correct value of the isothermal compressibility in the production run - by means of a barostat that generates the correct isothermal-isobaric ensemble - the actual value of the volume relaxation time can be calculated \emph{a posteriori} with a simple rescaling, as shown \cref{LJ_crystal}. When simulating solvated molecules, the common practice is to use as input the isothermal compressibility of the solvent.
    
    In the flexible formulation of the algorithm - but still assuming an isotropic external stress - \cref{berendsen_barostat_iso} is generalized to
    \begin{equation}\label{berendsen_aniso}
        \dot{\mathbf{h}} = -\frac{\beta_T}{3\tau_p} \left(P_0 \mathbf{I} - \mathbf{P}_{\text{int}}\right)\,\mathbf{h}\,,
    \end{equation}
    where $\mathbf{I}$ is the 3$\times$3 identity matrix. Equivalently, at each step the box matrix and the positions are rescaled by the matrix
    \begin{equation}
        \bm{\mu} = \mathbf{I} - \frac{\beta_T \Delta t}{3\tau_p}\left(P_0\mathbf{I} -\mathbf{P}_{\text{int}} \right)\,,
    \end{equation}
    namely by applying the matrix products $\mathbf{h}' = \bm{\mu} \mathbf{h}$ and $\mathbf{q}_i' = \bm{\mu}\mathbf{q}_i$. In the anisotropic implementation, the isothermal compressibility can be in principle generalized to a 3$\times$3 tensor $\bm{\beta}_T$, although this only affects the time constants with which the various components of $\mathbf{P}_{\text{int}}$ relax to the reference value $P_0$ \cite{berendsen}. For instance, in the GROMACS \cite{gromacs} implementation of the Berendsen barostat, the anisotropic box rescaling is performed by means of a matrix $\Tilde{\mu}$ with elements
    \begin{equation}
        \Tilde{\mu}_{\alpha\beta} = \delta_{\alpha\beta} - \frac{\Delta t}{3\tau_p}\beta_{T,\alpha\beta}\left(P_0 \delta_{\alpha\beta} - P_{\text{int},\alpha\beta} \right)\,.
    \end{equation}

    \subsection{Andersen barostat}\label{andersen_barostat}
    The Andersen barostat \cite{andersen} is a deterministic algorithm based on a second-order equation for the dynamics of the volume. Employing both rescaled coordinates as in \cref{rescaled_pos_iso} and rescaled momenta as in \cref{rescaled_mom_iso}, the 
    idea of the barostat is to treat the volume as a dynamical variable by adding to the system a new degree of freedom, which appears in the Hamiltonian as a new position-like variable $V$ and its conjugated momentum $p_V$:
    \begin{equation}\label{hamiltonian_andersen_barostat}
        \mathcal{H}\big(\{\mathbf{s}_i,\bm{\pi}_i\},V,p_V\big) \equiv \sum_i \frac{V^{-2/3}|\bm{\pi}_i|^2}{2 m_i} + U\big(\{V^{1/3}\mathbf{s}_i\}\big) + P_0 V +\frac{p_V^2}{2W}\,.
    \end{equation}
    The first two terms of the augmented Hamiltonian in \cref{hamiltonian_andersen_barostat} are the kinetic and potential energy as functions of the rescaled variables, while the third and fourth terms are respectively the potential and kinetic energy terms associated to the volume. The parameter $W$ is called \emph{barostat mass} and quantifies the inertia associated to $V$. The equations of motion derived from $\mathcal{H}$ are the following:
    \begin{subequations}\label[pluralequation]{andersen_barostat_eqs}
    \begin{align}
        \dot{\mathbf{s}}_i &= V^{-2/3}\frac{\bm{\pi}_i}{m_i}\,,\label{1st_eq_andersen}  \\
        \dot{\bm{\pi}}_i &= V^{1/3}\mathbf{F}_i\,,\label{2nd_eq_andersen} \\
        \dot{V} &= \frac{p_V}{W}\,\label{volume_eq_andersen} \\
        \dot{p}_V &= P_{\text{int}} - P_0\,, \label{p_V_eq_andersen}
    \end{align}
    \end{subequations}
    where $\mathbf{F}_i = -\frac{\partial U}{\partial \mathbf{q}_i}$ and $P_{\text{int}}$ is calculated as in \cref{internal_pressure}. Equivalently, applying the inverse transformations from the rescaled variables to the physical positions and momenta, \cref{1st_eq_andersen,2nd_eq_andersen} can also be written as: 
    \begin{subequations}\label{}
    \begin{align}
        \dot{\mathbf{q}}_i &= \frac{\mathbf{p}_i}{m_i} + \frac{1}{3}\frac{\dot{V}}{V}\mathbf{q}_i\,, \label{q_eq_andersen}\\
        \dot{\mathbf{p}}_i &= \mathbf{F}_i - \frac{1}{3}\frac{\dot{V}}{V}\mathbf{p}_i \label{p_eq_andersen}\,.
    \end{align}
    \end{subequations}
    Integrating these equations is equivalent to sample the distribution of an extended \emph{isoenthalpic-isobaric} ensemble, namely
    \begin{equation}
        \mathcal{P}_{NVH'}\Big(\{\mathbf{q}_i,\bm{p}_i\},V,p_V\Big) \propto \delta\Big(K + U + P_0 V + \frac{p_V^2}{2W} - E\Big)\,,
    \end{equation}
    where $H = K + U + P_0 V$ is called \emph{enthalpy} and the barostat kinetic energy $p_V^2/(2W)$ is an additional term with respect to the standard ensemble, leading to the conserved quantity
    \begin{equation}\label{conserved_enthalpy}
        H' = H + \frac{p_V^2}{2W}\,.
    \end{equation}
    However, according to the equipartition theorem the average contribution of this additional kinetic term is $\langle p_V^2/(2W)\rangle = k_B T/2$; as a consequence its effect becomes irrelevant when the number of degrees of freedom is large.
    Moreover, similarly to what discussed in \cref{hybrid_MC}, also the distribution
    \begin{equation}\label{distrib_andersen}
        \mathcal{P}_{N P_0 T}\big(\{\mathbf{q}_i,\bm{p}_i\},V,p_V\big) \propto \exp\left[-\frac{1}{k_BT}\left(K + U + P_0 V + \frac{p_V^2}{2W} \right)\right]\,
    \end{equation}
    is stationary with respect to \cref{andersen_barostat_eqs}. $\mathcal{P}_{NP_0 T}$ is exactly the target $NP_0 T$ distribution because the additional term depending on $p_V$ factorizes and can be integrated out, as $p_V$ is not a variable of the isothermal-isobaric ensemble. In order to sample $\mathcal{P}_{NP_0 T}$ in a ergodic way it is possible to exploit that $\mathcal{P}_{N P_0 T}$ is also stationary with respect to any thermostat generating the correct canonical ensemble, since also the Boltzmann distribution can be recovered from \cref{distrib_andersen} - in this case by integrating out both $V$ and $p_V$, which are not dynamical variables of the $NVT$ ensemble. Thus the ergodic sampling of the $NP_0 T$ distribution can be achieved by coupling the Andersen's equations with a thermostat, acting on the particles' degrees of freedom. Furthermore, the volume itself can be coupled with an external bath by means of a Langevin thermostat (see \cref{langevin_thermo}), in order to damp the volume oscillations and accelerate the equilibration of the system. With this additional coupling, \cref{volume_eq_andersen,p_V_eq_andersen} become 
    \begin{subequations}
    \begin{align}
        \dot{V} &= \frac{p_V}{W}\,, \\
        \dot{p}_V &= (P_{\text{int}} - P_0) -\gamma p_V + \sqrt{2\gamma W k_B T}\,\mathcal{R}(t)\,,
    \end{align}
    \end{subequations}
    where $\mathcal{R}(t)$ is a white noise (see Appendix \cref{appendix_stocdiffeq}). This method is typically referred to as \emph{Langevin piston} \cite{langevin_piston}.
    
    As all second-order algorithms, when the Andersen barostat is employed far from equilibrium the volume shows damped oscillations, which can decay slow if the barostat mass $W$ is not chosen properly. Nevertheless, setting the optimal value of $W$ is a system-dependent problem and there is not a general recipe to accomplish it. For this reason, the typical pipeline for a simulation in the isothermal-isobaric ensemble employs the Berendsen barostat for the equilibration phase, and a second-order barostat such as the Andersen one for the production run.

    \subsection{Parrinello-Rahman barostat}\label{pr_barostat}
    The Parrinello-Rahman method \cite{parr-rahman0,parr-rahman} extends the Andersen barostat to allow for changes in volume and shape of the system box, i.e. it generalizes the Andersen equations \cref{andersen_barostat_eqs} to the fully flexible $N\mathbf{S}T$ ensemble. This generalization includes a new definition of rescaled coordinates $\mathbf{s}_i$, namely
    \begin{equation}
        \mathbf{s}_i = \mathbf{h}^{-1}\mathbf{q}_i\,.
    \end{equation}
    Note that the matrix $\mathbf{h}$ is reasonably assumed to be invertible, since the three box vectors defining its columns are linearly independent. The augmented Hamiltonian is constructed by introducing nine new degrees of freedom, corresponding to the nine components of the box matrix $\mathbf{h}$:
    \begin{equation}
        \mathcal{H} = K + U + \frac{W}{2}\text{Tr}\left(\dot{\mathbf{h}}^T\dot{\mathbf{h}}\right) + P_0 \det \mathbf{h}\,.
    \end{equation}
    Considering as in \cite{parr-rahman0,parr-rahman} a pair-wise potential of the form 
    \begin{equation}
        U\left(\{ \mathbf{q}_i \}\right) = \frac{1}{2} \sum_{i,j} \Phi (q_{ij})\,,
    \end{equation}
    where $q_{ij} = |\mathbf{q}_i-\mathbf{q}_j|$, it is possible to derive the following equations of motion:
    \begin{subequations}
    \begin{align}
        \label{first_eq_PR}
        \ddot{\mathbf{s}}_i &= -\sum_{j\neq i}\frac{1}{m_i q_{ij}}\frac{\text{d}\Phi}{\text{d}q_{ij}} \left(\bm{s}_i-\bm{s}_j \right) - \mathbf{G}^{-1}\dot{\mathbf{G}}\,\dot{\mathbf{s}}_i\,, \\
        \ddot{\mathbf{h}} &= V W^{-1}\,\big(\mathbf{P} - P_0\mathbf{I}\big)\big( \mathbf{h}^{T}\big)^{-1}\,. \label{second_eq_PR}
    \end{align}
    \end{subequations}
    where $\mathbf{G} \equiv \mathbf{h}^T\mathbf{h}$ is called \emph{metric tensor}.
    \indent In case of a general anisotropic external stress, i.e. when $\mathbf{S}\neq P_0 \mathbf{I}$, the Hamiltonian is further augmented with the strain energy term defined in \cref{strain_energy}:
    \begin{equation}
        \mathcal{H}_S = \mathcal{H} + \frac{1}{2}\text{Tr}\left(\bm{\Sigma}\mathbf{G}\right)\,.
    \end{equation}
    As a consequence, \cref{first_eq_PR} remains unchanged while the dynamics of $\mathbf{h}$ is now described by
    \begin{equation}
        \ddot{\mathbf{h}} = V W^{-1} \big(\mathbf{P}_{\text{int}} - P_0\mathbf{I}\big)\big( \mathbf{h}^{T}\big)^{-1} - W^{-1}\mathbf{h}\bm{\Sigma}\,.
    \end{equation}
    \indent Even if in the original formulation of the Parrinello-Rahman barostat $W^{-1}$ is a scalar, in some implementations - such as in GROMACS \cite{gromacs} - it is treated as a 3$\times$3 symmetric tensor related to a tensorial expression of the isothermal compressibility. As already commented in \cref{berendsen_barostat}, this only affects the relaxation times of the various components of $\mathbf{P}_{\text{int}}$ towards the reference values of the external stress $\mathbf{S}$.

    \subsection{Martyna-Tobias-Klein barostat}\label{mtk_barostat}
    The pressure coupling method developed by Martyna, Tobias and Klein (MTK) \cite{mtk} develops a set of equations similar but not equivalent to the Andersen's ones, namely the Hoover's equations \cite{hoover,hoover2}
    \begin{subequations}
    \begin{align}
        \dot{\mathbf{q}}_i &= \frac{\mathbf{p}_i}{m_i} + \frac{p_\varepsilon}{W}\mathbf{q}_i\,, \label{pos_eq_mtk}\\
        \dot{\mathbf{p}}_i &= \mathbf{F}_i - \frac{p_\varepsilon}{W}\mathbf{p}_i\,, \label{mom_eq_mtk}\\
        \dot{V} &= \frac{3Vp_\varepsilon}{W}\,, \label{vol_eq_mtk} \\
        \dot{p}_\varepsilon &= 3 V (P_{\text{int}}-P_0) \label{peps_eq_mtk}\,,
    \end{align}
    \end{subequations}
    where the variables $\varepsilon$ and $p_{\varepsilon}$ are defined as:
    \begin{subequations}
    \begin{align}
        \varepsilon &= \frac{1}{3}\log\frac{V}{V_0}\,, \\
        p_{\varepsilon} &= W \dot{\varepsilon} = \frac{W\dot{V}}{3V}\,.
    \end{align}
    \end{subequations}
    It is easy to show that the first three equations above are equivalent to Andersen's \cref{q_eq_andersen,p_eq_andersen,volume_eq_andersen}, while the one for $p_\varepsilon$ is different from \cref{p_V_eq_andersen}. As a consequence, when coupled to a thermostat Hoover's equations actually generate a slightly different ensemble than the isothermal-isobaric one:
    \begin{equation}
        \mathcal{P}_{\text{Hoover}} \propto \frac{1}{V}\mathcal{P}_{NP_0 T}
    \end{equation}
    As an additional problem, in the extended phase space Hoover's equations do not satisfy anymore the incompressibility condition defined in \cref{incompressibility}, namely
    \begin{equation}
        \sum_{i,\alpha}\left(\frac{\partial \dot{q}_i^{\alpha}}{\partial q_i^{\alpha}} + \frac{\partial \dot{p}_i^{\alpha}}{\partial p_i^{\alpha}} \right) + \frac{\partial\dot{V}}{\partial V}\neq 0\,.
    \end{equation}
    In order to fix this problem without changing the conserved enthalpy defined in \cref{conserved_enthalpy} it is possible to modify \cref{mom_eq_mtk,peps_eq_mtk} with two corrections whose energy contributions cancel each other:
    \begin{subequations}
    \begin{align}
        \dot{\mathbf{p}}_i &= \tilde{\mathbf{F}}_i - \left(1+\frac{3}{N_f} \right)\frac{p_\varepsilon}{W}\mathbf{p}_i\,, \label{mom_eq_corr_mtk}\\
        \dot{p}_\varepsilon &= 3 V (P_{\text{int}}-P_0) + \frac{3}{N_f}\sum_i \frac{\mathbf{p}_i^2}{m_i}\,. \label{peps_eq_corr_mtk}
    \end{align}
    \end{subequations}
    In this case, $\tilde{\mathbf{F}}_i$ is the total force acting on atom $i$, including the contribution of constraints. The full set of MTK equations is obtained by coupling these modified equations with two Nosé-Hoover chains, one for the particles and one for the volume, to keep into account that positions and momenta thermalize at a considerably faster time than the volume \cite{tuckerman}. With this additional coupling, it is possible to show that the MTK equations just defined generate the correct (isotropic) isothermal-isobaric distribution defined in \cref{npt_isotropic}.
    
    The MTK method can be generalized to anisotropic cell fluctuations, promoting $V$ to the nine box variables in $\mathbf{h}$ and the conjugated momentum $p_\varepsilon$ to the 3$\times$3 matrix of box momenta $\mathbf{p}_g$, such that $\mathbf{p}_g / W_g = \dot{\mathbf{h}}\mathbf{h}^{-1}$. Then \cref{pos_eq_mtk,mom_eq_corr_mtk,vol_eq_mtk,peps_eq_corr_mtk} become respectively:
    \begin{subequations}\label[pluralequation]{mtk_aniso}
    \begin{align}
        \dot{\mathbf{q}}_i &= \frac{\mathbf{p}_i}{m_i} + \frac{\mathbf{p}_g}{W_g}\mathbf{q}_i\,, \\
        \dot{\mathbf{p}}_i &= \tilde{\mathbf{F}}_i - \frac{\mathbf{p}_g}{W_g}\mathbf{p}_i - \frac{1}{N_f}\frac{\text{Tr}(\mathbf{p}_g)}{W_g}\mathbf{p}_i \,, \\
        \dot{\mathbf{h}} &= \frac{\mathbf{p}_g}{W_g} \mathbf{h}\,, \\
        \dot{\mathbf{p}}_g &= \left(\det\mathbf{h}\right) \left(\mathbf{P}_{\text{int}}-P_0 \mathbf{I}\right) + \frac{1}{N_f}\sum_i \frac{\mathbf{p}_i^2}{m_i}\mathbf{I} \,.
    \end{align}
    \end{subequations}
    Also in this case, the full set of MTK equations is obtained by coupling particles and cell components with two separate Nosé-Hoover chains, resulting in the generation of the correct anisotropic $N\mathbf{S}T$ ensemble defined in \cref{NPT_anisotropic}. 
    
    Since the time-reversible integration scheme for the MTK equations has been derived by Tuckerman \emph{et al.}, the algorithm is also referred to as Martyna-Tuckerman-Tobias-Klein (MTTK) barostat.

    \subsection{Shinoda barostat}
    The equations developed by Shinoda \emph{et al.} \cite{shinoda} combine the hydrostatic MTK \cref{mtk_aniso} with the strain energy calculated as in \cref{strain_energy} within the Parrinello-Rahman barostat. Apart from the details in the time-reversible integration scheme, the only modification to the MTK method involves the equation for the matrix of box momenta $\mathbf{p}_g$:
    \begin{equation}
        \dot{\mathbf{p}}_g = \left(\det\mathbf{h}\right) \left(\mathbf{P}_{\text{int}}-P_0 \mathbf{I}\right) - \mathbf{h}\bm{\Sigma}\mathbf{h}^T + \frac{1}{N_f}\sum_i \frac{\mathbf{p}_i^2}{m_i}\mathbf{I}\,,
    \end{equation}
    with $\bm{\Sigma}$ defined as in \cref{sigma_matrix}.
    
    \subsection{Stochastic cell rescaling}\label{section:crescale_iso}
    Stochastic cell rescaling (SCR) \cite{crescale_iso} is a first-order stochastic barostat that generates the correct isothermal-isobaric ensemble when coupled to the Hamilton's equations and to a thermostat. SCR employs a Berendsen-like deterministic part and a suitable noise term, which is responsible for the correct volume fluctuations. In its isotropic version, the stochastic equation driving the dynamics of the volume is
    \begin{equation}\label{crescale_iso_V}
        \text{d}V = -\frac{\beta_T V}{\tau_p}\left(P_0 - P_{\text{int}} - \frac{k_B T}{V}\right)\text{d}t + \sqrt{\frac{2k_B T \beta_T V}{\tau_p}}\text{d}W\,,
    \end{equation}
    where $P_{\text{int}}$ is computed with the instantaneous kinetic energy as in \cref{internal_pressure} if momenta are rescaled, or with the average kinetic energy as in \cref{internal_pressure_avgkin} if they are not. The stationarity of the $NP_0 T$ distribution in \cref{npt_isotropic} can be proved by considering the associated Fokker-Planck equation (see Appendix \ref{appendix_fp}). As in the Berendsen barostat, the time constant $\tau_p$ defines the equilibration time of the volume and its autocorrelation time in equilibrium conditions. By defining the variables $\varepsilon = \log\left(V/V_0 \right)$, where $V_0$ is a reference volume, and $\lambda = \sqrt{V}$, it is possible to derive two equivalent ways of writing \cref{crescale_iso_V} by means of the It\^{o} chain rule (see Appendix \ref{appendix_stocdiffeq}):
    \begin{subequations}
    \begin{align}
        \text{d}\varepsilon &= -\frac{\beta_T}{\tau_p}\left(P_0 - P_{\text{int}} \right)\text{d}t + \sqrt{\frac{2 k_B T \beta_T}{V \tau_p}}\text{d}W\,, \label{eps_eq_SCR} \\
        \text{d}\lambda &= -\frac{\beta_T\lambda}{2\tau_p}\left(P_0 - P_{\text{int}} - \frac{k_B T}{2 V}\right)\text{d}t + \sqrt{\frac{k_B T \beta_T}{2\tau_p}}\text{d}W\,. \label{lambda_eq_SCR}
    \end{align}
    \end{subequations}
    In particular, \cref{lambda_eq_SCR} allows to write a time-reversible integrator for which the effective energy drift can be computed (see Appendix~\ref{appendix_effenergy}). Moreover, it is possible to derive \cref{eps_eq_SCR} as the high-friction limit of a Langevin piston algorithm (see \cref{andersen_barostat}) with a volume-dependent friction. 
    
    SCR has also been formulated in a semi-isotropic version, namely to generate the constant surface-tension ensemble $N P_0^\perp \gamma_0 T $ described by \cref{npt_constant_surface_tension}. In this case, by defining the variables $\varepsilon_{xy} = \log(A/A_0)$ and $\varepsilon_z = \log(L/L_0)$ the dynamics can be written in terms of two decoupled stochastic equations:
    \begin{subequations}\label[pluralequation]{crescale-semi-isotropic-eq}
    \begin{align}
        \text{d}\varepsilon_{xy} &= -\frac{2\beta_T}{3\tau_p}\left(P_0^\perp - \frac{\gamma_0}{L} - \frac{P_{\text{int},xx} + P_{\text{int},yy}}{2} \right)\text{d}t + \sqrt{\frac{4 k_B T \beta_T}{3 V \tau_p}}\text{d}W_{xy}\,, \label{eps_xy}\\
        \text{d}\varepsilon_z &= -\frac{\beta_T}{3\tau_p}\left(P_0^\perp - P_{\text{int},zz}\right)\text{d}t + \sqrt{\frac{2 k_B T \beta_T}{3 V \tau_p}}\text{d}W_z\,, \label{eps_z}
    \end{align}
    \end{subequations}
    where $W_{xy}$ and $W_z$ are two distinct and independent Wiener processes. As one might expect, by summing the two equations above with $\gamma_0 = 0$ it is possible to recover the isotropic \cref{eps_eq_SCR}, with $P_0 = P_0^\perp$. 
    
    This pressure coupling method preserves all the good properties of the Berendsen barostat, namely it is efficient in the equilibration phase (as it is not affected by instabilities or oscillations if the system is far from equilibrium), it allows to easily tune the relaxation time of the volume and it is easier to implement than second-order barostats. Moreover, since SCR generates the correct isothermal-isobaric ensemble it can be used also in production runs in place of second-order algorithms, replacing the typical pipeline for constant pressure simulations with a more efficient one, where a single algorithm is employed (see \cref{fig:pipeline}).



    
    \chapter{Fully flexible formulation of SCR}\label{chapter:formulation}
    The aim of this chapter is to derive and discuss the equations for the anisotropic version of SCR, which allows to generate the correct $N\mathbf{S}T$ ensemble both in presence of a hydrostatic external pressure and in case of a generic external stress. 
    
    \section{Derivation of the equations}\label{derivation_aniso}
    In order to formulate the anisotropic version of the stochastic cell rescaling method, starting from the case of a diagonal external stress $\mathbf{S} = P_0\mathbf{I}$, we look for a multidimensional It\^{o} equation (see Appendix \ref{appendix_stocdiffeq}) for the box matrix $\mathbf{h}$,
    \begin{equation}\label{SCR_flex_start}
        \text{d}\mathbf{h} =  \mathbf{A}(\mathbf{h})\, \text{d}t + \mathbf{B}(\mathbf{h})\,\text{d}\mathbf{W}  \,,
    \end{equation}
    such that the two following requirements are satisfied:
    \begin{enumerate}[label=(\roman*)]
        \item the anisotropic isothermal-isobaric distribution 
        \begin{equation}
            \mathcal{P}_{N\mathbf{S}T}\big(\{\mathbf{q}_i,\mathbf{p}_i\},\mathbf{h}\big) \propto \left(\det\mathbf{h}\right)^{-2}\exp\left[-\frac{1}{k_B T}\Big(K + U + P_0 \det\mathbf{h}\Big)\right]
        \end{equation}
        should be stationary with respect to the multidimensional FP equation (see Appendix \ref{appendix_fp}) corresponding to \cref{SCR_flex_start};\label{first_requirement}
        \item the deterministic part of these equations, $\text{d}\mathbf{h}^{\text{det}} = \mathbf{A}(\mathbf{h})\, \text{d}t$, should contain a Berendsen-like term as written in \cref{berendsen_aniso}.\label{second_requirement}
    \end{enumerate}
    Written explicitely, this multidimensional It\^{o} equation reads:
    \begin{equation}\label[pluralequation]{SCR_eq_A_B}
        \text{d}h_{\alpha i} = A_{\alpha i}(\mathbf{h})\, \text{d}t + \sum_{\beta j} B_{\alpha i \beta j}(\mathbf{h})\,\text{d}W_{\beta j} \,,
    \end{equation}
    where we remind that $\alpha = x,y,z$ and $i=1,2,3$.
    Note that $\mathbf{A}$ and $\mathbf{B}$ have already been defined with no explicit time dependence in order to satisfy \ref{first_requirement}.
    
    Instead of imposing the stationarity condition for $\mathcal{P}_{N\mathbf{S}T}$ it is easier to require the stricter condition of detailed balance, as defined in \cref{db_multidim}. As a starting point, the multidimensional FP equation corresponding to \cref{SCR_flex_start} reads:
    \begin{equation}
        \frac{\partial}{\partial t} \mathcal{P}(\mathbf{h},t) = -\sum_{\alpha i}\frac{\partial}{\partial h_{\alpha i}}\Big(A_{\alpha i}(\mathbf{h})\mathcal{P}(\mathbf{h},t)\Big) + \sum_{\alpha i \beta j} \frac{\partial^2}{\partial h_{\alpha i}\partial h_{\beta j}} \Big(D_{\alpha i \beta j} (\mathbf{h}) \mathcal{P}(\mathbf{h},t)\Big)\,,
    \end{equation}
    where $\mathbf{D} = \frac{1}{2}\mathbf{B}\mathbf{B}^T$, namely
    \begin{equation}\label{D_B_relation}
        D_{\alpha i \beta j} = \frac{1}{2}\sum_{\gamma k} B_{\alpha i \gamma k} B_{\beta j \gamma k}\,.
    \end{equation}
    Imposing the detailed balance condition for the generic distribution $\mathcal{P} = \mathcal{P}(\mathbf{h})$ means requiring that each component of the probability density tensor $\mathbf{J}$ vanishes, namely
    \begin{equation}
        J_{\alpha i} = A_{\alpha i}\,\mathcal{P} - \sum_{\beta j} \frac{\partial}{\partial h_{\beta j}} \Big(D_{\alpha i \beta j}\,\mathcal{P}\Big) = 0\,.
    \end{equation}
    By solving these equations with respect to the components of $\mathbf{A}$ one gets:
    \begin{equation}
        A_{\alpha i} = \sum_{\beta j} D_{\alpha i \beta j}\frac{\partial}{\partial h_{\beta j}}\Big(\log\big(D_{\alpha i \beta j}\,\mathcal{P}\big)\Big)\,.
    \end{equation}
    As a consequence, the condition \ref{first_requirement} is satisfied writing the target \cref{SCR_eq_A_B} as:
    \begin{equation}\label[pluralequation]{target_eqs}
        \text{d}h_{\alpha i} = \sum_{\beta j} D_{\alpha i \beta j} \frac{\partial}{\partial h_{\beta j}}\Big(\log\big(D_{\alpha i \beta j}\,\mathcal{P}_{N\mathbf{S}T}\big)\Big)\,\text{d}t + \sum_{\beta j} B_{\alpha i \beta j}\,\text{d}W_{\beta j}\,,
    \end{equation}
    \indent In order to satisfy the requirement \ref{second_requirement}, let's make the following \emph{ansatz} on the functional form of the diffusion tensor: 
    \begin{equation}
        D_{\alpha i\beta j} = \frac{\beta_T k_B T}{3V\tau_p}\delta_{\alpha\beta}\sum_\gamma h_{\gamma i}h_{\gamma j}\,.
    \end{equation}
    Leaving the complete calculations to Appendix \ref{appendix_derivation}, with this choice of $\mathbf{D}$ the deterministic part of \cref{target_eqs} becomes:
    \begin{equation}\label{dh_det_final}
        \text{d}h_{\alpha i}^{\text{det}} = -\frac{\beta_T}{3\tau_p} \Bigg[\sum_\beta \Big( P_0\,\delta_{\alpha\beta} - P_{\text{int},\alpha\beta} \Big)h_{\beta i} - \frac{k_B T}{V}h_{\alpha i}\Bigg]\,\text{d}t\,.
    \end{equation}
    The expression of the internal pressure tensor appearing in $\text{d}h_{\alpha i}^{\text{det}}$ is different if the propagation of $\mathbf{h}$ is performed at both constant rescaled positions $\mathbf{s}_i$ and rescaled momenta $\bm{\pi}_i$, or instead by keeping fixed rescaled positions and physical momenta. Representing physical positions $\mathbf{q}_i$ and momenta $\mathbf{p}_i$ as column vectors, the rescaled counterparts are defined via the following relations:
    \begin{subequations}\label[pluralequation]{rescaled_variables}
    \begin{align}
        \mathbf{q}_i &= \mathbf{h} \mathbf{s}_i \,,  \label{rescaled_positions}\\
        \mathbf{p}^T_i &=  \bm{\pi}^T_i \mathbf{h}^{-1}\,.\label{rescaled_momenta}
    \end{align}
    \end{subequations}
    As it is shown in Appendix \ref{appendix_derivation}, when both positions and momenta are rescaled the internal pressure tensor $\mathbf{P}_{\text{int}}$ is defined as in \cref{internal_pressure_tensor}, while in the formulation with only rescaled positions the kinetic energy tensor is replaced by an average contribution, namely
    \begin{equation}\label{internal_pressure_avgkineng}
        P^{\langle K\rangle}_{\text{int},\alpha\beta} = \frac{Nk_B T}{V}\delta_{\alpha\beta} + \frac{1}{V}\sum_{i=1}^N F_i^{\alpha}\,q_i^{\beta} \,. 
    \end{equation}
    Regardless of how the internal pressure is computed, the first part of \cref{dh_det_final} is exactly the anisotropic formulation of the Berendsen barostat; thus the initial choice of the diffusion tensor $\mathbf{D}$ appears meaningful in order to satisfy \ref{second_requirement}. 
    The additional term containing $k_B T/V$ can be seen as a correction that becomes more and more negligible as the system size increases, since the average kinetic contribution included in $\mathbf{P}_{\text{int}}$ is the dominant term when the number of atoms $N$ is large. Without considering the stochastic part of \cref{target_eqs}, it is obvious that this correction is not sufficient to generate the correct isothermal-isobaric ensemble.
    
    As a comment on the definition of the rescaled momenta $\bm{\pi}_i$, there are two reasons to multiply by the inverse box matrix $\mathbf{h}^{-1}$ \emph{on the right side}:
    \begin{itemize}
        \item this is the only way to have a consistency in the class of labels for physical and rescaled variables, which appear with latin and greek indices respectively if we write \cref{rescaled_variables} explicitely:
        \begin{subequations}\label[pluralequation]{rescaled_variables_explicit}
        \begin{align}
            q_i^{\alpha} &= \sum_{k=1}^3 h_{\alpha k}\, s_i^k \,, \\
            p_i^{\alpha} &= \sum_{k=1}^3 \pi_i^{k}\, h_{k\alpha}^{-1}\,;
        \end{align}
        \end{subequations}
        \item defined in this way, it is possible to show that the variables $\bm{\pi}_i$ are the actual \emph{conjugated momenta} of the rescaled coordinates $\mathbf{q}_i$, or equivalently that the transformation from physical to rescaled variables is \emph{canonical}, i.e. it preserves the form of Hamilton's equations \cite{tuckerman}. 
    \end{itemize}
    While the first motivation is necessary to be consistent with the definition of the box matrix $\mathbf{h}$, the second one is more of aesthetic nature, since the SCR method is not formulated within a Hamiltonian framework. As a comparison, in the MTK \cref{mtk_aniso} - which cannot be obtained from a Hamiltonian as well - positions and momenta are rescaled with a matrix multiplication \emph{on the same side}.
    
    Let's now consider the stochastic part of the target equations, $\text{d}\mathbf{h}^{\text{stoc}} = \mathbf{B}(\mathbf{h})\,\text{d}\mathbf{W}$. The tensor $\mathbf{B}$ that satisfies $\mathbf{D} = \frac{1}{2}\mathbf{B}\mathbf{B}^T$ has components
    \begin{equation}\label{B_tensor}
        B_{\alpha i \beta j} = \sqrt{\frac{2\beta_T k_B T}{3V\tau_p}}\,h_{\beta i}\,\delta_{\alpha j}\,;
    \end{equation}
    as a consequence, the stochastic term of the target \cref{target_eqs} becomes:
    \begin{equation}
        \text{d}h_{\alpha i}^{\text{stoc}} = \sqrt{\frac{2\beta_T k_B T}{3V\tau_p}} \sum_{\beta}h_{\beta i}\,\text{d}W_{\alpha\beta}\,.
    \end{equation}
    Putting together the deterministic and stochastic parts, we finally find the equations that generalize the SCR method to anisotropic cell fluctuations,
    \begin{equation}\label[pluralequation]{crescale_aniso}
        \text{d}h_{\alpha i} = -\frac{\beta_T}{3\tau_p} \Bigg[\sum_\beta \Big( P_0\delta_{\alpha\beta} - P_{\text{int},\alpha\beta} \Big)h_{\beta i} - \frac{k_B T}{V}h_{\alpha i}\Bigg]\,\text{d}t + \sqrt{\frac{2\beta_T k_B T}{3V\tau_p}} \sum_{\beta}h_{\beta i}\,\text{d}W_{\alpha\beta}\,,
    \end{equation}
    or in matrix notation:
    \begin{equation}
        \boxed{\text{d}\mathbf{h} = -\frac{\beta_T}{3\tau_p} \Big[ \big( P_0\mathbf{I} - \mathbf{P}_{\text{int}} \big) - \frac{k_B T}{V}\mathbf{I}\Big]\mathbf{h}\,\text{d}t + \sqrt{\frac{2\beta_T k_B T}{3V\tau_p}} \text{d}\mathbf{W}\,\mathbf{h}\,.\,}
    \end{equation}
    It is worth observing that the diffusion tensor $\mathbf{D}$ satisfying \ref{second_requirement} is not unique; for instance, it has been found that also the choice
    \begin{equation}
        \widetilde{D}_{\alpha i \beta j} = \frac{\beta_T k_B T}{3V\tau_p} h_{\alpha j} h_{\beta i}
    \end{equation}
    brings to the same expression for the Berendsen-like deterministic part of the equations. However, this choice has been discarded as it seems not possible to find an analytic expression for $\widetilde{\mathbf{B}}$ satisfying \cref{D_B_relation}.
    
    \subsection{Generic external stress}
    In case of a generic external stress, i.e. $\mathbf{S}\neq P_0\mathbf{I}$, the target distribution in \cref{NPT_anisotropic_shear} includes an additional term that only enters in the equations in an additive way, and that is independent on the choice of rescaling momenta or not. Indeed, as it is shown in Appendix \ref{derivation_strain}, the deterministic part of the equations has to be expanded with the contribution
    \begin{equation}\label{dh_strain_maintext}
        \text{d}h_{\alpha i}^{\text{strain}} = -\frac{\beta_T}{3V\tau_p}\big(\mathbf{h}\,\bm{\Sigma}\,\mathbf{h}^T \mathbf{h} \big)_{\alpha i}\,,
    \end{equation}
    where $\bm{\Sigma} = V_0\, \mathbf{h}_0^{-1}\big(\mathbf{S}-P_0\mathbf{I}\big)\left(\mathbf{h}_0^{-1}\right)^T$. Then the full equations for a generic external stress are the following:
    \begin{equation}\label[pluralequation]{crescale_eqs_strain}
        \boxed{\text{d}\mathbf{h} = -\frac{\beta_T}{3\tau_p} \Big[ \big( P_0\mathbf{I} - \mathbf{P}_{\text{int}} \big) - \frac{k_B T}{V}\mathbf{I} + \frac{1}{V}\mathbf{h}\,\bm{\Sigma}\,\mathbf{h}^T \Big]\,\mathbf{h}\,\text{d}t + \sqrt{\frac{2\beta_T k_B T}{3V\tau_p}} \text{d}\mathbf{W}\,\mathbf{h}\,.\,}
    \end{equation}

    \section{Properties}\label{properties}
    As a first important observation, both the isotropic and semi-isotropic formulations of SCR - namely \cref{crescale_iso_V} and \cref{crescale-semi-isotropic-eq} - can be derived from the anisotropic SCR equations (see Appendix \ref{appendix_flex2iso}), by changing the propagated variables with the multidimensional It\^{o} chain rule. Hence these equations represent a self-consistent generalization of the previous formulations of the method.
    
    A relevant feature of the anisotropic SCR \cref{crescale_eqs_strain} is that they are invariant under a redefinition of the box vectors leaving the Bravais lattice structure untouched. In MD simulations, the notion of Bravais lattice is used to periodically replicate the system in space, in order to minimize edge effects (see Appendix \ref{PBCs}). In general, if the three box vectors $\mathbf{a},\mathbf{b},\mathbf{c}$ (or $\mathbf{a}_1,\mathbf{a}_2,\mathbf{a}_3$) are thought as the primitive vectors of a Bravais lattice, the most general transformation leaving the Bravais lattice invariant is a subclass of the following mapping:
    \begin{equation}\label{change_cell_vectors}
        \mathbf{a}_i \longmapsto \mathbf{a}'_i = \sum_{j=1}^3 n_j^{(i)}\,\mathbf{a}_j\,
    \end{equation}
    where $n_j^{(i)}$ are integer numbers. The actual transformation is only a subclass of \cref{change_cell_vectors} because additional constraints should be imposed on the three integers, in order to obtain three vectors $\{\mathbf{a}'_i\}$ that are still linearly independent and that generate a cell with the same volume of the original one. Recalling that $\mathbf{a}_1,\mathbf{a}_2,\mathbf{a}_3$ define the columns of the box matrix $\mathbf{h}$, the transformation above can be also written as:
    \begin{equation}
        h_{\alpha i} \longmapsto h'_{\alpha i} = \sum_{j=1}^3 n_j^{(i)}\,h_{\alpha j}\,.
    \end{equation}
    By employing once again the multidimensional It\^{o} chain rule, it is possible to show that the anisotropic SCR equations assume the same form when written in terms of the transformed box matrix $\mathbf{h}'$ (see Appendix \ref{appendix_changecellvectors}). 
    
    \section{SCR as limit case of Parrinello-Rahman equations}\label{limit_PR}
    As already shown in the isotropic case \cite{crescale_iso}, also the anisotropic SCR equations can be derived as the high-friction limit of a second-order barostat, namely the Parrinello-Rahman equations coupled to a Langevin thermostat with a variable-dependent friction tensor $\bm{\gamma} = \bm{\gamma}(\mathbf{h})$. Writing the second-order Parrinello-Rahman \cref{second_eq_PR} as two first-order equations and adding both friction and noise terms, the equations of interest are
    \begin{subequations}\label[pluralequation]{PR_friction_eqs}
    \begin{align}
        \dot{\mathbf{h}} &= \mathbf{v}\,, \\
        \dot{\mathbf{v}} &= \frac{V}{W}\big(\mathbf{P}_{\text{int}} - P_0\mathbf{I}\big)\big(\mathbf{h}^{-1}\big)^T - \frac{1}{W}\bm{\gamma}\,\mathbf{v} + \frac{1}{W}\bm{\sigma}\,\bm{\eta}\,, \label{2nd_eq_PR}
    \end{align}
    \end{subequations}
    where $\bm{\eta}$ is a tensor of independent white noise processes and $\bm{\sigma}$ is a tensor satisfying the multidimensional \emph{fluctuation-dissipation theorem} \cite{high_friction_limit}:
    \begin{equation}\label{fluct_diss_multidim}
        \bm{\sigma}\bm{\sigma}^T = 2k_B T \bm{\gamma}\,.
    \end{equation}
    As shown in Appendix \ref{appendix_highfriction}, with a suitable choice of the tensor $\bm{\gamma}(\mathbf{h})$,
    \begin{equation}
        \gamma_{\alpha i\beta j} = \frac{3V\tau_p}{\beta_T}\delta_{\alpha\beta}\sum_{\eta}h_{i\eta}^{-1}\,h_{j\eta}^{-1}\,,
    \end{equation}
    the equations above bring to the anisotropic SCR \cref{crescale_aniso} in the high-friction limit described in \cite{high_friction_limit}, apart for a small correction in the deterministic part: 
    \begin{equation}
        \text{d}\mathbf{h}^{PR} = \frac{\beta_T}{3\tau_p}\frac{2 k_B T}{V}\mathbf{I}\,\text{d}t\,.
    \end{equation}
    This additional term comes from a slight difference in the $N\mathbf{S}T$ distribution sampled by the Parrinello-Rahman method, namely the absence of the factor $(\det\mathbf{h})^{-2}$ in the target distribution defined in \cref{NPT_anisotropic}. Anyway this term is negligible if the system includes a large number of atoms $N$, as in this case the main contribution in the deterministic part of the SCR equations comes from the internal pressure tensor. Indeed the average contribution of the kinetic energy included in $\mathbf{P}_{\text{int}}$ scales linearly in $N$, as it is clear from \cref{internal_pressure_tensor}. 
    
    \Cref{PR_friction_eqs} hold in case of a isotropic external stress ($\mathbf{S} = P_0\mathbf{I}$). The inclusion of a generic stress, namely of the additional term $\frac{1}{W}\bm{\Sigma}\mathbf{h}$ in the (RHS) of \cref{2nd_eq_PR}, brings to the general anisotropic SCR \cref{crescale_eqs_strain} in the same high-friction limit considered before. Also this generalization is better discussed in Appendix \ref{appendix_highfriction}.
     
    \section{Euler integrator}\label{euler_integrator}
    The simplest way to integrate \cref{crescale_eqs_strain} is to use the Euler method, which is a simple finite time step approximation:
    \begin{equation}\label[pluralequation]{Delta_h}
        \Delta\mathbf{h} = -\frac{\beta_T}{3\tau_p} \Big[ \big( P_0\mathbf{I} - \mathbf{P}_{\text{int}} \big) - \frac{k_B T}{V}\mathbf{I} + \frac{1}{V}\mathbf{h}\,\bm{\Sigma}\,\mathbf{h}^T \Big]\,\mathbf{h}\,\Delta t + \sqrt{\frac{2\beta_T k_B T\Delta t}{3V\tau_p}} \mathbfcal{R}\,\mathbf{h}
    \end{equation}
    Here $\mathbfcal{R}$ is a 3$\times$3 matrix of i.i.d. standard Gaussian numbers. The propagation $\mathbf{h}\mapsto\mathbf{h}+\Delta\mathbf{h}$ is equivalent to the rescaling $\mathbf{h}\mapsto\bm{\mu}\mathbf{h}$, which gives the name to the algorithm, where the rescaling matrix is
    \begin{equation}\label{mu_matrix}
        \bm{\mu} = \mathbf{I} -\frac{\beta_T}{3\tau_p} \Big[ \big( P_0\mathbf{I} - \mathbf{P}_{\text{int}} \big) - \frac{k_B T}{V}\mathbf{I} + \frac{1}{V}\mathbf{h}\,\bm{\Sigma}\,\mathbf{h}^T \Big]\,\Delta t + \sqrt{\frac{2\beta_T k_B T\Delta t}{3V\tau_p}} \mathbfcal{R}\,.
    \end{equation}
    The same matrix is also employed to rescale positions and momenta, according to the formulation chosen:
    \begin{subequations}\label[pluralequation]{pos_mom_rescaling}
    \begin{align}
        \mathbf{q}_i &\longmapsto \bm{\mu}\,\mathbf{q}_i\,, \\
        \mathbf{p}_i^T &\longmapsto \mathbf{p}_i^T\bm{\mu}^{-1}\,.
    \end{align}
    \end{subequations}
    The rescaling can be performed at each MD step or every $n_s$ steps in a multiple-time-step fashion \cite{multiple_timestep}, in order to speed up the simulation.
    
    Even if the continuous \cref{crescale_aniso} satisfy detailed balance, this condition is violated when they are integrated with a finite time step algorithm. To quantify this violation and find out if the time step and the other parameters were chosen correctly, in principle it is possible to compute a quantity called \emph{effective energy drift}, which can be interpreted as the work performed by the integration algorithm on the system (see Appendix \ref{appendix_effenergy}). However, as already observed in the isotropic case \cite{crescale_iso}, the effective energy drift has a "bad scaling" with the time step when - as in \cref{Delta_h} - the noise prefactor is variable-dependent. In other words, in such a situation variations of the effective energy appear not much sensitive to variations of the time step, and this makes the effective energy an unsuitable quantity to evaluate the quality of the integration. For this reason, no effective energy is computed for the simple Euler integrator in \cref{Delta_h}. In order to explain this behaviour in presence of a variable-dependent noise prefactor, it is possible to observe from \cref{effective_energy} that the effective energy is computed as the ratio between the probability of generating the forward move and the probability of generating the backward one, and these transition probabilities are dominated by the stochastic term when the time step is small, or equivalently when $\tau_p$ is large. Indeed, in this limit it is clear from the $\Delta t/\tau_p$-dependencies in \cref{Delta_h} that the deterministic contribution goes to zero faster than the stochastic one. Neglecting the deterministic term, the integration appears perfectly time-reversible if we suppose that the noise prefactor is constant; therefore this condition is expected to maximize the probability of the backward move at fixed time step.
    
    In the isotropic case, it is easy to find a change of variable - namely $\lambda = \sqrt{V}$ - that brings to a formulation with a constant noise prefactor, as shown in \cref{lambda_eq_SCR}. In the anisotropic case, instead, such a transformation appears not feasible, since the variable dependence in the stochastic term is notably complicated by a matrix product involving all the nine cell components. A possibility that has been taken into account but finally discarded is to perform a transformation such that one of the propagated variables is exactly $\lambda$, while the other eight variables are propagated according to equations with a $\lambda$-dependent noise prefactor, which is "symmetrized" with a geometric mean between consecutive steps in order to enhance the time-reversibility of the generated trajectory. More information about the attempt of constructing this integrator and its limitations are reported in Appendix \ref{appendix_epsvariables}.
    
    \section{Time-reversible integrator}\label{sec:TR}
    The Euler integrator of the previous section does not allow to use the effective energy drift (see Appendix \cref{appendix_effenergy}) to efficiently monitor the quality of the integration, since the volume dependence in the noise prefactor makes the box matrix update $\mathbf{h}\mapsto\mathbf{h}+\Delta\mathbf{h}$ non-reversible even when $\Delta t$ is small enough to allow neglecting the deterministic part of the move. In order to derive a time-reversible integration scheme in the limit of small $\Delta t$, let's first rewrite the anisotropic SCR \cref{crescale_eqs_strain} as
    \begin{equation}
        \text{d}\mathbf{h} = \left(\mathbf{A}+\frac{b^2}{2} \right)\,\mathbf{h}\,\text{d}t + b\,\text{d}\mathbf{W}\,\mathbf{h}\,,
    \end{equation}
    where $\mathbf{A}$ and $b$ are 
    \begin{subequations}
    \begin{align}
        \mathbf{A} &= -\frac{\beta_T}{3\tau_p}\Big(P_0\mathbf{I}-\mathbf{P}_{\text{int}} + \frac{1}{V}\mathbf{h}\,\bm{\Sigma}\,\mathbf{h}^T\Big)\,, \\
        b &= \sqrt{\frac{2\beta_T k_B T\Delta t}{3V\tau_p}}\,.
    \end{align}
    \end{subequations}
    Note that $\mathbf{A}$ depends on the full box matrix $\mathbf{h}$, while $b$ only depends on its determinant, which is the volume of the system.
    In terms of these quantities, the Euler integrator reads:
    \begin{equation}
        \mathbf{h}^{t+\Delta t} = \left(\mathbf{A}^t+\frac{\left(b^t\right)^2}{2} \right)\,\mathbf{h}^t\Delta t + b^t\Delta\mathbf{W}^t\,\mathbf{h}^t\,,
    \end{equation}
    where $\Delta\mathbf{W}^t = \sqrt{\Delta t}\, \bm{\mathcal{R}}^t$. Superscripts are referred to the MD time at which each quantity is computed. This expression can be seen as the first order approximation of
    \begin{equation}\label{eq:expmatrix}
        \mathbf{h}^{t+\Delta t} = \exp\big(\mathbf{A}^t\Delta t + b^t\Delta\mathbf{W}^t\big)\,\mathbf{h}^t\,,
    \end{equation}
    where the \emph{matrix exponential} of a generic matrix $\mathbf{M}$ is defined via the power series
    \begin{equation}\label{eq:exp_matrix_def}
        \exp\big(\mathbf{M}\big) = \sum_{k=0}^{\infty}\frac{1}{k!}M^k\,.
    \end{equation}
    Let's now decompose $\mathbf{A}$ and $\Delta \mathbf{W}$ as
    \begin{subequations}
    \begin{align}
        \mathbf{A} &= a_1\mathbf{I} + \mathbf{A}_2\,,\\
        \Delta\mathbf{W} &= \Delta W_1\mathbf{I} + \Delta \mathbf{W}_2\,,
    \end{align}
    \end{subequations}
    where the scalars $a_1$ and $\Delta W_1$ are the averages of the diagonal elements of the matrices $\mathbf{A}$ and $\Delta \mathbf{W}$ respectively:
    \begin{subequations}
    \begin{align}
        a_1 &= \frac{\text{Tr}\,\mathbf{A}}{3}\,, \\
        \Delta W_1 &= \frac{\text{Tr}\,\Delta\mathbf{W}}{3}\,.
    \end{align}
    \end{subequations}
    As a consequence, $\mathbf{A}_2$ and $\Delta \mathbf{W}_2$ are by construction traceless matrices. With this decomposition, \cref{eq:expmatrix} becomes:
    \begin{subequations}
    \begin{align}
        \mathbf{h}^{t+\Delta t} &= \exp\big(\left(a_{1}^t\mathbf{I} + \mathbf{A}_2^t\right)\Delta t + b^t\,\left(\Delta W_1^t\mathbf{I} + \Delta \mathbf{W}_2^t\right)\big)\,\mathbf{h}^t \\
        &= \exp\big(\left(a_1^t\Delta t + b^t\,\Delta W_1^t\right)\,\mathbf{I}+ \left(\mathbf{A}_2^t\,\Delta t + b^t\Delta \mathbf{W}_2^t\right)\big)\,\mathbf{h}^t \\
        &= \exp\big(a_1^t\Delta t + b^t\,\Delta W_1^t\big)\,\exp\big(\mathbf{A}_2^t\,\Delta t + b^t\Delta \mathbf{W}_2^t\big)\,\mathbf{h}^t\,, \label{eq:split_volume_shape}
    \end{align}
    \end{subequations}
    where in the last passage holds with no approximations, since the identity matrix $\mathbf{I}$ commutes with any other matrix and $\exp\big(c\mathbf{I}\big) = \exp(c)\, \mathbf{I}$ for any scalar $c$. Note that the matrix exponential in \cref{eq:split_volume_shape} has unitary determinant, as the matrix in the argument is traceless and the determinant of a matrix exponential is given by the well-known relation
    \begin{equation}
        \det\left[\exp\big(\mathbf{M}\big)\right] = e^{\text{Tr}\,\mathbf{M}}\,.
    \end{equation}
    As a consequence, \cref{eq:split_volume_shape} allows to separate in the overall rescaling two different contributions, the former related to the change of volume and given by the first (scalar) exponential $\exp\big(a_1^t\Delta t + b^t\,\Delta W_1^t\big)$, and the latter connected to the change of shape and given by the second (matrix) exponential $\bm{\mu}_s =\exp\big(\mathbf{A}_2^t\,\Delta t + b^t\Delta \mathbf{W}_2^t\big)$, which leaves the determinant of $\mathbf{h}_t$ untouched. In other words, the first operation propagates the isotropic degree of freedom, namely $V=\text{det}\mathbf{h}$, while the second rescaling evolves the remaining eight degrees of freedom, which are responsible for anisotropic box fluctuations and for global rotations. To obtain a time-reversible move, we can write the box matrix update with the following Trotterization:
    \begin{equation}\label{eq:trotterization}
        \mathbf{h}^{t+\Delta t} = \exp\left(a_1^t\frac{\Delta t}{2} + b^t\frac{\Delta W_1^t}{2}\right)\exp\left(\mathbf{A}_2^t\Delta t + b^{t+\frac{\Delta t}{2}}\Delta \mathbf{W}_2^t\right)\exp\left(a_1^t\frac{\Delta t}{2} + b^t\frac{\Delta W_1^t}{2}\right)\mathbf{h}^t
    \end{equation}
    The only difference with respect to \cref{eq:split_volume_shape} is that in the rescaling involving the matrix exponential, responsible for the change of shape of the box, the noise prefactor $b$ is computed after propagating the volume for half time step. Although not strictly necessary, this modification is expected to increase the reversibility of the integrator, since the noise term $\Delta \mathbf{W}_2^t$ will be scaled with the same prefactor in the forward and backward trajectories. 
    A scheme of the box matrix propagation according to \cref{eq:trotterization} is reported in \cref{alg:trotterized_integrator}.\\
    
    \begin{algorithm}[H]
    \SetAlgoLined
     propagate volume for $\Delta t/2$\;
     propagate box matrix shape for $\Delta t$\;
     propagate volume for $\Delta t/2$\;
     \caption{Time-reversible integrator for anisotropic SCR equations}
     \label{alg:trotterized_integrator}
    \end{algorithm}
    \vspace{0.4cm}
    
    The rescaling matrix $\bm{\mu}$ containing the information on both volume and shape update, used to rescale positions and momenta as shown in \cref{pos_mom_rescaling}, is computed after step 3 as
    \begin{equation}
        \bm{\mu} = \left(\frac{V^{t+\Delta t}}{V^t} \right)^{1/3}\bm{\mu}_s\,, 
    \end{equation}
    where $\bm{\mu}_s$ is evaluated at step 2. The matrix exponential $\bm{\mu}_s$ is computed by means of a \emph{Padé approximation} \cite{pade_approximation} that reproduces \cref{eq:exp_matrix_def} up to the sixth-order. Calling $\Delta\bm{\varepsilon} = \mathbf{A}_2^t\,\Delta t + b^t\Delta \mathbf{W}_2^t$, so that $\bm{\mu}_s =\exp\big(\Delta\bm{\varepsilon}\big)$, this approximation reads:
    \begin{equation}
        \bm{\mu}_s \simeq \Big(\mathbf{I}-\frac{1}{2}\Delta\bm{\varepsilon}+\frac{1}{10}\Delta\bm{\varepsilon}^2 - \frac{1}{120}\Delta\bm{\varepsilon}^3\Big)^{-1}\Big(\mathbf{I}+\frac{1}{2}\Delta\bm{\varepsilon}+\frac{1}{10}\Delta\bm{\varepsilon}^2 + \frac{1}{120}\Delta\bm{\varepsilon}^3\Big)\,.
    \end{equation}
    
    The isotropic moves at step 1 and 3 can be performed by propagating the variable $\lambda = \sqrt{V}$ instead of $V$, as this choice brings to a volume-independent noise prefactor and, as a consequence, to a "well behaved" effective energy (see \cref{section:crescale_iso}). 
    
    In principle, the random numbers employed for the isotropic propagations at step 1 and 3 depend on the "diagonal" random numbers extracted at step 2. Actually, it is easy to show that, given three i.i.d. Gaussian numbers $\mathcal{R}_{xx},\mathcal{R}_{yy},\mathcal{R}_{zz}$, 
    \begin{equation}
        \Big\langle \mathcal{R}_{\alpha\alpha}-\frac{\mathcal{R}_{xx}+\mathcal{R}_{yy}+\mathcal{R}_{zz}}{3}\Big\rangle = 0
    \end{equation}
    for any $\alpha=x,y,z$; as a consequence, the random numbers at step 1 and 3 can be extracted independently from the ones at step 2, with the only request that $\text{Tr}\,\Delta\mathbf{W}_2 = 0$.
    
    Note that steps 1+3 are equivalent to propagate $V$ for a full time step, and the split is only employed to compute the noise prefactor $b$ in an "intermediate" time at step 2. In other words, the update of $\lambda$ given by steps 1+3 is equivalent to
    \begin{equation}\label{eq:lambda_equivalent_1+3}
        \lambda \longmapsto \lambda 
        -\frac{\beta_T\lambda}{2\tau_P}\left(P_0-P_{\text{int}}+
        \frac{\text{Tr}(\mathbf{h}\mathbf{\Sigma}\mathbf{h}^T )}{3\lambda^2}-\frac{k_BT}{2\lambda^2}\right)\Delta t +\sqrt{\frac{k_BT\beta_T\Delta t}{2\tau_P}}\mathcal{R}\,,
    \end{equation}
    which is the Euler propagation of the isotropic \cref{lambda_eq_SCR} for a time $\Delta t$, except for the additional $\bm{\Sigma}$-dependent term, that is related to a possible deviatoric stress.
    As a consequence, the effective energy drift resulting from steps 1+3 is the same that one gets in the isotropic formulation of the barostat \cite{crescale_iso}, where $\lambda$ is propagated for a full time step with no further splitting:
    \begin{align}\label{eq:eff_eng_iso}
        \Delta \widetilde{H}_{1+3} =& \,\Delta K_{1+3} + \Delta U_{1+3} + \Delta E_{s,1+3} + P_0\Delta\lambda^2 - k_B T\Delta\log\lambda \nonumber\\
        &+ \Delta\lambda\left(\frac{f(\lambda^t)+f(\lambda^{t+\Delta t})}{2}\right) + \frac{\beta_T\Delta t}{16\tau_p}\Delta f^2,\,
    \end{align}
    where $f(\lambda) = -2\lambda\left(P_0-P_{\text{int}}-\frac{k_B T}{2\lambda^2} + \frac{\text{Tr}\left(\mathbf{h}\,\bm{\Sigma}\,\mathbf{h}^T\right)}{3\lambda^2}\right)$. Here $E_s$ is the strain energy defined in \cref{strain_energy}, and the terms $\Delta K_{1+3},\, \Delta U_{1+3}\,\text{and}\,\Delta E_{s,1+3}$ accumulate the energy increments due to the volume rescaling half-steps. In this case, the differences with respect to the effective energy drift reported in \cite{crescale_iso} are the presence of a term accounting for a possible anisotropic external stress in $f(\lambda)$ and the additional strain energy contribution.
    
    The contribution of the anisotropic step 2 is instead given by
    \begin{align}\label{eq:eff_eng_aniso}
        \Delta \widetilde{H}_{2} = \Delta K_{2} + \Delta U_{2} + \Delta E_{s,2} + \frac{k_B T}{2}\sum_{\alpha\beta}\left[\frac{\Delta t}{(b^t)^2}\Delta A_{2,\alpha\beta}^2 + \frac{2}{(b^{t})^2}\Delta\varepsilon_{\alpha\beta}\left( A_{2,\alpha\beta}^t + A_{2,\alpha\beta}^{t+\Delta t}\right)   \right]\,,
    \end{align}
    where the increments $\Delta K_{2},\, \Delta U_{2}\,\text{and}\,\Delta E_{s,2}$ account for energy differences due to the change of shape.
    For the full derivations of \cref{eq:eff_eng_iso} and \cref{eq:eff_eng_aniso}, see Appendix \ref{appendix:iso_effeng} and Appendix \ref{appendix:aniso_effeng}.

    \section{Elimination of box rotations}\label{eliminate_rotations}
    As mentioned in \cref{npt_theory}, three among the nine degrees of freedom in $\mathbf{h}$ only account for the global orientation of the box and their evolution describes overall rotations of the system, which are typically not of interest in MD simulations. At least two different strategies can be employed to eliminate box rotations.
    
    The first possibility, as suggested by Martyna \emph{et al.} \cite{mtk}, is to rescale the box matrix with a symmetric tensor, so that no torque is applied to the cell causing it to rotate. Note that a 3$\times$3 symmetric matrix only has six independent elements, coherently with the remaining degrees of freedom after eliminating three of them. In the specific case of the anisotropic MTK barostat (see \cref{mtk_barostat}), this is accomplished by symmetrizing the internal pressure tensor, which is the only source of possible asymmetries in the rescaling of the box matrix:
    \begin{equation}
        P_{\text{int},\alpha\beta} \longmapsto \frac{P_{\text{int},\alpha\beta} + P_{\text{int},\beta\alpha}}{2}\,.
    \end{equation}
    In our case, symmetrizing only the internal pressure tensor would be not sufficient, since by chance also the stochastic term in \cref{crescale_aniso} could be responsible for global rotations. Hence, a possibility is to symmetrize both $\mathbf{P}_{\text{int}}$ and the tensor $\text{d}\mathbf{W}$ containing nine independent Wiener processes:
    \begin{equation}
        W_{\alpha\beta} \longmapsto \frac{W_{\alpha\beta} + W_{\beta\alpha}}{2}\,.
    \end{equation}
    Equivalently, instead of performing two symmetrization it is possible to directly symmetrize the rescaling matrix $\mathbf{\mu}$ defined in \cref{mu_matrix}. Note that with this method the three redundant degrees of freedom are integrated out by means of the symmetrization procedure, but the number of propagated variables is still nine. As a final observation, using this approach the box will be in general \emph{triclinic}, (see Appendix \ref{PBCs}) since no constraints are imposed on its shape.
    
\begin{figure}[h!]
        \centering
        \includegraphics[width = 0.9\textwidth]{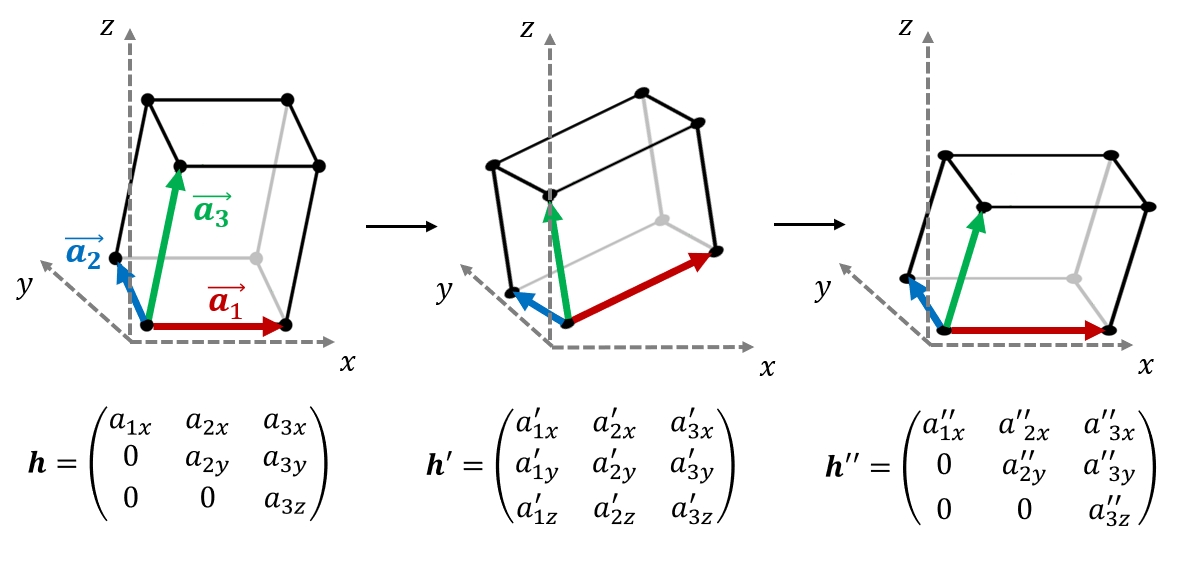}
        \caption{Graphical representation of the method applied to eliminate box rotations ($\mathbf{h}'\mapsto\mathbf{h}''$), after propagating all the nine degrees of freedom according to the anisotropic SCR \cref{crescale_aniso} ($\mathbf{h}\mapsto\mathbf{h}'$).}
        \label{fig:box_rot_elimination}
    \end{figure}

    The second possibility to eliminate rotations is to constrain $\mathbf{h}$ to be upper-triangular, so that the box vector $\mathbf{a}$ is always oriented along the $x$-axis, $\mathbf{b}$ lies in the $xy$-plane and only $\mathbf{c}$ is free to evolve in all its three components. It is important to underline that this constraint does \emph{not} imply any limitation on the shape of the box, since any triclinic box can be represented by an upper-triangular matrix $\mathbf{h}$ with a suitable choice of the chartesian frame. Hence, also this method allows to represent the box in the most general way. Since some of the most popular software for MD simulations employ such a representation for $\mathbf{h}$, this is also the method used for the anisotropic SCR algorithm, in all the implementations discussed in \cref{implementations}.  This constraint can be imposed either by evolving only the six degrees of freedom corresponding to the upper-triangular part of $\mathbf{h}$ (as for instance in the implementation of the MTK barostat in LAMMPS \cite{lammps}), or by evolving all the box matrix components and imposing the constraint afterwards (as for the Parrinello-Rahman and the anisotropic Berendsen barostats in GROMACS \cite{gromacs}). In this second implementation, at each step the matrix $\mathbf{h}' = \bm{\mu}\mathbf{h}$ obtained by evolving all the nine components,
    \begin{equation}
        \mathbf{h'}=
        \begin{pmatrix}
        a'_x & b'_x & c'_x\\
        a'_y & b'_y & c'_y\\
        a'_z & b'_z & c'_z
        \end{pmatrix}\,,
    \end{equation}
    has to be "rotated back" in order to eliminate the rotations acquired and restore the upper-triangular shape (see \cref{fig:box_rot_elimination}):
    \begin{equation}\label{backrotate}
        \mathbf{h''}= \mathbf{R}\mathbf{h'} = 
        \begin{pmatrix}
        a'_x & b'_x & c'_x\\
        0 & b'_y & c'_y\\
        0 & 0 & c'_z
        \end{pmatrix}\,.
    \end{equation}
    $\mathbf{R}$ is the rotation matrix the accounts for this operations, and it is applied on the left since it has to act on the columns of $\mathbf{h'}$, namely on the box vectors $\mathbf{a'},\mathbf{b'},\mathbf{c'}$. Note that \cref{backrotate} can be written as 
    \begin{equation}
        \mathbf{h''}= \mathbf{R}\mathbf{h'} = (\mathbf{R}\bm{\mu})\mathbf{h}\,;
    \end{equation}
    as a consequence, rotating back $\mathbf{h'}$ is equivalent to rescale the initial box matrix $\mathbf{h}$ with the rotated rescaling matrix $\bm{\mu'} = \mathbf{R}\bm{\mu}$, using the same rotation matrix $\mathbf{R}$ that makes $\mathbf{h''}$ upper-triangular. Clearly, this rotation also makes $\bm{\mu'}$ upper-triangular. The problem of determining $\bm{\mu'}$ knowing $\bm{\mu}$ is part of the so-called $\emph{QR factorization}$. In all the implementations of the anisotropic SCR algorithm, this problem is accomplished by considering a simple method equivalent to the Gram-Schmidt procedure, explained in Appendix \ref{appendix_rotations}. 
    
    \section{Equations with tensorial isothermal compressibility}
    The anisotropic SCR \cref{crescale_aniso} has been derived taking as a reference the Berendsen \cref{berendsen_aniso} for the deterministic part. In the derivation, the isothermal compressibility has been always considered as a scalar quantity, but the Berendsen algorithm also allows for a formulation where $\bm{\beta}_T$ is a 3$\times$3 tensor, which is multiplied by the pressure tensor in a element-wise manner (see \cref{berendsen_barostat}):
    \begin{equation}\label{berendsen_betat_tensor}
        \text{d}h_{\alpha i}^{\text{Ber}} = -\frac{1}{3\tau_p} \sum_\gamma \beta_{T,\alpha\gamma}\Big( P_0\,\delta_{\alpha\gamma} - P_{\text{int},\alpha\gamma} \Big)h_{\gamma i}\,\text{d}t\,.
    \end{equation}
    Moreover, the GROMACS software \cite{gromacs} employs exactly this formulation for the anisotropic Berendsen barostat, by considering in its input six independent components of $\bm{\beta}_T$, which is always constructed as a symmetric matrix. Therefore, even if such a formulation should only affect the relaxation times for the different components of $\mathbf{h}$ and $\mathbf{P}_{\text{int}}$, it is worth giving a theoretical formulation of the SCR method with a tensorial expression of $\bm{\beta}_T$, which is treated on the model of \cref{berendsen_betat_tensor}.
    
    The derivation is completely equivalent to the one described in \cref{derivation_aniso}, but starting from a different \emph{ansatz} for the diffusion tensor,
    \begin{equation}
        D_{\alpha i\beta j}' = \frac{k_B T}{3V\tau_p}\delta_{\alpha\beta}\sum_\gamma \beta_{T,\alpha\gamma}\,h_{\gamma i}\,h_{\gamma j}\,,
    \end{equation}
    and as a consequence employing a different tensor $\mathbf{B}'$ for the stochastic term:
    \begin{equation}
        B'_{\alpha i\beta j} = \sqrt{\frac{2\beta_{T,\alpha\beta}k_B T}{3V\tau_p}}h_{\beta i}\,\delta_{\alpha j}\,.
    \end{equation}
     With calculations similar to those reported in Appendix \ref{appendix_derivation}, the equations that one obtains with these two tensors are the following:
     \begin{align}
         \text{d}h_{\alpha i} = -&\frac{1}{3\tau_p} \Bigg[\sum_\gamma \beta_{T,\alpha\gamma}\Big( P_0\delta_{\alpha\gamma} - P_{\text{int},\alpha\gamma} \Big)h_{\gamma i} - \beta_{T,\alpha\alpha}\frac{k_B T}{V}h_{\alpha i}\Bigg]\,\text{d}t \nonumber\\
         &+ \sqrt{\frac{2\beta_T k_B T}{3V\tau_p}} \sum_{\gamma}\sqrt{\beta_{T,\alpha\gamma}}\,h_{\gamma i}\,\text{d}W_{\alpha\gamma}\,,
    \end{align}
    Including the correction for a generic external stress, the equations become:
    \begin{align}
         \text{d}h_{\alpha i} = -&\frac{1}{3\tau_p} \Bigg[\sum_\gamma \beta_{T,\alpha\gamma}\Big( P_0\delta_{\alpha\gamma} - P_{\text{int},\alpha\gamma} \Big)h_{\gamma i} - \beta_{T,\alpha\alpha}\frac{k_B T}{V}h_{\alpha i} + \frac{1}{V}\big(\bm{\beta}_T\odot\mathbf{h}\,\bm{\Sigma}\,\mathbf{h}^T \mathbf{h} \big)_{\alpha i}\Bigg]\,\text{d}t \nonumber\\
         &+ \sqrt{\frac{2 k_B T}{3V\tau_p}} \sum_{\gamma}\sqrt{\beta_{T,\alpha\gamma}}\,h_{\gamma i}\,\text{d}W_{\alpha\gamma}\,, 
    \end{align}
    where the notation $\mathbf{A}\odot\mathbf{B}$ is used for the element-wise (or \emph{Hadamard}) product, namely
    \begin{equation}
        \big(\mathbf{A}\odot\mathbf{B}\big)_{ij} = A_{ij}\,B_{ij}\,.
    \end{equation}

    \chapter{Implementations and tests}\label{implementations}
    The anisotropic SCR method developed in \cref{chapter:formulation} was tested with three different implementations, built by modifying the MD software SimpleMD, GROMACS 2021.2 \cite{gromacs} and LAMMPS (2 July 2021 release) \cite{lammps}. All the implementations employ the Euler integrator described in \cref{euler_integrator} and eliminate box rotations by constraining $\mathbf{h}$ to be a triclinic upper triangular matrix, first by propagating all the nine box components, and then by rotating back the cell vectors using the procedure described in Appendix \ref{appendix_rotations}.
    
    The SimpleMD program is an educational code to perform MD simulations of a Lennard-Jones (LJ) system, namely using a simple pair potential of the form
    \begin{equation}
        V_{LJ}(q_{ij}) = 4\varepsilon\left[\left(\frac{\sigma}{q_{ij}}\right)^{12} - \left(\frac{\sigma}{q_{ij}}\right)^{6} \right]\,,
    \end{equation}
    where $q_{ij} = |\mathbf{q}_i - \mathbf{q}_j|$.
    The modified code includes both the isotropic and the anisotropic implementations of the SCR algorithm. The available temperature coupling methods are the Langevin and the SVR thermostats, discussed in \cref{langevin_thermo,SVR} respectively. The scheme applied to propagate the SCR equations together with Hamilton's equations and the selected thermostat is the one referred to as \emph{Trotter-based integrator} in \cite{crescale_iso}:\\
    
    \begin{algorithm}[H]\label{trotter_based_integrator}
    \SetAlgoLined
     apply thermostat for $\Delta t/2$\;
     propagate momenta for $\Delta t/2$: $\mathbf{p}_i \gets \mathbf{p}_i + \mathbf{F}_i\Delta t/2$\;
     \label{modulo}
     \If{$\big(\text{n}_{\text{MD}}\text{ \% n}_s =0\big)$} {
        apply SCR barostat to compute $\bm{\mu}$\; \label{line3}
        rescale box matrix: $\mathbf{h} \gets \bm{\mu}\mathbf{h}$\;
        rescale and propagate positions with (rescaled) momenta: \cref{rescale_propagate_positions}\; \label{line6}
        (rescale momenta: $\mathbf{p}_i^T \gets \mathbf{p}_i^T\bm{\mu}^{-1}$)\; \label{line7}
    }
     recompute forces\;
     propagate momenta for $\Delta t/2$: $\mathbf{p}_i \gets \mathbf{p}_i + \mathbf{F}_i\Delta t/2$\;
     apply thermostat for $\Delta t/2$\;
     \caption{Trotter-based integrator implemented in SimpleMD}
    \end{algorithm}
    \vspace{0.4cm}
    \noindent The symbol \% at \cref*{modulo} stands for the modulo operation, meaning that the barostat is applied once every $n_s$ steps, and the brackets at \cref*{line6,line7} refer to the formulation where both positions and momenta are rescaled. Step \ref*{line6} is performed with a further Trotter splitting, where positions are first propagated with momenta for half time step, then rescaled and finally propagated again for half time step. Depending on the formulation chosen, this splitting results in: 
    \begin{subequations}\label[pluralequation]{rescale_propagate_positions}
    \begin{align}
        \mathbf{q}_i &\gets \bm{\mu}\,\mathbf{q}_i +  \frac{\bm{\mu}\mathbf{p}_i}{2m_i} \Delta t + \frac{\mathbf{p}_i^T\bm{\mu}^{-1}}{2m_i} \Delta t \hspace{0.4cm}\text{(if momenta are rescaled)}\,, \\
        \mathbf{q}_i &\gets \bm{\mu}\,\mathbf{q}_i + \left(\bm{\mu}+\mathbf{I}\right)\frac{\mathbf{p}_i}{2m_i} \Delta t \hspace{0.54cm}\text{(if momenta are \emph{not} rescaled)}\,.
    \end{align}
    \end{subequations}
    The SimpleMD implementation allows to apply the anisotropic SCR barostat both with and without cell rotations, and also includes the time-reversible integration scheme outlined in \cref{sec:TR}. 
    A comparison between the main features of the three implementations of the anisotropic SCR algorithm is reported in \cref{tab:implementations}.
    
    \begin{table}[h!]
    \centering
    \caption{Comparison between the SimpleMD\footnote{\href{https://github.com/bussilab/crescale.git}{https://github.com/bussilab/crescale.git}}, GROMACS\footnote{\href{https://github.com/bussilab/crescale-gromacs.git}{https://github.com/bussilab/crescale-gromacs.git}} and LAMMPS\footnote{\href{https://github.com/bussilab/crescale-lammps.git}{https://github.com/bussilab/crescale-lammps.git}} implementations of the anisotropic SCR algorithm (top lines), with additional software specifities (bottom lines).}
    \begin{threeparttable}
    \label{tab:implementations}
        \begin{tabular}{l c c c}
        \toprule[0.5pt]\toprule[0.5pt]
        {} & \small\textbf{SimpleMD}       & \small\textbf{GROMACS}   & \small\textbf{LAMMPS} \\\midrule
        Number of propagated variables      & 9 & 9 & 9 \\
        Multiple time step                  & yes  & yes & no \\
        Rotations or not                    & both  & no  & no \\
        Rescaling of momenta or not         & both  & yes  & yes \\
        Time-reversible implementation      & yes  &  no  & no \\
        Isotropic implementation            &  yes  &  yes  & yes \\
        Semi-isotropic implementation       &  no  & yes & yes \\
        Coupling of arbitrary components    & no  & no & yes \\\midrule
        Electrostatics                      & no  & yes & yes \\
        Constraints                         & no  & yes & yes \\
        Potentials for solid-state materials & no & no & yes \\
        Parrinello-Rahman barostat          & no & yes & no \\
        MTTK barostat                       & no & yes\tnotex{tnote:MTTK_gromacs} & yes
        \\\bottomrule[0.5pt]\bottomrule[0.5pt] 
        \end{tabular}
        \begin{tablenotes}
          \item\label{tnote:MTTK_gromacs}Only isotropic version and in absence of constraints
        \end{tablenotes}
    \end{threeparttable}
    \end{table}
    
    The tests discussed in the following are performed on a variety of crystal systems, including a Lennard-Jones (LJ) solid, ice, gypsum (chemical formula: $\text{(CaSO)}_\text{4}\cdot\text{2(H}_\text{2}\text{O)}$) and gold (Au). Since in all the simulations the reference distributions for $\mathbf{h}$ are unknown, the results are validated against the reference barostats available in the MD software employed. 
    
    \section{Lennard-Jones crystal}\label{LJ_crystal}
    The first tests were performed with the modified SimpleMD program on a LJ crystal with $N = 256$ particles arranged in a face-centered-cubic (fcc) lattice. For each choice of the external parameters ($\tau_p$ and $n_s$), the system was simulated for $10^6$ steps with time step $\Delta t = 0.005$, external hydrostatic pressure $P_0=1$ and temperature $T=0.1$, using a SVR thermostat with relaxation time $\tau_T=0.05$. All the parameters are reported here in reduced LJ units. The input isothermal compressibility  was set to $\beta_T = 0.3$, as estimated for the same system in the liquid phase at $T=1.5$ \cite{crescale_iso}. The simulations were carried out by using a cut-off distance $r_{\text{cut}}=2.5$ for the interactions and accumulating statistics at each step. 
    
\begin{figure}[h!]
    \centering
    \begin{subfigure}[b]{0.45\textwidth}
        \centering
        \includegraphics[height = 4.1 cm]{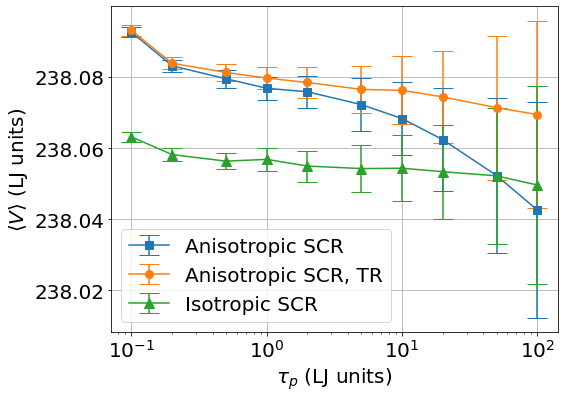}\\[1 ex]
        \caption{\RaggedLeft Volume average vs. $\tau_p\hspace{0.5cm}$}
        \label{fig:volume_taups_LJ}
    \end{subfigure}
    \hfil
    \begin{subfigure}[b]{0.45\textwidth}
        \centering
        \includegraphics[height = 4.1cm]{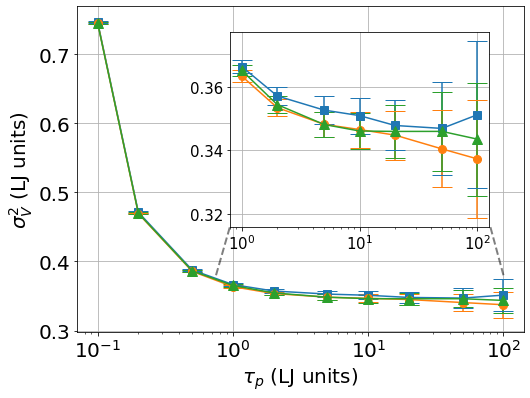}\\[1 ex]
        \caption{\RaggedLeft Volume variance vs. $\tau_p\hspace{0.7cm}$}
        \label{fig:std_taups_LJ}
    \end{subfigure}
    \caption{Results from the simulations of a LJ crystal system in the $N\mathbf{S}T$ ensemble, as a function of $\tau_p$. Error bars were computed by means of block bootstrap analysis (see Appendix \ref{appendix_block_bootstrap}). Blue and orange lines refer to the anisotropic simulations using respectively the Euler and the time-reversible (TR) integrators in SimpleMD.}
    \label{fig:taups_LJ}
    \end{figure}

    First, the behaviour of the volume distribution was studied using different values of $\tau_p$ at fixed barostat stride $n_s = 1$. Results were validated by comparing the volume distributions generated by the anisotropic barostat, both in the Euler and the time-reversible (TR) implementations, with the ones generated by the isotropic barostat, employing the same input parameters (see \cref{fig:taups_LJ}). All the analysis were carried out by discarding the first $10^5$ steps of the simulations.
    
    \noindent As \cref{fig:volume_taups_LJ} shows, the average volumes obtained with the isotropic and the anisotropic SCR barostats converge by increasing $\tau_p$, although this behaviour is more accentuated in the Euler case. The reason of this behaviour is that integration errors become more relevant when $\tau_p$ is small, since decreasing this parameter is equivalent to increase the time step at fixed relaxation time. In this regime, the integration errors on the average volume appear identical for the Euler and the TR integrators. As expected, the average volume appears less dependent on $\tau_p$ for the distributions generated by the isotropic barostat; indeed, integration errors in the anisotropic case generate larger deviations from the exact volume distribution, as they result from the propagation of nine variables instead of one. These deviations have a systematic nature and do not enter in the error bars, which only depend on the autocorrelation time of the volume. As shown in \cref{fig:acfs_volume_LJ} and discussed in detail in Appendix \ref{appendix_relaxation_autocorrelation}, this autocorrelation time is completely dictated by the barostat relaxation time, namely it can be effectively identified with $\tau_p$. Since the uncertainty of an average value is known to increase with the autocorrelation time of the series from which it is computed (see Appendix \ref{appendix_autocorrelation}), the behaviour of the error bars of $\langle V\rangle$ is meaningful. Looking at \cref{fig:std_taups_LJ}, the volume fluctuations generated by the anisotropic SCR method appear cosnistent with the ones obtained with the isotropic barostat, regardless of the relaxation time employed. The deviations of the volume variance due to integrations errors in the small-$\tau_p$ regime are more evident for the Euler integrator, suggesting an increased accuracy in the TR scheme. 
    
    Only for the anisotropic integrators, a further comparison was carried out between some relevant distributions extracted from the box matrix components, namely for the squared moduli of the cell vectors $|\mathbf{a}|^2,\,|\mathbf{b}|^2,\,|\mathbf{c}|^2$ and their three scalar products $\mathbf{a}\cdot\mathbf{b}\,, \mathbf{a}\cdot\mathbf{c}\,, \mathbf{b}\cdot\mathbf{c}$. Also in this case (see \cref{fig:box_comp_LJ,fig:box_vars_LJ} in the Appendices) the performances of the Euler and the TR integrators appear almost identical.
    
    For the same simulations, the volume autocorrelation function (ACF) was computed for each value of $\tau_p$ (see \cref{fig:acfs_volume_LJ}). Note that, since the input isothermal compressibility is the one in the liquid phase, the real $\beta_T$ for the crystal system is larger than the input one. As a consequence, the values of $\tau_p$ are smaller than the actual volume relaxation times. Anyway, the correct values can be calculated \emph{a posteriori} with the simple rescaling
    \begin{equation}
        \tau_{p,\text{exp}} = \frac{\beta_{T,\text{exp}}}{\beta_{T,\text{input}}}\tau_{p,\text{input}}\,,
    \end{equation}
    after estimating the real isothermal compressibility according to \cref{beta_T_formula}, resulting in $\beta_{T,\text{exp}}\simeq 0.015$. 
    
    \noindent It is also possible to study how fast an estimate of the volume variance decorrelates as a function of $\tau_p$, since this quantity is essential to calculate physical observables such as the isothermal compressibility. Note that in principle it is \emph{not} possible to define the ACF of the variance, since this quantity is a global property of the trajectory and is not calculated as the average of consecutive "instantaneous variances". If we assume that the average volume $\overbar{V}$ is known exactly, however, the variance can be computed as the mean value over the time series $\{(V_1-\overbar{V})^2),...,(V_n-\overbar{V})^2 \}$, namely 
    \begin{equation}
        \sigma_V^2 = \frac{1}{n}\sum_{j=1}^n\Big(V_j-\overbar{V}\Big)^2\,,
    \end{equation}
    where $j$ is the index for the MD step and $n$ is the total number of samplings. In the unbiased estimator of the variance $n$ should be replaced by $n-1$, but this correction is negligible for a large number of samplings. The ACFs of the volume variance are shown \cref{fig:acfs_var_LJ}; for each $\tau_p$, the value of $\overbar{V}$ is taken as the average over the entire volume trajectory. 
    \begin{figure}[h!]
    \centering
    \begin{subfigure}[b]{0.45\textwidth}
        \centering
        \includegraphics[height = 4.4 cm]{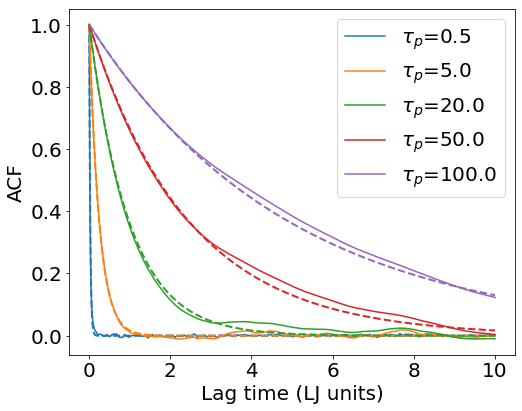}\\[1 ex]
        \caption{\RaggedLeft Volume ACFs$\hspace{1.4cm}$}
        \label{fig:acfs_volume_LJ}
    \end{subfigure}
    \hfil
    \begin{subfigure}[b]{0.45\textwidth}
        \centering
        \includegraphics[height = 4.4 cm]{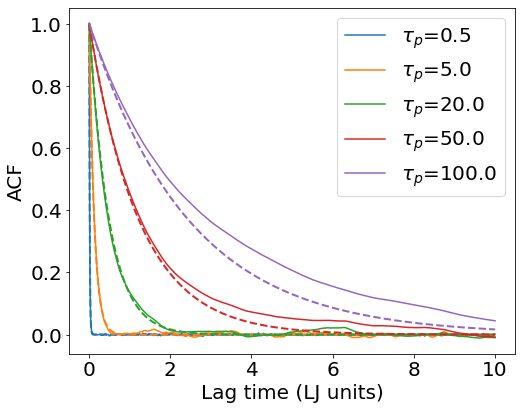}\\[1 ex]
        \caption{\RaggedLeft ACFs of volume variance$\hspace{0.6cm}$}
        \label{fig:acfs_var_LJ}
    \end{subfigure}
    \caption{Left panel: ACFs of the volume time series from the simulations of a LJ crystal with the anisotropic SCR barostat, using the Euler integrator in SimpleMD. The values of $\tau_p$ reported in the legend are the input ones, while the dashed lines represent the exponentially decaying functions $e^{-t/\tau_{p,\text{exp}}}$, where the correct values of the relaxation times are computed \emph{a posteriori}. Right panel: variance ACFs computed under the assumption of known volume averages. The dashed lines are in this case the exponential functions $e^{-2t/\tau_{p,\text{exp}}}$, which represent the anlytical ACFs of the variance in the limit case of a Gaussian-distributed volume following a Langevin dynamics (see Appendix \ref{appendix_relaxation_autocorrelation}).}
    \label{fig:ACFs_LJ}
    \end{figure}
    
    \noindent Note that these ACFs decay faster than the volume ones (ideally, with halved characteristic time); as a consequence, a calculation of the variance of the volume converges faster than a calculation of its average when the SCR barostat is employed. Similar results were obtained in the isotropic case. The ACFs shown in \cref{fig:ACFs_LJ} are obtained with the anisotropic barostat in the Euler integration scheme; employing the TR integrator, the deviations of the variance ACFs in the large-$\tau_p$ regime are no present anymore (see \cref{fig:ACFs_LJ_TR} in the Appendices). 
    
    For each tested value of $\tau_p$, the ACFs of the squared moduli and the scalar products of the box vectors were also computed, resulting qualitatively in the same behaviour shown in \cref{fig:ACFs_LJ}. These ACFs go to zero slower than the volume ACFs, but their limiting analytical behaviour cannot be easily predicted. As an example, see for instance \cref{fig:ACFs_mod2_LJ} in the Appendices.
    
\begin{figure}[h!]
        \centering
        \includegraphics[width = 0.55\textwidth]{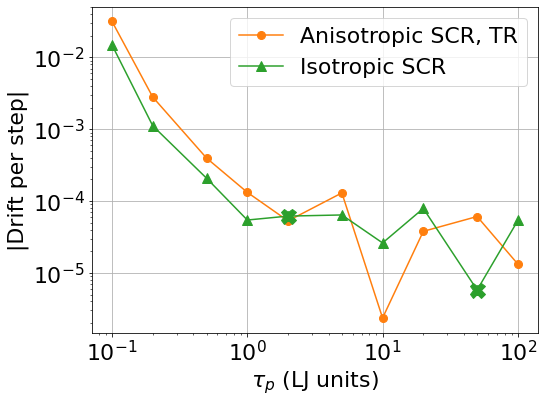}
        \caption{Absolute value of the effective energy drift per step from the simulations of the LJ crystal in SimpleMD. Each value is the slope of a line interpolating the effective energy drift on the entire trajectory. The green $\times$ markers are actually negative values, for which the absolute value is taken in order to represent all data on a logarithmic scale. The drift consistently increase in the small-$\tau_p$ regime, i.e. when integration errors have a larger impact. For $\tau_p \gtrapprox 1$, the energy drift appears irrelevant on the time scale of the trajectories, resulting in noisy estimates of the slope above and around zero.}
        \label{fig:slopes_LJ}
    \end{figure}

    For the same simulations, the effective energy drift is calculated along the trajectories generated by the isotropic and the TR anisotropic implementations (see \cref{fig:slopes_LJ}). 
    
    Both \cref{fig:taups_LJ,fig:ACFs_LJ} are obtaind from simulations where momenta are rescaled and, in the anisotropic case, rotations are eliminated. However, all the possible four combinations of these two options were tested, showing equivalent results to the ones discussed here. 
    
\begin{figure}[h!]
    \centering
    \begin{subfigure}[b]{0.45\textwidth}
        \centering
        \includegraphics[height = 4.2 cm]{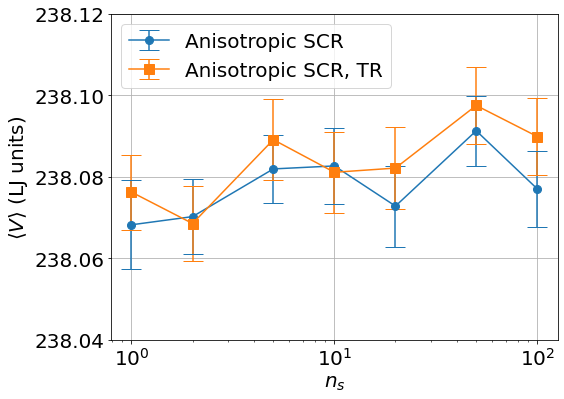}\\[1 ex]
        \caption{\RaggedLeft Volume average vs. $n_s\hspace{0.5cm}$}
        \label{fig:avg_vs_stride}
    \end{subfigure}
    \hfil
    \begin{subfigure}[b]{0.45\textwidth}
        \centering
        \includegraphics[height = 4.2 cm]{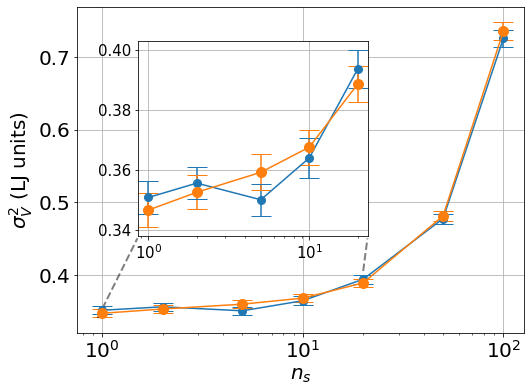}\\[1 ex]
        \caption{\RaggedLeft Volume variance vs. $n_s\hspace{0.5cm}$}
        \label{fig:std_vs_stride}
    \end{subfigure}
    \caption{Average and variance of volume distributions as a function of the barostat stride $n_s$, using the anisotropic SCR barostat in SimpleMD. Error bars are computed with block bootstrap analysis. The simulations were performed with $\tau_p = 10$.}
    \label{fig:volume_vs_stride}
    \end{figure}

    Additional simulations were performed at fixed volume relaxation time ($\tau_p = 10$) and different barostat strides $n_s$, in order to validate the application of a multiple-time-step approach to the anisotropic SCR barostat. The results are reported in \cref{fig:volume_vs_stride}.
    Interestingly, the average volume appears to be well reproduced even for large barostat strides (up to $n_s = 100)$, while systematic errors for the volume fluctuations become evident for $n_s \gtrapprox 20$. Also in this test, the results obtained with the Euler and the TR integration scheme appear equivalent.
    
    Similar simulations were also performed with the modified version of GROMACS 2021.2, using a larger crystal with $N=1000$ Argon atoms, employing the GROMOS 54A7 force field and setting $T=5$ K and $P_0 = 1$ bar. The crystal structure was obtained first by equilibrating the system in the $NVT$ ensemble for 500 ps, and then with a simulated annealing protocol in the $NP_0 T$ ensemble from $T = 80$ K to $T=5$ K, resulting in an hexagonal-close-packed (hcp) structure with defects. Scanning different values of $\tau_p$, production runs of $10$ ns were carried out using both the SCR barostat and the other pressure coupling methods available in GROMACS (see \cref{tab:implementations}), except for the Berendsen one, which is known to generate wrong volume fluctuations. Temperature was controlled by a SVR thermostat with a relaxation time $\tau_T = 0.01$ ps. Both thermostat and barostat were applied every 10 steps, and statistics were accumulated every 20 steps (40 fs). The time step was set to $\Delta t = 2$ fs, the cut-off distance for the LJ interactions to 1 nm and the input isothermal compressibility to $\beta_T = 3.53\times 10^{-5}$ bar$^{-1}$, using a rough estimation on a preliminary run. Instead of using the standard GROMACS leap frog integrator, the simulations with the Parrinello-Rahman and the MTTK barostats were carried out with a velocity Verlet scheme, which is more accurate for these coupling methods and partially reduces (but does not eliminate) the pathological behaviours discussed in the following. All the analysis were carried out by discarding the first 2.5$\times$10$^5$ steps of the simulations.
    
\begin{figure}[h!]
    \centering
    \begin{subfigure}[b]{0.45\textwidth}
        \centering
        \includegraphics[height = 4.9 cm]{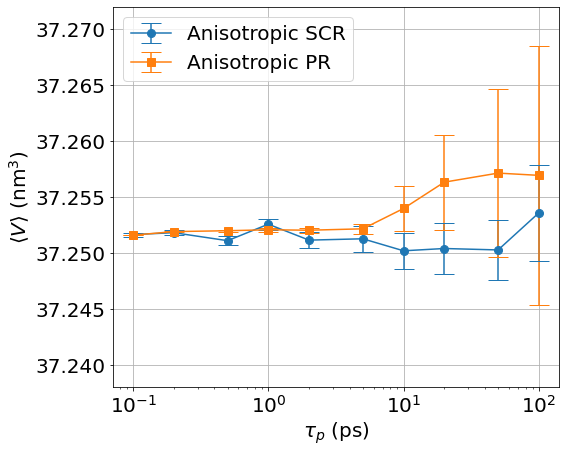}\\[1 ex]
        \caption{\RaggedLeft Anisotropic barostats$\hspace{0.5cm}$}
        \label{fig:avg_aniso_vs_taup_LJ_GMX}
    \end{subfigure}
    \hfil
    \begin{subfigure}[b]{0.45\textwidth}
        \centering
        \includegraphics[height = 4.9 cm]{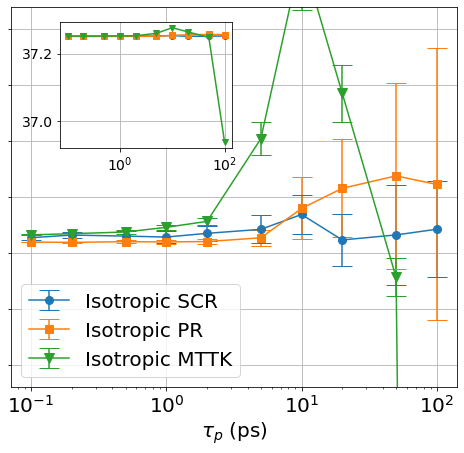}\\[1 ex]
        \caption{\centering Isotropic barostats}
        \label{fig:avg_iso_vs_taup_LJ_GMX}
    \end{subfigure}
    \caption{Average of volume distributions as a function of the barostat relaxation time $\tau_p$, using the barostats available in GROMACS 2021.2 (SCR: Stochastic Cell Rescaling, PR: Parrinello-Rahman, MTTK: Martyna-Tuckerman-Tobias-Klein). In the PR and MTTK methods, $\tau_p$ is related to the barostat mass $W$ as $W\propto \tau_p^2$.}
    \label{fig:avg_vs_taup_LJ_GMX}
    \end{figure}

\begin{figure}[h!]
    \centering
    \begin{subfigure}[b]{0.45\textwidth}
        \centering
        \includegraphics[height = 5.1 cm]{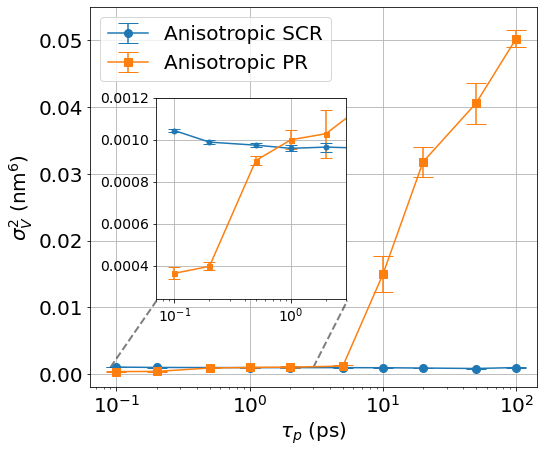}\\[1 ex]
        \caption{\RaggedLeft Anisotropic barostats$\hspace{0.5cm}$}
        \label{fig:std_aniso_vs_taup_LJ_GMX}
    \end{subfigure}
    \hfil
    \begin{subfigure}[b]{0.45\textwidth}
        \centering
        \includegraphics[height = 5.1 cm]{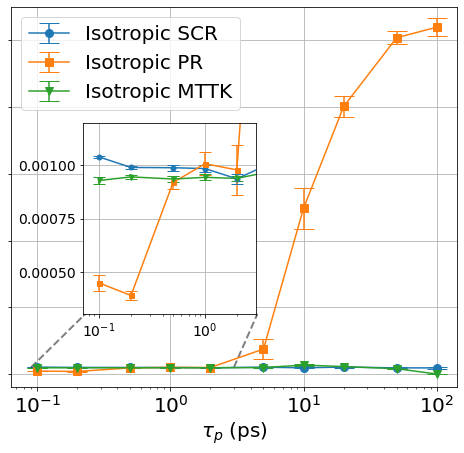}\\[1 ex]
        \caption{\centering Isotropic barostats}
        \label{fig:std_iso_vs_taup_LJ_GMX}
    \end{subfigure}
    \caption{Variance of volume distributions as a function of the barostat relaxation time $\tau_p$. The zoomed regions of the plots refer to the range of $\tau_p$ values where the PR and the MTTK methods do not produce pathological volume distributions.}
    \label{fig:std_vs_taup_LJ_GMX}
    \end{figure}

    Averages and variance of the volume distributions from the GROMACS simulations are shown in \cref{fig:avg_vs_taup_LJ_GMX,fig:std_vs_taup_LJ_GMX}. 
    
    \noindent As a necessary observation, the time constant $\tau_p$ is not equivalently defined for the different barostats employed; as a consequence, comparing the distributions generated by different methods for a given value of $\tau_p$ is not completely meaningful. However, it is useful to study how the volume distributions generated by different barostats are affected by equivalent variations of $\tau_p$. As \cref{fig:avg_aniso_vs_taup_LJ_GMX} shows, the average volume reproduced by the anisotropic SCR barostat appears perfectly in agreement with the one obtained from the anisotropic PR method, as well as in the case of isotropic volume fluctuations (see \cref{fig:avg_iso_vs_taup_LJ_GMX}). Moreover, in both the cases the SCR method shows a reduced sensitivity to the input parameter $\tau_p$ than the PR barostat, for which sampling problems for large $\tau_p$ are more evident. Using the MTTK method, which works only in the isotropic case in GROMACS 2021.2, these sampling problems are already present from $\tau_p \gtrapprox 2$ ps. In general, above $\tau_p\approx 5$ the volume trajectories generated by the PR and the MTTK barosats do not equilibrate properly and show non-stationary behaviours (see Appendix \cref{appendix_pathological_distributions}). As a consequence, the results of these simulations are reliable and actually comparable with the SCR method only up to $\tau_p \approx 2-5$. Looking at \cref{fig:std_vs_taup_LJ_GMX}, the SCR method appears more robust than the PR barostat in reproducing the correct volume fluctuations against variations of $\tau_p$. Moreover, the PR method is more affected by accuracy problems for small relaxation times. It has not to be excluded that the errors observed for the MTTK and the PR barostats - especially in the anisotropic case - could depend on the technical details of the GROMACS implementation. 
    
    \begin{figure}[h!]
        \centering
        \includegraphics[width = 0.95\textwidth]{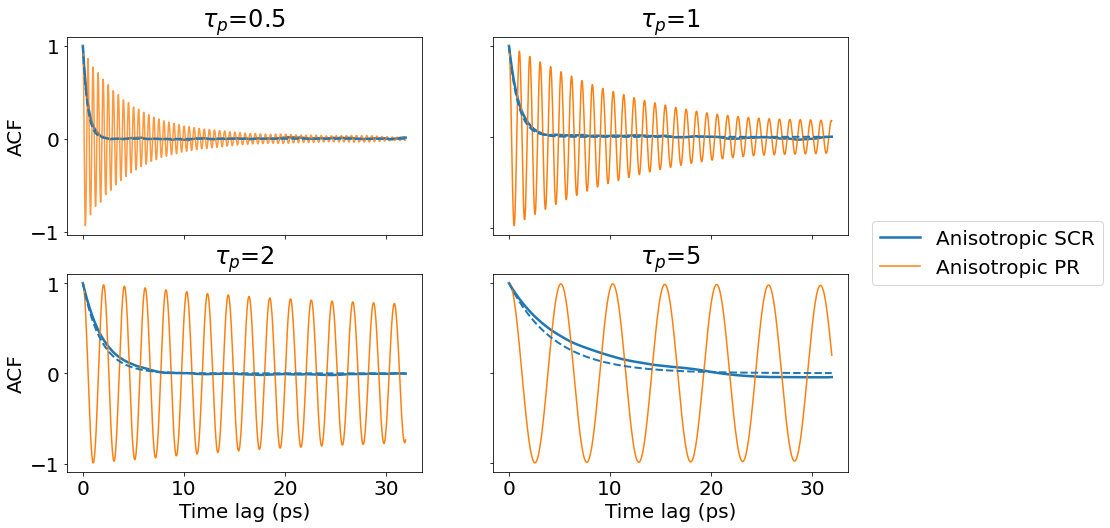}
        \caption{Comparison of the volume ACFs obtained with the anisotropic SCR barostat and the PR method in GROMACS. The dashed lines represent the decaying functions $\text{exp}(-t/\tau_{p,\text{exp}})$, where $\tau_{p,\text{exp}}$ is computed from the correct value of the isothermal compressibility.}
        \label{fig:ACFs_volume_LJ_GMX}
    \end{figure}
    
\begin{figure}[h!]
    \centering
    \begin{subfigure}[b]{0.44\textwidth}
        \centering
        \includegraphics[height = 3.9 cm]{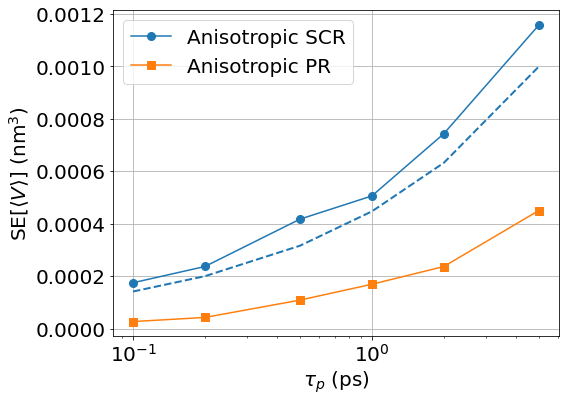}\\[1 ex]
        \caption{\RaggedLeft Errors of volume average$\hspace{0.3cm}$}
        \label{fig:SE_avg_LJ_GMX}
    \end{subfigure}
    \hfil
    \begin{subfigure}[b]{0.44\textwidth}
        \centering
        \includegraphics[height = 3.9 cm]{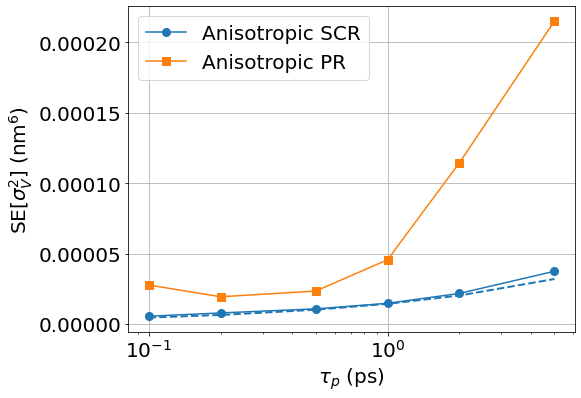}\\[1 ex]
        \caption{\RaggedLeft Errors of volume variance$\hspace{0.2cm}$ }
        \label{fig:SE_std_LJ_GMX}
    \end{subfigure}
    \caption{Standard errors of volume average and variance computed with block bootstrap analysis (see Appendix \ref{appendix_block_bootstrap}). The dashed lines represent the expected behaviour as a function of $\tau_p$, computed with \cref{standard_error_autocorr}.}
    \label{fig:SEs_LJ_GMX}
    \end{figure}

    For the anisotropic SCR barostat, the ACFs of the volume show the same behaviour observed in the SimpleMD simulations, with autocorrelation times scaling linearly with the input values of $\tau_p$. 
    \Cref{fig:ACFs_volume_LJ_GMX} shows a comparison between the volume ACFs obtained with the anisotropic SCR and PR barostats, considering values of $\tau_p$ within the reliable range previously discussed. At fixed $\tau_p$, the anisotropic SCR barostat appears more efficient in terms of decorrelation speed. However, this does \emph{not} imply that the uncertainty associated to an estimate of the average volume is larger with the PR method. Indeed, the actual autocorrelation time related to the statistical uncertainty of $\langle V\rangle$ is defined as the integral of the ACF (see Appendix \ref{appendix_autocorrelation}); as a consequence, if the damped oscillations of the ACF are symmetric with respect to zero, the integral can be very small even if the envelope of this function does not go to zero as fast as in the SCR case. This is the reason why the standard errors on the average volume, estimated with a block bootstrap analysis, are smaller from the trajectories generated by the PR barostat for $\tau_p< 5$ (see \cref{fig:SE_avg_LJ_GMX}). On the other hand, the ACFs of the volume variance from the PR trajectories do not show the same symmetry (see \cref{fig:ACFs_var_LJ_GMX} in the Appendices), resulting in an integrated autocorrelation time that is larger than in the SCR trajectories. As a consequence, the standard error on the estimate of volume fluctuations is smaller in the SCR simulations (see \cref{fig:SE_std_LJ_GMX}).
    
    \section{Crystal Ice I$_h$}\label{IceIh_analysis}
\begin{figure}[h!]
        \centering
        \includegraphics[width = 0.34\textwidth]{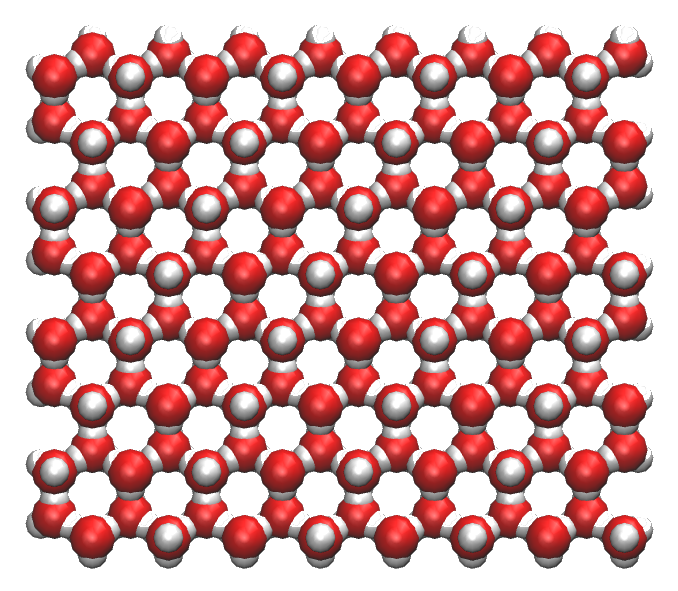}
        \caption{\centering Orthographic perspective of the simulated hexagonal crystal ice I$_h$.}
        \label{fig:IceIh}
    \end{figure}

\begin{figure}[h!]
    \centering
    \begin{subfigure}[b]{0.44\textwidth}
        \centering
        \includegraphics[height = 4.3 cm]{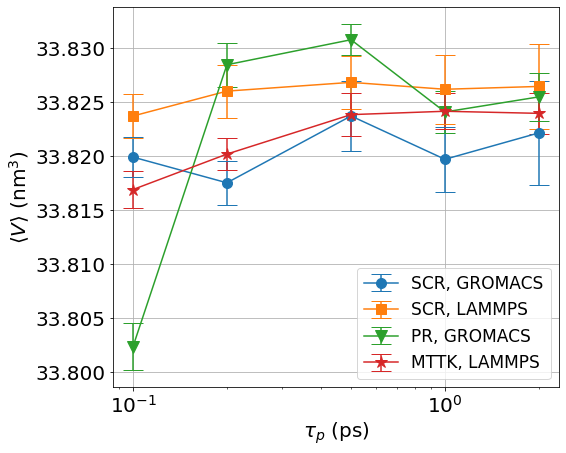}\\[1 ex]
        \label{fig:avg_IceIh}
    \end{subfigure}
    \hfil
    \begin{subfigure}[b]{0.44\textwidth}
        \centering
        \includegraphics[height = 4.3 cm]{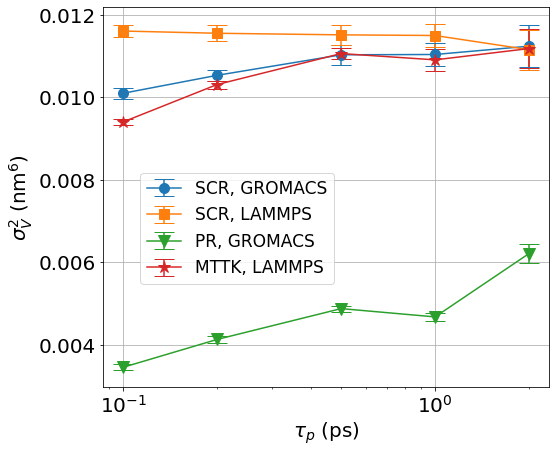}\\[1 ex]
        \label{fig:std_IceIh}
    \end{subfigure}
    \caption{Results from the simulations of the Ice$_{h}$ system. Average (left panel) and variance (right panel) of volume distributions are shown as functions of the barostat relaxation time $\tau_p$, comparing the anisotropic SCR barostat with the anisotropic pressure coupling methods available in GROMACS 2021.2 and LAMMPS (2 July 2021 release).}
    \label{fig:vol_IceIh}
    \end{figure}

    In order to test the performance of the anisotropic SCR barostat in presence of constraints, simulations were performed on a system of Ice I$_h$ (see \cref{fig:IceIh}) composed of $N=3072$ atoms, using the TIP4P/Ice model \cite{tip4pice}. 
    In GROMACS, for each $\tau_p$ the crystal system was simulated for 4 $ns$ with a time step of $1$ fs, accumulating statistics every 10 steps (10 fs) and applying both barostat and thermostat at each step. The simulations were carried out with external hydrostatic pressure $P_0 = 1$ bar and external temperature $T = 270$ K, employing a SVR thermostat with relaxation time $\tau_T = 0.1$ ps. The input isothermal compressibility, estimated over a preliminary run of $1$ ns, was set to $\beta_T = 9.49\times 10^{-6}$ bar$^{-1}$. All the simulations were performed by employing as integrator of Hamilton's equations a modified version of the velocity Verlet algorithm, where the kinetic energy is determined as the average of the two half step kinetic energies. The use of this integrator appears to reduce the pathological behaviour of some distributions generated by the PR method, as discussed in \cref{LJ_crystal}. In all the simulations, electrostatics was treated with a Particle-Mesh Ewald (PME) approach. The same system was also simulated in LAMMPS with equivalent MD options, and performing a comparison with the anisotropic MTTK barostat. The behaviour of average and variance of the volume distributions, computed by discarding the first 4$\times 10^4$ steps, are shown in \cref{fig:vol_IceIh} for different $\tau_p$ values.
    
    \noindent Both in GROMACS and in LAMMPS, the SCR barostat generates volume distributions that are consistent with the MTTK ones. Moreover, the LAMMPS implementation of the anisotropic SCR method is the one showing the smallest sensibility to variations of the relaxation time. The anisotropic PR barostat in GROMACS, instead, generates volume fluctuations that are not consistent with the other pressure coupling methods, and it is the one showing the largest deviation for small $\tau_p$. Similar results are obtained by studying how the squared moduli of the box vectors and their scalar products are distributed (see \cref{fig:box_comp_ice,fig:box_vars_ice} in the Appendices). 
    
\begin{figure}[h!]
        \centering
        \includegraphics[width = 0.95\textwidth]{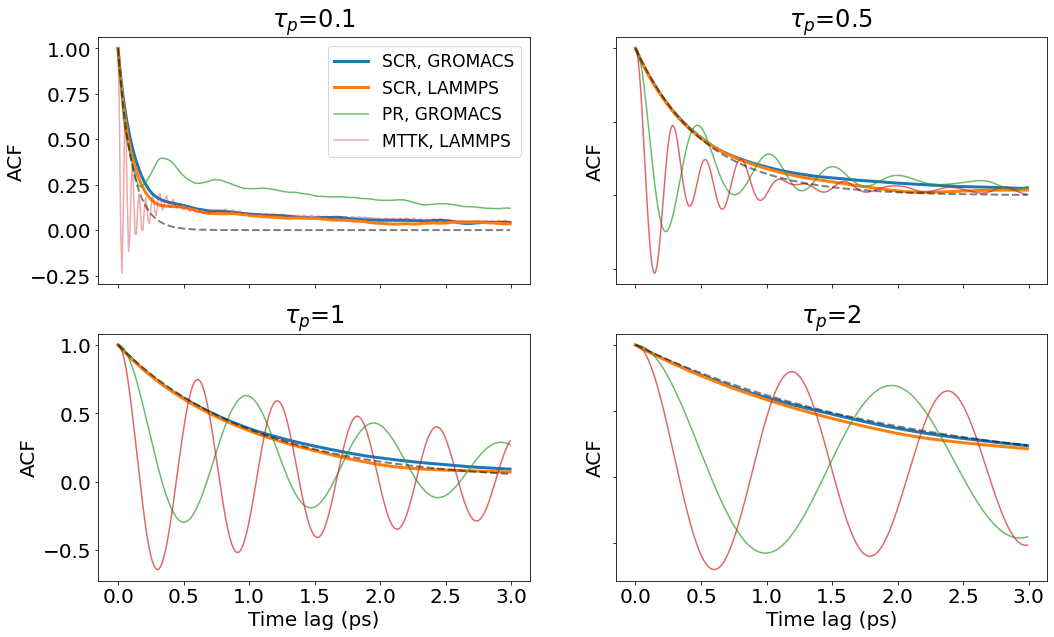}
        \caption{ACFs of the volume for different values of $\tau_p$, compared among the pressure coupling methods employed in the simulations of the Ice I$_h$ crystal.}
        \label{fig:ACFs_avgs_Ice}
    \end{figure}

    The ACFs of the volume for the SCR simulations are consistent with the expected exponential behaviour as a function of $\tau_p$ (see \cref{fig:ACFs_avgs_Ice}). The largest deviation is shown for $\tau_p = 0.1$ ps, when the observed decaying is slower than the reference exponential one. This slowdown of the first-order relaxation occurs when $\tau_p$ is of the same order or smaller than the timescale in the rearrangement of atoms, which then becomes the bottleneck for volume dynamics. 
    
    \noindent Also the ACFs of the volume variance match the expected decaying functions (see \cref{fig:ACFs_vars_Ice} in the Appendices). In this case, statistical errors associated to average and variance of volume distributions are larger for the SCR trajectories, at fixed $\tau_p$ (see \cref{fig:SE_Ice} in the Appendices). However, as already commented for the simulations of the Argon crystal, the results obtained from different barostats are not directly comparable for the same value of the relaxation time, since it does not exist a clear mapping between the corresponding definitions of $\tau_p$. Moreover, since the SCR method performs well up to $\tau_p=0.1$, this analysis suggests that small values of the relaxation time could be used to reduce the statistical error without introducing systematic ones.
    
\begin{figure}[h!]
    \centering
    \begin{subfigure}[b]{0.44\textwidth}
        \centering
        \includegraphics[height = 3.8 cm]{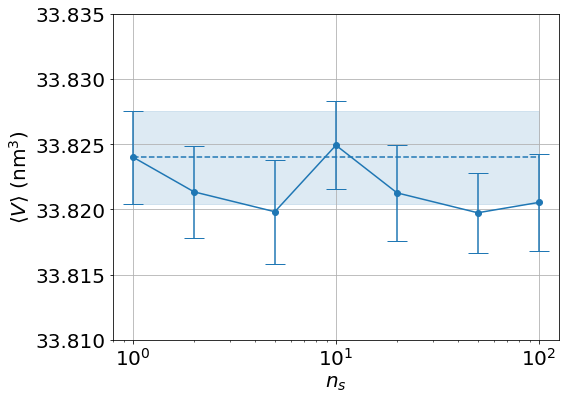}\\[1 ex]
    \end{subfigure}
    \hfil
    \begin{subfigure}[b]{0.44\textwidth}
        \centering
        \includegraphics[height = 3.8 cm]{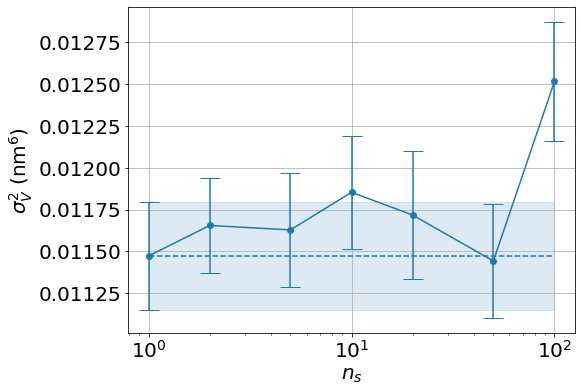}\\[1 ex]
    \end{subfigure}
    \caption{Results from GROMACS simulations of the ice I$_h$ system, employing the anisotropic SCR barostat at fixed relaxation time $\tau_p=1$ ps. Average (left panel) and variance (right panel) of volume distributions are shown as functions of the barostat stride $n_s$. The error region, computed with block bootstrap analysis, is associated to the smallest stride tested ($n_s = 1$).}
    \label{fig:variable_stride_ice}
    \end{figure}

    Using the GROMACS implementation, additional simulations were performed by varying the barostat stride $n_s$ at fixed relaxation time (see \cref{fig:variable_stride_ice}). The value selected is $\tau_p = 1$ ps, for which the SCR barostat in GROMACS appears consistent with the other methods. With the only exception of $n_s=100$, average and variance of the volume distributions are consistent with the results obtained by applying the barostat at each step.

    \section{Gypsum crystal}
\begin{figure}[h!]
        \centering
        \includegraphics[width = 0.55\textwidth]{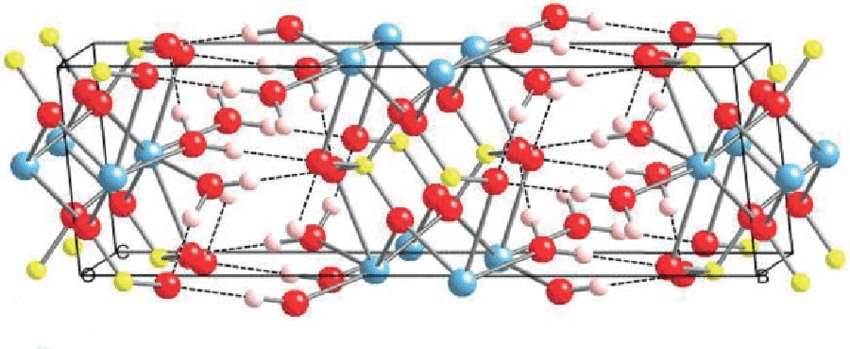}
        \caption{Crystal structure of $\text{(CaSO)}_\text{4}\cdot\text{2(H}_\text{2}\text{O)}$, commonly known as gypsum. White, red, blue and yellow beads represent H, O, Ca and S atoms respectively. Hydrogen bonds are shown as dashed lines. Source: \cite{gypsum_img}.}
        \label{fig:gypsum}
    \end{figure}

    Additional tests of the anisotropic SCR barostat in LAMMPS were performed by simulating a gypsum crystal (see \cref{fig:gypsum}) composed of $N=3456$ atoms, setting $P_0 = 1$ bar and $T=270$ K. 
    Simulations of $4$ ns were carried out using the SCR and the MTTK barostats, both in the the anisotropic implementations, accumulating statistics every $10$ steps (4 ps). Barostat and thermostat relaxation times were both set to $\tau_p = \tau_T = 0.1$ ps. For the simulations performed with the SCR barostat, the input isothermal compressibility was fixed to $\beta_T = 2.3\times 10^{-6}$ bar$^{-1}$. The statistics of interest are reported in \cref{tab:gypsum_lammps_aniso} for a comparison with the MTTK pressure coupling method. For additional results related to the single components of the box matrix $\mathbf{h}$ and the internal pressure tensor $\mathbf{P}_{\text{int}}$, see \cref{tab:gypsum_lammps_aniso_box} in the Appendices.
    \begin{table}[h!]
    \centering
    \caption{Results from the simulations of the gypsum crystal in LAMMPS. The pairs of raws refer respectively to the distributions of volume, internal pressure, density, potential energy and temperature, for which average and standard deviations are reported.}
    \begin{threeparttable}
    \label{tab:gypsum_lammps_aniso}
        \begin{tabular}{l c c c c}
        \toprule[0.5pt]\toprule[0.5pt]
        {} & \small\textbf{Anisotropic SCR}       & \small\textbf{Anisotropic MTTK}  \\\midrule
        $\langle V\rangle\,$ (nm$^3$)                     & $36.8682\pm0.0008$ & $36.8691\pm0.0010$ \\
        $\sigma_V\,$                & $0.0624\pm0.0003$ & $0.0622\pm0.0002$  \\ \midrule
        $\langle P_{\text{int}}\rangle\,$ (bar)            & $-7\pm 5$ & $1.0\pm 1.4$  \\
        $\sigma_{P_{\text{int}}}\,\,$          & $946\pm3$ & $940\pm 10$  \\ \midrule
        $\langle\rho\rangle\,$ (kg/cm$^3$)                 & $2233.44\pm0.05$ & $2233.38\pm0.06$  \\
        $\sigma_\rho\,\,$                  & $3.78\pm0.02$ &  $3.765\pm0.014$ \\ \midrule
        $\langle U\rangle\,$ (kJ/mol)            & $-789541.4\pm1.4$ & $-789542.3\pm1.4$  \\
        $\sigma_U\,\,$         & $171.0\pm0.7$ & $170.5\pm0.7$ \\ \midrule
        $\langle T\rangle\,$ (K)                      & $270.00\pm0.03$ &  $270.00\pm0.03$ \\
        $\sigma_T\,\,$                       & $3.762\pm0.012$ & $3.759\pm0.012$
        \\\bottomrule[0.5pt]\bottomrule[0.5pt] 
        \end{tabular}
    \end{threeparttable}
    \end{table}
    
\begin{figure}[h!]
    \centering
    \begin{subfigure}[b]{0.44\textwidth}
        \centering
        \includegraphics[height = 4. cm]{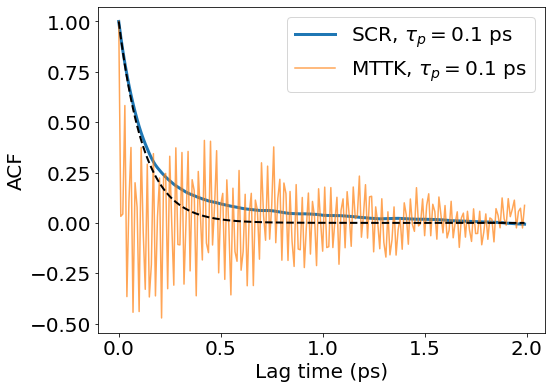}\\[1 ex]
    \end{subfigure}
    \hfil
    \begin{subfigure}[b]{0.44\textwidth}
        \centering
        \includegraphics[height = 4. cm]{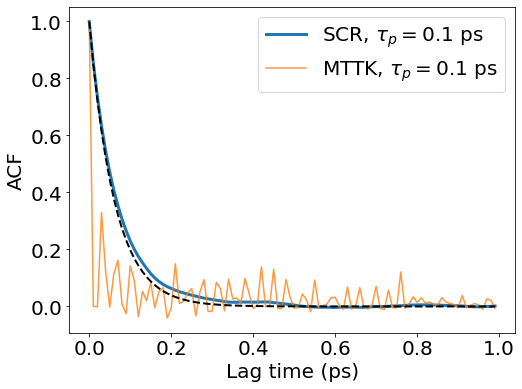}\\[1 ex]
    \end{subfigure}
    \caption{ACFs of the volume (left panel) and its variance (right panel) from the simulations of the gypsum crystal system in LAMMPS. The dashed lines in the left and right panels are respectively the exponential functions $e^{-t/\tau_{p,\text{exp}}}$ and $e^{-2t/\tau_{p,\text{exp}}}$.}
    \label{fig:ACFs_gypsum}
    \end{figure}

    \noindent The distributions generated by the two barostats appear consistent, with the usual \emph{caveat} that errors cannot be directly compared for the same choice of $\tau_p$. As already discussed in \cref{LJ_crystal} for the PR barostat, the symmetric oscillations in the MTTK volume ACF can result in a shorter autocorrelation time than the one obtained with the SCR method, due to a cancellation effect between positive and negative contributions when the volume ACF is integrated. When this happens, as in the case of the Ice I$_h$ simulations, the MTTK barostat achieves smaller statistical errors on the average of volume-related quantities. However, when a small enough $\tau_p$ is employed (as in this case, i.e. $\tau_p = 0.1$ ps) the MTTK volume ACF (see \cref{fig:ACFs_gypsum}) can suffer from a damping that makes the cancellation effect less efficient, resulting in statistical errors on volume-related averages that are comparable with the SCR method or even larger (see for instance $\langle V\rangle$ and $\langle \rho\rangle$ in the table above). 
    \newpage
    Also for this system, the the ACFs of the volume and its variance show an exponential decay in agreement with the expected behaviour in the SCR simulation (see \cref{fig:ACFs_gypsum}).

    \section{Au crystal}
\begin{figure}[h!]
        \centering
        \includegraphics[width = 0.75\textwidth]{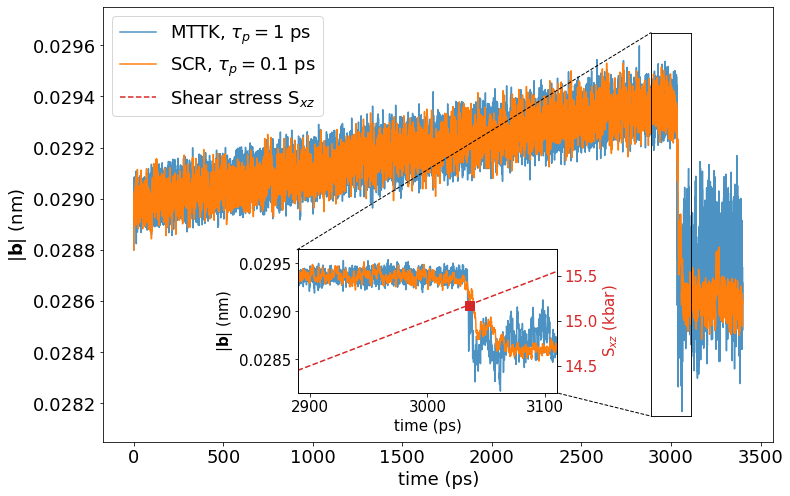}
        \caption{Results from the simulations of the gold crystal systems in LAMMPS, employing the MTTK and SCR anisotropic barostats in presence of a linearly increasing shear stress $S_{xz}$. The plot shows the trajectory for the modulus of the cell vector $\mathbf{b}$, which appears as a good variable to describe the crystal breaking.}
        \label{fig:gold_slip_plot}
    \end{figure}

    In order to test the anisotropic SCR method in presence of a generic external stress, namely for $\mathbf{S}_{\text{dev}}\neq 0$ (see \cref{npt_theory}), two simulations of a gold (Au) crystal system with $N=4000$ atoms were carried out in LAMMPS, applying the SCR and the MTTK barostats at each step and employing the modified embedded-atom model (EAM) described in \cite{embedded_atom}. In both the runs the crystal, intialized to an FCC structure, was simulated with time step $\Delta t=1$ fs, keeping the external hydrostatic pressure constant to $P_0 = 1$ bar and increasing linearly the $xz$ shear stress (and its symmetric $zx$) of 0.05 bar at each step, starting from $S_{xz} = S_{zx} = 0$ bar. The purpose of the simulations is to identify the extreme shear that the crystal can bear before its breaking. The external temperature was set to $T=298.15$ K and controlled with a SVR thermostat with relaxation time $\tau_T = 0.1$ ps. The barostat relaxation time was fixed to $\tau_p = 1$ ps for the MTTK barostat and $\tau_p = 0.1$ ps for the SCR run, using as input bulk modulus $k_T = \beta_T^{-1} = 1.7\times 10^6$ bar. Statistics were saved every 100 steps ($0.1$ ps). In the two simulations, the breaking of the crystal structure occurs consistently at $t\approx 3.03-3.05$ ns, when the external shear stress is $S_{xz}\approx 15.1-15.3$ kbar (see \cref{fig:gold_slip_plot}). This result is in agreement with the value predicted by \cite{slip_gold_reference}, performing static calculations with the GULP software \cite{GULP}.
    
\begin{figure}[h!]
    \centering
    \begin{subfigure}[b]{0.44\textwidth}
        \RaggedLeft
        \includegraphics[height = 3.5 cm]{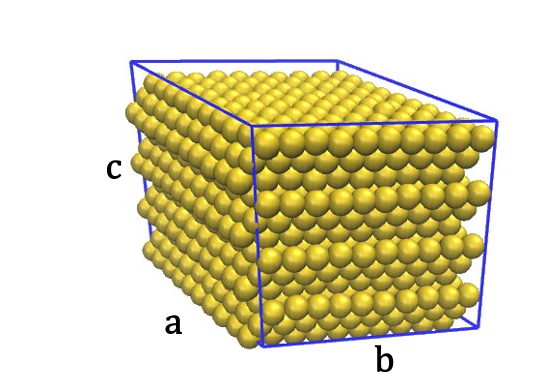}\\[1 ex]
        \caption{\RaggedLeft $\,t=3.035$ ns $\hspace{0.5cm}$}
    \end{subfigure}
    \hfil
    \begin{subfigure}[b]{0.44\textwidth}
        \RaggedRight
        \includegraphics[height = 3.5 cm]{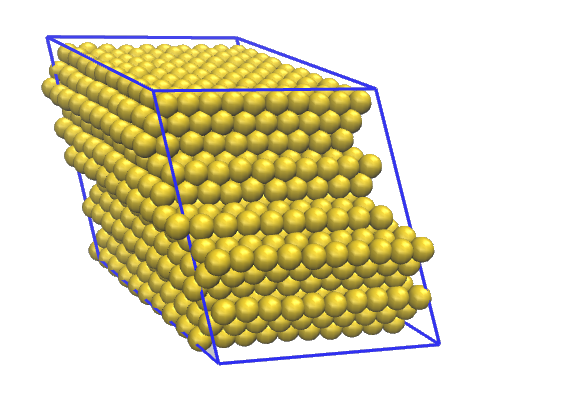}\\[1 ex]
        \caption{\RaggedLeft $\,t=3.260$ ns$\hspace{1.5cm}$}
    \end{subfigure}
    \vfil
    \vspace{0.3cm}
    \begin{subfigure}[b]{0.44\textwidth}
        \RaggedLeft
        \includegraphics[height = 3.5 cm]{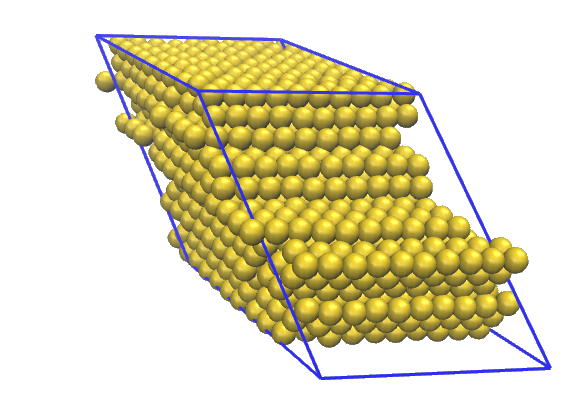}\\[1 ex]
        \caption{\RaggedLeft $\,t=3.515$ ns $\hspace{0.2cm}$}
    \end{subfigure}
    \hfil
    \begin{subfigure}[b]{0.44\textwidth}
        \RaggedRight
        \includegraphics[height = 3.5 cm]{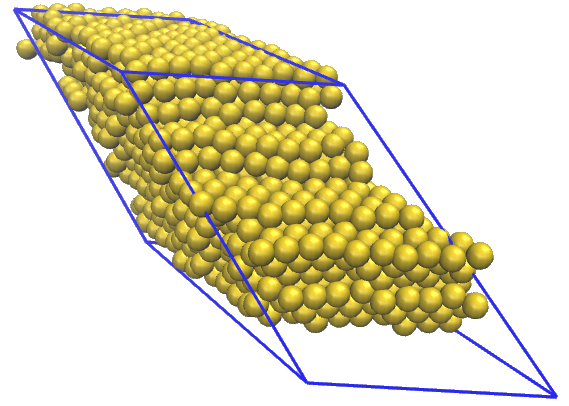}\\[1 ex]
        \caption{\RaggedLeft $\,t=3.517$ ns$\hspace{0.8cm}$}
    \end{subfigure}
    \caption{Frames from the simulations of the gold crystal in LAMMPS, using the anisotropic SCR barostat.}
    \label{fig:gold_slip_vmd}
    \end{figure}

    A visualization of the crystal breaking, which occurs through the slipping of the crystallographic planes (1\,1\,1), is reported in \cref{fig:gold_slip_vmd}.

    \chapter{Conclusions}
    In this work, the stochastic cell rescaling (SCR) barostat \cite{crescale_iso} is generalized to anisotropic volume fluctuations, allowing the system box to change its shape during MD simulations in the isothermal-isobaric ensemble. This pressure coupling method is formulated in terms of a nine-dimensional and first-order stochastic differential equation, whose deterministic part resembles the equation of the anisotropic Berendsen barostat \cite{berendsen}, except for a corrective term that becomes negligible in the thermodynamic limit. Unlike second-order methods, the SCR barostat can be used effectively during the equilibration phase. When coupled with a thermostat, the algorithm generates the correct anisotropic isothermal-isobaric ensemble, both when a isotropic external pressure is applied and in the case of a deviatoric external stress, i.e. with anisotropic external conditions. The anisotropic SCR method is shown to be equivalent to the high-friction limit of a second-order barostat, namely the Parrinello-Rahman barostat coupled with a Langevin thermostat. 
    
    Two equivalent formulations of the method are presented, where the components of the system box are propagated by keeping constant either the physical or the rescaled momenta. These two possibilities result respectively in the use of the average and instantaneous kinetic energy to compute the virial. 
    
    As in the isotropic case and in the Berendsen barostat, the algorithm has effectively a single input parameter, namely the ratio between the system isothermal compressibility $\beta_T$ and the barostat relaxation time $\tau_p$. If the accurate value of the isothermal compressibility is unknown \emph{a priori}, it is possible to use an estimate over a short preliminary run for its input, re-computing \emph{a posteriori} the correct values of both $\beta_T$ and $\tau_p$. 
    
    For the integration of the anisotropic SCR equations, two schemes are proposed and tested. The first method employs a simple Euler propagation of the box matrix components, generating trajectories that are not reversible. The second one is built on a conceptual separation between the propagation of the isotropic degree of freedom, i.e. the volume, and the evolution of the remaining eight degrees of freedom, which are responsible for the change of shape of the box. The two operations are applied in a way that allows generating time-reversible trajectories. As a consequence, a quantity called \emph{energy drift} \cite{langevin_bussi}, which behaves as a constant of motion is the limit of small time step, can be defined and computed for this integrator, so that its variations along the trajectory can be used to monitor the violation of detailed balance.  
    
    The algorithm has been tested with three different implementations, using the SimpleMD, GROMACS 2021.2 and LAMMPS (2 July 2021 release) MD softwares. Consistently among all the implementations, box rotations are eliminated after propagating all the nine box variables, with an orthogonal transformation that compensates for the three redundant degrees of freedom. Simulations were carried out on a variety of solid-state systems, both in presence and in absence of inter-molecular interactions and constraints. In all the tests, results appear consistent with the ones from the reference barostats employed in the comparison, especially as regards the distributions of the volume and the ones extracted from the single box matrix components. In terms of accuracy, the two integration schemes proposed (Euler and time-reversible) show an equivalent performance. The generated distributions appear to be stable in the entire range of $\tau_p$ values tested, covering 3 orders of magnitude in the LJ crystal simulations and 1-2 orders of magnitude in the Ice I$_h$ tests. In some cases, the method appears to be more robust than other second-order pressure coupling methods against systematic sampling errors in the small-$\tau_p$ regime. For this reason, the inaccuracy in the input value of $\tau_p$, due to a possible error in the input isothermal compressibility, is considered as a minor drawback of the method. The algorithm was also tested in a multiple-time-step fashion, showing stable results up to a stride $n_s = 10-20$. 
    
    In all the simulations, the autocorrelation functions of the volume and of its variance match the expected exponential decays as functions of the corrected relaxation time $\tau_p$, namely $\exp(-t/\tau_p)$ and $\exp(-2t/\tau_p)$. In other words, the input parameter $\tau_p$ can be used to estimate \emph{a priori} how fast the statistical error of any volume-dependent quantity approaches zero with the length of the trajectory. The current formulation of the algorithm does not allow to control the autocorrelation times of the single box components with the same accuracy, as the their limiting behaviours cannot be analytically predicted. A further formulation of the algorithm that could allow for this additional feature, employing a tensorial expression for the isothermal compressibility, is proposed in this work in view of future refinements of the method.
    
    As other anisotropic pressure coupling methods, the proposed barostat could find applications in MD simulation of two classes of systems, namely in the presence of internal anisotropies, as in the case of crystal systems, or in the case of an external anisotropic stress, that can result in modifications of the crystal structure and eventually in conformational phase transitions \cite{crystal_phase_trans,crystal_phase_trans2}. In this work, an example of this realistic applications is shown by simulating a gold crystal system in presence of a variable external shear stress, identifying the limiting stress causing the breaking of the structure.

    \begin{appendices}

    \chapter{Stochastic differential equations}\label{appendix_stocdiffeq}
    Stochastic differential equations (SDEs) allow to model and describe stochastic processes, but can also employed to construct non-deterministic sampling algorithms. Considering a generic one-dimensional variable $x$, the most common SDE that one finds in the literature is the first-order equation
    \begin{equation}\label{Langevin_colored}
        \dot{x}(t) = a(x,t) + b(x,t) \mathcal{\eta}(t)\,,
    \end{equation}
    where $\eta(t)$ is a stochastic process called \emph{Gaussian white noise}, namely it follows a zero-mean and unit-variance Gaussian distribution at any fixed time $t$ and it has an instantaneously decaying autocorrelation:
    \begin{subequations}\label[pluralequation]{gaussian_white_noise}
    \begin{align}
        \eta(t) &\sim \mathcal{N}(0,1)\,, \\
        \langle \eta(t) \eta(t') \rangle &= \delta (t-t')\,.
    \end{align}
    \end{subequations}
    \Cref{Langevin_colored} is a generalized Langevin equation including a \emph{multiplicative noise}, since in the (RHS) the white noise $\eta(t)$ is multiplied by a function of the dynamical variable $x$. A more rigorous way of writing this equation is
    \begin{equation}\label{ito_1d}
        \text{d}x(t) = a(x,t)\text{d}t + b(x,t) \text{d}W\,,
    \end{equation}
    where $W(t)$ satisfies $\frac{\text{d}W}{\text{d}t} = \eta(t)$. The stochastic process $W(t)$ is called \emph{Wiener noise} and it is characterized by independent and Gaussian-distributed time increments,
    \begin{equation}\label{wiener_incement}
        \Delta W (t_0,\Delta t) = W(t_0 + \Delta t) - W(t_0) \sim \mathcal{N}(0,\Delta t)\,,
    \end{equation}
    where $\sigma^2 = \Delta t$ is the variance of the distribution and \cref{wiener_incement} holds for any initial time $t_0$ and any finite increment $\Delta t$. 
    \noindent The formal solution of \cref{ito_1d} is
    \begin{equation}
        x(t) = x(0) + \int_0^t \text{d}t' a\big(x(t'),t'\big) + \underbrace{\int_0^t \text{d}W(t') b\big(x(t'),t'\big)}_{= I}
    \end{equation}
    and it depends on how the \emph{stochastic integral} $I$ is computed. The result of this integral can be defined through an approach similar to Riemann integration, starting from a partition $\{t_0\equiv0,t_1,...,t_n,t_{n+1}\equiv t\}$ of the time interval $\left[0,t\right]$. Given this partition, $I$ can be approximated as the sum of $n$ (signed) rectangular areas:
    \begin{equation}
        I_n = \sum_{i=1}^n \big[W(t_{i+1}) -W(t_i)\big]b\big(x(\tau_i),\tau_i\big)\,,
    \end{equation}
    with the condition $\tau_i \in \left[t_i,t_{i+1}\right]$. Due to the stochastic nature of $W(t)$, the classical limit of the Riemann sums $I_n$ is replaced by the notion of \emph{mean-square convergence}:
    \begin{equation}
        I = \mathop{\text{ms-lim}}_{n\rightarrow\infty} I_n \hspace{0.3cm}\Longleftrightarrow \hspace{0.3cm} \lim_{n\rightarrow\infty} \langle \left(I-S_n\right)^2\rangle = 0\,.
    \end{equation}
    The main difference with respect to Riemann integration is that the result depends on where the points $\tau_i$ are located in the partition $\{t_i\}$, and different conventions give rise to different rules of \emph{stochastic calculus}. Among all the possible choices, the two limiting cases are $\tau_i = t_i$ and $\tau_i = (t_i + t_{i+1})/2$ are what define the \emph{It\^{o} convention} and the \emph{Stratonovich convention} respectively. In this work, all the SDEs are interpreted \emph{à la} It\^{o}.
    
    A relevant result in \emph{It\^{o} calculus} \cite{gardiner} is the so-called \emph{It\^{o}'s lemma} (or chain rule), which allows to write the differential of a function of a stochastic process. 
    \begin{ito_lemma}
    Let $y = y(x)$ be a function of a stochastic process $x$ described by the It\^{o} equation \cref{ito_1d}. Then:
    \begin{equation}\label{ito_chain}
        \text{d}y(x) = \left[\frac{\text{d}y}{\text{d}x}a(x,t) + \frac{\text{d}^2 y}{\text{d}x^2}\frac{b^2(x,t)}{2}\right]\text{d}t + \frac{\text{d}y}{\text{d}x} b(x,t) \text{d}W\,.
    \end{equation}
    \end{ito_lemma}
    \noindent Hence, if the transformation $y = y(x)$ is invertible it is possible to obtain a new It\^{o} SDE for $y$ by substituting $x = x(y)$ in the expression above.
    
    It\^{o} \cref{ito_1d} can be generalized to the following multidimensional SDE for the variable $\mathbf{x}^T = (x_1,...,x_n)$:
    \begin{equation}\label{ito_multidim}
        \text{d}\mathbf{x}(t) = \mathbf{a}(\mathbf{x},t)\text{d}t + \mathbf{b}(\mathbf{x},t) \text{d}\mathbf{W}\,,
    \end{equation}
    where now $\mathbf{a}$ and $\text{d}\mathbf{W}$ are vectors while $\mathbf{b}$ is a matrix. Equivalently, written for the single component $i$:
     \begin{equation}
        \text{d}x_i(t) = a_i(x,t)\text{d}t + \sum_{j=1}^n b_{ij}(x,t) \text{d}W_j\,.
    \end{equation}
    The multidimensional Wiener process $\text{d}\mathbf{W}$ is simply composed by independent scalar Wiener processes. In the derivation of the anisotropic SCR \cref{crescale_aniso}, a further generalization is considered by promoting $\mathbf{a}$,$\mathbf{b}$ and $\text{d}\mathbf{W}$ to tensors. The corresponding generalization of It\^{o}'s lemma is the following.
    \begin{ito_lemma_multidim}
    Let $y = y(\mathbf{x})$ be a function of a multidimensional stochastic process $\mathbf{x}$ described by the It\^{o} \cref{ito_multidim}. Then
    \begin{equation}\label{ito_chain_multidim}
        \text{d}y(x) = \left\{\big(\nabla_{\mathbf{x}}y\big)^T \mathbf{a} + \frac{1}{2}\text{Tr}\left[\mathbf{b}^T \mathbf{H}_{\mathbf{x}}(y)\mathbf{b}\right]\right\}\text{d}t + (\nabla_{\mathbf{x}}y)^T \mathbf{b}\, \text{d}\mathbf{W}\,,
    \end{equation}
    where $\nabla_{\mathbf{x}}y$ and $\mathbf{H}_{\mathbf{x}}(y)$ are the gradient and the Hessian matrix of y respectively:
    \begin{subequations}
    \begin{align}
        \big(\nabla_{\mathbf{x}}y\big)_i &= \frac{\partial y}{\partial x_i}\,, \\
        \big(\mathbf{H}_{\mathbf{x}}(y)\big)_{ij} &= \frac{\partial^2 y}{\partial x_i \partial x_j}\,.
    \end{align}
    \end{subequations}
    \end{ito_lemma_multidim}

    \chapter{Fokker-Planck equations}\label{appendix_fp}
    Given a stochastic variable $x$ that evolves in time according to It\^{o} \cref{ito_1d}, it is possible to give an equivalent description of its stochastic dynamics in terms of a time-dependent probability density $\mathcal{P} = \mathcal{P}(x,t)$, such that
    \begin{equation}
        \mathcal{P}(x,t)\,\text{d}x\text{d}t
    \end{equation}
    is the probability of finding the system in $(x,x+dx)$ within the time interval $(t,t+\text{d}t)$. 
    
    \noindent The partial differential equation that defines the time evolution of $\mathcal{P}$ is called Fokker-Planck (FP) equation:
    \begin{equation}\label{fp_1d}
        \frac{\partial}{\partial t }\mathcal{P}(x,t) = -\frac{\partial}{\partial x}\Big(a(x,t)\mathcal{P}(x,t)\Big) + \frac{\partial^2}{\partial x^2}\Big(D(x,t)\mathcal{P}(x,t)\Big)\,.
    \end{equation}
    Here the \emph{drift coefficient} $a(x,t)$ is the same function appearing in \cref{ito_1d}, while the \emph{diffusion coefficient} $D(x,t)$ is given by:
    \begin{equation}
        D(x,t) = \frac{b^2(x,t)}{2}\,.
    \end{equation}
    In this context, the stationarity (or \emph{balance}) condition reads $\frac{\partial}{\partial t }\mathcal{P}(x,t) = 0$. A stronger condition - namely \emph{detailed balance} - can be defined by writing \cref{fp_1d} as the continuity equation
    \begin{equation}
        \frac{\partial}{\partial t }\mathcal{P}(x,t) = -\frac{\partial}{\partial x} J(x,t)
    \end{equation}
    and requiring the \emph{probability current} $J$ to be zero at any time $t$:
    \begin{equation}
        J(x,t) = a(x,t)\mathcal{P}(x,t) - \frac{\partial}{\partial x}\Big(D(x,t)\mathcal{P}(x,t)\Big) = 0\,.
    \end{equation}
    The relation between the It\^{o} and FP equations can be shown by considering a generic function $y=y(x)$ and employing It\^{o}'s lemma. Fixing the time $t$ and taking the average over all the possible configurations of $x$, \cref{ito_chain} becomes:
    \begin{equation}\label{proof_fp}
        \big\langle \text{d}y \big\rangle = \Big\langle a(x,t)\frac{\text{d}y}{\text{d}x}\Big\rangle\,\text{d}t + \Big\langle\frac{b^2(x,t)}{2}\frac{\text{d}^2 y}{\text{d}x^2}\Big\rangle\,\text{d}t + \Big\langle \frac{\text{d}y}{\text{d}x} b(x,t) \text{d}W\Big\rangle\,.
    \end{equation}
    In the third term of the (RHS), $\text{d}W = W(t+\text{d}t) - W(t)$ is independent on the remaining part - that only depends on the time $t$ but not on consecutive times - and as a consequence:
    \begin{equation}
        \Big\langle \frac{\text{d}y}{\text{d}x} b(x,t) \text{d}W\Big\rangle = \Big\langle \frac{\text{d}y}{\text{d}x} b(x,t)\Big\rangle \big\langle \text{d}W\big\rangle = 0\,.
    \end{equation}
    Therefore \cref{proof_fp} can be rewritten as:
    \begin{equation}\label{proof_fp2}
        \underbrace{\frac{\text{d}}{\text{d}t} \big\langle y \big\rangle}_{(A)} = \underbrace{\Big\langle a(x,t)\frac{\text{d}y}{\text{d}x}\Big\rangle}_{(B)} + \underbrace{\Big\langle\frac{b^2(x,t)}{2}\frac{\text{d}^2 y}{\text{d}x^2}\Big\rangle}_{(C)}
    \end{equation}
    The three terms appearing in this equation can be written separately as 
    \begin{subequations}
    \begin{align}
        (A) &= \frac{\text{d}}{\text{d}t} \int \text{d}x\, \mathcal{P}(x,t) y(x) = \int \text{d}x\, \frac{\partial\mathcal{P}(x,t)}{\partial t} y(x) \mathcal{P}(x,t)\,, \\
        (B) &= \int \text{d}x\,\mathcal{P}(x,t)a(x,t)\frac{\text{d}y}{\text{d}x} 
        = -\int \text{d}x\,\frac{\partial}{\partial x}\left(\mathcal{P}(x,t)a(x,t)y(x)\right)\,, \label{(B)}\\
        (C) &= \int \text{d}x\,\mathcal{P}(x,t)\frac{b^2(x,t)}{2}\frac{\text{d}^2 y}{\text{d}x^2} 
        = \int \text{d}x\,\frac{\partial^2}{\partial x^2}\left(\mathcal{P}(x,t)\frac{b^2(x,t)}{2}y(x)\right)\,,\label{(C)}
    \end{align}
    \end{subequations}
    where the results in \cref{(B),(C)} come from integrating by parts and assuming that $P(x,t)$ vanishes at the boundaries, which is a necessary condition in order to have a normalizable probability distribution. The initial statement immediately follows by substituting $(A)$, $(B)$ and $(C)$ in \cref{proof_fp2}, since this condition must hold for any $y=y(x)$.
    
    A similar reasoning can be followed starting from the multidimensional It\^{o} \cref{ito_multidim}; in this case, the Fokker-Planck equation that one obtains for the probability density $\mathcal{P} = \mathcal{P}(\mathbf{x},t)$ is
    \begin{equation}\label{fp_multidim}
        \frac{\partial}{\partial t} \mathcal{P}(\mathbf{x},t) = -\sum_{i=1}^n \frac{\partial}{\partial x_i}\Big(a_i(\mathbf{x},t)\mathcal{P}(\mathbf{x},t)\Big) + \sum_{i,j=1}^n \frac{\partial^2}{\partial x_i \partial x_j} \Big( D_{ij}(\mathbf{x},t)\mathcal{P}(\mathbf{x},t)\Big)\,,
    \end{equation}
    where here the \emph{diffusion matrix} $\mathbf{D}$ is related to $\mathbf{b}$ via the relation
    \begin{equation}
        \mathbf{D} = \frac{1}{2}\mathbf{b}\mathbf{b}^T\,,
    \end{equation}
    or equivalently, written for the single component $ij$: 
    \begin{equation}
    D_{ij} = \frac{1}{2}\sum_k b_{ik}b_{jk}.
    \end{equation}
    By introducing a probability current vector $\mathbf{J}$, also \cref{fp_multidim} can be written as a continuity equation,
    \begin{equation}
        \frac{\partial}{\partial t} \mathcal{P}(\mathbf{x},t) = \sum_{i=1}^n \frac{\partial}{\partial x_i} J_i (\mathbf{x},t)\,,
    \end{equation}
    and in this case the detailed balance condition reads, for any component $i$:
    \begin{equation}\label{db_multidim}
        J_i (\mathbf{x},t) = a_i(\mathbf{x},t)\mathcal{P}(\mathbf{x},t) - \sum_{j=1}^n \frac{\partial}{\partial x_j } \Big( D_{ij}(\mathbf{x},t)\mathcal{P}(\mathbf{x},t)\Big) = 0\,.
    \end{equation}
    All these relations are generalized to a tensorial formulation in the derivation of the anisotropic SCR \cref{crescale_aniso} (see \cref{derivation_aniso}).

    \chapter{Complete derivations}
    This appendix includes calculations and mathematical technicalities that are not reported in the main text.
    \section{Full derivation of the anisotropic SCR equations}\label{appendix_derivation}
    Starting from the case of a isotropic external stress, we report here the calculations to derive the deterministic part of the anisotropic SCR equations, namely the expression in \cref{dh_det_final}. 
    
    Let's consider for the moment the formulation where both positions and momenta are rescaled according to \cref{rescaled_variables_explicit}. The Jacobians of the two changes of variable are respectively
    \begin{subequations}
    \begin{align}
        J(\mathbf{q}_i\mapsto\mathbf{s}_i)_{\alpha k} = \frac{\partial q_i^\alpha}{\partial s_i^k} = h_{\alpha k} \,, \label{J_qs} \\
        J(\mathbf{p}_i\mapsto\bm{\pi}_i)_{\alpha k} = \frac{\partial p_i^{\alpha}}{\partial \pi_i^k} = h^{-1}_{k\alpha}\,, \label{J_ppi}
    \end{align}
    \end{subequations}
    and the $N\mathbf{S}T$ distribution as a function of $\{\mathbf{s}_i,\bm{\pi}_i\}$ gains no additional prefactor, as
    \begin{subequations}
    \begin{align}
        \text{d}\mathbf{q}_i\,\text{d}\mathbf{p}_i &= \det \big(\mathbf{J}(\mathbf{q}_i\mapsto\mathbf{s}_i)\big) \text{d}\mathbf{s}_i\, \det\big(\mathbf{J}(\mathbf{p}_i\mapsto\bm{\pi}_i)\big) \text{d}\bm{\pi}_i \\
        &= \big(\det \mathbf{h}\big)\big(\det \mathbf{h}^{-1}\big) \text{d}\mathbf{s}_i\, \text{d}\bm{\pi}_i \\
        &= \left(\det \mathbf{h h}^{-1}\right) \text{d}\mathbf{s}_i\, \text{d}\bm{\pi}_i \\ &= \text{d}\mathbf{s}_i\, \text{d}\bm{\pi}_i\,.
    \end{align}
    \end{subequations}
    Hence, using the expression of the $N\mathbf{S}T$ distribution as reported in \cref{NPT_anisotropic}, the deterministic part of \cref{target_eqs} can be decomposed as it follows:
    \begin{equation}\label{dh_det_terms}
        \text{d}h_{\alpha i}^{\text{det}} = \sum_{\beta j} D_{\alpha i\beta j}\Bigg[\underbrace{\frac{\partial \log D_{\alpha i\beta j}}{\partial h_{\beta j}}}_{(a)} + \underbrace{\frac{\partial \log (V^{-2})}{\partial h_{\beta j}}}_{(b)} -\frac{1}{k_B T}\Big(\underbrace{\frac{\partial K}{\partial h_{\beta j}}}_{(c)}  +\underbrace{\frac{\partial U}{\partial h_{\beta j}}}_{(d)} +\underbrace{P_0\frac{\partial V}{\partial h_{\beta j}}}_{(e)} \Big)\Bigg]\,\text{d}t
    \end{equation}
    Let's evaluate these terms one by one.
    \begin{subequations}
    \begin{align}
        (a) &= \sum_{\beta j} D_{\alpha i\beta j}\frac{\partial \log D_{\alpha i\beta j}}{\partial h_{\beta j}} = \sum_{\beta j} \frac{\partial D_{\alpha i\beta j}}{\partial h_{\beta j}}\\ 
        &= \frac{\beta_T k_B T}{3 \tau_p}\sum_{\beta j \eta} 
        \delta_{\alpha \beta} \frac{\partial }{\partial h_{\beta j}} \left( \frac{1}{V}h_{\eta i}h_{\eta j}\right) \\
        &= \frac{\beta_T k_B T}{3 \tau_p}\sum_{j \eta} 
        \left(-\frac{1}{V^2}\frac{\partial V}{\partial h_{\beta j}}h_{\eta i}h_{\eta j}  +\frac{1}{V}\delta_{\alpha \eta}\,\delta_{ji}h_{\eta j} +\frac{1}{V} h_{\eta i}\,\delta_{\alpha \eta}\delta_{jj}\right)  \\
        &= \frac{\beta_T k_B T}{3 \tau_p} 
        \left( -\frac{1}{V^2}\sum_{j \eta}V\,h_{j\alpha}^{-1}\,h_{\eta i}h_{\eta j}  +\frac{1}{V}h_{\alpha j} +\frac{3}{V} h_{\alpha i}\right)  \\
        &= \frac{\beta_T k_B T}{3V \tau_p} \left(-h_{\alpha j} + h_{\alpha j} + 3h_{\alpha i}\right) 
        = \frac{3\beta_T k_B T}{3V \tau_p}h_{\alpha i}\,;
    \end{align}
    \end{subequations}
    \begin{subequations}
    \begin{align}
        (b) &= \sum_{\beta j} D_{\alpha i\beta j}\frac{(-2)}{V}\frac{\partial V}{\partial h_{\beta j}} = -\frac{2\beta_T k_B T}{3V^2 \tau_p}\sum_{\beta j \eta}\delta_{\alpha\beta}\,h_{\eta i}\,h_{\eta j}\,\text{cof}(\mathbf{h})_{\beta j} \\
        &= -\frac{2\beta_T k_B T}{3V^2 \tau_p}\sum_{\eta}h_{\eta i}\,\left(\mathbf{h}\,\text{cof}(\mathbf{h})^T\right)_{\eta\alpha} = -\frac{2\beta_T k_B T}{3V^2 \tau_p}\sum_{\eta}h_{\eta i}\left(\det\mathbf{h}\right)\delta_{\eta\alpha} \\
        &= -\frac{2\beta_T k_B T}{3V \tau_p}h_{\alpha i}\,. \label{term_b_derivation}
    \end{align}
    \end{subequations}
    Both to evaluate $(a)$ and $(b)$, Jacobi's formula for the derivative of a matrix determinant has been used:
    \begin{equation}\label{jacobi}
        \frac{\partial V}{\partial h_{\beta j}} = \frac{\partial \det\mathbf{h}}{\partial h_{\beta j}} = \text{cof}(\mathbf{h})_{\beta j} = (\det\mathbf{h})\,h^{-1}_{j\beta}\,,
    \end{equation}
    where $\text{cof}(\mathbf{h})$ is the cofactor matrix of $\mathbf{h}$. In order to evaluate $(c)$, let's focus on the derivative of the kinetic energy, which is a function of the rescaled momenta $\bm{\pi}_i$ in the formulation that we are considering:
    \begin{subequations}
    \begin{align}
            \frac{\partial K}{\partial h_{\beta j}} &= \frac{\partial }{\partial h_{\beta j}} \sum_{k=1}^N \frac{1}{2m_k}\sum_{\alpha}\big(p_k^\alpha(\bm{\pi}_k)\big)^2 \\
            &= \frac{\partial }{\partial h_{\beta j}} \sum_{k=1}^N \frac{1}{2m_k}\sum_{\alpha l m} h^{-1}_{l\alpha}\,h^{-1}_{m\alpha}\,\pi_k^{l}\,\pi_l^m \\
            &= \sum_{k=1}^N \frac{1}{2m_k}\sum_{\alpha l m}\Big[ \frac{\partial h^{-1}_{l\alpha}}{\partial h_{\beta j}}\,h^{-1}_{m\alpha}\,\pi_k^{l}\,\pi_l^m + h^{-1}_{l\alpha}\,\frac{\partial{h^{-1}_{m\alpha}}}{\partial h_{\beta j}}\,\pi_k^{l}\,\pi_l^m \Big] \label{identical_terms}\\
            &= \sum_{k=1}^N \frac{1}{m_k}\sum_{\alpha l m} \frac{\partial h^{-1}_{l\alpha}}{\partial h_{\beta j}}\,h^{-1}_{m\alpha}\,\pi_k^{l}\,\pi_l^m \,.\label{kineng_derivative}
    \end{align}
    \end{subequations}
    The last passage is possible since the two terms in \cref{identical_terms} are identical, as it is possible to observe by exchanging the summed indices $i$ and $j$ in the second one. The derivative of the inverse box matrix can be evaluated by using the following property, that holds in general for square matrices:
    \begin{equation}
        \frac{\partial h^{-1}_{l\alpha}}{\partial h_{\beta j}} = -\sum_{\gamma k} h^{-1}_{l\gamma}\frac{\partial h_{\gamma k}}{\partial h_{\beta j}}h^{-1}_{k\alpha} = -\sum_{\gamma k} h^{-1}_{l\gamma}\,\delta_{\gamma\beta}\,\delta_{k j}\,h^{-1}_{k\alpha} = -h^{-1}_{l\beta}\,h^{-1}_{j\alpha}\,.
    \end{equation}
    Then, by substituting this last expression in \cref{kineng_derivative}, applying the Kronecker deltas and recomposing the physical momenta one gets:
    \begin{equation}
        \frac{\partial K}{\partial h_{\beta j}} = -\sum_{k = 1}^N \frac{1}{m_k} p^{\beta}_k \sum_\gamma h^{-1}_{j\gamma}\,p_k^{\gamma}\,.
    \end{equation}
    We can now evaluate the term $(c)$ appearing in \cref{dh_det_terms}:
    \begin{subequations}
    \begin{align}
        (c) &= \sum_{\beta j}\frac{ D_{\alpha i\beta j}}{k_B T}\frac{\partial K}{\partial h_{\beta j}} = \frac{\beta_T}{3V \tau_p} \sum_{\beta j \eta k \gamma}\delta_{\alpha\beta}\,h_{\eta i}\,h_{\eta j}\,\frac{1}{m_k} p^{\beta}_k\, h^{-1}_{j\gamma}\,p_k^{\gamma} \\
        &= \frac{\beta_T}{3V \tau_p} \sum_{\eta k \gamma} h_{\eta i}\,\delta_{\eta \gamma}\,\frac{1}{m_k} p^{\alpha}_k\,p_k^{\gamma} = \frac{\beta_T}{3V \tau_p} \sum_{\gamma}\Big(\sum_k \frac{1}{m_k} p^{\alpha}_k\,p_k^{\gamma}\Big)h_{\gamma i} \\
        &= \frac{\beta_T}{3\tau_p} \sum_{\gamma}P_{\text{int},\alpha \gamma}^{(p)}\,h_{\gamma i}\,
    \end{align}
    \end{subequations}
    where $\mathbf{P}_{\text{int}}^{(p)}$ is the kinetic part of the internal pressure tensor. Let's now focus on term $(d)$, starting from the calculation of the derivative of the potential energy. For the sake of simplicity we restrict the derivation to the case of a two-body potential:
    \begin{equation}
        U = \frac{1}{2}\sideset{}{'}\sum_{l,m =1}^{N} u(|\mathbf{q}_{lm}|) = \frac{1}{2}\sideset{}{'}\sum_{l,m =1}^N u(q_{lm})\,,
    \end{equation}
    where $\mathbf{q}_{lm} = \mathbf{q}_l - \mathbf{q}_m = \mathbf{h}\,\mathbf{s}_{lm}$ and the symbol ' in the sum is a notation for the constraint $l\neq m$. Then we have
    \begin{subequations}
    \begin{align}
        \frac{\partial U}{\partial h_{\beta j}} &= \frac{1}{2}\sideset{}{'}\sum_{l,m =1}^N \sum_\gamma \frac{\partial q_{lm}^\gamma}{\partial h_{\beta j}}\frac{\partial u({q}_{lm})}{\partial q_{lm}^\gamma} 
        = \frac{1}{2}\sideset{}{'}\sum_{l,m =1}^N \sum_\gamma \left(\frac{\partial }{\partial h_{\beta j}}\sum_k h_{\gamma k} s_{lm}^k\right) \frac{\partial u({q}_{lm})}{\partial q_{lm}^\gamma} \\
        &= \frac{1}{2}\sideset{}{'}\sum_{l,m =1}^N s_{lm}^j \frac{\partial u({q}_{lm})}{\partial q_{lm}^\beta} = 
        \frac{1}{2}\sideset{}{'}\sum_{l,m =1}^N \left(\sum_\gamma h_{j \gamma}^{-1}\,q_{lm}^\gamma \right)\left(-F_{lm}^\beta\right) \\
        &= - V\sum_\gamma h_{j \gamma}^{-1} \Bigg(\frac{1}{2V}\sideset{}{'}\sum_{l,m =1}^N q_{lm}^\gamma \,F_{lm}^\beta\Bigg) = -V\sum_\gamma h_{j \gamma}^{-1}\,P_{\text{int},\beta\gamma}^{(q)} \label{virial_variant} \,,
    \end{align}
    \end{subequations}    
    where $\mathbf{F}_{lm} = \mathbf{F}_{l} - \mathbf{F}_{m} = - \frac{\partial u({q}_{lm})}{\partial \mathbf{q}_{lm}}$ and $\mathbf{P}_{\text{int}}^{(q)}$ is the virial part of the internal pressure tensor, since the expression in \cref{virial_variant} can be shown to be equivalent to the second term in \cref{internal_pressure_avgkineng}. We can finally evaluate the term $(d)$ of \cref{dh_det_terms}:
    \begin{subequations}
    \begin{align}
        (d) &= \sum_{\beta j}\frac{D_{\alpha i \beta j}}{k_B T}\frac{\partial U}{\partial h_{\beta j}} = -\frac{\beta_T}{3\tau_p} \sum_{\beta j \eta \gamma}\delta_{\alpha\beta}\,h_{\eta i}\,h_{\eta j}\,h_{j\gamma}^{-1}\,P_{\text{int},\beta\gamma}^{(q)} \\
        &= -\frac{\beta_T}{3\tau_p} \sum_{\eta \gamma}\,\delta_{\eta\gamma}\,h_{\eta i}\,P_{\text{int},\alpha\gamma}^{(q)} = -\frac{\beta_T}{3\tau_p} \sum_{ \gamma}\,P_{\text{int},\alpha\gamma}^{(q)}\,h_{\gamma i}\,.
    \end{align}
    \end{subequations}
    We are left with the last term of \cref{dh_det_terms}:
    \begin{subequations}
    \begin{align}
        (e) &= \sum_{\beta j}\frac{D_{\alpha i \beta j}}{k_B T}P_0\frac{\partial V}{\partial h_{\beta j}} = \frac{\beta_T P_0}{3V\tau_p}\sum_{\beta j\eta}\delta_{\alpha\beta}\,h_{\eta i}\,h_{\eta j}\,\text{cof}(h)_{\beta j} \\
        &= \frac{\beta_T P_0}{3V\tau_p}\sum_{\eta}h_{\eta i}\left(\sum_j h_{\eta j}\,V\,h_{j\alpha}^{-1}\right) = \frac{\beta_T P_0}{3\tau_p}\sum_{\eta}h_{\eta i}\,\delta_{\eta\alpha} \\
        &= \frac{\beta_T P_0}{3\tau_p}h_{\alpha i}\,,
    \end{align}
    \end{subequations}
    where we have used once again the property in \cref{jacobi}. Putting all these terms together, namely
    \begin{equation}
        \text{d}h_{\alpha i}^{\text{det}} = \big[(a) + (b) - (c) - (d) - (e)\big]\,\text{d}t\,,
    \end{equation}
    the result that one gets is the one in \cref{dh_det_final}.
    
    Let's now consider the formulation where the components of $\mathbf{h}$ are propagated at constant rescaled positions $\mathbf{s}_i$ and physical momenta $\mathbf{p}_i$. The $N\mathbf{S}T$ distribution as a function of this variables acquires in this case an additional prefactor, since the Jacobian in \cref{J_ppi} is no more present and it does not cancel the one in \cref{J_qs}. Hence the distribution to be considered is:
    \begin{subequations}
    \begin{align}
        \mathcal{P}_{N\mathbf{S}T}\big(\{\mathbf{s}_i,\mathbf{p}_i \},V \big) &= \left(\prod_{i=1}^N \det \mathbf{J}(\mathbf{q}_i\mapsto\mathbf{s}_i)\right)\, \mathcal{P}_{N\mathbf{S}T}\big(\{\mathbf{q}_i(\mathbf{s}_i),\mathbf{p}_i \},V \big) \\
        &= V^N\,\mathcal{P}_{N\mathbf{S}T}\big(\{\mathbf{q}_i(\mathbf{s}_i),\mathbf{p}_i \},V \big) 
    \end{align}
    \end{subequations}
    Then the derivative of the kinetic energy - i.e. the term $(c)$ in the previous derivation - is now zero, while an additional term $(f)$ appears in \cref{dh_det_terms} as a consequence of the new factor $V^N$:
    \begin{subequations}
    \begin{align}
        (f) &= \sum_{\beta j} D_{\alpha i \beta j} \frac{\partial\log\left(V^N\right)}{\partial h_{\beta j}} = \sum_{\beta j} \frac{\beta_T k_B T}{3V\tau_p}\delta_{\alpha\beta}\sum_\eta h_{\eta i}\,h_{\eta j} \frac{N}{V} \frac{\partial V}{\partial h_{\beta j}} \\
        &=  \frac{N \beta_T k_B T}{3V^2\tau_p} \sum_{j\eta}h_{\eta i}\,h_{\eta j}\,\text{cof}(\mathbf{h})_{\alpha j} = \frac{N \beta_T k_B T}{3V^2\tau_p} \sum_{\eta}h_{\eta i}\sum_j h_{\eta j}\,V\,h_{j\alpha}^{-1} \\
        &= \frac{N \beta_T k_B T}{3V\tau_p} \sum_{\eta}h_{\eta i}\,\delta_{\eta\alpha} = \frac{N \beta_T k_B T}{3V\tau_p} h_{\alpha i}\,.
    \end{align}
    \end{subequations}
    As a result, \cref{dh_det_final} remains the same except for the expression of the internal pressure $\mathbf{P}_{\text{int}}$, where the kinetic energy tensor $\mathbf{K}$ is replaced by the diagonal tensor $\frac{N k_B T}{V}\mathbf{I}$, as stated in \cref{derivation_aniso}.
    
    \subsection*{Derivation with a generic external stress}\label{derivation_strain}
    If we consider the $N\mathbf{S}T$ dsitribution defined in \cref{NPT_anisotropic_shear}, the additional strain energy brings the following contribution to the deterministic part of the equations:
    \begin{equation}\label{dh_strain}
        \text{d}h_{\alpha i}^{\text{strain}} = -\frac{1}{k_B T}\sum_{\beta j} D_{\alpha i \beta j} \frac{\partial}{\partial h_{\beta j}}\left( \frac{1}{2}\text{Tr}(\bm{\Sigma}\,\mathbf{G})\right)\,\text{d}t\,,
    \end{equation}
    where we remind that $\mathbf{G}$ and $\bm{\Sigma}$ are defined as:
    \begin{subequations}
    \begin{align}
        \mathbf{G} &= \mathbf{h}^T\mathbf{h} \,,\\
        \bm{\Sigma} &= V_0\,\mathbf{h}_0^{-1}\big(\mathbf{S}-P_0\mathbf{I}\big)\left(\mathbf{h}_0^{-1}\right)^T\,.
    \end{align}
    \end{subequations}
    Let's focus on the derivative:
    \begin{subequations}
    \begin{align}
        \frac{\partial}{\partial h_{\beta j}}\Bigg( &\frac{1}{2}\text{Tr}(\bm{\Sigma} \,\mathbf{G})\Bigg) = \frac{1}{2}\frac{\partial}{\partial h_{\beta j}} \sum_{l m \gamma} \Sigma_{l m}\,h_{\gamma l}\,h_{\gamma m} \\
        &= \frac{1}{2}\sum_{l m \gamma}\left( \Sigma_{l m}\,\delta_{\gamma \beta}\,\delta_{l j}\,h_{\gamma m} + \Sigma_{l m}\,h_{\gamma l}\,\delta_{\gamma\beta}\,\delta_{m j}\right) \\
        &= \frac{1}{2}\left(\sum_{m}\Sigma_{j m}\,h_{\beta m} + \sum_{l}\Sigma_{l j}\,h_{\beta l}\right) \\
        &= \sum_{m}\Sigma_{j m}\,h_{\beta m}\,.
    \end{align}
    \end{subequations}
    To perform the last passage we renamed the summed index $l$ as $m$ in the second sum, and we used the fact that $\bm{\Sigma}$ is a symmetric tensor ($\Sigma_{j m} = \Sigma_{m j}$) as $\mathbf{S}$ is symmetric as well. By substituting in \cref{dh_strain} with the explicit expression of the diffusion tensor we get:
    \begin{subequations}
    \begin{align}
        \text{d}h_{\alpha i}^{\text{strain}} &= -\frac{\beta_T}{3V\tau_p}\sum_{\beta j \eta m} \delta_{\alpha\beta}\,h_{\eta i}\,h_{\eta j}\,\Sigma_{j m}\,h_{\beta m}\,\text{d}t\\
        &= -\frac{\beta_T}{3V\tau_p}\sum_{j \eta m}\,h_{\eta i}\,h_{\eta j}\,\Sigma_{j m}\,h_{\alpha m}\,\text{d}t\\ 
        &= -\frac{\beta_T}{3V\tau_p} \big(\mathbf{h}^T\mathbf{h}\,\bm{\Sigma}\,\mathbf{h}^T \big)_{i\alpha}\,\text{d}t = -\frac{\beta_T}{3V\tau_p} \big(\mathbf{h}\,\bm{\Sigma}\,\mathbf{h}^T\mathbf{h} \big)_{\alpha i}\,\text{d}t\,,
    \end{align}
    \end{subequations}
    where in the last line we used the property $\big(\mathbf{A}\,\mathbf{B}\big)^T = \mathbf{B}^T\mathbf{A}^T$ and once again the symmetry of $\bm{\Sigma}$. This proves the expression of $\text{d}h_{\alpha i}^{\text{strain}}$ given in \cref{dh_strain_maintext}.

    \section{Self-consistency with isotropic SCR equations}\label{appendix_flex2iso}
    Let's show the first feature of the anisotropic SCR equations stated in \cref{properties}, namely that they are consistent with the isotropic \cref{crescale_iso_V}. By applying the multidimensional It\^{o}'s lemma in \cref{ito_chain_multidim} to the anisotropic \cref{crescale_aniso} with respect to the variable $y = V(\mathbf{h}) = \det\mathbf{h}$, we get:
    \begin{equation}\label{ito_lemma_volume}
        \text{d}V = \Big[\underbrace{\left(\nabla_{\mathbf{h}}V\right)^T\mathbf{A}}_{(a)} + \underbrace{\frac{1}{2}\text{Tr}\big(\mathbf{B}^T\,\mathbf{H}_{\mathbf{h}}(V)\,\mathbf{B}\big)}_{(b)}\Big]\,\text{d}t + \underbrace{\left(\nabla_{\mathbf{h}}V\right)^T \mathbf{B}\, \text{d}\mathbf{W}}_{(c)}\,.
    \end{equation}
    Using Jacobi's formula in \cref{jacobi} the gradient and the Hessian matrix turn out to be:
    \begin{subequations}
    \begin{align}
        \left(\nabla_{\mathbf{h}}V\right)_{\alpha i} &= \frac{\partial V}{\partial h_{\alpha i}} = V\,h_{i \alpha}^{-1}\,,\\
        \big(\mathbf{H}_{\mathbf{h}}(V)\big)_{\alpha i\beta j} &= \frac{\partial V}{\partial h_{\alpha i}\partial h_{\beta j}} = V\left(h_{i \alpha}^{-1}\,h_{j \beta}^{-1} - h_{j \alpha}^{-1}\,h_{i \beta}^{-1}\right)\,.
    \end{align}
    \end{subequations}
    Let's evaluate the three terms of \cref{ito_lemma_volume} separately, starting from $(a)$:
    \begin{subequations}
    \begin{align}
        (a) &= -\frac{\beta_T}{3\tau_p}\sum_{\alpha i} V\,h_{i \alpha}^{-1}\Big[\sum_\beta \Big( P_0\delta_{\alpha\beta} - P_{\text{int},\alpha\beta} \Big)h_{\beta i} - \frac{k_B T}{V}h_{\alpha i}\Big] \\
        &= -\frac{\beta_T V}{3\tau_p}\sum_{\alpha} \Big[\sum_\beta \Big( P_0\delta_{\alpha\beta} - P_{\text{int},\alpha\beta} \Big)\,\delta_{\alpha\beta} - \frac{k_B T}{V}\,\delta_{\alpha\alpha}\Big] \\
        &= -\frac{\beta_T V}{\tau_p} \left(P_0 - P_{\text{int}} - \frac{k_B T}{V}\right)\,.
    \end{align}
    \end{subequations}
    In the last passage we used that $P_{\text{int}} = \text{Tr}\left(\mathbf{P}_{\text{int}}\right)$. The second term of \cref{ito_lemma_volume} is actually zero, in fact:
    \begin{subequations}
    \begin{align}
        (b) &= \frac{1}{2} \frac{2\beta_T k_B T}{3V\tau_p}\sum_{\alpha i\beta j\gamma k}B_{\alpha i\gamma k}\,\big(\mathbf{H}_{\mathbf{h}}(V)\big)_{\alpha i\beta j}\,B_{\beta j \gamma k} \\
        &= \frac{\beta_T k_B T}{3V\tau_p}\sum_{\alpha i\beta j\gamma k}h_{\gamma i}\,\delta_{\alpha k}\left(h_{i\alpha}^{-1}\,h_{j\beta}^{-1} - h_{j\alpha}^{-1}\,h_{i\beta}^{-1} \right)h_{\gamma j}\,\delta_{\beta k} \\
        &= \frac{\beta_T k_B T}{3V\tau_p}\sum_{\alpha i\gamma j} \left(h_{\gamma i}\,h_{i\alpha}^{-1}\,h_{j\alpha}^{-1}\,h_{\gamma j} - h_{\gamma i}\,h_{i\alpha}^{-1}\,h_{j\alpha}^{-1}\,h_{\gamma j}\right) = 0\,.
    \end{align}
    \end{subequations}
    Finally we can evaluate the term $(c)$ in \cref{ito_lemma_volume}:
    \begin{subequations}
    \begin{align}
        (c) &= \sqrt{\frac{2\beta_T k_B T}{3V\tau_p}}\sum_{\alpha i \beta j} V\,h_{i\alpha}^{-1}\,h_{\beta i}\,\delta_{\alpha j}\,\text{d}W_{\beta j} \\
        &= \sqrt{\frac{2\beta_T k_B T V}{3\tau_p}}\sum_{\alpha \beta j} \delta_{\alpha\beta}\,\delta_{\alpha j}\,\text{d}W_{\beta j} \\
        &= \sqrt{\frac{2\beta_T k_B T V}{\tau_p}}\frac{\text{Tr}(\text{d}\mathbf{W})}{\sqrt{3}}\,.
    \end{align}
    \end{subequations}
    Since $\text{\text{d}}\mathbf{W}$ is a Wiener noise whose variance is three times that of the single components, $\text{\text{d}}\mathbf{W}/\sqrt{3} = \eta(t)\text{d}t$ where $\eta(t)$ is a Gaussian white noise. Hence we can do the replacement
    \begin{equation}
        \text{d}W = \frac{\text{Tr}(\text{d}\mathbf{W})}{\sqrt{3}}\,,
    \end{equation}
    where $\text{d}W$ is the standard Wiener noise as defined in Appendix \ref{appendix_stocdiffeq}.
    Putting together the terms $(a)$ and $(b)$ we obtain the isotropic SCR \cref{crescale_iso_V}.

    \section{Self-consistency with semi-isotropic SCR equations}\label{appendix_flex2semi-iso}
    Also this derivation employs the multidimensional It\^{o} chain rule in \cref{ito_chain_multidim}. The starting equations are actually a modified version of the anisotropic \cref{crescale_aniso}, namely the equations that one obtains with the derivation in Appendix \ref{appendix_derivation} but using the of the $NP_0^\perp \gamma_0 T$ distribution as a function of the box matrix components,
    \begin{equation}\label{semi-isotropic-target}
        \mathcal{P}_{NP_0^\perp \gamma_0 T}\big(\{\mathbf{q}_i,\mathbf{p}_i\},\mathbf{h}\big) \propto \left(\det \mathbf{h}\right)^{-2} \exp\left[-\frac{1}{k_B T}\Big(K + U + P_0^\perp \det \mathbf{h} - \gamma_0\,A\Big)\right]\,.
    \end{equation}
    Although the $NP_0^\perp \gamma_0 T$ was introduced considering an orthorhombic box in \cref{npt_theory}, it is possible to consider here a slightly more general case with two additional nonzero off-diagonal elements:
    \begin{equation}
        \mathbf{h}=
        \begin{pmatrix}
        h_{x1} & h_{x2} & 0\\
        h_{y1} & h_{y2} & 0\\
        0 & 0 & h_{z3}
        \end{pmatrix}\,.
    \end{equation}
    Obviously $A = h_{x1}\,h_{y2} - h_{x2}\,h_{y1}$ and $L = h_{z3}$. By repeating the derivation of \cref{derivation_aniso} with \cref{semi-isotropic-target} as target distribution, the same equations are obtained but with two differences, namely the hydrostatic pressure $P_0$ is substituted by the normal pressure $P_0^\perp$ and an additional term with the reference surface tension $\gamma_0$ appears:
    \begin{equation}
        \text{d}h_{\alpha i}^{\gamma_0} = - \sum_{\beta j}\frac{D_{\alpha i \beta j}}{k_B T}\,\left(-\gamma_0\frac{\partial A}{\partial h_{\beta j}}\right)\text{d}t\,. \\
    \end{equation}
    Observing that the derivative of $A$ can be written as
    \begin{equation}\label{dh_gamma0}
        \frac{\partial A}{\partial h_{\beta j}} =  \delta_{\beta x}\,\delta_{j1}\,h_{y2} + h_{x1}\,\delta_{\beta y}\,\delta_{j2} - \delta_{\beta x}\,\delta_{j2}\,h_{y1} - h_{x2}\,\delta_{\beta y}\,\delta_{j1}\,,
    \end{equation}
    we obtain, by substituting this expression in \cref{dh_gamma0} and applying the Kronecker deltas:
    \begin{subequations}
    \begin{align}
        \text{d}h_{\alpha i}^{\gamma_0} &= \frac{\beta_T \gamma_0}{3V\tau_p} \Big[\left(\delta_{\alpha x}\,h_{xi} + \delta_{\alpha y}\,h_{yi}\right)\left(h_{x1}\,h_{y2} - h_{x2}\,h_{y1}\right) \Big] \\
        &= \frac{\beta_T \gamma_0 A}{3V\tau_p} \left(\delta_{\alpha x}\,h_{xi} + \delta_{\alpha y}\,h_{yi}\right) =\frac{\beta_T \gamma_0}{3L\tau_p} h_{\alpha i}\left(1-\delta_{\alpha z}\right)\,.
    \end{align}
    \end{subequations}
    Then the modified anisotropic equations for the $NP_0^\perp \gamma_0 T$ ensemble read:
    \begin{align}
        \text{d}h_{\alpha i} = -&\frac{\beta_T}{3\tau_p} \Bigg[\sum_\beta \Big( P_0^\perp\delta_{\alpha\beta} - P_{\text{int},\alpha\beta} \Big)h_{\beta i} - \frac{k_B T}{V}h_{\alpha i} - \frac{\gamma_0}{L}h_{\alpha i}\left(1-\delta_{\alpha z}\right)\Bigg]\,\text{d}t \nonumber \\
        &+ \sqrt{\frac{2\beta_T k_B T}{3V\tau_p}} \sum_{\beta}h_{\beta i}\,\text{d}W_{\alpha\beta}\,.
    \end{align}
    Starting from these equations, let's derive the semi-isotropic ones for the variables $\varepsilon_{xy} = \log A/A_0$ and $\varepsilon_z = \log L/L_0$, namely \cref{crescale-semi-isotropic-eq}. The equation for $L$ is simply the one for $h_{3z}$:
    \begin{equation}
        \text{d}L = \underbrace{-\frac{\beta_T L}{3\tau_p} \left( P_0^\perp - P_{\text{int},zz} - \frac{k_B T}{V}\right)}_{a(L)}\,\text{d}t + \underbrace{\sqrt{\frac{2\beta_T k_B T}{3V\tau_p}}\,L}_{b(L)}\,\text{d}W_{zz}\,.
    \end{equation}
    Then, by applying It\^{o}'s lemma for $\varepsilon_z$, we find the following SDE:
    \begin{equation}
        \text{d}\varepsilon_z = \left(\frac{\partial \varepsilon_z}{\partial L}\,a + \frac{b^2}{2}\frac{\partial^2 \varepsilon_z}{\partial L^2} \right)\text{d}t + b\frac{\partial \varepsilon_z}{\partial L}\,\text{d}W \\
    \end{equation}
    Using that $\frac{\partial \varepsilon_z}{\partial L} = \frac{1}{L}$ and $\frac{\partial^2 \varepsilon_z}{\partial L^2} = -\frac{1}{L^2}$ we finally obtain
    \begin{equation}
        \text{d}\varepsilon_z = -\frac{\beta_T}{3\tau_p}\left(P_0^\perp - P_{\text{int},zz}\right)\text{d}t + \sqrt{\frac{2 k_B T \beta_T}{3 V \tau_p}}\text{d}W_z\,,
    \end{equation}
    which is exactly \cref{eps_xy}. In order to obtain the second equation, we can first isolate the four equations that evolve the box matrix components entering in $A$:
    \begin{equation}
        \text{d}\mathbf{h}^{(xy)} = \left(\mathbf{A}^{(xy)} + \frac{\gamma_0\beta_T}{3L\tau_p}\mathbf{I}_{2\times2}\right)\,\mathbf{h}^{(xy)}\,\text{d}t + \mathbf{B}^{(xy)}\,\text{d}\mathbf{W}^{(xy)}\,.
    \end{equation}
    Here the upperscript $(xy)$ identifies the 2$\times$2 upper-left submatrices, and $A$ can be written as $A = \det \mathbf{h}^{(xy)}$. The equation for $A$ can be obtained by applying the multidimensional It\^{o} chain rule to \cref{ito_chain_multidim}, performing the same calculations shown in Appendix \ref{appendix_flex2iso} to derive the equation for $V$, but in two dimensions instead of three, and with the additional diagonal term containing $\gamma_0$. The equation that one obtains is:
    \begin{equation}
        \text{d}A = -\frac{2A\beta_T}{3\tau_p}\left(P_0^\perp - \frac{P_{\text{int},xx}+P_{\text{int},yy}}{2} - \frac{k_B T}{V} - \frac{\gamma_0}{L}\right)\,\text{d}t + A\sqrt{\frac{4 k_B T \beta_T}{3 V \tau_p}}\,\text{d}W\,,
    \end{equation}
    where here the Wiener noise comes from $\text{d}W = (\text{d}W_{xx} + \text{d}W_{yy})/\sqrt{2}$. By applying It\^{o}'s lemma for the variable $\varepsilon_{xy}$ as already done for $\varepsilon_z$, we finally get
    \begin{equation}
        \text{d}\varepsilon_{xy} = -\frac{2\beta_T}{3\tau_p}\left(P_0^\perp - \frac{\gamma_0}{L} - \frac{P_{\text{int},xx} + P_{\text{int},yy}}{2} \right)\text{d}t + \sqrt{\frac{4 k_B T \beta_T}{3 V \tau_p}}\text{d}W_{xy}\,,
    \end{equation}
    namely \cref{eps_xy}.

    \section{Change of box vectors}\label{appendix_changecellvectors}
    We show here that the anisotropic SCR equations are invariant under the transformation 
    \begin{equation}\label{change_box}
        h_{\alpha i} \longmapsto h'_{\alpha i} = \sum_{j=1}^3 n_j^{(i)}\,h_{\alpha j}\,.
    \end{equation}
    As shown in \cref{derivation_aniso}, the equations can be written as
    \begin{equation}
        \text{d}\mathbf{h} = \mathbf{A}(\mathbf{h})\, \text{d}t + \mathbf{B}(\mathbf{h})\,\text{d}\mathbf{W}\,,
    \end{equation}
    where $\mathbf{B}$ is defined in \cref{B_tensor} and $\mathbf{A}$ is given by:
    \begin{equation}
        \mathbf{A} = -\frac{\beta_T}{3\tau_p} \Big[ \big( P_0\mathbf{I} - \mathbf{P}_{\text{int}} \big) - \frac{k_B T}{V}\mathbf{I} \Big]\,\mathbf{h}\,.
    \end{equation}
    Once again the demonstration is based on the multidimensional It\^{o}'s chain rule, applied to the equations above with respect to each transformed variable $h'_{\alpha i}$:
    \begin{equation}\label{dh_prime}
        \text{d}h'_{\alpha i} = \Big[\underbrace{\left(\nabla_{\mathbf{h}}h'_{\alpha i}\right)^T\mathbf{A}}_{(a)} + \underbrace{\frac{1}{2}\text{Tr}\Big(\mathbf{B}^T\,\mathbf{H}_{\mathbf{h}}\left(h'_{\alpha i}\right)\,\mathbf{B}\Big)}_{(b)}\Big]\,\text{d}t + \underbrace{\left(\nabla_{\mathbf{h}}h'_{\alpha i}\right)^T \mathbf{B}\, \text{d}\mathbf{W}}_{(c)}\,,
    \end{equation}
    Since the transformation \cref{change_box} is linear, the first derivatives are 
    \begin{equation}
        \left(\nabla_{\mathbf{h}}h'_{\alpha i}\right)_{\beta j} = \frac{\partial h'_{\alpha i}}{\partial h_{\beta j}} = n_j^{(i)}\delta_{\alpha\beta} \,,
    \end{equation}
    while the second derivatives are zero:
    \begin{equation}
        \big(\mathbf{H}_{\mathbf{h}}(h'_{\alpha i})\big)_{\beta j\gamma k} = \frac{\partial h'_{\alpha i}}{\partial h_{\beta j}\partial h_{\gamma k}} = 0\,.
    \end{equation}
    As a consequence, $(b) = 0$. Let's evaluate the other two terms:
    \begin{align}
        (a) &= -\frac{\beta_T}{3\tau_p} \sum_{\beta j} n_j^{(i)}\,\delta_{\alpha\beta}\, \Bigg[\sum_\gamma \Big( P_0\delta_{\beta\gamma} - P_{\text{int},\beta\gamma} \Big)h_{\gamma j} - \frac{k_B T}{V}h_{\beta j}\Bigg]\nonumber \\
        &= -\frac{\beta_T}{3\tau_p} \left[\sum_{\gamma } \Big( P_0\delta_{\alpha\gamma} - P_{\text{int},\alpha\gamma} \Big)\left(\sum_{j}n_j^{(i)}\,h_{\gamma j}\right) - \frac{k_B T}{V}\sum_{j}n_j^{(i)}h_{\alpha j}\right] \nonumber \\
        &= -\frac{\beta_T}{3\tau_p} \Bigg[\sum_\gamma \Big( P_0\delta_{\alpha\gamma} - P_{\text{int},\alpha\gamma} \Big)h'_{\gamma i} - \frac{k_B T}{V}h'_{\alpha i}\Bigg] 
    \end{align}
    \begin{align}
        (c) &= \sum_{\beta j\gamma k} n_j^{(i)}\,\delta_{\alpha\beta} \,B_{\beta j\gamma k}\,\text{d}W_{\gamma k} = \sum_{j}n_j^{(i)}\,\sqrt{\frac{2\beta_T k_B T}{3V\tau_p}} \sum_\gamma h_{\gamma j}\,\text{d}W_{\alpha\gamma} \nonumber\\
        &= \sqrt{\frac{2\beta_T k_B T}{3V\tau_p}} \sum_\gamma\left(\sum_{j}n_j^{(i)}\, h_{\gamma j}\right)\text{d}W_{\alpha\gamma} =\sqrt{\frac{2\beta_T k_B T}{3V\tau_p}} \sum_\gamma h'_{\gamma j}\text{d}W_{\alpha\gamma}  
    \end{align}
    Then the transformation $\mathbf{h}\mapsto\mathbf{h}'$ does not change the form of the anisotropic SCR equations.

    \section{Method for QR factorization}\label{appendix_rotations}
    Given a rescaling matrix $\bm{\mu}$ with all non-zero elements, obtained by propagating nine degrees of freedom according to \cref{mu_matrix}, let's show the procedure to rotate the columns of $\mu$ in order to obtain an upper-triangular matrix $\bm{\mu'}$:
    \begin{equation}
        \bm{\mu'}= \mathbf{R}\bm{\mu} = 
        \begin{pmatrix}
        \mu^{\prime}_{xx} & \mu^{\prime}_{xy} & \mu^{\prime}_{xz}\\
        0 & \mu^{\prime}_{yy} & \mu^{\prime}_{yz}\\
        0 & 0 & \mu^{\prime}_{zz}
        \end{pmatrix}\,.
    \end{equation}
    The rotation of the first column is simply achieved by imposing 
    \begin{subequations}\label{rot_column1}
    \begin{align}
        \mu^{\prime\,2}_{xx} &= \mu_{xx}^2 + \mu_{yx}^2 + \mu_{zx}^2\,, \\
        \mu^{\prime\,2}_{yx} &= \mu^{\prime\,2}_{zx} = 0\,.
    \end{align}  
    \end{subequations}
    To rotate the second column of $\bm{\mu}$ we have to take into account the invariance both of the norm and of the scalar product with the first column:
    \begin{subequations}\label{rot_column2}
    \begin{align}
        \mu^{\prime\,2}_{xy} + \mu^{\prime\,2}_{yy} &= \mu_{xy}^2 + \mu_{yy}^2 + \mu_{zy}^2\,, \\
        \mu^{\prime}_{xy}\,\mu^{\prime}_{xx} &= \mu_{xy}\,\mu_{xx} + \mu_{yy}\,\mu_{yx} + \mu_{zy}\,\mu_{zx}\,, \\
        \mu^{\prime}_{zy} &= 0\,.
    \end{align}  
    \end{subequations}
    Finally, the conditions to rotate the third column include the invariance of the scalar product both with the first and the second column:
    \begin{subequations}\label{rot_column3}
    \begin{align}
        \mu^{\prime\,2}_{xz} + \mu^{\prime\,2}_{yz} + \mu^{\prime\,2}_{zz}  &= \mu_{xz}^2 + \mu_{yz}^2 + \mu_{zz}^2\,, \\
        \mu^{\prime}_{xz}\,\mu^{\prime}_{xx} &= \mu_{xz}\,\mu_{xx} + \mu_{yz}\,\mu_{yx} + \mu_{zz}\,\mu_{zx}\,, \\
        \mu^{\prime}_{xz}\,\mu^{\prime}_{xy} + \mu^{\prime}_{yz}\,\mu^{\prime}_{yy} &= \mu_{xz}\,\mu_{xy} + \mu_{yz}\,\mu_{yy} + \mu_{zz}\,\mu_{zy}\,. 
    \end{align}  
    \end{subequations}
    By solving together \cref{rot_column1,rot_column2,rot_column3} one gets the six non-zero elements of $\bm{\mu'}$:
    \begin{subequations}
    \begin{align}
        \mu^{\prime}_{xx} &= \sqrt{\mu_{xx}^2 + \mu_{yx}^2 + \mu_{zx}^2} \,, \\
        \mu^{\prime}_{xy} &= \frac{\mu_{xy}\,\mu_{xx} + \mu_{yy}\,\mu_{yx} + \mu_{zy}\,\mu_{zx}}{\mu^{\prime}_{xx}}\,, \\
        \mu^{\prime}_{yy} &= \sqrt{\mu_{xy}^2 + \mu_{yy}^2 + \mu_{zy}^2 - \mu^{\prime\,2}_{xy}}\,, \\
        \mu^{\prime}_{xz} &= \frac{\mu_{xz}\,\mu_{xx} + \mu_{yz}\,\mu_{yx} + \mu_{zz}\,\mu_{zx}}{\mu^{\prime}_{xx}}\,, \\
        \mu^{\prime}_{yz} &= \frac{\mu_{xz}\,\mu_{xy} + \mu_{yz}\,\mu_{yy} + \mu_{zz}\,\mu_{zy} - \mu^{\prime}_{xz}\,\mu^{\prime}_{xy}}{\mu^{\prime}_{yy}}\,, \\
        \mu^{\prime}_{zz} &= \sqrt{\mu_{xz}^2 + \mu_{yz}^2 + \mu_{zz}^2 -\mu^{\prime\,2}_{xz} - \mu^{\prime\,2}_{yz}}\,.
    \end{align}  
    \end{subequations}
    As a final observation, in the GROMACS \cite{gromacs} implementation of the anisotropic Berendsen barostat, the same operation is performed using a first order approximation of the equations above, namely 
    \begin{equation}
        \bm{\mu'}=
        \begin{pmatrix}
        \mu_{xx} & \mu_{xy} + \mu_{yx} & \mu_{xz} + \mu_{zx}\\
        0 & \mu_{yy} & \mu_{yz} +\mu_{zy}\\
        0 & 0 & \mu_{zz}
        \end{pmatrix}\,.
    \end{equation}

    \section{High-friction limit of Parrinello-Rahman equations}\label{appendix_highfriction}
    We report here the main calculations to derive the anisotropic SCR \cref{crescale_aniso} from the extended Parrinello-Rahman \cref{PR_friction_eqs}.
    The limit taken into account is known as \emph{Smoluchowski-Kramers limit} in the field of SDEs, and it is performed sending the friction to infinity and the mass to zero, such that their product stays finite. A detailed discussion of this limit in case of a variable-dependent friction is reported in \cite{high_friction_limit}. To apply the limit, we first rewrite the extended Parrinello-Rahman equations as
    \begin{subequations}\label[pluralequation]{PR_plusfriction}
    \begin{align}
        \text{d}\mathbf{h} &= \mathbf{v}\,\text{d}t\,, \\
        \text{d}\mathbf{v} &= \left(\frac{\mathbf{F}(\mathbf{h})}{W} - \frac{\bm{\gamma}(\mathbf{h})}{W}\mathbf{v} \right)\,\text{d}t + \frac{\bm{\sigma}(\mathbf{h})}{W}\,\,\text{d}\mathbf{W}\,,
    \end{align}
    \end{subequations}
    where $\mathbf{F}(\mathbf{h}) = V\big(\mathbf{P}_{\text{int}} - P_0\mathbf{I}\big)\big(\mathbf{h}^{-1}\big)^T$. In case of a variable-dependent friction \cite{high_friction_limit}, the result of the limit is
    \begin{equation}\label{hf_limit}
        \text{d}\mathbf{h} = \Big(\underbrace{\bm{\gamma}^{-1}(\mathbf{h})\,\mathbf{F}(\mathbf{h})}_{(a)} + \underbrace{\mathbf{T}(\mathbf{h})}_{(b)}\Big)\,\text{d}t + \underbrace{\bm{\gamma}^{-1}(\mathbf{h})\,\bm{\sigma}(\mathbf{h})\,\text{d}\mathbf{W}}_{(c)}\,,
    \end{equation}
    with the \emph{noise-induced drift tensor} $\mathbf{T}(\mathbf{h})$ determined via the following relations:
    \begin{subequations}
    \begin{align}
        &T_{\alpha i} = \sum_{\beta j\gamma k}\left(\frac{\partial}{\partial h_{\beta j}}\gamma_{\alpha i\gamma k}^{-1} \right)\,J_{\gamma k \beta j}\,, \\
        &\mathbf{J}\bm{\gamma}^T + \bm{\gamma}\mathbf{J} = \bm{\sigma}\bm{\sigma}^T\,. \label{lyapunov}
    \end{align}
    \end{subequations}
    \Cref{lyapunov} is called \emph{Lyapunov equation} for $\mathbf{J}$. Let's start by considering the term $(a)$:
    \begin{subequations}
    \begin{align}
        (a)_{\alpha i} &= \sum_{\beta j}\gamma^{-1}_{\alpha i\beta j}\,F_{\beta j} \\
        &= \sum_{\beta j}\gamma^{-1}_{\alpha i\beta j}\,V\sum_{\gamma}\left(P_{\text{int},\beta\gamma} - P_0\,\delta_{\beta\gamma}\right)h_{j\gamma}^{-1} 
    \end{align}
    \end{subequations}
    We can now arbitrary set the following functional form for $\bm{\gamma}^{-1}$:
    \begin{equation}\label{gamma_inverse}
        \gamma_{\alpha i\beta j}^{-1} = \frac{\beta_T}{3V\tau_p}\delta_{\alpha\beta}\sum_{\eta}h_{\eta i}\,h_{\eta j}\,.
    \end{equation}
    In spirit, this \emph{ansatz} is similar to the one for the diffusion tensor in the derivation of \cref{derivation_aniso}, namely it breaks the generality of the equations in order to reproduce a Berendsen-like deterministic term, but without affecting the sampled distribution. 
    Substituting $\bm{\gamma}^{-1}$ in the previous expression we get:
    \begin{subequations}
    \begin{align}
        (a)_{\alpha i} &= \frac{\beta_T V}{3V\tau_p}\sum_{\beta j \gamma \eta}\delta_{\alpha\beta}h_{\eta i}\,h_{\eta j}\,h_{j\gamma}^{-1}\left(P_{\text{int},\beta\gamma} - P_0\,\delta_{\beta\gamma}\right) \\
        &= -\frac{\beta_T}{3\tau_p}\sum_{\gamma} \left(P_0\,\delta_{\alpha\gamma} - P_{\text{int},\alpha\gamma}\right)h_{\gamma i} \,.
    \end{align}
    \end{subequations}
    Hence we have recovered the first term in the deterministic part of the anisotorpic SCR equations. Let's now focus on the term (b). To compute the drift tensor $\mathbf{T}$ we first need to solve the Lyapunov equation \cref{lyapunov}, where both $\bm{\gamma}$ and $\bm{\sigma}$ appear. The tensor $\bm{\gamma}$ is obtained by inverting \cref{gamma_inverse}:
    \begin{equation}\label{gamma}
        \gamma_{\alpha i\beta j} = \frac{3V\tau_p}{\beta_T}\delta_{\alpha\beta}\sum_{\eta}h_{i\eta}^{-1}\,h_{j\eta}^{-1}\,.
    \end{equation}
    The tensor $\bm{\sigma}$ is obtained instead by imposing the fluctuation-dissipation theorem in \cref{fluct_diss_multidim}, resulting in:
    \begin{equation}
        \sigma_{\alpha i\beta j} = \sqrt{\frac{6 V\tau_p k_B T}{\beta_T}}\,h_{i\beta}^{-1}\delta_{\alpha j}\,.
    \end{equation}
    With these expressions for $\bm{\gamma}$ and $\bm{\sigma}$, the solution of the Lyapunov equation turns out to be
    \begin{equation}
        J_{\alpha i \beta j} = k_B T \delta_{\alpha\beta}\,\delta_{ij}\,.
    \end{equation}
    We can now evaluate the term $(b)$:
    \begin{subequations}\label{almost_finished}
    \begin{align}
        (b)_{\alpha i} &= T_{\alpha i} = \sum_{\beta j\gamma k}\left(\frac{\partial}{\partial h_{\beta j}}\gamma_{\alpha i\gamma k}^{-1} \right)\,J_{\gamma k \beta j} = k_B T\sum_{\gamma k}\left(\frac{\partial}{\partial h_{\gamma k}}\gamma_{\alpha i\gamma k}^{-1} \right)\,.
    \end{align}
    \end{subequations}
    Performing the calculations with Jacobi's formula in \cref{jacobi} one finds for the derivative
    \begin{equation}
        \frac{\partial}{\partial h_{\gamma k}}\gamma_{\alpha i\gamma k}^{-1} = \frac{1}{V}\left( -h_{k\gamma}^{-1}\,\delta_{\alpha\gamma}\sum_\eta h_{\eta i}\,h_{\eta k} + \delta_{\alpha\gamma}\,\delta_{k i}\,h_{\gamma k} + \delta_{\alpha\gamma}\,\delta_{kk}\,h_{\gamma i}\right)   \,,
    \end{equation}
    and substituting in \cref{almost_finished}:
    \begin{equation}
        (b)_{\alpha i} = \frac{\beta_T}{3\tau_p}\frac{3 k_B T}{V}\,h_{\alpha i}\,.
    \end{equation}
    Finally, the last term is:
    \begin{subequations}
    \begin{align}
        (c)_{\alpha i} &= \sum_{\beta j\gamma k}\gamma_{\alpha i\beta j}^{-1}\,\sigma_{\beta j\gamma k}\,\text{d}W_{\gamma k} \\
        &= \sum_{\beta j\gamma k}\frac{\beta_T}{3V\tau_p}\delta_{\alpha\beta}\sum_\eta h_{\eta i}\,h_{\eta j}\sqrt{\frac{6V\tau_p k_B T}{\beta_T}}\,h_{\gamma j}^{-1}\,\delta_{\beta k}\,\text{d}W_{\gamma k} \\
        &= \sqrt{\frac{2\beta_T k_B T}{3V\tau_p}}\sum_\gamma h_{\gamma i}\,\text{d}W_{\alpha \gamma}\,.
    \end{align}
    \end{subequations}
    Evaluating $\left[(a) + (b)\right]\,\text{d}t + (c)$ we obtain the equations
    \begin{equation}
    \text{d}h_{\alpha i} = -\frac{\beta_T}{3\tau_p} \Bigg[\sum_\beta \Big( P_0\delta_{\alpha\beta} - P_{\text{int},\alpha\beta} \Big)h_{\beta i} - \frac{3 k_B T}{V}h_{\alpha i}\Bigg]\,\text{d}t + \sqrt{\frac{2\beta_T k_B T}{3V\tau_p}} \sum_{\beta}h_{\beta i}\,\text{d}W_{\alpha\beta}\,,
    \end{equation}
    which perfectly match the anisotropic SCR \cref{crescale_aniso} except for the additional term 
    \begin{equation}
        \text{d}\mathbf{h}^{PR} = \frac{\beta_T}{3\tau_p}\frac{2 k_B T}{V}\mathbf{I}\,\text{d}t\,,
    \end{equation}
    as claimed in \cref{limit_PR}. Note that these equations can be obtained with the same derivation of the anisotropic SCR equations outlined in Appendix \ref{appendix_derivation}, but employing as target distribution 
    \begin{equation}
        \mathcal{P}'_{N\mathbf{S}T}\big(\{\mathbf{q}_i,\mathbf{p}_i\},\mathbf{h}\big) \propto \exp\left[-\frac{1}{k_B T}\Big(K + U + P_0 \det \mathbf{h}\Big)\right]\,,
    \end{equation}
    namely neglecting the factor $\left(\det \mathbf{h}\right)^{-2}$ that was instead included in \cref{NPT_anisotropic}. In fact, neglecting this factor is equivalent to omit the term $(b)$ computed in \cref{term_b_derivation}, which is exactly the additional term appearing in this derivation but changed of sign.
    
    In summary, the anisotropic SCR barostat can be seen as the Parrinello-Rahman barostat plus a Langevin thermostat applied to the components of $\mathbf{h}$, with a $\mathbf{h}$-dependent friction tensor $\bm{\gamma} = \bm{\gamma}(\mathbf{h},W)$ defined as in \cref{gamma} and in the high-friction and zero-mass limit described in \cite{high_friction_limit}.
    
    \subsection*{High-friction limit with a generic external stress}
    If we consider the most general case $\mathbf{S}\neq P_0\mathbf{I}$, the tensor $\mathbf{F}(\mathbf{h})$ in \cref{PR_plusfriction} gains an additional term, namely it has to be substituted by
    \begin{equation}
        \tilde{\mathbf{F}}(\mathbf{h}) = \mathbf{F}(\mathbf{h}) - \mathbf{h}\,\bm{\Sigma}\,.
    \end{equation}
    Taking the high-friction limit, this term contributes only to the term $(a)$ in \cref{hf_limit}:
    \begin{subequations}
    \begin{align}
        \text{d}h^{\text{strain}}_{\alpha i} &= -\sum_{\beta j}\gamma_{\alpha i\beta j}^{-1}\left(\mathbf{h}\bm{\Sigma}\right)_{\beta j}\,\text{d}t \\
        &= -\sum_{\beta j}\frac{\beta_T}{3V\tau_p}\delta_{\alpha\beta}\sum_{\gamma}h_{\gamma i}\,h_{\gamma j}\sum_k h_{\beta k}\,\Sigma_{kj}\,\text{d}t \\
        &= -\frac{\beta_T}{3V\tau_p}\sum_{\gamma j k},h_{\alpha k}\,\Sigma_{kj}\,h_{\gamma j}\,h_{\gamma i}\,\text{d}t \\
        &= -\frac{\beta_T}{3V\tau_p}\left(\mathbf{h}\,\bm{\Sigma}\,\mathbf{h}^T \mathbf{h} \right)_{\alpha i}\,\text{d}t\,.
    \end{align}
    \end{subequations}
    This is exactly the additional term that appears in \cref{crescale_eqs_strain} in presence of a generic external stress.

    \chapter{Effective energy drift}\label{appendix_effenergy}
    Let's consider a sampling algorithm based on a differential equation that satisfies the detailed balance condition with respect to the distribution $\mathcal{P}(x)$:
    \begin{equation}
        \mathcal{P}(x)\Pi(x\rightarrow x') = \mathcal{P}(x')\Pi(x'\rightarrow x)\,.
    \end{equation}
    Here, $\Pi(x\rightarrow x')$ is the transition probability of moving to $x'$ starting from $x$.
    If the variable $x$ is a point in phase space, this condition has to be substituted with the generalized detailed balance described in \cref{hybrid_MC}.
    When the equation satisfying this condition is integrated approximately - using for instance a finite time step propagation of the variable $x$ - detailed balance is violated. The amount of this violation, namely how much the ratio
    \begin{equation}
        \frac{\mathcal{P}(x')\Pi(x'\rightarrow x)}{\mathcal{P}(x)\Pi(x\rightarrow x')}
    \end{equation}
    moves away from $1$, can be used to evaluate the quality of the integration,  in order to understand on the fly if the time step or other parameters of the integration algorithm were chosen correctly. The quantity that can be introduced with this purpose is the \emph{effective energy} $\widetilde{H}$ \cite{langevin_bussi}, defined through its finite increments:
    \begin{equation}\label{effective_energy}
        \widetilde{H}(t+\Delta t) - \widetilde{H}(t) = -k_B T \log\frac{\mathcal{P}(x')\Pi(x'\rightarrow x)}{\mathcal{P}(x)\Pi(x\rightarrow x')}\,.
    \end{equation}
    Clearly the transition probabilities embedded in $\Pi$ depend on the specific sampling algorithm.
    Summing consecutive increments $\Delta \tilde{H}$ one typically observes a stationary \emph{effective energy drift}, whose slope increases with the time step. Hence, the effective energy plays the role of a conserved quantity whose conservation law is violated for any finite time step, and that can be monitored to detect problems in the simulation. As an example, considering the microcanonical ensemble and the velocity Verlet integrator, the effective energy is just the total energy of the system. 
    
    The effective energy variations $\Delta \widetilde{H}$ can alternatively be used to implement accept-reject algorithms using the Metropolis-Hastings rule, where the acceptance $\alpha$ is computed as
    \begin{equation}
        \alpha = \text{min}\left[1,\text{exp}(-\Delta \widetilde{H})\right]\,;
    \end{equation}
    in this way, finite step errors are by construction corrected without the need of changing the time step. This scheme defines what are typically called \emph{Metropolized integrators} \cite{metropolized_integrator}.

    \section{Derivation of isotropic energy drift for SCR}\label{appendix:iso_effeng}
    We discuss here the isotropic contribution to the effective energy drift associated to the time-reversible integrator of \cref{sec:TR}. The derivation is the same reported in \cite{crescale_iso}, except for the additional strain energy $E_s = \frac{1}{2}\text{Tr}(\bm{\Sigma}\mathbf{G})$ included in the target distribution. Indeed, in the limit of small $\Delta t$ it is possible to show that \cref{eq:lambda_equivalent_1+3} samples the volume distribution
    \begin{equation}
        \mathcal{P}_{1+3}(V) \propto\exp\left[-\frac{1}{k_B T}\Big(K + U + P_0 V + E_s\Big)\right]\,,
    \end{equation}
    or equivalently, as a function of $\lambda$:
    \begin{equation}\label{eq:distr_1+3}
        \mathcal{P}_{1+3}(\lambda) \propto\lambda \exp\left[-\frac{1}{k_B T}\Big(K + U + P_0 \lambda^2 + E_s\Big)\right]\,.
    \end{equation}
    In \cref{eq:distr_1+3} the additional factor $\lambda$ comes from the Jacobian of the change of variable. By condensing the two half-steps in a single step of size $\Delta t$, according to \cref{eq:lambda_equivalent_1+3}, we can write the forward and backward moves as
    \begin{subequations}
    \begin{align}
        &\lambda^{i+1} = \lambda^i + \frac{D_\lambda}{k_B T}f({\lambda}^i) \Delta t + \sqrt{2D_\lambda \Delta t}\,\mathcal{R}^i\,,\label{eq:lambda_forward}\\
        &\lambda^{i} = \lambda^{i+1} + \frac{D_\lambda}{k_B T}f({\lambda}^{i+1}) \Delta t + \sqrt{2D_\lambda \Delta t}\,\mathcal{R}^{i+1}\,,\label{eq:lambda_backward}
    \end{align}
    \end{subequations}
    where $D_\lambda = \frac{k_B T\beta_T}{4\tau_p}$ and $f(\lambda) = -2\lambda\left(P_0 - P_{\text{int}} - \frac{k_BT}{2\lambda^2} + \frac{\text{Tr}\left(\mathbf{h}\,\bm{\Sigma}\,\mathbf{h}^T\right)}{3\lambda^2}\right)$. Then, the isotropic contribution to the effective energy drift is given by:
    \begin{equation}\label{eq:1+3_effective_energy}
        \Delta\widetilde{H}_{i\rightarrow i+1} = -k_B T \log\frac{\mathcal{P}_{1+3}(\lambda^{i+1})\Pi(\lambda^{i+1}\rightarrow \lambda^i)}{\mathcal{P}_{1+3}(\lambda^i)\Pi(\lambda^i\rightarrow \lambda^{i+1})}\,.
    \end{equation}
    The part due to the $\mathcal{P}_{1+3}$ probabilities is computed as:
    \begin{subequations}
    \begin{align}
        \Delta\widetilde{H}_{i\rightarrow i+1}^\mathcal{P} = -k_B T \log\frac{\mathcal{P}_{1+3}(\lambda^{i+1})}{\mathcal{P}_{1+3}(\lambda^i)} 
        = \Delta K + \Delta U + \Delta E_s + P_0\Delta\lambda^2 - k_B T\Delta\log\lambda \,.
    \end{align}
    \end{subequations}
    Recalling that $R^i$ and $R^{i+1}$ are zero-mean and unit-variance Gaussian numbers, the forward and backward transition probabilities have the following expressions: 
    \begin{subequations}
    \begin{align}
        \Pi(\lambda^i\rightarrow \lambda^{i+1}) &= \frac{1}{\sqrt{2\pi}}e^{-  \frac{(\mathcal{R}^i)^2}{2}}\frac{\text{d}\mathcal{R}^i}{\text{d}\lambda^{i+1}}\,, \label{eq:lambda_forward_pi}\\
        \Pi(\lambda^{i+1}\rightarrow \lambda^i) &= \frac{1}{\sqrt{2\pi}}e^{-  \frac{(\mathcal{R}^{i+1})^2}{2}}\frac{\text{d}\mathcal{R}^{i+1}}{\text{d}\lambda^{i}}\,,\label{eq:lambda_backward_pi}
    \end{align}
    \end{subequations}
    where the derivatives come from the changes of variable $\mathcal{R}^i\mapsto\lambda^{i+1}$, for the forward move, and $\mathcal{R}^{i+1}~\mapsto~\lambda^{i}$, for the backward one.
    By inverting \cref{eq:lambda_forward,eq:lambda_backward} with respect to $R^i$ and $R^{i+1}$ and substituting their expressions in \cref{eq:lambda_forward_pi,eq:lambda_backward_pi}, it is straightforward to show that the contribution to the effective energy drift given by the transition probabilities is:
    \begin{subequations}
    \begin{align}
        \Delta\widetilde{H}_{i\rightarrow i+1}^\Pi = -k_BT\log\frac{\Pi(\lambda^{i+1}\rightarrow \lambda^i)}{\Pi(\lambda^{i}\rightarrow \lambda^{i+1})} 
        = \Delta\lambda\left(\frac{f(\lambda^i)+f(\lambda^{i+1})}{2}\right) + \frac{\beta_T\Delta t}{16\tau_p}\Delta f^2\,.
    \end{align}
    \end{subequations}
    Then, by summing $\Delta\widetilde{H}_{i\rightarrow i+1}^{\mathcal{P}}$ and $\Delta\widetilde{H}_{i\rightarrow i+1}^\Pi$ one recovers the result written in \cref{eq:eff_eng_iso}.

    \section{Derivation of anisotropic energy drift for SCR}\label{appendix:aniso_effeng}
    In this section we report the derivation of the anisotropic contribution to the effective energy drift, within the time-reversible integration scheme of \cref{sec:TR}. For this purpose, let's rewrite the second step of \cref{alg:trotterized_integrator}, namely the rescaling
    \begin{equation}
        \mathbf{h}^{i+1} = \exp\big(\mathbf{A}_2^i\,\Delta t + b\,\Delta \mathbf{W}_2^i\big)\,\mathbf{h}^{i}\,,
    \end{equation}
    by introducing an auxiliary momentum variable $\bm{\alpha}$ in a time-reversible fashion, shown in \cref{alg:TR_alpha}.\\
    
    \begin{algorithm}[H]
    \SetAlgoLined
     $\bm{\alpha}\gets\bm{\mathcal{R}}_2$, where $\bm{\mathcal{R}}_2 = \bm{\mathcal{R}} - \text{Tr}\left(\bm{\mathcal{R}}\right)/3$\;
     $\bm{\alpha}\gets\bm{\alpha}+\mathbf{A}_2\sqrt{\Delta t}/b$\;
     $\mathbf{h} \gets \exp\big(\bm{\alpha}b\sqrt{\Delta t} \big)\,\mathbf{h}$\;
     recompute $\mathbf{A}_2$\;
     $\bm{\alpha}\gets\bm{\alpha}+\mathbf{A}_2\sqrt{\Delta t}/b$\;
     $\bm{\alpha}\gets\bm{\mathcal{R}}_2'$, where $\bm{\mathcal{R}}_2' = \bm{\mathcal{R}}' - \text{Tr}\left(\bm{\mathcal{R}}'\right)/3$\;
     \caption{Change of box shape with the auxiliary variable $\bm{\alpha}$.}
     \label{alg:TR_alpha}
    \end{algorithm}
    \vspace{2.5cm}
    We recall that $b$ only depends on $\mathbf{h}$ through its determinant, that is untouched in the rescaling at step 3. Therefore, in this context $b$ can be treated as a constant and its time index is omitted; for the same reason, only $\mathbf{A}_2$ is recomputed at step 4.\\
    \noindent In steps 1 and 6, $\bm{\mathcal{R}}$ and $\bm{\mathcal{R}}'$ are 3$\times$3 matrices of i.i.d. zero-mean and unit-variance Gaussian numbers. The effective energy drift is formally defined according to \cref{effective_energy}, which here takes the form
    \begin{equation}\label{eq:aniso_drift}
        \Delta \widetilde{H}_{i\rightarrow i+1} = -k_B T \log\frac{\mathcal{P}\big(-\bm{\alpha}^{i+1},\mathbf{h}^{i+1}\big)\,\Pi\big((-\bm{\alpha}^{i+1},\mathbf{h}^{i+1})\rightarrow (-\bm{\alpha}^i,\mathbf{h}^i)\big)}{\mathcal{P}\big(\bm{\alpha}^i,\mathbf{h}^i\big)\,\Pi\big((\bm{\alpha}^i,\mathbf{h}^i)\rightarrow(\bm{\alpha}^{i+1},\mathbf{h}^{i+1})\big)}\,.
    \end{equation}
    Numerator and denominator in the argument of the logarithm can be evaluated by considering respectively the \emph{forward} and \emph{backward} moves in the scheme of \cref{alg:TR_alpha}, where the momentum-like variables $\bm{\alpha}$ gain a minus sign in the backward trajectory, exactly as discussed in the context of generalized detailed balance in \cref{hybrid_MC}. The forward move ($\bm{\alpha}^i,\mathbf{h}^i$)$\,\mapsto\,$($\bm{\alpha}^{i+1},\mathbf{h}^{i+1}$) can be written as
    \begin{subequations}\label[pluralequation]{forward_traj}
    \begin{align}
        \bm{\alpha}^{i_+} &= \bm{\mathcal{R}}_{2F}^i + \mathbf{A}_2^i\frac{\sqrt{\Delta t}}{b}\,,\label{1st_eq_forward}\\
        \mathbf{h}^{i+1} &= \exp\big(b\sqrt{\Delta t}\,\bm{\alpha}^{i_+}\big)\,\mathbf{h}^i\,, \label{2nd_eq_forward}\\
        \bm{\alpha}^{i+1} &= \bm{\mathcal{R}}_{2F}^{i+1}\,,\label{3rd_eq_forward}
    \end{align}
    \end{subequations}
    where the superscript $i_+$ identifies an intermediate step between $i$ and $i+1$. Similarly, calling $(i+1)_-$ an intermediate step in the backward trajectory, the move ($-\bm{\alpha}^{i+1},\mathbf{h}^i$)$\,\mapsto\,$($-\bm{\alpha}^{i},\mathbf{h}^{i}$) reads:
    \begin{subequations}\label[pluralequation]{backward_traj}
    \begin{align}
        -\bm{\alpha}^{(i+1)_-} &= \bm{\mathcal{R}}_{2B}^{i+1} + \mathbf{A}_2^{i+1}\frac{\sqrt{\Delta t}}{b}\,,\\
        \mathbf{h}^{i+1} &= \exp\big(-b\sqrt{\Delta t}\,\bm{\alpha}^{(i+1)_-}\big)\,\mathbf{h}^i\,,\label{2nd_eq_backward}\\
        -\bm{\alpha}^{i} &= \bm{\mathcal{R}}_{2B}^{i}\,.
    \end{align}
    \end{subequations}
    Note that by inverting \cref{2nd_eq_forward,2nd_eq_backward} with respect to $\bm{\alpha}^{i_+}$ and $\bm{\alpha}^{(i+1)_-}$ we obtain respectively:
    \begin{subequations}
    \begin{align}
        \bm{\alpha}^{i_+} &= \frac{1}{b\sqrt{\Delta t}}\log\big(\mathbf{h}^{i+1}(\mathbf{h}^{i})^{-1}\big)\,,\\
        \bm{\alpha}^{(i+1)_-} &= -\frac{1}{b\sqrt{\Delta t}}\log\big(\mathbf{h}^{i}(\mathbf{h}^{i+1})^{-1}\big) = \frac{1}{b\sqrt{\Delta t}}\log\big(\mathbf{h}^{i+1}(\mathbf{h}^{i})^{-1}\big)\,,\label{matrix_log_prop}
    \end{align}
    \end{subequations}
    where the \emph{matrix logarithm} appearing in the two equations is defined as the inverse operation of the matrix exponential. In the last passage of \cref{matrix_log_prop} we have used the properties 
    \begin{subequations}\label[pluralequation]{generic_properties}
    \begin{align}
        -\log\big(\mathbf{M}_1\big) &= \log\big(\mathbf{M}_1^{-1}\big)\,, \\
        \big(\mathbf{M}_1\,\mathbf{M}_2\big)^{-1} &= \mathbf{M}_2^{-1}\,\mathbf{M}_1^{-1}\,,
    \end{align}
    \end{subequations}
    holding for two generic matrices $\mathbf{M}_1,\,\mathbf{M}_2$. As a consequence, $\bm{\alpha}^{i_+}$ and $\bm{\alpha}^{(i+1)_-}$, are actually the same matrix, which will be called $\bm{\alpha}^{i+1/2}$ in the following. Let's now evaluate the forward transition probability $\Pi_F = \Pi\big((\bm{\alpha}^i,\mathbf{h}^i)\rightarrow(\bm{\alpha}^{i+1},\mathbf{h}^{i+1})\big)$ appearing in \cref{eq:aniso_drift}. To write $\Pi_F$ correctly, we recall that $\bm{\mathcal{R}}_{2F}^i$ and $\bm{\mathcal{R}}_{2F}^{i+1}$ are obtained
    as 
    \begin{subequations}
    \begin{align}
        \bm{\mathcal{R}}_{2F}^i &= \bm{\mathcal{R}}_{F}^i - \text{Tr}\big(\bm{\mathcal{R}}_{F}^i\big)/3\,,\\
        \bm{\mathcal{R}}_{2F}^{i+1} &= \bm{\mathcal{R}}_{F}^{i+1} - \text{Tr}\big(\bm{\mathcal{R}}_{F}^{i+1}\big)/3\,,
    \end{align}
    \end{subequations}
    where $\bm{\mathcal{R}}_{F}^i$ and $\bm{\mathcal{R}}_{F}^{i+1}$ are 3$\times$3 matrices of i.i.d. zero-mean and unit-variance Gaussian numbers. As a consequence, it is easy to show that the diagonal elements of $\bm{\mathcal{R}}_{2F}^i$ and $\bm{\mathcal{R}}_{2F}^{i+1}$ are Gaussian numbers with zero-mean and variance equal to 2/3, satisfying the constraints $\text{Tr}\big(\bm{\mathcal{R}}_{2F}^i\big) = \text{Tr}\big(\bm{\mathcal{R}}_{2F}^{i+1}\big) = 0$. In other words, $\bm{\mathcal{R}}_{2F}^i$ and $\bm{\mathcal{R}}_{2F}^{i+1}$ are random matrices containing only 8 independent elements, and a transformation followed by a marginalization over the redundant degree of freedom is necessary to write their joint probability distribution. With this regard, we introduce a linear transformation $\mathcal{G}$ that maps a generic 3$\times$3 matrix $\mathbf{M}$ into a 9-dimensional vector $\widetilde{\mathbf{M}}$ such that
    \begin{subequations}\label[pluralequation]{eq:m0m1m2}
    \begin{align}
        \widetilde{M}_0 &= \text{Tr}\big(\mathbf{M}\big)/2\,, \label{first_tilde_var}\\
        \widetilde{M}_1 &= \big(M_{xx} + M_{yy} - M_{zz}\big) / \sqrt{2}\,,\\
        \widetilde{M}_2 &= \big(M_{xx} - M_{yy} + M_{zz}\big) / \sqrt{2}\,,
    \end{align}
    \end{subequations}
    and the remaining six components $\widetilde{M}_j$ ($j = 3,...,8$) correspond to the off-diagonal elements of $\widetilde{\mathbf{M}}$, in an order that is irrelevant for the following reasoning. If we apply this transformation to $\bm{\mathcal{R}}_{2F}^i$ and $\bm{\mathcal{R}}_{2F}^{i+1}$, we can observe that the choice of the prefactors $1/\sqrt{2}$ in \cref{eq:m0m1m2} are such that the components $\widetilde{{\mathcal{R}}}_{2F,j}^i$ and $\widetilde{{\mathcal{R}}}_{2F,j}^{i+1}$ from $j=1$ to $j=8$ are zero-mean and unit-variance Gaussian numbers, and that in these new variables the constraints reported above simply read $\widetilde{{\mathcal{R}}}_{2F,0}^i = \widetilde{{\mathcal{R}}}_{2F,0}^{i+1} = 0$. Then, since the Jacobian of the transformation $\mathcal{G}$ is $1$ in absolute value, we can write the the forward transition probability as
    \begin{equation}\label{Pi_f}
        \Pi_F = \frac{1}{(2\pi)^9}\exp\left[-\frac{1}{2}\sum_{j=1}^8 \left(\left(\widetilde{{\mathcal{R}}}_{2F,j}^i\right)^2 +\left(\widetilde{{\mathcal{R}}}_{2F,j}^{i+1}\right)^2 \right)\right]\,\Big|\text{det}\,\mathbf{J}_F \Big|\,,
    \end{equation}
    where $\mathbf{J}_F$ is the Jacobian of the transformation $(\bm{\mathcal{R}}_{2F}^i,\bm{\mathcal{R}}_{2F}^{i+1})\mapsto(\bm{\alpha}^{i+1},\mathbf{h}^{i+1})$. Since $\bm{\mathcal{R}}_{2F}^i$ only depends on $\mathbf{h}^{i+1}$ and $\bm{\mathcal{R}}_{2F}^{i+1}$ only depends on $ \bm{\alpha}^{i+1}$, this matrix is block-diagonal and its determinant factorizes as 
    \begin{equation}
        \text{det}\,\mathbf{J}_F = \left(\text{det}\,\mathbf{J}_{\bm{\mathcal{R}}_{2F}^i\mapsto\mathbf{h}^{i+1}} \right) \left(\text{det}\,\mathbf{J}_{\bm{\mathcal{R}}_{2F}^{i+1}\mapsto\bm{\alpha}^{i+1}} \right)\,.
    \end{equation}
    From \cref{3rd_eq_forward} we observe that the second determinant is simply 1, while from \cref{1st_eq_forward} we can write the components of the first Jacobian tensor as 
    \begin{equation}
        \left(\mathbf{J}_{\bm{\mathcal{R}}_{2F}^i\mapsto\mathbf{h}^{i+1}}\right)_{\gamma k\alpha\beta} = \frac{\partial\, \mathcal{R}_{2F,\alpha\beta}^i}{\partial\, h_{\gamma k}^{i+1}} = \frac{1}{b\sqrt{\Delta t}}\frac{\partial}{\partial h^{i+1}_{\gamma k}}\left(\log \mathbf{h}^{i+1}(\mathbf{h}^{i})^{-1} \right)_{\alpha\beta}
    \end{equation}
    Performing the same calculations on the time-reversed trajectory in \cref{backward_traj}, the expression that one finds for the backward transition probability $\Pi_B = \Pi\big((-\bm{\alpha}^{i+1},\mathbf{h}^{i+1})\rightarrow (-\bm{\alpha}^i,\mathbf{h}^i)\big)$ reads
    \begin{equation}\label{Pi_b}
        \Pi_B = \frac{1}{(2\pi)^9}\exp\left[-\frac{1}{2}\sum_{j=1}^8 \left(\left(\widetilde{{\mathcal{R}}}_{2B,j}^i\right)^2 +\left(\widetilde{{\mathcal{R}}}_{2B,j}^{i+1}\right)^2 \right)\right]\,\Big|\text{det}\,\mathbf{J}_B \Big|\,,
    \end{equation}
   and in this case the non-trivial part of the Jacobian is:
    \begin{equation}
        \left(\mathbf{J}_{\bm{\mathcal{R}}_{2B}^{i+1}\mapsto\mathbf{h}^{i}}\right)_{\gamma k\alpha\beta} = \frac{\partial\, \mathcal{R}_{2B,\alpha\beta}^{i+1}}{\partial\, h_{\gamma k}^{i}} = \frac{1}{b\sqrt{\Delta t}}\frac{\partial}{\partial h^{i}_{\gamma k}}\left(\log \mathbf{h}^{i}(\mathbf{h}^{i+1})^{-1} \right)_{\alpha\beta}\,.
    \end{equation}
    Therefore the contribution of the two Jacobians in the ratio $\Pi_B/\Pi_F$ in \cref{eq:aniso_drift} can be manipulated as it follows:
    \begin{subequations}
    \begin{align}
        \frac{\text{det}\,\mathbf{J}_B }{\text{det}\,\mathbf{J}_F } &= \frac{\text{det}\left[\frac{\partial}{\partial \,\mathbf{h}^{i}}\log \mathbf{h}^{i}(\mathbf{h}^{i+1})^{-1} \right]}{\text{det}\left[\frac{\partial}{\partial\, \mathbf{h}^{i+1}}\log \mathbf{h}^{i+1}(\mathbf{h}^{i})^{-1} \right]} = \\
        &= \text{det}\left[\left(\frac{\partial}{\partial \,\mathbf{h}^{i}}\log \mathbf{h}^{i}(\mathbf{h}^{i+1})^{-1}\right)\left(\frac{\partial}{\partial\, \mathbf{h}^{i+1}}\log \mathbf{h}^{i+1}(\mathbf{h}^{i})^{-1}\right)^{-1} \right] \label{eq:d17b}\\
        &= \text{det}\left[\left(\frac{\partial}{\partial \,\mathbf{h}^{i}}\log \mathbf{h}^{i}(\mathbf{h}^{i+1})^{-1}\right)\left(\frac{\partial\mathbf{h}^{i+1}}{\partial\, \log \mathbf{h}^{i+1}(\mathbf{h}^{i})^{-1}}\right) \right]\,. \label{eq:d17c}
    \end{align}
    \end{subequations}
    \Cref{eq:d17b} is obtained using the well-known property $(\text{det}\,\mathbf{M})^{-1} = \text{det}\big(\mathbf{M}^{-1}\big)$, while \cref{eq:d17c} relies on the fact that 
    \begin{equation}
        \frac{\partial \mathbf{X}}{\partial\mathbf{Y}} = \left(\frac{\partial\mathbf{Y}}{\partial \mathbf{X}}\right)^{-1}
    \end{equation}
    if $\mathbf{Y} = \mathbf{Y}(\mathbf{X})$ is continuous and differentiable in $X_{ij}$, if it is invertible and its inverse $\mathbf{X} = \mathbf{X}(\mathbf{Y})$ has the same properties with respect to $Y_{ij}$. In fact if these hypothesis hold, as in the case of $\mathbf{Y}=\log \mathbf{h}^{i+1}(\mathbf{h}^{i})^{-1}$ and $\mathbf{X}=\mathbf{h}^{i+1}$, then
    \begin{subequations}
    \begin{align}
        \sum_{kl}\left(\frac{\partial \mathbf{Y}}{\partial\mathbf{X}} \right)_{ijkl}\left(\frac{\partial \mathbf{X}}{\partial\mathbf{Y}} \right)_{klmn}  = \sum_{kl} \frac{\partial {Y}_{kl}}{\partial{X}_{ij}}\frac{\partial {X}_{mn}}{\partial{Y}_{kl}} = \delta_{mi}\,\delta_{jn}\,.
    \end{align}
    \end{subequations}
    Starting again from \cref{eq:d17c} we can write:
    \begin{subequations}
    \begin{align}
        \frac{\text{det}\,\mathbf{J}_B }{\text{det}\,\mathbf{J}_F } &= \text{det}\left[\left(-\frac{\partial\,\log \mathbf{h}^{i+1}(\mathbf{h}^{i})^{-1}}{\partial \,\mathbf{h}^{i}}\right)\left(\frac{\partial\mathbf{h}^{i+1}}{\partial\, \log \mathbf{h}^{i+1}(\mathbf{h}^{i})^{-1}}\right) \right] \\
        &= \text{det}\left[-\left(\frac{\partial\,\log \mathbf{h}^{i+1}(\mathbf{h}^{i})^{-1}}{\partial\, \log \mathbf{h}^{i+1}(\mathbf{h}^{i})^{-1}}\right)\left(\frac{\partial\mathbf{h}^{i+1}}{\partial \,\mathbf{h}^{i}}\right) \right] \\
        &= \text{det}\left[-\left(\frac{\partial\mathbf{h}^{i+1}}{\partial \,\mathbf{h}^{i}}\right) \right]\,.
    \end{align}
    \end{subequations}
    Here we have used both the properties in \cref{generic_properties} and the fact that 
    \begin{equation}
        \frac{\partial\mathbf{A}}{\partial\mathbf{B}}\frac{\partial\mathbf{C}}{\partial\mathbf{A}} = \frac{\partial\mathbf{C}}{\partial\mathbf{B}}
    \end{equation}
    if these matrix derivatives are well defined. In fact, writing explicitly the product:
    \begin{subequations}
    \begin{align}
        \left(\frac{\partial\mathbf{A}}{\partial\mathbf{B}}\frac{\partial\mathbf{C}}{\partial\mathbf{A}}\right)_{ijmn} &= \sum_{kl}\left(\frac{\partial\mathbf{A}}{\partial\mathbf{B}}\right)_{ijkl}\left(\frac{\partial\mathbf{C}}{\partial\mathbf{A}}\right)_{klmn} = \sum_{kl}\frac{\partial{A}_{kl}}{\partial{B}_{ij}}\frac{\partial{C}_{mn}}{\partial{A}_{kl}} = \frac{\partial{C}_{mn}}{\partial{B}_{ij}}\,.
    \end{align}
    \end{subequations}
    Finally, the remaining derivative can be easily computed using \cref{2nd_eq_forward}:
    \begin{equation}
        \frac{\partial\mathbf{h}^{i+1}}{\partial\,\mathbf{h}^{i}} = \frac{\partial}{\partial\,\mathbf{h}^{i}}\exp\big(b\sqrt{\Delta t}\,\bm{\alpha}^{i+1/2}\big)\,\mathbf{h}^i = \exp\big(b\sqrt{\Delta t}\,\bm{\alpha}^{i+1/2}\big)\,.
    \end{equation}
    Since this exponential matrix has determinant equal to 1, as discussed in \cref{sec:TR}, the ratio of the Jacobians in $\Pi_B/\Pi_F$ is simply 
    \begin{equation}
        \left| \frac{\text{det}\,\mathbf{J}_B }{\text{det}\,\mathbf{J}_F }\right| = 1\,.
    \end{equation}
    Then, using \cref{Pi_f,Pi_b} the contribution of the transitions probabilities to the anisotropic drift reads:
    \begin{subequations}
    \begin{align}
        \Delta \widetilde{H}_{i\rightarrow i+1}^{\Pi} &= -k_B T\log\frac{\Pi_B}{\Pi_F} \\
        &=  \frac{k_B T}{2}\sum_{j=1}^8\left[\left(\widetilde{\mathcal{R}}_{2B,j}^{i+1}\right)^2+\left(\widetilde{\mathcal{R}}_{2B,j}^{i}\right)^2 - \left(\widetilde{\mathcal{R}}_{2F,j}^{i+1}\right)^2 -\left(\widetilde{\mathcal{R}}_{2F,j}^{i}\right)^2\right]
    \end{align}
    \end{subequations}
    Substituting the expressions of these four random vectors in terms of the $\mathcal{G}$-transformed quantities appearing in \cref{forward_traj} and \cref{backward_traj} one finds:
    \begin{equation}\label{deltaH_Pi_tilde}
        \Delta \widetilde{H}_{i\rightarrow i+1}^{\Pi} = \frac{k_B T}{2}\sum_{j=1}^8 \left[\frac{\Delta t}{b^2}\Delta \widetilde{A}_{2j}^{\,2} -\Delta\widetilde{\alpha}_{j}^2 + 2\widetilde{\alpha}_j^{i+1/2}\frac{\sqrt{\Delta t}}{b}\left(\widetilde{A}_{2j}^{\,i} + \widetilde{A}_{2j}^{\,i+1} \right)\right]\,.
    \end{equation}
    Note that we can include for each term in the sum also the component $j=0$ as defined in \cref{first_tilde_var}, since all the matrices considered here are traceless. As a consequence, all the terms in \cref{deltaH_Pi_tilde} can be interpreted as squared moduli or scalar products of 9-dimensional vectors obtained from the transformation $\mathcal{G}$. Since this transformation is unitary and then preserves all scalar products, it is possible to map back \cref{deltaH_Pi_tilde} to the original matrix quantities by applying the inverse transformation $\mathcal{G}^{-1}$:
    \begin{equation}
        \Delta \widetilde{H}_{i\rightarrow i+1}^{\Pi} = \frac{k_B T}{2}\sum_{\alpha\beta} \left[\frac{\Delta t}{b^2}\Delta {A}_{2,\alpha\beta}^{\,2} -\Delta{\alpha}_{\alpha\beta}^2 + 2{\alpha}_{\alpha\beta}^{i+1/2}\frac{\sqrt{\Delta t}}{b}\left({A}_{2,\alpha\beta}^{\,i} + {A}_{2,\alpha\beta}^{\,i+1} \right)\right]\,.
    \end{equation}
    Let's now evaluate the remaining contribution to \cref{eq:aniso_drift}, starting from the following ratio:
    \begin{equation}
        \frac{\mathcal{P}\big(-\bm{\alpha}^{i+1},\mathbf{h}^{i+1}\big)}{\mathcal{P}\big(\bm{\alpha}^{i},\mathbf{h}^{i}\big)} = \frac{\mathcal{P}\big(-\bm{\alpha}^{i+1}\big)\,\mathcal{P}\big(\mathbf{h}^{i+1}\big)}{\mathcal{P}\big(\bm{\alpha}^{i}\big)\,\mathcal{P}\big(\mathbf{h}^{i}\big)}\,.
    \end{equation}
    The equality holds because the distributions of $\mathbf{h}$ and $\bm{\alpha}$ are independent. According to the $N\mathbf{S}T$ distribution in \cref{NPT_anisotropic}, the contribution from the ratio of the $\mathbf{h}$-distributions results in a sum of energy increments, only due to the change of shape: 
    \begin{equation}
        \Delta \widetilde{H}_{i\rightarrow i+1}^{\mathcal{P}(\mathbf{h})} = -k_B T \log\frac{\mathcal{P}\big(\mathbf{h}^{i+1}\big)}{\mathcal{P}\big(\mathbf{h}^{i}\big)} = \Delta K + \Delta U + \Delta E_s\,.
    \end{equation}
    The only remaining term is 
    \begin{subequations}\label{last_effort}
    \begin{align}
         \Delta \widetilde{H}_{i\rightarrow i+1}^{\mathcal{P}(\bm{\alpha})} = -k_B T \log\frac{\mathcal{P}\big(-\bm{\alpha}^{i+1}\big)}{\mathcal{P}\big(\bm{\alpha}^{i}\big)}\,.
    \end{align}
    \end{subequations}
    Recalling from \cref{forward_traj} and \cref{backward_traj} that $\bm{\alpha}^{i+1} = \bm{\mathcal{R}}_{2F}^{i+1}$ and $\bm{\alpha}^{i} = -\bm{\mathcal{R}}_{2B}^{i}$, it is possible to evaluate the distribution of $\alpha$ and the ratio in \cref{last_effort} using exactly the same procedure shown before to compute the transition probabilities. The final result,
    \begin{equation}
        \Delta \widetilde{H}_{i\rightarrow i+1}^{\mathcal{P}(\bm{\alpha})} = \frac{k_B T}{2}\sum_{\alpha\beta}\Delta\alpha_{\alpha\beta}^2\,,
    \end{equation}
    cancels the second term in \cref{deltaH_Pi_tilde}. Then, calling $\Delta\bm{\varepsilon}$ the argument of the exponential matrix responsible for the change of shape, namely $\Delta\bm{\varepsilon} = b\sqrt{\Delta t}\bm{\alpha}^{i+1/2}$, the sum $\Delta \widetilde{H}_{i\rightarrow i+1}^{\Pi} + \Delta \widetilde{H}_{i\rightarrow i+1}^{\mathcal{P}(\mathbf{h})} + \Delta \widetilde{H}_{i\rightarrow i+1}^{\mathcal{P}(\bm{\alpha})}$ results in the expression reported in \cref{eq:eff_eng_aniso}.

    \chapter{PBCs and Bravais lattices}\label{PBCs}
\begin{figure}[h!]
        \centering
        \includegraphics[width = 0.35\textwidth]{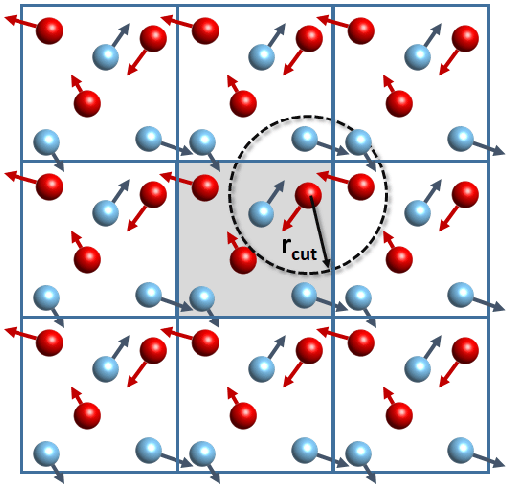}
        \caption{Graphical representation of PBCs in a two-dimensional system within a square box. Source: \cite{pbcs_img}.}
        \label{fig:pbcs}
    \end{figure}

    Periodic boundary conditions (PBCs) are widely used in MD simulations to eliminate effects due to the boundaries, where the physical behaviour of the system could be substantially different than in the bulk. The implementation of these conditions is based on a geometric construction called \emph{Bravais lattice}, which is built using as \emph{primitive vectors} the box vectors $\mathbf{a},\mathbf{b},\mathbf{c}$. This periodic construction is mathematically defined as the infinite set of points $\{\mathbf{r}_{n_1 n_2 n_3}\}$ that one can generate by means of integer linear combinations of the primitive vectors:
    \begin{equation}
        \mathbf{r}_{n_1 n_2 n_3} = n_1 \mathbf{a} + n_2 \mathbf{b} + n_3 \mathbf{c}\,,
    \end{equation}
    where $n_1,\,n_2$ and $n_3$ are integer numbers. The points $\{\mathbf{r}_{n_1 n_2 n_3}\}$ define a periodic structure throughout the space where the fundamental unit, called \emph{unit cell}, has the geometry of the box defined by the three vectors $\mathbf{a},\mathbf{b},\mathbf{c}$. If we duplicate and translate every atom according to all the Bravais lattice vectors $\{\mathbf{r}_{n_1 n_2 n_3}\}$, all the space will be filled by periodic copies of the system of interest (see \cref{fig:pbcs}).  
    Then, PBCs can be applied by making each atom interact with all the periodic copies of the remaining ones. Since it is not possible to deal with infinite interactions, a scheme called \emph{minimal image convention} is typically employed: the (short-range) interactions are considered only within a cut-off $r_{\text{cut}}$ such that, with a suitable choice of the box dimensions, each atom turns out to interact with only the nearest copy of each other atom. In particular, the necessary condition to apply this convention is
    \begin{equation}
        r_{\text{cut}} < \frac{1}{2}\text{min}\left(|\mathbf{a}|,|\mathbf{b}|,|\mathbf{c}| \right)\,.
    \end{equation}
    
\begin{figure}[h!]
        \centering
        \includegraphics[width = 0.45\textwidth]{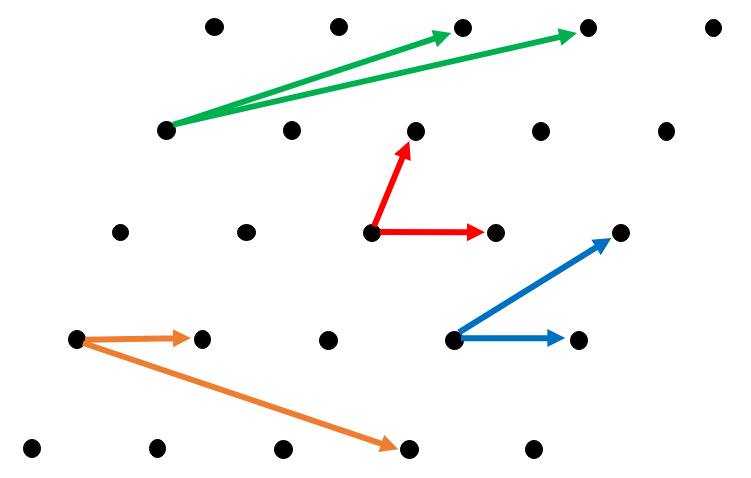}
        \caption{Equivalent choices of the primitive vectors for a two-dimensional Bravais lattice.}
        \label{fig:bravais_lattice_equiv}
    \end{figure}

    It is relevant to observe that the box vectors generating a given Bravais lattice structure are not unique (see \cref{fig:bravais_lattice_equiv}), and independent choices are connected by a transformation of the form \cref{change_cell_vectors}. Since the results of any MD simulation should be independent on this choice, this explains the check for the invariance of the anisotropic SCR equations under a redefinition of the box vectors, and the reason why this property should be satisfied by any equation describing anisotropic volume fluctuations in finite systems. 
    
    We finally report in \cref{tab:boxes_def} some of the most employed box shapes in MD simulations, using the convention of upper triangular box matrices. 
    \begin{table}[h!]
    \centering
    \caption{Definitions of common box shapes.}
    \label{tab:boxes_def}
    \begin{tabular}{p{5cm} p{4cm} c}
        \toprule[0.5pt]\toprule[0.5pt]
        \textbf{Box type}       & \textbf{Box matrix} $\mathbf{h}$  & \textbf{Volume} \\\midrule
        Cubic                   & $\begin{pmatrix}                  
                                        d & 0 & 0\\
                                        0 & d & 0\\
                                        0 & 0 & d
                                \end{pmatrix}$                      & $d^3$ \\
        Orthorhombic            & $\begin{pmatrix}                  
                                        d_x & 0 & 0\\
                                        0 & d_y & 0\\
                                        0 & 0 & d_z
                                \end{pmatrix}$                      & $d_x\,d_y\,d_z$ \\
        Rhombic dodecahedron    & $\begin{pmatrix}                  
                                        d & 0 & d/2\\
                                        0 & d & d/2\\
                                        0 & 0 & \sqrt{2}\,d/2
                                \end{pmatrix}$                      & $\frac{\sqrt{2}}{2}d^3 \simeq 0.71\,d^3$ \\
        Truncated octahedron    & $\begin{pmatrix}                  
                                        d & d/3 & -d/3\\
                                        0 & 2\sqrt{2}\,d/3 & \sqrt{2}\,d/3\\
                                        0 & 0 & \sqrt{2}\,d/3
                                \end{pmatrix}$                      & $\frac{4\sqrt{3}}{9}d^3 \simeq 0.77\,d^3$ \\\bottomrule[0.5pt]\bottomrule[0.5pt]                          
        \end{tabular}
    \end{table}
    
    \noindent Note that all these shapes are particular cases of \emph{triclinic} boxes, where $\mathbf{h}$ contains six independent non-zero elements, resulting in a parallelepiped with generic edge lengths $|\mathbf{a}|,|\mathbf{b}|,|\mathbf{c}|$ and generic angles $\alpha,\beta,\gamma$,
    
    \chapter{First attempt for a time-reversible integrator}\label{appendix_epsvariables}
    We discuss here an attempt to construct an integrator of \cref{crescale_aniso} with a time-reversible behaviour in the limit of small time steps (i.e. only considering the stochastic part of the equations). Let's introduce nine new variables as the components of a matrix $\bm{\varepsilon}$ defined as:
    \begin{equation}
        \mathbf{h} = e^{\bm{\varepsilon}}\,\mathbf{h}_0\,.
    \end{equation}
    Here $\mathbf{h}_0$ is a reference box matrix and $e^{\bm{\varepsilon}}$ is the \emph{matrix exponential} of $\bm{\varepsilon}$, defined by the power series
    \begin{equation}
        e^{\bm{\varepsilon}} = \sum_{k= 0}^\infty \frac{1}{k!}\,\bm{\varepsilon}^k\,.
    \end{equation}
    Note that $\bm{\varepsilon}$ can be seen as the generalization of the logarithmic volume $\varepsilon = \log(V/V_0)$ defined in the isotropic case. Since the expansion above contains infinite terms, $\exp(\bm{\varepsilon})$ cannot be practically computed in general, but in case of small matrix increments $\exp(\text{d}\bm{\varepsilon})$ can be approximated efficiently by truncating the expansion after a few terms. 
    With a relevant \emph{caveat} highlighted below, the equations for $\bm{\varepsilon}$ can be obained by means of It\^{o}'s chain rule. Then, using the well-known property $\text{det}\big(\text{exp}(\bm{\varepsilon})\big) = \text{exp}\big(\text{Tr}(\bm{\varepsilon})\big)$ the isotropic degree of freedom can be embedded in the variable 
    \begin{equation}\label{det_of_exp}
        \lambda = \sqrt{V_0}\exp\left(\frac{1}{2}\text{Tr}(\bm{\varepsilon})\right) = \sqrt{V}\,,
    \end{equation}
    which is decoupled from the other degrees of freedom and can be propagated as in the isotropic case with \cref{lambda_eq_SCR}, where the noise prefactor is constant. The other eight variables $\Phi_k$ ($k=1,...,8$) can be chosen as independent linear combinations of the box matrix components and they are propagated according to eight SDEs with a $\lambda$-dependent noise prefactor $b(\lambda)$, which can be symmetrized by means of a geometric mean between the values of $\lambda$ at the current and the next steps:
    \begin{equation}
        b(\lambda_{t}) \longmapsto b\left(\sqrt{\lambda_{t}\,\lambda_{t+\Delta t}}\,\right)\,.
    \end{equation}
    This operation, which requires $\lambda$ to be propagated before the other eight variables at each step, should enhance the time-reversibility of the generated trajectory, and as a consequence the "good scaling" of the effective energy drift with the integration time step. However, this integrator actually appears to work worse than the Euler one (see \cref{euler_integrator}) for at least two reasons:
    \begin{itemize}
        \item The derivation of the equations for $\bm{\varepsilon}$ requires to know the derivatives 
        \begin{equation}
            \frac{\partial \varepsilon_{\alpha\beta}}{\partial h_{\gamma i}} = \frac{\partial }{\partial h_{\gamma i}}\log(\mathbf{h}\,\mathbf{h}_0^{-1})_{\alpha\beta}\,,
        \end{equation}
        where the \emph{matrix logarithm} appearing above is the inverse of the matrix exponential previously defined. Unfortunately, these derivatives do not admit any closed-form solution and they can only be computed in the (wrong) hypothesis that the following commutator is zero:
        \begin{equation}
            \left[\frac{\partial }{\partial h_{\gamma i}}(\mathbf{h}\,\mathbf{h}_0^{-1}), (\mathbf{h}\,\mathbf{h}_0^{-1}) \right] = 0\,.
        \end{equation}
        As a consequence, the equations for $\bm{\varepsilon}$ can only be obtained within a certain degree of approximation, which cannot be clearly quantified.
        \item After computing the increments of the variables $\{\lambda,\Phi_k\}$ and mapping them back to the increments of the $\bm{\varepsilon}$ variables, namely $\text{d}\bm{\varepsilon}$, the rescaling matrix $\bm{\mu}$ should be computed as
        \begin{equation}
            \bm{\mu} = \mathbf{h}^{-1}\left(\mathbf{h}+\text{d}\mathbf{h} \right) = e^{-\bm{\varepsilon}}\left(e^{\bm{\varepsilon}\,+\,\text{d}\bm{\varepsilon}} \right) \simeq e^{\text{d}\bm{\varepsilon}}\,.
        \end{equation}
        The last passage includes a relevant approximation, coming from the fact that the matrix $\bm{\varepsilon}$ and its increment do not commute in general: $\left[\bm{\varepsilon},\text{d}\bm{\varepsilon} \right]\neq 0$. However, this approximation is unavoidable to implement the algorithm, as only the matrix exponential $\text{exp}(\text{d}\bm{\varepsilon})$ can be computed with a sufficient accuracy. As a consequence, the rescaling matrix is $\bm{\mu}$ necessarily obtained with a further error.
    \end{itemize}
    Because of these issues, the tests of this time-reversible integrator have shown a bad behaviour of the effective energy drift, together with additional problems related to the ACFs of the box matrix components. A well-behaved time-reversible integrator for the anisotropic SCR equations is discussed in \cref{sec:TR}.

    \chapter{Integration to simulation analysis}
    
    \section{Error estimation}\label{appendix_block_bootstrap}
    In the analysis reported in \cref{implementations}, errors are computed with \emph{block-bootstrap analysis}, which is a technique to estimate the uncertainty of statistics calculated over time series, namely by employing correlated samplings. To illustrate the procedure, let's consider a stationary time series $\mathbf{x}=\{x_1,...,x_n\}$, from which a statistics of interest $s(\mathbf{x})$ has been computed. If the samplings were uncorrelated, the error associated to $s(\mathbf{x})$ could be estimated with the \emph{bootstrap} approach \cite{efron}, which consists in the following steps:
    \begin{itemize}
        \item $B$ new time series $\mathbf{x}^{\prime\,b}$ of length $n$ are generated by resampling with replacement the original series;
        \item the statistics of interest is recomputed over each bootstrap sample $\mathbf{x}^{\prime\, b}$;
        \item the standard error of $s(\mathbf{x})$ is calculated as the standard deviation of the bootstrap estimates, namely as 
        \begin{equation}
            \text{SE}\left[s(\mathbf{x}) \right] = \sqrt{\frac{1}{B}\sum_{b=1}^B\Big(s(\mathbf{x}^{\prime\,b })-\bar{s}\Big)^2}\,,
        \end{equation}
    \end{itemize}
    where $\bar{s} = \sum_b s(\mathbf{x}^{\prime\,b })/B$. However, this method only works under the assumption of i.i.d. samplings, and brings to underestimate the actual error when this condition is not satisfied.
    
\begin{figure}[h!]
        \centering
        \includegraphics[width = 0.75\textwidth]{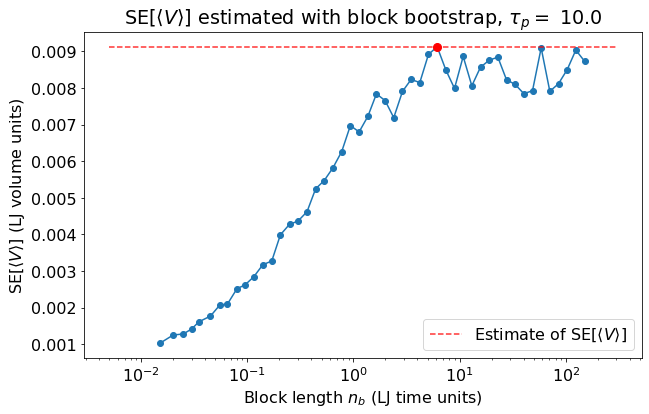}
        \caption{Example of block bootstrap analysis, applied to the standard error of the average volume from a simulation performed with SimpleMD. In the bootstrap procedure, the number of resampled trajectories is $B=200$. The optimal value of $\text{SE}\left[\langle V\rangle\right])$ is determined conservatively  using the largest estimate in the plateau region.}
        \label{fig:block_bootstrap_example}
    \end{figure}

    A possibility to circumvent this problem is to divide the time series in $N_B$ non-ovelapping blocks, each one containing $n_b$ samplings, and to apply the bootstrap resampling on the blocks, studying the standard error obtained as a function of $n_b$. Each bootstrap trajectory is composed in this way by a sequence of blocks coming from the original series, each one preserving its internal order. For large values of $n_b$ the standard error is expected to saturate, and the value of the plateau gives a meaningful estimate of $\text{SE}\left[s(\mathbf{x}) \right]$. \Cref{fig:block_bootstrap_example} shows an example of this systematic procedure applied to a volume time series, produced in one of the the simulations illustrated in \cref{LJ_crystal}.
    
    \section{Autocorrelation time and error of the sample mean}\label{appendix_autocorrelation}
    We show here that, if a statistics is estimated as the average over a finite-length time series, the uncertainty of the estimate depends both on the length of the series and on its autocorrelation. We consider the case of a continuous time series, but the same reasoning can be applied to discretized MD trajectories just replacing integrals with sums. If $x(t)$ is a stationary stochastic process, its autocorrelation time $\tau_x$ is defined as
    \begin{equation}
        \tau_x = \int_0^\infty \text{d}t\, C_x(t)\,,
    \end{equation}
    where $C_x(t)$ is the (normalized) ACF of the process:
    \begin{equation}
        C_x(t) = \frac{\langle x(t)x(0) \rangle - \langle x\rangle^2}{\langle x^2 \rangle - \langle x\rangle^2} = \frac{\langle x(t)x(0)\rangle -\langle x\rangle^2}{\sigma^2_x}\,.
    \end{equation}
    Note that $\langle x\rangle$ and $\langle x^2\rangle$ are time-independent, since the process is assumed to be stationary. Let's now consider the average value of $x$ estimated in a trajectory of finite length $T$, where the system is supposed to be already equilibrated at time $t=0$:
    \begin{equation}
        \overbar{x} = \frac{1}{T}\int_0^T \text{d}t\,x(t)\,.
    \end{equation}
    The variance of $\overbar{x}$ as an estimator of the real average value $\langle x \rangle$ can be written as:
    \begin{equation}
        \sigma_{\overbar{x}}^2 = \langle \overbar{x}^2\rangle - \langle \overbar{x}\rangle^2\,,  
    \end{equation}
    where these averages can be thought over infinite simulations of length $T$ of the process. Then:
    \begin{subequations}
    \begin{align}
        \sigma_{\overbar{x}}^2 &= \Big\langle \frac{1}{T^2}\int_0^T \text{d}t\,x(t)\int_0^T \text{d}t'\,x(t')\Big\rangle - \langle x\rangle^2 \\
        &= \frac{1}{T^2}\int_0^T\int_0^T \text{d}t\,\text{d}t'\,\langle x(t) x(t')\rangle - \langle x\rangle^2 \label{intx2}\\
        &= \frac{2}{T^2}\int_0^T\int_0^t \text{d}t\,\text{d}t'\,\langle x(t) x(t')\rangle - \langle x\rangle^2 \,.
    \end{align}
    \end{subequations}
    The last passage is performed observing that the integrand function is symmetric with respect to the exchange $x\leftrightarrow x'$, and the integral in \cref{intx2} contains for each pair of integration points $(t_1,t_2)$ its symmetric $(t_2,t_1)$. With other manipulations we can wite:
    \begin{subequations}
    \begin{align}
         \sigma_{\overbar{x}}^2 &= \frac{2}{T^2}\int_0^T\int_0^t \text{d}t\,\text{d}t'\,\Big[\langle x(t) x(t')\rangle - \langle x\rangle^2\Big] \\
         &= \frac{2\sigma_x^2}{T^2}\int_0^T\int_0^t \text{d}t\,\text{d}t'\,C_x(t-t') \\
         &\simeq \frac{2\sigma_x^2}{T^2}\int_0^T \text{d}t \,\tau_x\,,
    \end{align}
    \end{subequations}
    where the last approximation is meaningful if $T\gg\tau_x$. Then the final result reads
    \begin{equation}\label{standard_error_autocorr}
        \sigma_{\overbar{x}}^2 \simeq \frac{2\tau_x}{T}\sigma_x^2\,,
    \end{equation}
    namely the uncertainty of an average $\overbar{x}$ computed over a correlated time series:
    \begin{itemize}
        \item decreases with the length of the series;
        \item increases with its autocorrelation time.
    \end{itemize} 
    \vfil
    
    \section{Reference ACFs of volume and its variance}\label{appendix_relaxation_autocorrelation}
    The aim of this section is to justify the dashed lines in \cref{fig:ACFs_LJ}, 
    \begin{subequations}\label[pluralequation]{ref_acfs}
    \begin{align}
        C_{V}(t) &= e^{-t/\tau_p}\,, \\
        C_{\sigma_V^2}(t) &= e^{-2t/\tau_p}\,,
    \end{align}
    \end{subequations}
    which are the analytical ACFs of the volume and its variance in the limit case of a Gaussian-distributed volume following a Langevin dynamics:
    \begin{equation}
        \text{d}V = -\frac{1}{\tau_p}\left(V-\overbar{V}\right) \text{d}t + \sqrt{\frac{2\sigma_V^2}{\tau_p}}\,\text{d}W\,.
    \end{equation}
    Note that the SCR dynamics for the volume, given by \cref{crescale_iso_V}, is a first-order stochastic dynamics that resembles to the above Langevin equation if we neglect the additional dependencies on $V$ in the internal pressure and in the noise prefactor; moreover, due to the central limit theorem the volume distribution is expected to approach a Gaussian in the thermodynamic limit, unless the system is in a critical point where different phases coexist. As a consequence, \cref{ref_acfs} represent the reference behaviours for the ACFs of interest when $N$ is large.

    In order to derive \cref{ref_acfs}, let's first apply It\^{o}'s lemma to rewrite the Langevin equation in terms of $\Delta V = V-\overbar{V}$:
    \begin{equation}
        \text{d}\,\Delta V = -\frac{\Delta V}{\tau_p} \text{d}t + \sqrt{\frac{2\sigma_V^2}{\tau_p}}\,\text{d}W\,.
    \end{equation}
    Using a different formalism, this equation can be equivalently written as
    \begin{equation}
        \frac{\text{d}\Delta V}{\text{d}t} = -\frac{\Delta V}{\tau_p} + \sqrt{\frac{2\sigma_V^2}{\tau_p}}\,\eta(t) \,,
    \end{equation}
    where $\eta(t)$ is a Gaussian white noise satisfying \cref{gaussian_white_noise}. The formal solution of this equation, decomposed as the sum of the general homogeneous solution and a particular solution of the inhomogenous problem, can be written as
    \begin{equation}
        \Delta V(t) = \Delta V(0)e^{-t/\tau_p} + \sqrt{\frac{2\sigma_V^2}{\tau_p}}\int_0^t \text{d}t'\, e^{-(t-t')/\tau_p}\,\eta(t')\,.
    \end{equation}
    Representing with $\langle \cdot\rangle$ the average over different realization of the stochastic process $\Delta V(t)$, we can evaluate the following correlation:
    \begin{subequations}\label{deltaVdeltaV0}
    \begin{align}
        \langle \Delta V(t)\Delta V(0)\rangle = \langle \Delta V(0)^2\rangle e^{-t/\tau_p} + 
        \sqrt{\frac{2\sigma_V^2}{\tau_p}}\int_0^t \text{d}t'\, e^{-(t-t')/\tau_p}\,\langle \Delta V(0)\eta(t')\rangle\,.
    \end{align}
    \end{subequations}
    We can now observe that:
    \begin{itemize}
        \item $\langle \Delta V(0)^2\rangle = \sigma_V^2$, since the process is stationary;
        \item $\langle \Delta V(0)\eta(t')\rangle = \langle \Delta V(0)\rangle \langle\eta(t')\rangle = 0$, since $\Delta V(t)$ is a non-anticipating function of $\eta(t)$ and is then independent on future realizations of the noise.
    \end{itemize}
    As a consequence, \cref{deltaVdeltaV0} becomes
    \begin{equation}
        \langle \Delta V(t)\Delta V(0)\rangle = \sigma_V^2\,e^{-t/\tau_p}\,,
    \end{equation}
    and the ACF of the volume can be simply computed as:
    \begin{subequations}
    \begin{align}
        C_V(t) &= \frac{\langle V(t) V(0)\rangle - \langle V\rangle^2 }{\langle \Delta V^2\rangle} =\frac{\langle \Delta V(t)\Delta V(0)\rangle}{ \sigma_V^2} = e^{-t/\tau_p}\,.
    \end{align}
    \end{subequations}
    
    In order to compute the ACF of the volume variance, let's start by evaluating the following correlation:
    \begin{subequations}
    \begin{align}
        \langle \Delta V(t)^2 &\Delta V(0)^2\rangle = \langle \Delta V(0)^4\rangle e^{-2t/\tau_p} \\
        &+ 2\langle \Delta V(0)^3\rangle \sqrt{\frac{2\sigma_V^2}{\tau_p}}e^{-t/\tau_p}\int_0^t \text{d}t'\, e^{-(t-t')/\tau_p}\,\langle\eta(t')\rangle \nonumber\\
        &+ \frac{2\sigma_V^2}{\tau_p}\langle \Delta V(0)^2\rangle \int_0^t \text{d}t'\int_0^t \text{d}t''\, e^{-(t-t')/\tau_p}e^{-(t-t'')/\tau_p}\,\langle\eta(t')\eta(t'')\rangle\,. \nonumber
    \end{align}
    \end{subequations}
    Using the properties of $\eta(t)$ in \cref{gaussian_white_noise} we get:
    \begin{equation}
        \langle \Delta V(t)^2 \Delta V(0)^2\rangle = \langle \Delta V^4\rangle e^{-2t/\tau_p} + \sigma_V^4 \left( 1-e^{-2t/\tau_p}\right)\,.
    \end{equation}
    Hence, the ACF of the volume variance can be written as:
    \begin{subequations}
    \begin{align}
        C_{\sigma_V^2}(t) &= \frac{\langle \Delta V(t)^2 \Delta V(0)^2\rangle - \langle \Delta V^2\rangle^2}{\langle \Delta V^4\rangle - \langle \Delta V^2\rangle^2} \\
        &= \frac{\langle \Delta V^4\rangle e^{-2t/\tau_p} + \sigma_V^4 \left( 1-e^{-2t/\tau_p}\right)-\sigma_V^4}{\langle \Delta V^4\rangle - \sigma_V^4} \\
        &= \frac{\left(\langle \Delta V^4\rangle - \sigma_V^4 \right)e^{-2t/\tau_p}}{\langle \Delta V^4\rangle - \sigma_V^4} = e^{-2t/\tau_p}\,.
    \end{align}
    \end{subequations}
    This concludes the derivation of the ACFs in \cref{ref_acfs}.

    \section{Pathological volume distributions in GROMACS}\label{appendix_pathological_distributions}
\begin{figure}[h]
    \centering
    \begin{subfigure}[b]{0.45\textwidth}
        \centering
        \includegraphics[height = 4.4 cm]{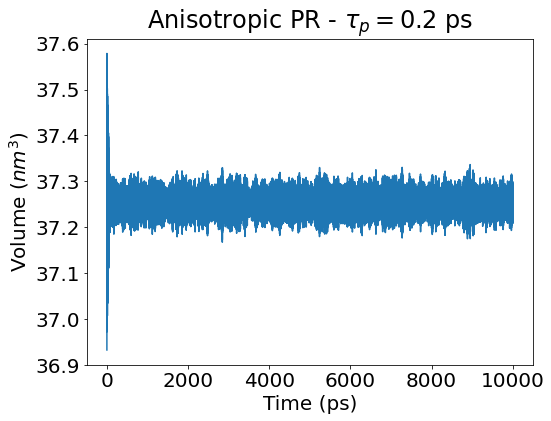}\\[1 ex]
        \label{fig:PR_aniso_ok}
    \end{subfigure}
    \hfil
    \begin{subfigure}[b]{0.45\textwidth}
        \centering
        \includegraphics[height = 4.4 cm]{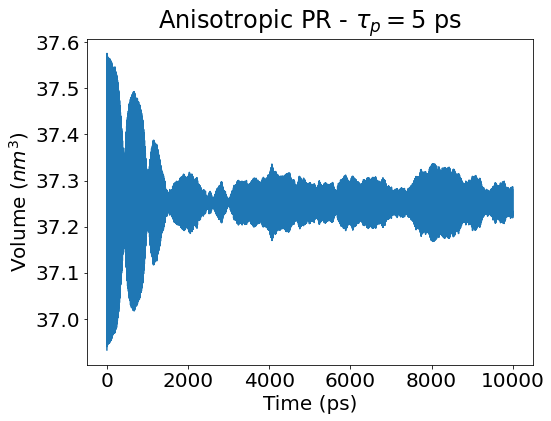}\\[1 ex]
        \label{fig:PR_aniso_notok}
    \end{subfigure}
    \caption{Volume distributions generated by the Parrinello-Rahman (PR) anisotropic barostat in GROMACS.}
    \label{fig:PR_aniso_prob}
    \end{figure}

\begin{figure}[h]
    \centering
    \begin{subfigure}[b]{0.45\textwidth}
        \centering
        \includegraphics[height = 4.4 cm]{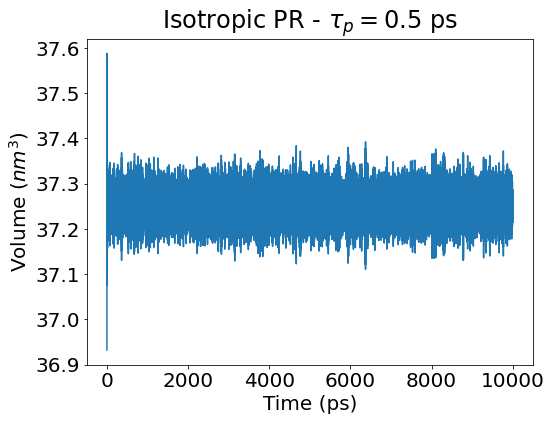}\\[1 ex]
        \label{fig:PR_iso_ok}
    \end{subfigure}
    \hfil
    \begin{subfigure}[b]{0.45\textwidth}
        \centering
        \includegraphics[height = 4.4 cm]{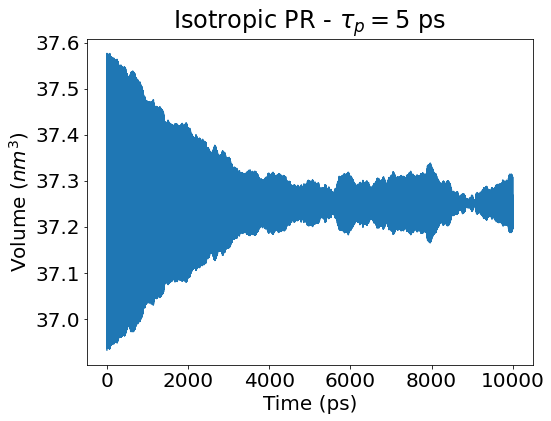}\\[1 ex]
        \label{fig:PR_iso_notok}
    \end{subfigure}
    \caption{Volume distributions generated by the Parrinello-Rahman (PR) isotropic barostat in GROMACS.}
    \label{fig:PR_iso_prob}
    \end{figure}

\begin{figure}[h!]
    \centering
    \begin{subfigure}[b]{0.45\textwidth}
        \centering
        \includegraphics[height = 4.4 cm]{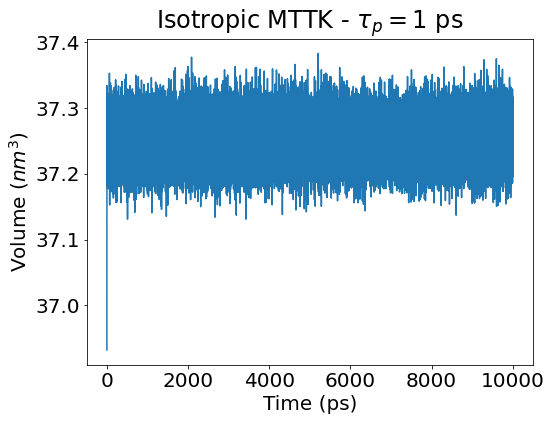}\\[1 ex]
        \label{fig:MTTK_ok}
    \end{subfigure}
    \hfil
    \begin{subfigure}[b]{0.45\textwidth}
        \centering
        \includegraphics[height = 4.4 cm]{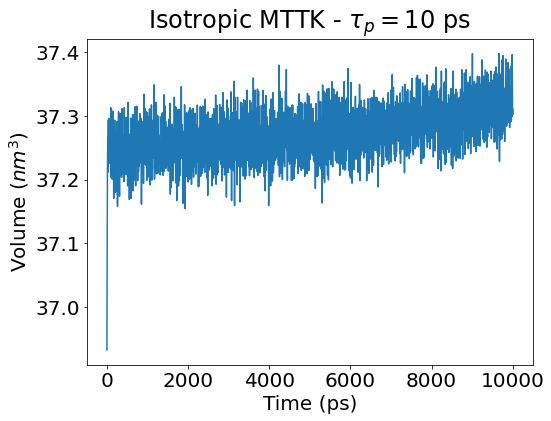}\\[1 ex]
        \label{fig:MTTK_notok}
    \end{subfigure}
    \caption{Volume distributions generated by the Martyna-Tuckerman-Tobias-Klein (MTTK) isotropic barostat in GROMACS.}
    \label{fig:MTTK_prob}
    \end{figure}

    We show in this section some examples of problematic volume distributions obtained with the barostats available in GROMACS 2021.2. \Cref{fig:PR_aniso_prob,fig:PR_iso_prob,fig:MTTK_prob} show respectively three comparisons between "well behaving" and "pathological" volume distributions obtained from the simulations of the Argon crystal system described in \cref{LJ_crystal}.
    

    \section{Supplementary results}\label{appendix_supplementaryresults}
    
    \begin{figure}[h!]
        \centering
        \includegraphics[width = 1.\textwidth]{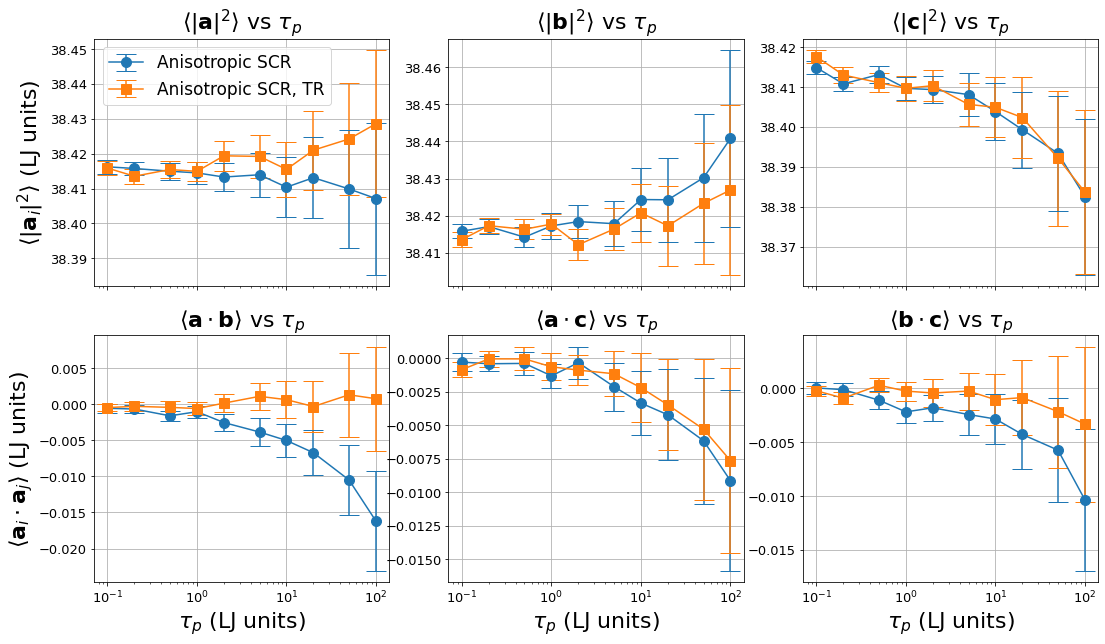}
        \caption{Results from the simulations of the LJ crystal in SimpleMD. The average squared moduli of the cell vectors $\mathbf{a},\mathbf{b},\mathbf{c}$ and their average scalar products are shown respectively in the upper and lower plots, as functions of the relaxation time $\tau_p$. Deviations in the large-$\tau_p$ regime are related to the large statistical errors of the estimates, due to trajectories that are too short with respect to the autocorrelation times of the analyzed quantities. Note that these deviations are not so evident when observing the volume distributions (see \cref{fig:taups_LJ}), because the autocorrelation time of the single box matrix components is typically larger than $\tau_p$ (see for instance \cref{fig:ACFs_mod2_LJ,fig:ACFs_mod2_IceIh}).} 
        \label{fig:box_comp_LJ}
    \end{figure}
    \begin{figure}[h!]
        \centering
        \includegraphics[width = 1.\textwidth]{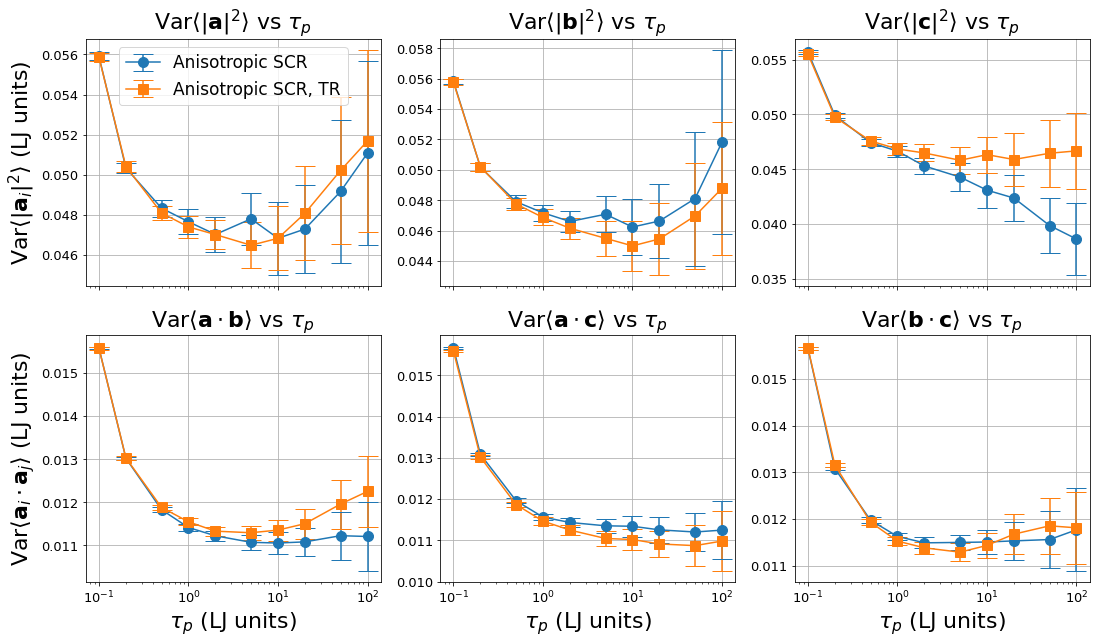}
        \caption{Results from the simulations of the LJ crystal in SimpleMD. The variances of the distributions considered in \cref{fig:box_comp_LJ} are shown as functions of $\tau_p$.}
        \label{fig:box_vars_LJ}
    \end{figure}
    
    \begin{figure}[h!]
    \centering
    \begin{subfigure}[b]{0.45\textwidth}
        \centering
        \includegraphics[height = 4.4 cm]{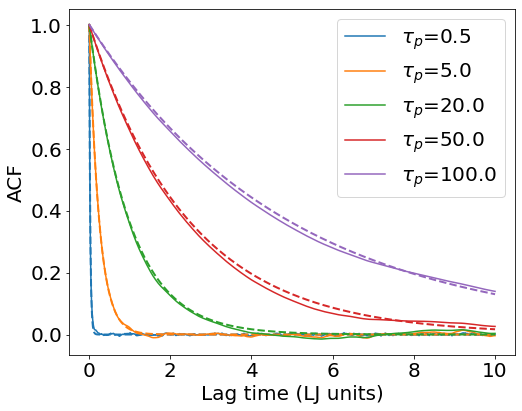}\\[1 ex]
        \caption{\RaggedLeft Volume ACFs$\hspace{1.4cm}$}
    \end{subfigure}
    \hfil
    \begin{subfigure}[b]{0.45\textwidth}
        \centering
        \includegraphics[height = 4.4 cm]{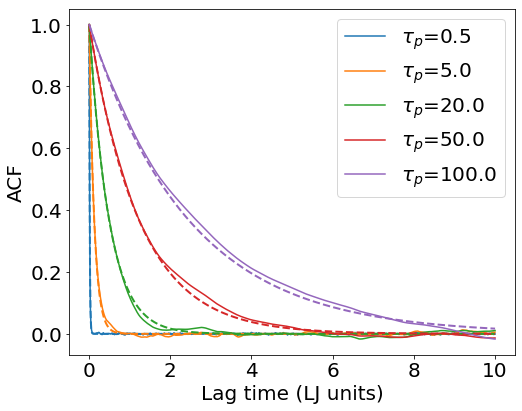}\\[1 ex]
        \caption{\RaggedLeft ACFs of volume variance$\hspace{0.6cm}$}
    \end{subfigure}
    \caption{ACFs of the volume (left panel) and its variance (right panel) from time series from the simulations of a LJ crystal with the anisotropic SCR barostat, using the time-reversible integrator in SimpleMD. The dashed lines are the same exponential functions reported in \cref{fig:ACFs_LJ}.}
    \label{fig:ACFs_LJ_TR}
    \end{figure}
    
    \begin{figure}[h!]
        \centering
        \includegraphics[width = 1.0\textwidth]{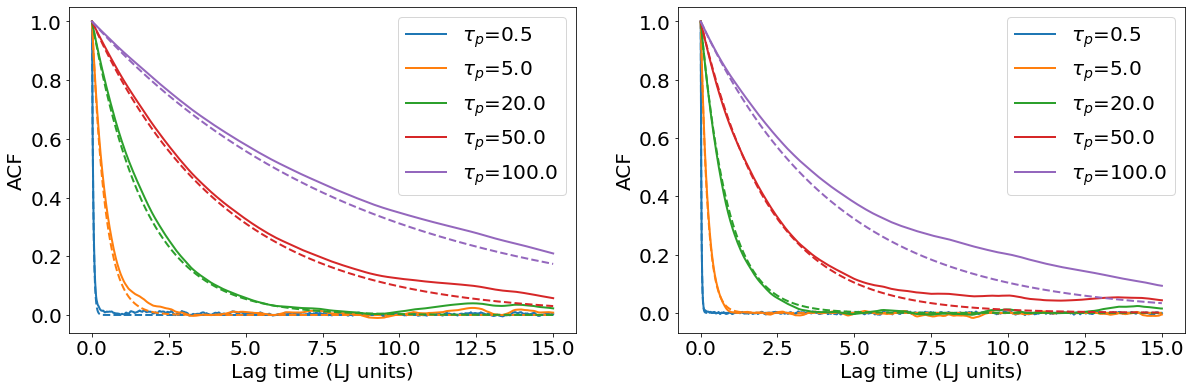}
        \caption{Results from the simulation of the LJ crystal system in SimpleMD, using the anisotropic SCR barostat. The left panel shows the ACF of the squared modulus of the cell vector $\mathbf{a}$; the right panel shows the ACF of its variance. The dashed lines are exponential functions $e^{-t/(c\tau_{p,\text{exp}})}$, where the coefficient $c$ is set by hand ($c=1.75$ for the left plot, $c=0.9$ for the right one) to show that the autocorrelation time of interest appears linearly dependent on $\tau_p$, as in the case of the volume and its variance. However, extending this analysis to the other systems discussed in \cref{implementations}, the values of $c$ appear system-dependent and cannot be exactly predicted $\emph{a priori}$ (see for instance \cref{fig:ACFs_mod2_IceIh}). In the case of the LJ system, identical behaviours are observed for the ACFs of the remaining squared moduli and the ACFs of the scalar products, with the same parameters $c$ in both the classes of functions.}
        \label{fig:ACFs_mod2_LJ}
    \end{figure}
    
    \begin{figure}[h!]
        \centering
        \includegraphics[width = 1.0\textwidth]{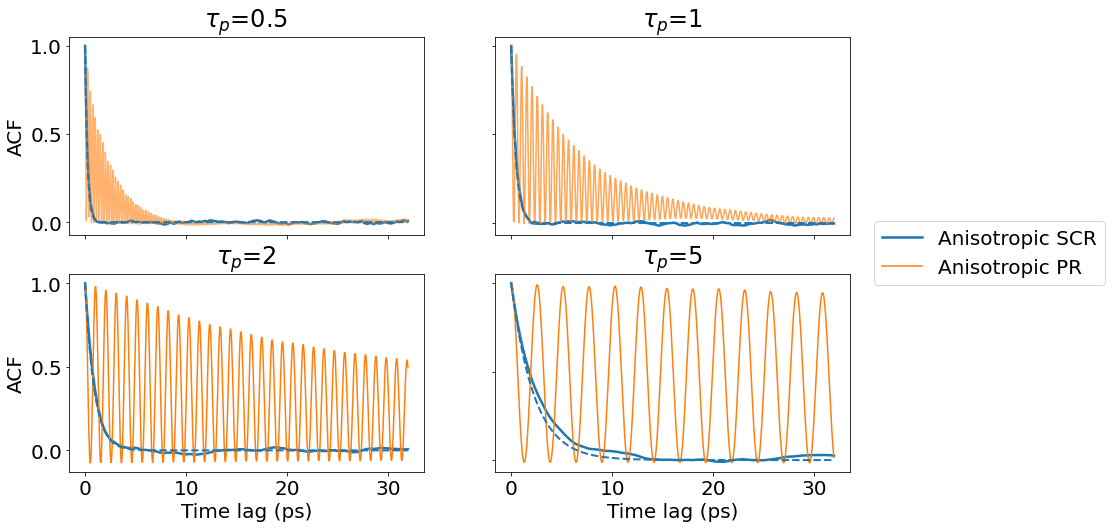}
        \caption{Comparison between the ACFs of the volume variance from GROMACS simulations of the Argon crystal system, using the anisotropic SCR and PR barostats. Variance ACFs are computed as explained in \cref{LJ_crystal}, and dashed curves represent the exponentially decaying functions $\exp\left(-2t/\tau_{p,\text{exp}} \right)$. Comparing these ACFs with the volume ones in \cref{fig:ACFs_volume_LJ_GMX}, it is possible to conclude that a calculation of the volume variance converges faster than a calculation of the average volume using the SCR method. The ACFs obtained with the PR barostats show no symmetry with respect to zero, resulting in an integrated autocorrelation time that is larger than the one from SCR simulations.}
        \label{fig:ACFs_var_LJ_GMX}
    \end{figure}
    
    \begin{figure}[h!]
        \centering
        \includegraphics[width = 1.\textwidth]{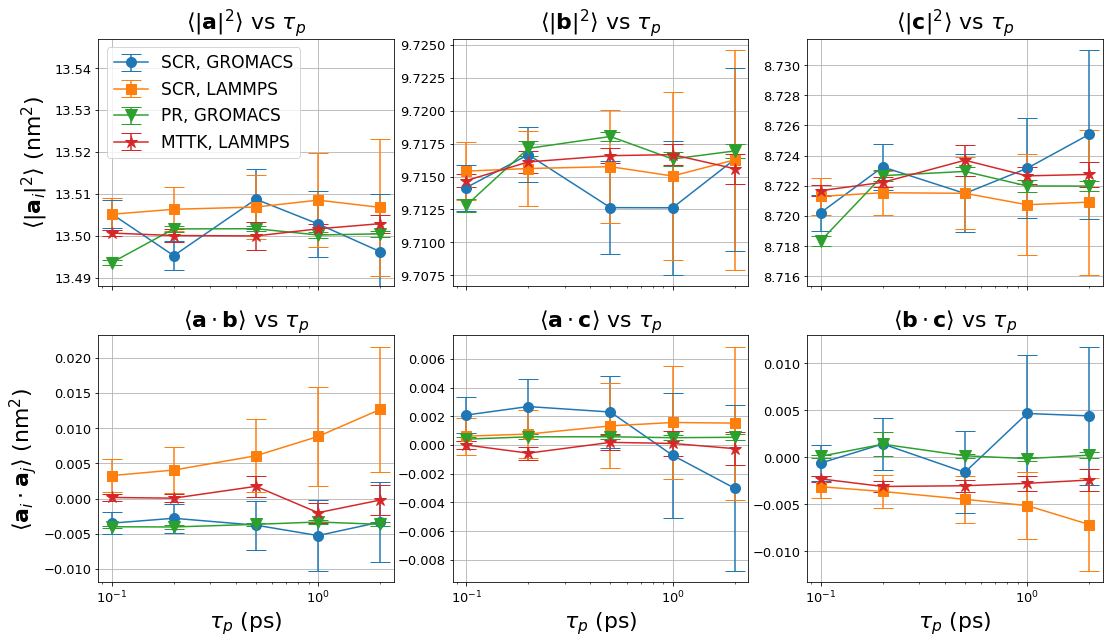}
        \caption{Results from the simulations of the ice I$_h$ crystal. The average squared moduli of the cell vectors $\mathbf{a},\mathbf{b},\mathbf{c}$ and their average scalar products are shown respectively in the upper and lower plots, as functions of the relaxation time $\tau_p$.} 
        \label{fig:box_comp_ice}
    \end{figure}
    \begin{figure}[h!]
        \centering
        \includegraphics[width = 1.\textwidth]{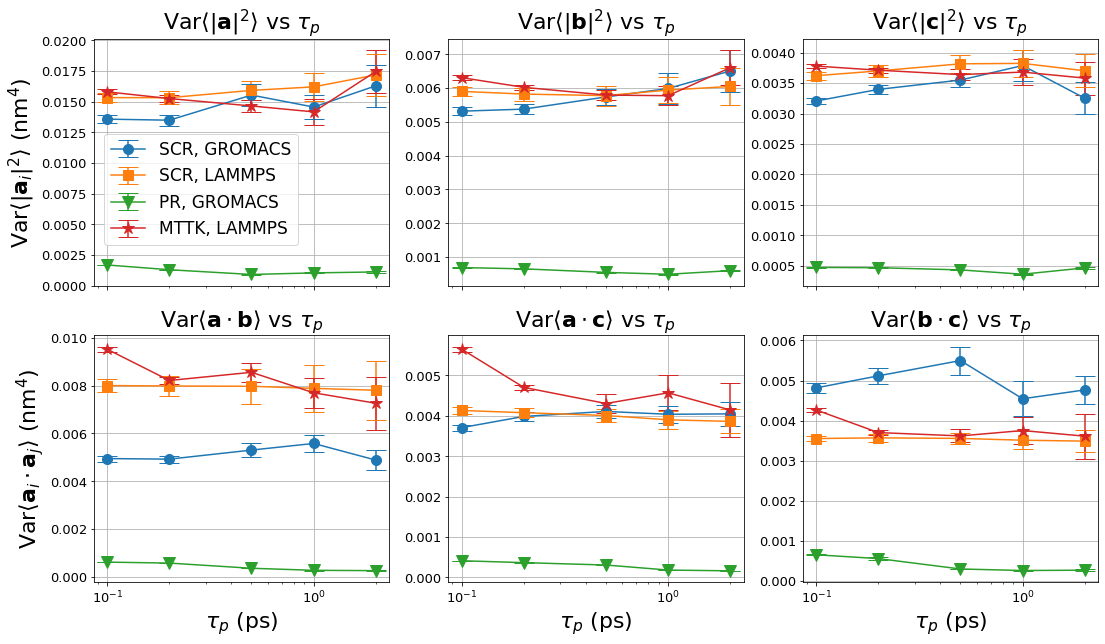}
        \caption{Results from the simulations of the ice I$_h$ crystal. The variances of the distributions considered in \cref{fig:box_comp_ice} are shown as functions of $\tau_p$.}
        \label{fig:box_vars_ice}
    \end{figure}
    
    \begin{figure}[h!]
        \centering
        \includegraphics[width = 0.9\textwidth]{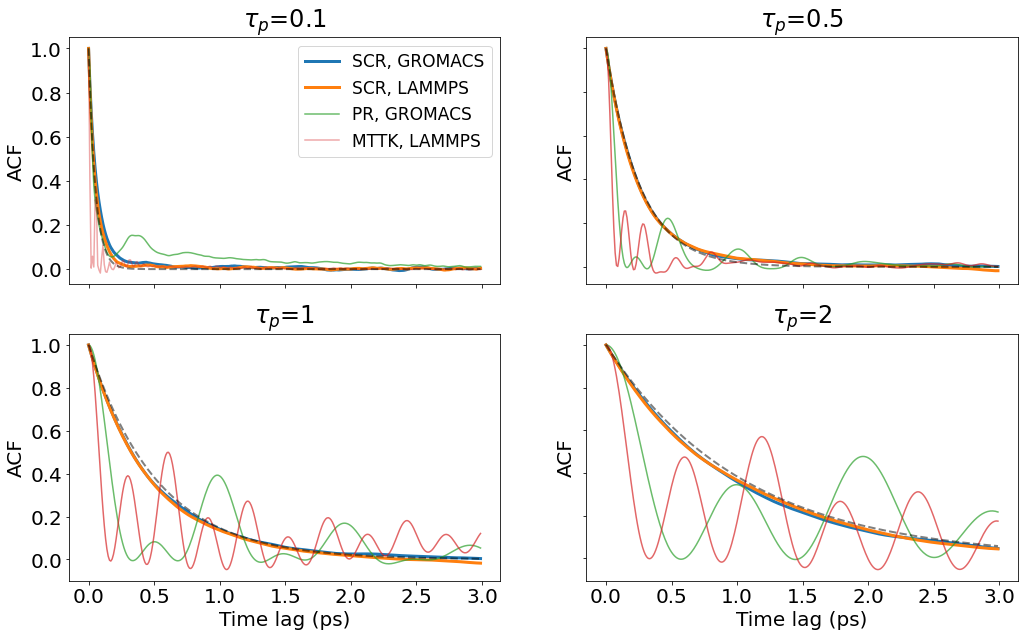}
        \caption{ACFs of volume variance from the simulations of the Ice I$_h$ system. Dashed lines represent the expected decay $\exp\left(-2t/\tau_p\right)$.}
        \label{fig:ACFs_vars_Ice}
    \end{figure}
    
    \begin{figure}[h!]
        \centering
        \includegraphics[width = 1.0\textwidth]{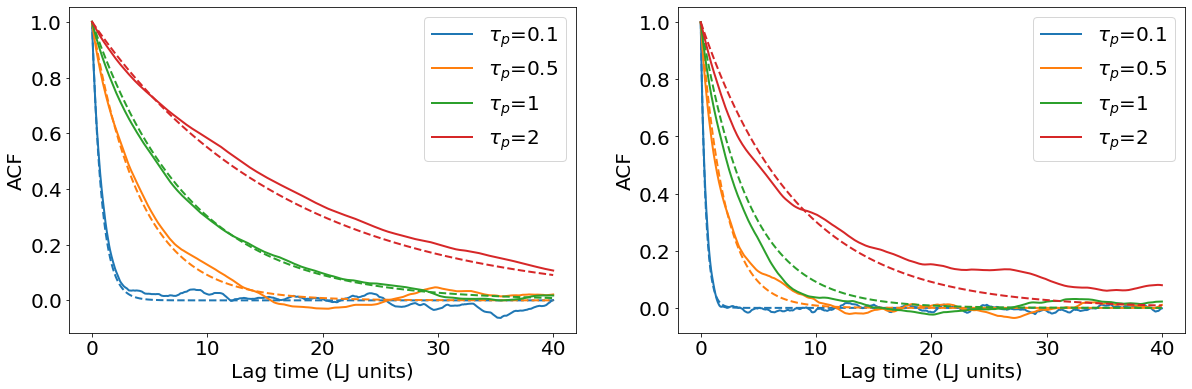}
        \caption{Results from the simulation of the Ice I$_h$ system in GROMACS, using the anisotropic SCR barostat. Left and right panels show respectively the ACF of the squared modulus of the cell vector $\mathbf{a}$ and the ACF of its variance. The dashed lines are the exponential functions $e^{-t/(c\tau_{p,\text{exp}})}$ with $c = 8$ and $c=4$ respectively, showing as a first approximation a linear scaling between the autocorrelation time of the quantity of interest and $\tau_p$. The ACFs of the other squared moduli and of the three scalar products between the cell vectors follow a similar behaviour, but with different parameters $c$.}
        \label{fig:ACFs_mod2_IceIh}
    \end{figure}
    
    \begin{table}[h!]
    \centering
    \caption{Results from the simulations of the gypsum crystal in LAMMPS, with relaxation time $\tau_p = 1$ ps for both the SCR and the MTTK barostats. The first six raws are related to the distributions of the squared moduli and the scalar products of the three cell vectors $\mathbf{a},\mathbf{b},\mathbf{c}$, while the last six raws concern the distributions of the independent components of the internal pressure tensor $\mathbf{P}_{\text{int}}$. }
    \begin{threeparttable}
    \label{tab:gypsum_lammps_aniso_box}
        \begin{tabular}{l c c c c}
        \toprule[0.5pt]\toprule[0.5pt]
        {} & \small\textbf{Anisotropic SCR}       & \small\textbf{Anisotropic MTTK}  \\\midrule
        $\langle|\mathbf{a}|^2\rangle\,$ (nm$^2$)                     & $11.3993\pm0.0005$ & $11.39903\pm0.00010$ \\
        $\sigma_{|\mathbf{a}|^2}\,$              & $0.0309\pm0.0002$ & $0.03080\pm0.00009$ \\ \midrule
        $\langle|\mathbf{b}|^2\rangle\,$ (nm$^2$)                     & $9.3914\pm0.0007$ & $9.3915\pm0.0002$\\
        $\sigma_{|\mathbf{b}|^2}\,$                & $0.0290\pm0.0003$ & $0.02925\pm0.00009$  \\ \midrule
        $\langle|\mathbf{c}|^2\rangle\,$ (nm$^2$)                    & $16.5031\pm0.0015$ &  $16.5041\pm0.0007$ \\
        $\sigma_{|\mathbf{c}|^2}\,$                & $0.0691\pm0.0007$ & $0.0694\pm 0.0004$ \\ \midrule
        $\langle\mathbf{a}\cdot\mathbf{b}\rangle\,$ (nm$^2$)                     & $0.00106\pm0.00010$ & $0.00001\pm0.00016$ \\
        $\sigma_{\mathbf{a}\cdot\mathbf{b}}\,$                & $0.0366\pm0.0005$ & $0.03350\pm0.00019$  \\ \midrule
        $\langle\mathbf{a}\cdot\mathbf{c}\rangle\,$ (nm$^2$)                     & $-6.5865\pm0.0007$ & $-6.58668\pm0.00014$ \\
        $\sigma_{\mathbf{a}\cdot\mathbf{c}}\,$                & $0.0338\pm0.0003$ & $0.03409\pm0.00014$  \\ \midrule
        $\langle\mathbf{b}\cdot\mathbf{c}\rangle\,$ (nm$^2$)                     & $-0.0007\pm0.0012$  & $-0.0000\pm0.0003$ \\
        $\sigma_{\mathbf{b}\cdot\mathbf{c}}\,$                & $0.0480\pm0.0006$ & $0.0454\pm0.0005$  \\\midrule
        $\langle P_{\text{int},xx} \rangle\,$ (bar)                     & $-14\pm10$ & $1\pm2$ \\
        $\sigma_{P_{\text{int},xx}}\,$              & $1490\pm5$ & $1473\pm10$ \\ \midrule
        $\langle P_{\text{int},yy}\rangle\,$ (bar)                     & $-8\pm10$ & $2\pm2$\\
        $\sigma_{P_{\text{int},yy}}\,$                & $1545\pm4$ & $1534\pm12$  \\ \midrule
        $\langle P_{\text{int},zz}\rangle\,$ (bar)                    & $-0.0 \pm9.6$ &  $0.6\pm1.1$ \\
        $\sigma_{P_{\text{int},zz}}\,$                & $1139\pm4$ & $1144\pm 4$ \\ \midrule
        $\langle P_{\text{int},xy}\rangle\,$ (bar)                     & $5.112\pm6.57$ & $-0.3\pm 1.3$ \\
        $\sigma_{P_{\text{int},xy}}\,$                & $887\pm3$ & $883\pm5$  \\ \midrule
        $\langle P_{\text{int},xz}\rangle\,$ (bar)                     & $-2\pm8$ & $-0.3\pm 1.3$ \\
        $\sigma_{P_{\text{int},xz}}\,$                & $850\pm3$ & $851\pm3$  \\ \midrule
        $\langle P_{\text{int},yz}\rangle\,$ (bar)              & $-7\pm7$  & $-0.2\pm1.3$ \\
        $\sigma_{P_{\text{int},yz}}\,$                & $816\pm3$ & $813\pm6$
        \\\bottomrule[0.5pt]\bottomrule[0.5pt] 
        \end{tabular}
    \end{threeparttable}
    \end{table}
    
    \begin{figure}[h!]
    \centering
    \begin{subfigure}[b]{0.44\textwidth}
        \centering
        \includegraphics[height = 3.9 cm]{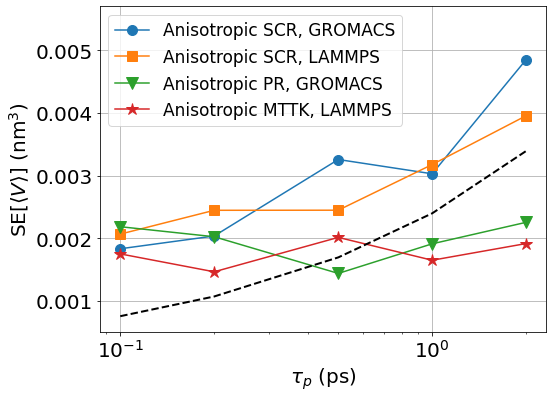}\\[1 ex]
        \caption{\RaggedLeft Errors of volume average$\hspace{0.3cm}$}
    \end{subfigure}
    \hfil
    \begin{subfigure}[b]{0.44\textwidth}
        \centering
        \includegraphics[height = 3.9 cm]{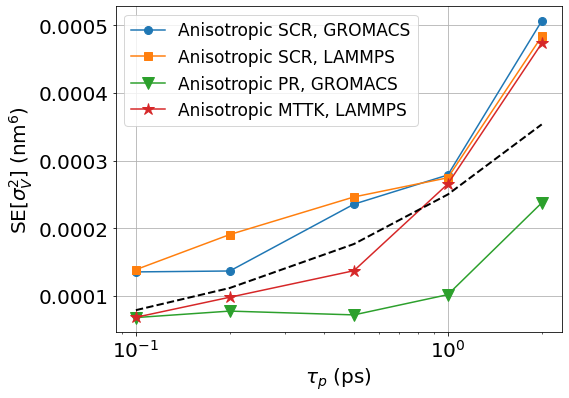}\\[1 ex]
        \caption{\RaggedLeft Errors of volume variance$\hspace{0.2cm}$ }
    \end{subfigure}
    \caption{Standard errors of volume average (left panel) and variance (right panel) computed with block bootstrap analysis, referred to the simulations of the Ice I$_h$ crystal. Dashed lines represent the expected behaviour of the errors as a function of $\tau_p$, computed with \cref{standard_error_autocorr}. For the SCR barostat, errors of the average volume show deviations from the expected behaviour when $\tau_p$ is small, resulting in an effective autocorrelation time that is larger than the barostat relaxation time. As commented for the volume ACFs in \cref{fig:ACFs_avgs_Ice}, this limit in the volume dynamics is given by the autocorrelation time in the rearrangement of atoms.}
    \vspace{12cm}
    \label{fig:SE_Ice}
    \end{figure}

    \end{appendices}
    
    \pagestyle{plain}

    \renewcommand*{\bibfont}{\small}
    \printbibheading
    \addcontentsline{toc}{chapter}{Bibliography}
    \printbibliography[heading = none]


\end{document}

%% file: core/preamble/document_settings.tex
\newif\ifprintVersion   
\newif\iffancyTheorems  
\newif\ifboldNumberSets 
\newif\ifbachelorThesis 

\printVersionfalse
\fancyTheoremstrue
\boldNumberSetstrue
\bachelorThesistrue

\newcommand*{\printTitle}{}
\newcommand*{\printGermanTitle}{}
\newcommand*{\myTitle}[2]{\renewcommand*{\printTitle}{#1}\renewcommand*{\printGermanTitle}{#2}}
\newcommand*{\printTitleBold}{\textbf{\printTitle}}

\newcommand*{\printAuthor}{}
\newcommand*{\myName}[1]{\renewcommand*{\printAuthor}{#1}}

\newcommand*{\printProgram}{}
\newcommand*{\myProgram}[1]{\renewcommand*{\printProgram}{#1}}

\newcommand*{\printDateReceived}{}
\newcommand*{\dateOfHandingIn}[1]{\renewcommand*{\printDateReceived}{#1}}

\newcommand*{\printSubject}{}
\newcommand*{\mySubject}[1]{\renewcommand*{\printSubject}{#1}}

\newcommand*{\printKeywords}{}
\newcommand*{\myKeywords}[1]{\renewcommand*{\printKeywords}{#1}}

\newcommand*{\printNameOfSupervisor}{}
\newcommand*{\nameOfMySupervisor}[1]{\renewcommand*{\printNameOfSupervisor}{#1}}

\newcommand*{\printAdditionalExaminers}{}
\newcommand*{\additionalExaminers}[1]{\renewcommand*{\printAdditionalExaminers}{#1}}

\newlength{\extraborderlength}
\newcommand*{\extraBorder}[1]{\setlength{\extraborderlength}{#1}}

\newlength{\mybindingcorrection}
\newcommand*{\bindingCorrection}[1]{\setlength{\mybindingcorrection}{#1}}

%% file: core/preamble/format.tex
\usepackage[utf8]{inputenc} 
\usepackage[T1]{fontenc}    
\usepackage
[
    ngerman,         
    main = american, 
]
{babel}                     

\usepackage{calc} 

\newlength{\myparindent}
\newlength{\myparskip}
\setfootnoterule{0 cm}

\deffootnote[1.2 em]{1.2 em}{0 em}{\makebox[1.4 em][l]{\textbf{\thefootnotemark}}}

\makeatletter%
    \@removefromreset{footnote}{chapter}%
\makeatother

\usepackage[dvipsnames]{xcolor} 

\definecolor{stroke1}{HTML}{2574A9} 

\colorlet{captionlabel}{black}
\colorlet{footerpagenr}{black}
\colorlet{footerchapter}{stroke1}
\colorlet{footerchaptername}{black}
\colorlet{footersection}{stroke1}
\colorlet{footersectionname}{black}
\colorlet{chapternumber}{stroke1}

\newcommand*{\magicratio}{1 * \ratio{193 mm}{210 mm}}

\newlength{\mypaperwidth}
\newlength{\mypaperheight}
\newlength{\mybodywidth}
\newlength{\mybodyheight}
\newlength{\myoutermargin}
\ifprintVersion
\else
\fi

\newlength{\mytopmargin}
\ifprintVersion
\else
\fi

\newlength{\myinnermargin}
\ifprintVersion
\else
\fi

\newlength{\mybottommargin}
\ifprintVersion
\else
\fi

\newcommand{\goldenratio}{1.618}

\newlength{\myheadsep} 
\ifprintVersion
\else
\fi

\newlength{\myfootskip} 
\ifprintVersion
\else
\fi

\newlength{\mymargininnersep} 
\newlength{\mymarginoutersep} 
\ifprintVersion
\else
\fi

\newlength{\mymarginwidth} 
\newlength{\mymarginwidthwithinnersep} 
\usepackage
[
    \ifprintVersion
        paperwidth = \mypaperwidth + 2\extraborderlength + \mybindingcorrection,
        paperheight = \mypaperheight + 2\extraborderlength,
    \else
        paperwidth = \mypaperwidth,
        paperheight = \mypaperheight,
    \fi
    textwidth = \mybodywidth,
    textheight = \mybodyheight,
    outer = \myoutermargin,
    top = \mytopmargin,
    headsep = \myheadsep,
    footskip = \myfootskip,
    marginparsep = \mymargininnersep,
    marginparwidth = \mymarginwidth,
]
{geometry} 

\usepackage
[
]
{scrlayer-scrpage} 

\KOMAoptions
{%
    headwidth = \textwidth + \mymarginwidthwithinnersep,%
    footwidth = \myoutermargin : \textwidth,%
}

\automark[chapter]{chapter}
\automark*[section]{}

\lehead%
{%
    \begin{minipage}[b]{\mymarginwidth}%
        \small\raggedleft\normalfont\textsf{\textbf{\color{footerchapter}\chaptername\ \thechapter}}
    \end{minipage}
}
\cehead{\hspace*{\mymarginwidthwithinnersep}\parbox{\textwidth}{\raggedright\leftmark}}

\rohead%
{%
    \Ifstr{\rightmark}{\leftmark}%
    {%
        \begin{minipage}[b]{\mymarginwidth}%
            \small\raggedright\normalfont\textsf{\textbf{\color{footersection}Chapter\ \thechapter}}%
        \end{minipage}%
    }%
    {%
        \begin{minipage}[b]{\mymarginwidth}%
            \small\raggedright\normalfont\textsf{\textbf{\color{footersection}Section\ \thesection}}%
        \end{minipage}%
    }%
}
\cohead{\hspace*{-\mymarginwidthwithinnersep}\parbox{\textwidth}{\raggedleft\rightmark}}

\lefoot*%
{%
    \vspace*{1 ex}%
    {\color{stroke1}\rule{\myoutermargin - \mymargininnersep}{0.5 mm}}\\
    \begin{minipage}[b]{\myoutermargin - \mymargininnersep}%
        \raggedleft\normalfont\color{footerpagenr}\textbf{\thepage}%
    \end{minipage}%
}
\rofoot*%
{%
    {\color{stroke1}\rule{\myoutermargin - \mymargininnersep}{0.5 mm}}\\
    \begin{minipage}[b]{\myoutermargin - \mymargininnersep}%
        \raggedright\normalfont\color{footerpagenr}\textbf{\thepage}%
    \end{minipage}%
}

\usepackage{caption}
\usepackage{tikz} 

\newlength{\mytmpa}
\newlength{\mytmpb}

\renewcommand*{\partlineswithprefixformat}[3]%
{%
    #2
    \thispagestyle{empty}
    \ifprintVersion
        \setlength{\mytmpa}{0.618\mypaperwidth + \mybindingcorrection + \extraborderlength}%
        \setlength{\mytmpb}{0.382\mypaperheight + \extraborderlength}%
    \else
        \setlength{\mytmpa}{0.618\mypaperwidth}%
        \setlength{\mytmpb}{0.382\mypaperheight}%
    \fi
    \begin{tikzpicture}[overlay, remember picture]%
        \node [inner sep = 0, outer sep = 0, anchor = north] at (current page.north west)%
        {%
            \begin{tikzpicture}[overlay, remember picture]%
            \draw[color = stroke1, line width = 0.7 mm] (\mytmpa, 0) -- (\mytmpa, -\mytmpb);%
            \end{tikzpicture}%
        };%
        \node (align) [align = right, below = \mytmpb - 2 ex, inner sep = 0, outer sep = 0, anchor = north west] at (current page.north west)%
        {%
            \hspace{\mytmpa}\hspace{0.5 em}\partname\ \thepart\\[1 ex]
            \color{stroke1}#3%
        };%
    \end{tikzpicture}%
}
\RedeclareSectionCommand%
[%
    font = \normalfont\Huge\sffamily,
    prefixfont = \normalfont\Huge\sffamily,
]
{part}

\usepackage{etoolbox}

\newbool{chapterHasANumber}
\newbool{chapterHasAStar}
\renewcommand*{\chapterlinesformat}[3]%
{%
    \Ifnumbered{#1}{\setbool{chapterHasANumber}{true}}{\setbool{chapterHasANumber}{false}}%
    \Ifstr{#2}{}{\setbool{chapterHasAStar}{true}}{\setbool{chapterHasAStar}{false}}%
    \ifboolexpr{bool{chapterHasANumber} and not bool{chapterHasAStar}}%
    {%
        \begin{tikzpicture}[overlay, remember picture]%
            \node [right = \myinnermargin, below = \mytopmargin, inner sep = 0, outer sep = 0, anchor = north west] (numbernode) at (current page.north west)%
            {%
                \hspace{\myinnermargin}%
                \sffamily\fontsize{60}{60}\selectfont%
                \color{chapternumber}%
                \thechapter%
            };%
            \node [inner sep = 0, outer sep = 0, anchor = north west] at (numbernode.south west)%
            {%
                \begin{tikzpicture}[overlay, remember picture]%
                    \draw[color = stroke1, line width = 0.7 mm] (\myinnermargin, -1 ex) -- (\paperwidth, -1 ex);%
                \end{tikzpicture}%
            };%
            \node (align) [text width = \textwidth - 2 cm, align = right, right = \myinnermargin + \mybodywidth, inner sep = 0, outer sep = 0, anchor = east] at (numbernode.west)%
            {%
                #3%
            };%
        \end{tikzpicture}%
    }%
    {%
        \begin{tikzpicture}[overlay, remember picture]%
            \node [right = \myinnermargin, below = \mytopmargin, inner sep = 0, outer sep = 0, anchor = north west] (numbernode) at (current page.north west)%
            {%
                \hspace{\myinnermargin}%
                \sffamily\fontsize{60}{60}\selectfont%
                \color{white}%
                \thechapter%
            };%
            \node [inner sep = 0, outer sep = 0, anchor = north west] at (numbernode.south west)%
            {%
                \begin{tikzpicture}[overlay, remember picture]%
                    \draw[color = stroke1, line width = 0.7 mm] (\myinnermargin, -1 ex) -- (\paperwidth, -1 ex);%
                \end{tikzpicture}%
            };%
            \node (align) [align = left, right = \myinnermargin, inner sep = 0, outer sep = 0, anchor = south west] at (numbernode.south west)%
            {%
                #3%
            };%
        \end{tikzpicture}%
    }%
}
\RedeclareSectionCommand%
[%
    font = \color{stroke1}\normalfont\huge\sffamily,
    afterskip = 20 pt,
]
{chapter}

\BeforeStartingTOC[toc]{\pagestyle{plain}}
\AfterStartingTOC{\thispagestyle{plain}}

%% file: core/bibliography/bibliography_format.tex
\usepackage
[
    sortcites,              
    sorting = none,
    style = numeric-comp,     
    defernumbers,           
    safeinputenc,           
    backref = true,         
    backrefstyle = three,   
    hyperref = true,        
    maxbibnames = 99,       
    maxcitenames = 2,       
]
{biblatex} 

\renewbibmacro{in:}%
{%
    \ifentrytype{article}{}{\printtext{\bibstring{in}\intitlepunct}}%
}

\renewbibmacro*{volume+number+eid}%
{%
    \printfield{volume}%
    \iffieldundef{number}{}{\addcolon}%
    \printfield{number}%
    \setunit*{\addcomma\space}%
    \printfield{eid}%
}

\DefineBibliographyStrings{english}%
{%
    backrefpage  = {\lowercase{s}ee page}, 
    backrefpages = {\lowercase{s}ee pages} 
}

\DeclareFieldFormat[article]{title}{\textbf{\color{stroke1}#1}}
\DeclareFieldFormat[inproceedings]{title}{\textbf{\color{stroke1}#1}}
\DeclareFieldFormat[thesis]{title}{\textbf{\color{stroke1}#1}}
\DeclareFieldFormat[book]{title}{\textbf{\color{stroke1}#1}}
\DeclareFieldFormat[unpublished]{title}{\textbf{\color{stroke1}#1}}
\DeclareFieldFormat[report]{title}{\textbf{\color{stroke1}#1}}
\DeclareFieldFormat[inbook]{chapter}{\textbf{\color{stroke1}#1}}
\DeclareFieldFormat[inbook]{title}{#1}
\DeclareFieldFormat{pages}{#1}

\newtoggle{authorend}
\togglefalse{authorend}

\DeclareBibliographyDriver{article}%
{%
  \usebibmacro{bibindex}%
  \usebibmacro{begentry}%
  \iftoggle{authorend}{}{\usebibmacro{author/translator+others}}%
  \setunit{\labelnamepunct}\newblock
  \usebibmacro{title}%
  \newunit
  \printlist{language}%
  \newunit\newblock
  \usebibmacro{byauthor}%
  \newunit\newblock
  \usebibmacro{bytranslator+others}%
  \newunit\newblock
  \printfield{version}%
  \newunit\newblock
  \usebibmacro{in:}%
  \usebibmacro{journal+issuetitle}%
  \newunit
  \usebibmacro{byeditor+others}%
  \newunit
  \printfield{pages}%
  \newunit\newblock
  \iftoggle{bbx:isbn}
  {\printfield{issn}}
  {}%
  \newunit\newblock
  \newunit\newblock
  \usebibmacro{addendum+pubstate}%
  \setunit{\bibpagerefpunct}\newblock
  \newunit\newblock
  \iftoggle{bbx:related}
  {\usebibmacro{related:init}%
    \usebibmacro{related}}
  {}%
  \usebibmacro{finentry}%
  \iftoggle{authorend}{\usebibmacro{author/translator+others}}{}%
}

\DeclareBibliographyDriver{inbook}%
{%
  \usebibmacro{bibindex}%
  \usebibmacro{begentry}%
  \iftoggle{authorend}{}{\usebibmacro{author/translator+others}}%
  \setunit{\labelnamepunct}\newblock
  \usebibmacro{chapter+pages}%
  \newunit
  \printlist{language}%
  \newunit\newblock
  \usebibmacro{byauthor}%
  \newunit\newblock
  \usebibmacro{in:}%
  \usebibmacro{bybookauthor}%
  \newunit\newblock
  \usebibmacro{maintitle+booktitle}%
  \newunit\newblock
  \usebibmacro{byeditor+others}%
  \newunit\newblock
  \printfield{edition}%
  \newunit
  \iffieldundef{maintitle}
  {\printfield{volume}%
    \printfield{part}}
  {}%
  \newunit
  \printfield{volumes}%
  \newunit\newblock
  \usebibmacro{series+number}%
  \newunit\newblock
  \printfield{note}%
  \newunit\newblock
  \usebibmacro{publisher+location+date}%
  \newunit\newblock
  \newunit\newblock
  \iftoggle{bbx:isbn}
  {\printfield{isbn}}
  {}%
  \newunit\newblock
  \usebibmacro{doi+eprint+url}%
  \newunit\newblock
  \usebibmacro{addendum+pubstate}%
  \setunit{\bibpagerefpunct}\newblock
  \usebibmacro{pageref}%
  \newunit\newblock
  \iftoggle{bbx:related}
  {\usebibmacro{related:init}%
    \usebibmacro{related}}
  {}%
  \usebibmacro{finentry}%
  \iftoggle{authorend}{\usebibmacro{author/translator+others}}{}%
}

\DeclareBibliographyDriver{inproceedings}%
{%
  \usebibmacro{bibindex}%
  \usebibmacro{begentry}%
  \iftoggle{authorend}{}{\usebibmacro{author/translator+others}}%
  \setunit{\labelnamepunct}\newblock
  \usebibmacro{title}%
  \newunit
  \printlist{language}%
  \newunit\newblock
  \usebibmacro{byauthor}%
  \newunit\newblock
  \usebibmacro{in:}%
  \usebibmacro{maintitle+booktitle}%
  \newunit\newblock
  \usebibmacro{event+venue+date}%
  \newunit\newblock
  \usebibmacro{byeditor+others}%
  \newunit\newblock
  \iffieldundef{maintitle}
  {\printfield{volume}%
    \printfield{part}}
  {}%
  \newunit
  \printfield{volumes}%
  \newunit\newblock
  \usebibmacro{series+number}%
  \newunit\newblock
  \printfield{note}%
  \newunit\newblock
  \printlist{organization}%
  \newunit
  \usebibmacro{publisher+location+date}%
  \newunit\newblock
  \usebibmacro{chapter+pages}%
  \newunit\newblock
  \iftoggle{bbx:isbn}
  {\printfield{isbn}}
  {}%
  \newunit\newblock
  \usebibmacro{doi+eprint+url}%
  \newunit\newblock
  \usebibmacro{addendum+pubstate}%
  \setunit{\bibpagerefpunct}\newblock
  \usebibmacro{pageref}%
  \newunit\newblock
  \iftoggle{bbx:related}
  {\usebibmacro{related:init}%
    \usebibmacro{related}}
  {}%
  \usebibmacro{finentry}%
  \iftoggle{authorend}{\usebibmacro{author/translator+others}}{}%
}

\DeclareBibliographyDriver{thesis}%
{%
  \usebibmacro{bibindex}%
  \usebibmacro{begentry}%
  \iftoggle{authorend}{}{\usebibmacro{author}}%
  \setunit{\labelnamepunct}\newblock
  \usebibmacro{title}%
  \newunit
  \printlist{language}%
  \newunit\newblock
  \usebibmacro{byauthor}%
  \newunit\newblock
  \printfield{note}%
  \newunit\newblock
  \printfield{type}%
  \newunit
  \usebibmacro{institution+location+date}%
  \newunit\newblock
  \usebibmacro{chapter+pages}%
  \newunit
  \printfield{pagetotal}%
  \newunit\newblock
  \iftoggle{bbx:isbn}
  {\printfield{isbn}}
  {}%
  \newunit\newblock
  \usebibmacro{doi+eprint+url}%
  \newunit\newblock
  \usebibmacro{addendum+pubstate}%
  \setunit{\bibpagerefpunct}\newblock
  \newunit\newblock
  \iftoggle{bbx:related}
  {\usebibmacro{related:init}%
    \usebibmacro{related}}
  {}%
  \usebibmacro{finentry}%
  \iftoggle{authorend}{\usebibmacro{author}}{}%
}

\providebool{bbx:subentry}
\newbibmacro*{citenum}%
{
  \printtext[bibhyperref]{
    \printfield{prefixnumber}%
    \printfield{labelnumber}%
    \ifbool{bbx:subentry}
    {\printfield{entrysetcount}}
    {}}%
}

\DeclareCiteCommand{\conline}[\mkbibbrackets]
{\usebibmacro{prenote}}
{\usebibmacro{citeindex}%
  \usebibmacro{citenum}}
{\multicitedelim}
{\usebibmacro{postnote}}

%% file: packages_and_commands/standard_packages.tex
\usepackage
[
    babel = true, 
]
{microtype}           
\usepackage{csquotes} 

\usepackage{amsmath}
\usepackage{amssymb}
\usepackage{amsthm}
\usepackage{thmtools}
\usepackage{mathtools}
\usepackage{thm-restate}
\usepackage{dsfont}        
\usepackage{braceMnSymbol} 

\usepackage
[
    ttscale = 0.85, 
]
{libertine} 
\usepackage
[
    libertine,    
    slantedGreek, 
    vvarbb,       
    libaltvw,     
]
{newtxmath} 
\usepackage{url} 
\usepackage{bm}  

\usepackage{graphicx} 
\usepackage
[
    subrefformat = simple, 
    labelformat = simple,  
]
{subcaption}         
\usepackage{wrapfig} 

\columnsep = \mymargininnersep

\usepackage{array}     
\usepackage{booktabs}  
\usepackage{longtable} 
\usepackage{pdflscape} 

\usepackage{enumitem} 

\usepackage
[
    ruled,         
    vlined,        
    linesnumbered, 
]
{algorithm2e} 

\usepackage{xspace}   
\usepackage
[
    shortcuts, 
]
{extdash}             
\usepackage{setspace} 

\usepackage{xparse}    
\usepackage{footnote}  
\usepackage{afterpage} 
\usepackage
[
    textsize = scriptsize, 
]
{todonotes}            


\usepackage
[
    bookmarks = true,                 
    bookmarksopen = false,            
    bookmarksnumbered = true,         
    pdfstartpage = 1,                 
    pdftitle = {{\printTitle}},       
    pdfauthor = {{\printAuthor}},     
    pdfsubject = {{\printSubject}},   
    pdfkeywords = {{\printKeywords}}, 
    breaklinks = true,                
    \ifprintVersion
        hidelinks,                    
    \else
    colorlinks = true,            
    allcolors = stroke1,          
    \fi
]
{hyperref} 

\usepackage
[
    nameinlink, 
]
{cleveref} 

%% file: packages_and_commands/general_commands.tex
\ifbachelorThesis

\fi

\makeatletter
    \def\IfEmptyTF#1%
    {%
        \if\relax\detokenize{#1}\relax%
            \expandafter\@firstoftwo%
        \else%
            \expandafter\@secondoftwo%
        \fi%
    }
\makeatother

\NewDocumentCommand{\mathOrText}{m}
{%
    \ensuremath{#1}\xspace%
}

\let\originalleft\left
\let\originalright\right
\renewcommand{\left}{\mathopen{}\mathclose\bgroup\originalleft}
\renewcommand{\right}{\aftergroup\egroup\originalright}

\makeatletter
    \DeclareRobustCommand{\bfseries}%
    {%
        \not@math@alphabet\bfseries\mathbf%
        \fontseries\bfdefault\selectfont%
        \boldmath%
    }
\makeatother

\xspaceaddexceptions{]\}}

\allowdisplaybreaks

\crefname{ineq}{inequality}{inequalities}
\creflabelformat{ineq}{#2{\upshape(#1)}#3} 

\crefname{term}{term}{terms}
\creflabelformat{term}{#2{\upshape(#1)}#3}

\let\oldfootnote\footnote

\newlength{\spaceBeforeFootnote} 
\newlength{\spaceAfterFootnote}  

\RenewDocumentCommand{\footnote}{o o o m}%
{%
    \IfNoValueTF{#1}%
    {%
        \oldfootnote{#4}%
    }%
    {%
        \setlength{\spaceBeforeFootnote}{\IfEmptyTF{#1}{0}{#1} em}%
        \IfNoValueTF{#2}%
        {%
            \hspace*{\spaceBeforeFootnote}\oldfootnote{#4}%
        }%
        {%
            \setlength{\spaceAfterFootnote}{\IfEmptyTF{#2}{0}{#2} em}%
            \hspace*{\spaceBeforeFootnote}\IfNoValueTF{#3}{\oldfootnote{#4}}{\oldfootnote[#3]{#4}}\hspace*{\spaceAfterFootnote}%
        }%
    }%
}

\makesavenoteenv{figure}
\makesavenoteenv{table}
\makesavenoteenv{tabular}

\iffancyTheorems
    \declaretheoremstyle
    [
        spaceabove = \topsep,
        spacebelow = \topsep,
        headfont = \bfseries,
        headformat = \textcolor{stroke1}{$\blacktriangleright$} \NAME~\NUMBER \NOTE,
        notefont = \bfseries,
        notebraces = {(}{)},
        bodyfont = \normalfont,
        postheadspace = 0.5 em,
        qed = \textcolor{stroke1}{\bfseries$\blacktriangleleft$},
    ]
    {myTheoremStyle}
    
    \declaretheorem
    [
        style = myTheoremStyle,
        name = Conjecture,
        numberwithin = chapter,
    ]
    {conjecture}
    \declaretheorem
    [
        style = myTheoremStyle,
        name = Proposition,
        sharenumber = conjecture,
    ]
    {proposition}
    \declaretheorem
    [
        style = myTheoremStyle,
        name = Claim,
        sharenumber = conjecture,
    ]
    {claim}
    \declaretheorem
    [
        style = myTheoremStyle,
        name = Lemma,
        sharenumber = conjecture,
    ]
    {lemma}
    \declaretheorem
    [
        style = myTheoremStyle,
        name = Corollary,
        sharenumber = conjecture,
    ]
    {corollary}
    \declaretheorem
    [
        style = myTheoremStyle,
        name = Theorem,
        sharenumber = conjecture,
    ]
    {theorem}
    \declaretheorem
    [
        style = myTheoremStyle,
        name = Definition,
        sharenumber = conjecture,
    ]
    {definition}
    \declaretheorem
    [
        style = myTheoremStyle,
        name = Example,
        sharenumber = conjecture,
    ]
    {example}
    \declaretheorem
    [
        style = myTheoremStyle,
        name = Remark,
        sharenumber = conjecture,
    ]
    {remark}
\else
    \theoremstyle{plain}

\fi

\NewDocumentCommand{\functionTemplate}{m m m m o}%
{%
    \IfNoValueTF{#5}%
    {%
        \mathOrText{#1\left#2{#4}\right#3}%
    }%
    {%
        \mathOrText{#1#5#2{#4}#5#3}%
    }%
}

\newcommand*{\leftBracketType}{(}
\newcommand*{\rightBracketType}{)}

\NewDocumentCommand{\createFunction}{m m o o}%
{%
    \renewcommand*{\leftBracketType}{\IfNoValueTF{#3}{(}{#3}}%
    \renewcommand*{\rightBracketType}{\IfNoValueTF{#4}{)}{#4}}%
    \NewDocumentCommand{#1}{o o}%
    {%
        \IfNoValueTF{##1}%
        {%
            \mathOrText{#2}%
        }%
        {%
            \functionTemplate{#2}{\leftBracketType}{\rightBracketType}{##1}[##2]%
        }%
    }%
}

\DeclareDocumentCommand{\probabilisticFunctionTemplate}{m m O{} o}
{%
    \functionTemplate{#1}%
    {\lbrack}%
    {\rbrack}%
    {#2\IfEmptyTF{#3}{}{\ \IfNoValueTF{#4}{\left}{#4}\vert\ \vphantom{#2}#3\IfNoValueTF{#4}{\right.}{}}}%
    [#4]%
}

\ifboldNumberSets

    \newcommand*{\indicatorFunctionSymbol}{\mathbf{1}}
\else

    \newcommand*{\indicatorFunctionSymbol}{\mathds{1}}
\fi

\RenewDocumentCommand{\Pr}{m O{} o}%
{%
    \probabilisticFunctionTemplate{\mathrm{Pr}}{#1}[#2][#3]%
}

\NewDocumentCommand{\E}{m O{} o}%
{%
    \probabilisticFunctionTemplate{\mathrm{E}}{#1}[#2][#3]%
}

\NewDocumentCommand{\Var}{m O{} o}%
{%
    \probabilisticFunctionTemplate{\mathrm{Var}}{#1}[#2][#3]%
}

\DeclareDocumentCommand{\bigO}{m o}%
{%
    \functionTemplate{\mathrm{O}}{(}{)}{#1}[#2]%
}

\DeclareDocumentCommand{\smallO}{m o}%
{%
    \functionTemplate{\mathrm{o}}{(}{)}{#1}[#2]%
}

\DeclareDocumentCommand{\bigTheta}{m o}%
{%
    \functionTemplate{\upTheta}{(}{)}{#1}[#2]%
}

\DeclareDocumentCommand{\bigOmega}{m o}%
{%
    \functionTemplate{\upOmega}{(}{)}{#1}[#2]%
}

\DeclareDocumentCommand{\smallOmega}{m o}%
{%
    \functionTemplate{\upomega}{(}{)}{#1}[#2]%
}

\DeclareDocumentCommand{\eulerE}{o}%
{%
    \mathOrText{\mathrm{e}\IfNoValueTF{#1}{}{^{#1}}}%
}

\DeclareDocumentCommand{\poly}{m o}%
{%
    \functionTemplate{\mathrm{poly}}{(}{)}{#1}[#2]%
}

\createFunction{\id}{\mathrm{id}}

\NewDocumentCommand{\ind}{m o o}%
{%
    \IfNoValueTF{#2}%
    {%
        \mathOrText{\indicatorFunctionSymbol_{#1}}%
    }%
    {%
        \functionTemplate{\indicatorFunctionSymbol_{#1}}{(}{)}{#2}[#3]%
    }%
}

\DeclareDocumentCommand{\dom}{m o}%
{%
    \functionTemplate{\mathrm{dom}}{(}{)}{#1}[#2]%
}

\DeclareDocumentCommand{\rng}{m o}%
{%
    \functionTemplate{\mathrm{rng}}{(}{)}{#1}[#2]%
}

\DeclareDocumentCommand{\d}{o}%
{%
    \mathrm{d}\IfNoValueTF{#1}{}{^{#1}}%
}

\DeclareDocumentCommand{\set}{m m o}%
{%
    \mathOrText{\IfNoValueTF{#3}{\left}{#3}\{#1\ \IfNoValueTF{#3}{\left}{#3}\vert\ \vphantom{#1}#2\IfNoValueTF{#3}{\right.}{}\IfNoValueTF{#3}{\right}{#3}\}}%
}

%% file: core/title_page/title_page.tex

\ifprintVersion
    \newgeometry
    {
        textwidth = 134 mm,
        textheight = 220 mm,
        top = 38 mm - (297 mm - 297 mm * \magicratio) * \real{0.5} + \extraborderlength,
        inner = (\paperwidth - 134 mm - 2\extraborderlength - \mybindingcorrection) * \real{0.5} + \mybindingcorrection + \extraborderlength,
    }
\else
    \newgeometry
    {
        textwidth = 134 mm,
        textheight = 220 mm,
        top = 38 mm - (297 mm - 297 mm * \magicratio) * \real{0.5},
        inner = (\mypaperwidth - 134 mm) * \real{0.5},
    }
\fi

\begin{titlepage}
    \sffamily
    \begin{center}
        \Large
        \textsc{University of Trento}\\
	    \vspace{0.6cm}
        \includegraphics[height = 3.2 cm]{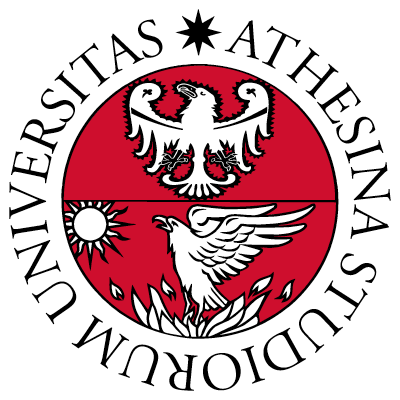} \\
        \vspace{0.3cm}
        \textsc{Department of Physics} \\
        \vspace{0.2cm}
        \textsc{Master Degree in Physics} \\
        \vspace{1cm}
        {\LARGE
            \rule[1 ex]{\textwidth}{1. pt}
            \onehalfspacing\printTitleBold\\[1 ex]
            \rule[-1 ex]{\textwidth}{1. pt}
        }
        \vfil
        \vspace{2.5cm}
    	\begin{tabular}{ l c c c c c c c c  r} 
			\emph{Candidate}: &&&&&&&&& \emph{Supervisor}: \\ 
			Vittorio Del Tatto &&&&&&&&& Prof. Dr. Giovanni Bussi \\ 
			                   &&&&&&&&&  \\  
			                   &&&&&&&&&  \\  
			                   &&&&&&&&& \emph{Co-supervisor}: \\
			                   &&&&&&&&& Prof. Dr. Raffaello Potestio
		\end{tabular}\\
        \vspace{2.5cm}
        \Large
        \vspace{0.2cm}
        \textsc{Academic Year 2020/2021} \\
    \end{center}
    \vfil
\end{titlepage}

\restoregeometry

%% file: preface/acknowledgments.tex
I am extremely thankful to my supervisor in SISSA, Giovanni Bussi, for his constant guidance and support during this work. Your passion and your intuitive approach to complex problems are a source of inspiration for me. 

I would also like to thank Professor Paolo Raiteri from Curtin University, for his valuable help with the implementation and test of the algorithm using the LAMMPS software.  

A sincere thank you to Mattia Bernetti, post-doctoral researcher in SISSA, for introducing me to the world of GROMACS and for his support during my first simulations.

I would also like to thank my co-supervisor at the University of Trento, Professor Raffaello Potestio, for his useful suggestions and for his inspiring course in Soft Matter Physics, which introduced me to the field of Biophysics.

Thank you to my friends in Trieste, Trento and Milano, who are an endless source of motivation and cheerfulness.

A special thank goes to my parents and my family, for their unfailing support and continuous encouragement over the years.